\documentclass[11pt]{article}
\usepackage{bm}
\usepackage{algorithm}
\usepackage{algpseudocode}
\usepackage[numbers]{natbib}
\usepackage[colorlinks,citecolor=blue,linkcolor=blue,urlcolor=blue]{hyperref}
\usepackage{latexsym, epsfig, amssymb, amsmath, amsthm, graphicx, mathrsfs, booktabs}
\usepackage{rotating, tabularx, setspace,paralist}
\usepackage{color}
\usepackage{graphicx}
\usepackage[OT1]{fontenc}
\usepackage{lscape}
\usepackage{multirow,multicol}
\usepackage{booktabs}
\usepackage{anysize}

\marginsize{1.1in}{1.1in}{0.55in}{0.8in} %

\usepackage{amsmath,amsfonts,amssymb,amsthm,color,hyperref,url,authblk,setspace,natbib,todonotes}
\usepackage{verbatim}
\usepackage{titlesec}
\usepackage{enumitem}

\newcommand{\norm}[1]{\|#1\|}
\newcommand{\vect}[1]{\text{vec}(#1)}
\newcommand{\wh}[1]{\widehat{#1}}
\newcommand{\wt}[1]{\widetilde{#1}}

\newcommand{\diag}[1]{\text{diag}\{#1\}}

\newcommand{\abs}[1]{\vert #1 \vert}

\newcommand{\der}[2]{\frac{\partial{#1}}{\partial{#2}}}

\newcommand{\derr}[3]{\frac{\partial^2{#1}}{\partial{#2}\partial{#3}}}

\newcommand{\bDelta}{\boldsymbol{\Delta}}
\newcommand{\bdelta}{\boldsymbol{\delta}}
\newcommand{\bSigma}{\boldsymbol{\Sigma}}

\newcommand{\bbeta}{\boldsymbol{\beta}}
\newcommand{\balpha}{\boldsymbol{\alpha}}
\newcommand{\bTheta}{\boldsymbol{\Theta}}
\newcommand{\btheta}{\boldsymbol{\theta}}
\newcommand{\bepsilon}{\boldsymbol{\epsilon}}
\newcommand{\bmu}{\boldsymbol{\mu}}

\newcommand{\boldeta}{\boldsymbol{\eta}}
\newcommand{\bxi}{\boldsymbol{\xi}}
\newcommand{\N}{\mathcal{N}}

\newtheorem{theorem}{Theorem}
\newtheorem{lemma}{Lemma}
\newtheorem{condition}{Condition}

\newtheorem{proposition}{Proposition}
\newtheorem{definition}{Definition}

\def\trans{^{T}}
\def\strans{^{* T}}

\def\R{{\mathbb R}}

\def\C{{\bf C}}
\def\X{{\bf X}}
\def\Y{{\bf Y}}
\def\U{{\bf U}}
\def\V{{\bf V}}
\def\D{{\bf D}}
\def\I{{\bf I}}
\def\E{{\bf E}}

\def\A{{\bf A}}
\def\B{{\bf B}}
\def\W{{\bf W}}
\def\L{{\bf L}}

\def\Q{{\bf Q}}
\def\T{{\bf T}}
\def\M{{\bf M}}

\def\x{\mathbf{x}}

\def\e{\boldsymbol{e}}

\def\u{\boldsymbol{u}}
\def\v{\boldsymbol{v}}

\def\a{\boldsymbol{a}}
\def\b{\boldsymbol{b}}

\def\w{\boldsymbol{w}}

\def\r{\boldsymbol{r}}

\def\l{\boldsymbol{l}}

\def\0{\boldsymbol{0}}
\def\1{\boldsymbol{1}}

\renewcommand{\N}{N}
\renewcommand{\ldots}{\cdots}
\renewcommand{\trans}{^T}

\begin{document}

\title{SOFARI-R: High-Dimensional Manifold-Based Inference for Latent Responses%
\thanks{
Zemin Zheng is Professor, Department of Statistics and Finance, School of Management, University of Science and Technology of China, China (E-mail: zhengzm@ustc.edu.cn). %
Xin Zhou is Postdoctor, Department of Statistics and Finance, School of Management, University of Science and Technology of China, China (E-mail: zx1120@mail.ustc.edu.cn). %
Jinchi Lv is Kenneth King Stonier Chair in Business Administration and Professor, Data Sciences and Operations Department, Marshall School of Business, University of Southern California, Los Angeles, CA 90089 (E-mail: \textit{jinchilv@marshall.usc.edu}). %
This work was supported in part by NSF grant DMS-2324490.
}
\date{April 24, 2025}
\author{Zemin Zheng$^1$, Xin Zhou$^1$ and Jinchi Lv$^2$
\medskip\\
University of Science and Technology of China$^1$ and University of Southern California$^2$
\\
} %
}

\maketitle

\begin{abstract}
Data reduction with uncertainty quantification plays a key role in various multi-task learning applications, where large numbers of responses and features are present. To this end, a general framework of high-dimensional manifold-based SOFAR inference (SOFARI) was introduced recently in Zheng, Zhou, Fan and Lv (2024) for interpretable multi-task learning inference focusing on the left factor vectors and singular values exploiting the latent singular value decomposition (SVD) structure. Yet, designing a valid inference procedure on the latent right factor vectors is not straightforward from that of the left ones and can be even more challenging due to asymmetry of left and right singular vectors in the response matrix. To tackle these issues, in this paper we suggest a new method of high-dimensional manifold-based SOFAR inference for latent responses (SOFARI-R), where two variants of SOFARI-R are introduced. The first variant deals with strongly orthogonal factors by coupling left singular vectors with the design matrix and then appropriately rescaling them to generate new Stiefel manifolds. The second variant handles the more general weakly orthogonal factors by employing the hard-thresholded SOFARI estimates and delicately incorporating approximation errors into the distribution. Both variants produce bias-corrected estimators for the latent right factor vectors that enjoy asymptotically normal distributions with justified asymptotic variance estimates. We demonstrate the effectiveness of the newly suggested method using extensive simulation studies and an economic application. 
\end{abstract}

\textit{Running title}: SOFARI-R


\section{Introduction}

Statistical estimation and inference for large-scale multi-task learning are crucial in modern big data applications such as brain memory encoding and autonomous driving. Multi-task learning often leverages multi-response regression models, where each task corresponds to a response. Over the past decade, many existing works have focused on recovering high-dimensional latent association network structures revealed by multi-response regression models through joint penalization methods \citep{Bunea2012, chen2012sparse, chendong2013, zou2022estimation}, or via sparse singular value decomposition (SVD) based approaches \citep{mishra2017, uematsu2017sofar, zheng2019scalable, chen2022stagewise}. An advantage of the latter is that different sparse SVD components provide natural interpretation of a sparse latent factor model, as exemplified by the discovered latent pathways in yeast eQTL data analyses from \cite{uematsu2017sofar} and \cite{chen2022stagewise}. Recently, \cite{sofari} proposed a general framework of high-dimensional manifold-based SOFAR inference (SOFARI) for interpretable multi-task learning to make statistical inference on the sparse left factor vectors and singular values of the latent SVD structure. Despite the success of this manifold-based inference approach, a valid procedure for inferring the latent right factor vectors remains unavailable. The goal of this paper is to develop an approach for inferring the latent sparse right factor vectors, thereby enabling inference on the entire high-dimensional SVD structure and broadening the applicability.

In this paper, we propose and investigate a new method of high-dimensional manifold-based SOFAR inference for latent responses, named as the SOFARI-R, for statistical inference on the sparse right singular vectors of the latent SVD structure in multi-response regression models. Different from the SOFARI inference \citep{sofari} targeting at the latent left factor vectors that correspond to feature (covariate) selection, our suggested SOFARI-R inference on the right singular vectors is from the prediction point of view and designed for response selection. As a special case of our applications, when the design matrix reduces to an identity matrix, the right singular vectors become the principal components of the response matrix, which are frequently used in practice.

For instance, in the context of high-dimensional principal component analysis (PCA), \cite{wang2017asymptotics} derived the asymptotic distributions of spiked eigenvalues and eigenvectors under the spiked covariance structures, and \cite{Yan2024} extended the results to the settings of spiked covariance models with missing data and heteroskedastic noises. As their spiked covariance structures essentially require that the maximum eigenvalue or the eigengap diverges, \cite{jankova2021biased} developed a debiased sparse PCA estimator for constructing confidence intervals and hypothesis tests on the first eigenvector and the largest eigenvalue even when the maximum eigenvalue and the eigengap are bounded. However, whether a similar inference procedure is applicable to deal with the remaining principal components is unclear due to the noise accumulation from the previous layers. 
Motivated by the SOFARI procedure for inferring the latent left singular vectors in \cite{sofari}, we form our inference approach using the similar idea of drawing on the Neyman near-orthogonality inference \citep{doubleML2018} while incorporating the Stiefel manifold structure imposed by the latent SVD constraints. That is, constructing a debiased estimator based on a modified score function that is locally insensitive to the nuisance parameters when they are constrained to be on the manifolds induced by the SVD constraints. Then a natural idea is to construct the Neyman near-orthogonal score function when the left singular vectors viewed as the nuisance parameters are constrained to be on the corresponding manifolds. However, this naive approach \textit{fails} to work because of the asymmetry of left and right singular vectors in the response matrix.

In fact, inference for the latent right factor vectors is \textit{not} straightforward from that of the left factor vectors and can be even \textit{more challenging}. The key difficulty lies in that when we target at the right factor vectors, the manifold induced by the SVD constraints on the left singular vectors would not help as the signals of different layers in the response matrix are generally not orthogonal to each other from the side of the left singular vectors, but are instead \textit{correlated} via the design matrix. This results in the inseparability of nuisance parameters from other layers and prevents the manifold of left singular vectors from yielding a valid modified score function to save degrees of freedom under the SVD constraints.

To overcome such intrinsic difficulty, we propose to couple left singular vectors with the design matrix and then rescale these coupled left factors to have a unit length, which will provide new Stiefel manifold structures on them. When the rescaled coupled left factors are constrained to be on such new Stiefel manifolds, we can show that the nuisance parameters from other layers can be separated from the target layer such that a valid score function satisfying the Neyman near-orthogonality can be constructed. Based on our \textit{coupling and rescaling} strategy, under the case of strongly orthogonal factors which includes the aforementioned high-dimensional PCA as a special application, the suggested debiased estimator SOFARI-R$_s$ generated from the new score function will admit the asymptotic normality.

Compared to the debiased sparse PCA approach \citep{jankova2021biased} which focuses on the asymptotic distribution of the first eigenvector, our manifold-based SOFARI-R$_s$ procedure can provide asymptotic distributions for all significant eigenvectors of the data matrix. Moreover, both the debiased approach in \cite{jankova2021biased} and the SOFARI estimator in \cite{sofari} need to approximately inverse the Hessian matrix or the precision matrix, and require certain sparsity conditions on them. In contrast, our SOFARI-R$_s$ procedure requires \textit{neither} estimation of the inverse Hessian matrix or the precision matrix, \textit{nor} sparsity constraints on them.

Another substantial difference of SOFARI-R from SOFARI lies in the more general case of weakly orthogonal factors, where latent factors can have stronger correlations among each other than for the case of strongly orthogonal factors so as to accommodate a wider range of multi-response applications. In the weakly orthogonal case of SOFARI, the intrinsic bias induced by correlations between latent factors can be controlled with the aid of significant gaps between nonzero singular values. However, such eigengap will \textit{not} assist in reducing the intrinsic bias of SOFARI-R, which depends solely on the correlations between latent factors. To alleviate the intrinsic bias in SOFARI-R, we propose to subtract the latent factors corresponding to other layers from the response matrix, which can be achieved by employing the SOFARI debiased estimates for left factor vectors with an additional hard-thresholding step to prevent noise accumulation. In this way, after delicately incorporating the approximation errors of SOFARI estimates into the distribution, we can show that even under the case of weakly orthogonal factors, the suggested SOFARI-R estimator for inference on the latent right factor vectors enjoys asymptotic normality with justified asymptotic variance estimates. Table \ref{tab:my_label} provides an overview of the major differences between SOFARI-R and SOFARI from different perspectives; see Section \ref{sec:method} for detailed descriptions.

The rest of the paper is laid out as follows. Section \ref{sec:method} introduces the SOFARI-R method under both  strong and weak orthogonality constraints. We establish asymptotic normalities of the SOFARI-R debiased estimates in Section \ref{sec4:theory}. In Section  \ref{new.Sec.5}, we conduct simulation studies to demonstrate the finite-sample performance of the suggested method. In Section \ref{new.Sec.6}, we apply our method to an economic forecasting data set. Section \ref{sec:discu} discusses some implications and extensions of our work. We provide the proofs of main results and some additional details in the Supplementary Material.

\begin{table}
    \centering
    \caption{A comparison of SOFARI-R and SOFARI}
    \label{tab:my_label}
    \resizebox{\textwidth}{!}{
    \begin{tabular}{c|c|c}
        \hline
         & SOFARI-R & SOFARI\\
         \hline
         Goal & Response selection & Covariate selection \\[5pt]
         \hline
       Inference  & $  (\l_i\trans \wh{\bSigma}\l_i)^{1/2} d_i \r_i$  & $d_i \l_i$ \text{and} $d_i^2$ \\[5pt]
       \hline
       \multirow{2}{*}{Scaling}  & $\u_i =  (\l_i\trans \wh{\bSigma}\l_i)^{-1/2} n^{-1/2} \X \l_i,$ & $\u_i = d_i \l_i,$ \\ 
        & $\v_i = (\l_i\trans \wh{\bSigma}\l_i)^{1/2} d_i \r_i$ & $\v_i = \r_i$ \\[5pt] 
        \hline
        Need the approximate inverse of & \multirow{2}{*}{No}  & \multirow{2}{*}{Yes} \\
        $\wh{\bSigma} = n^{-1}\X\trans\X$&  &  \\
        \hline
        Techniques for strongly orthogonal factors  & Manifold on rescaled $\u$ & Manifold on $\r$ \\
        \hline
        Techniques for &Debiased estimate of SOFARI& Eigengap\\
        weakly orthogonal factors&   and hard thresholding  &  and weak correlation\\
         \hline
    \end{tabular}}
\end{table}

\section{High-dimensional manifold-based SOFAR inference for latent responses}\label{sec:method}

\subsection{SOFARI-R\texorpdfstring{$_s$}{s} under strongly orthogonal factors}\label{sec:vs}

To better motivate the idea of high-dimensional manifold-based SOFAR inference for latent responses (SOFARI-R), we will first investigate the case of strongly orthogonal factors which gives rise to the basic form SOFARI-R$_s$. Consider the following multi-response regression model given a sample of $n$ independent observations 
\begin{align}\label{model}
\Y = \X\C^* + \E,
\end{align}
where $\Y\in\R^{n\times q}$ is the response matrix from $q$ possibly related tasks, $\X\in\R^{n\times p}$ is the fixed design matrix with $p$ features, $\C^* \in \R^{p\times q}$ is the unknown population regression coefficient matrix, and $\E\in\R^{n\times q}$ stands for the random noise matrix. Here, we allow for the high-dimensional setup so that both feature dimensionality $p$ and response dimensionality $q$ can exceed the sample size $n$. Moreover, to characterize the dependence structure between responses and features via latent pathways, we exploit the singular value decomposition (SVD) of $\C^*$ that
$\C^* =  \sum_{i=1}^{r^*} d_i^*\l_i^*\r_i\strans$, where $\l_i^*$ and $\r_i^*$ represent the $i$th left and right singular vectors, respectively, $d_i^*$ denotes the $i$th singular value, and $r^* \geq 1$ is the true rank.

In our fixed design setup, the columns of $\X$ are assumed to have a common $L_2$-norm $\sqrt{n}$. Denote by $\wh{\bSigma}=n^{-1}\X\trans\X$. By the SVD of $\C^*$, we write
$n^{-1/2} \X\C^* = \sum_{i=1}^{r^*} (n^{-1/2}\X\l_i^*)(d_i^* \r_i^*)\trans $ $= \sum_{i=1}^{r^*} \u_i^*\v_i\strans$, 
where the left and right factor vectors $\u_i^*$ and $\v_i^*$ are defined as
\begin{align}\label{uvtransform}
    \u_i^* =  (\l_i\strans \wh{\bSigma}\l_i^*)^{-1/2} n^{-1/2} \X \l_i^* \ \text{ and } \ \v_i^* =   (\l_i\strans \wh{\bSigma}\l_i^*)^{1/2} d_i^* \r_i^*,
\end{align}
respectively. Based on this decomposition, we observe that
$ \u_i\strans\u_i^* = 1, \ \v_i\strans\v_i^* =  d_i^{*2}  \l_i\strans \wh{\bSigma}\l_i^*,$ and $\v_i\strans\v_j^* = 0 \ \text{for} \ i \neq j. $

The key difficulty of our inference problem lies in that when we target at the right factor vectors, the signals of different layers $d_i^*\X\l_i^*\r_i\strans$ in the response matrix are generally not orthogonal to each other from the side of the left singular vectors in view of $\l_i\strans \wh{\bSigma}\l_j^* \neq 0$. This prevents the manifold induced by the SVD constraints on the original left singular vectors $\l_i^*$ from yielding a valid score function to save degrees of freedom under SVD constraints.
Thus, we couple $\l_i^*$ with $\X$ and rescale them to have a unit length, which induces new Stiefel manifold structures on $\u_i^*$. When $\u_i^*$ are constrained to be on such manifolds, we can show that nuisance parameters from other layers can be separated from the target layer such that a valid score function satisfying the Neyman near-orthogonality can be constructed.

For a given layer $k$, $\v_k^*$ is our inference target and the corresponding nuisance parameter vector is $\boldeta_k^* = \left(\u_1\strans,\ldots, \u_{r^*}\strans, \v_1\strans, \ldots, \v_{k-1}\strans, \v_{k+1}\strans, \ldots, \v_{r^*}\strans\right)\trans$. To alleviate the impacts of nuisance parameter vector $\boldeta_k^*$, we will make use of the Neyman orthogonality scores \citep{Neyman59,doubleML2018} and construct a vector $\wt{\psi}_k(\v_k,\boldeta_k)$ of score functions for $\v_k$ that is approximately insensitive to $\boldeta_k = \left(\u_1\trans,\ldots, \u_{r^*}\trans, \v_1\trans, \ldots, \v_{k-1}\trans, \v_{k+1}\trans, \ldots, \v_{r^*}\trans\right)\trans$ when evaluated at $(\v_k^*, \boldeta_k^*)$ locally. We start with the following constrained least-squares loss function
\begin{align}
	&L(\v_k,\boldeta_k) = (2n)^{-1}\norm{\Y - \sum_{i=1}^{r^*} \sqrt{n} \u_i\v_i\trans}_F^2, \label{constraint00}\\
	&\text{subject to } ~\u_i\trans\u_i = 1 ~ \text{and} ~ \v_i\trans\v_j = 0. \label{SVDc}
\end{align}

Since $L(\v_k,\boldeta_k)$ is sensitive to nuisance parameter vector $\boldeta_k$ due to its non-vanishing score function $\der{L}{\v_k}$ at $\boldeta_k^*$, we define a modified score function for $\v_k$ as
\begin{align*}
	\wt{\psi}_k(\v_k,\boldeta_k) = \der{L}{\v_k} - \M^{(k)}\der{L}{\boldeta_k},
\end{align*}
where matrix $\M^{(k)}=[\mathbf{M}_{1}^{u}, \ldots, \mathbf{M}_{r^*}^{u}, $ $\mathbf{M}_{1}^{v}, \ldots, \mathbf{M}_{k-1}^{v}, \mathbf{M}_{k+1}^{v}, \ldots,  \mathbf{M}_{r^*}^{v}]$  will be chosen such that $\wt{\psi}_k(\v_k,\boldeta_k)$ is approximately insensitive to $\boldeta_k$ under SVD constraint \eqref{SVDc}. Here, submatrices $\M_i^{u}\in\R^{q\times p}$ and $\M_j^{v}\in\R^{q\times q}$ correspond to nuisance parameter vectors $\u_i$ and $\v_j$, respectively. Note that these submatrices depend essentially on $k$, but we make such dependence implicit whenever no confusion. 

Similar to \cite{sofari}, we define a bias-corrected function for $\v_k^*$ as 
\begin{align*}
	\psi_k(\v_k,\boldeta_k) = \v_k - \W_k\wt{\psi}_k(\v_k,\boldeta_k),
\end{align*}
where $\W_k \in \R^{q \times q}$ is constructed to correct the bias in some initial estimate. In this paper, we exploit the initial estimate as the SOFAR estimator $\wt{\C}$ with SVD components $(\widetilde{\L}, \wt{\D}, \widetilde{\mathbf{R}})$, to be formally defined in Definition \ref{defi:sofar}. It  further provides estimates of the left and right factor matrices based on \eqref{uvtransform}. Denote by $\wt{\U}$ and $\wt{\V}$ with different factor vectors $ \wt{\u}_i =  (\wt{\l}_i\trans \wh{\bSigma}\wt{\l}_i)^{-1/2} n^{-1/2} \X \wt{\l}_i,$ $ \wt{\v}_i =   (\wt{\l}_i\trans \wh{\bSigma}\wt{\l}_i)^{1/2} \wt{d}_i \wt{\r}_i$ for $i = 1, \ldots, r^*$, and $\wt\boldeta_k$ the corresponding estimated nuisance parameter vector.

To specify the constructions of $\M^{(k)}$ and $\W_k$, it is essential to first gain some insights into the explicit structure of the score function vector $\wt{\psi}_k$.
In view of Proposition \ref{prop:strong1:psi} in Section \ref{sec:proof:prop1} of the Supplementary Material, we can see that $\wt{\psi}_k$ at the true parameter values $(\v_k^*, \boldeta^*_k)$ is
	$\wt{\psi}_k(\v_k^*,\boldeta^*_k)
	= \sum_{j \neq k}  \M_j^v \sum_{i \neq j} \v_i^*\u_i\strans\u_j^* -  \sum_{i \neq k} \v_i^*\u_i\strans \u_k^* + \bepsilon_k^*$, 
where $\bepsilon_k^*$ is obtained by substituting $\wt{\v}_k$ with $\v_k^*$ in $\bepsilon_k$ of Proposition \ref{prop:strong1:psi}.
It is easy to see that the expectation of $\bepsilon_k^*$ equals zero. Then the remaining term above is an intrinsic bias associated with this inference problem, induced by the correlations between different layers $\u_i\strans\u_j^*$ which are propositional to $\l_i\strans \wh{\bSigma} \l_j^*$. Hence, to design a valid inference procedure, we impose some orthogonality between the latent factors so that
    \begin{align}\label{intribias}
    \|  \sum_{j \neq k}  \M_j^v \sum_{i \neq j} \v_i^*\u_i\strans\u_j^* -  \sum_{i \neq k} \v_i^*\u_i\strans \u_k^*\|_{\infty}
    = o(n^{-1/2}). 
    \end{align}
In this section, we consider the strong orthogonality (to be formally specified in Condition \ref{con4:strong}) between latent factors, where $\sum_{j \neq k} |\l_j\strans\wh{\bSigma}\l_k^*| = o(n^{-1/2})$ is imposed for each given $k$, $1 \leq k \leq r^*$. Then the intrinsic bias can be secondary and the SOFARI-R$_s$ procedure will be suggested to attain the asymptotic distribution.

Then we continue to provide valid constructions of $\M^{(k)}$ and $\W_k$ to generate the bias-corrected estimator $\psi_k(\wt\u_k,\wt\boldeta_k)$. Due to the SVD constraints on $\boldeta_k$, similar to SOFARI \citep{sofari}, we utilize a manifold-based inference framework. Specifically, we need only the local insensitiveness to hold on the manifolds induced by the SVD constraints, rather than requiring that $\wt{\psi}_k$ be locally insensitive to the nuisance parameters on the full Euclidean space. To this end, we provide in the following proposition the gradient of $\wt{\psi}_k$ on the corresponding manifolds.

\begin{proposition}\label{prop:deri2}
Under constraint \eqref{SVDc}, the
 orthonormal vectors ${\u}_i$ with $1 \leq i \leq r^*$ belong to the Stiefel manifold $\operatorname{St}(1,n) = \{\u \in \mathbb{R}^n : \u\trans\u = 1   \}$.
 The gradient of $\wt{\psi}_k$ on the manifold is $\Q\big(\der{\wt{\psi}_k}{\boldeta_k}\big)$, where $\Q = \diag{\I_n - \u_1\u_1\trans, \dots, \I_n - \u_{r^*}\u_{r^*}\trans, \I_{q(r^* - 1)}}$  and $\der{\wt{\psi}_k}{\boldeta_k}$ is the regular derivative vector on the Euclidean space.
\end{proposition}

Proposition \ref{prop:deri2} shows that we can make $\wt{\psi}_k$ approximately insensitive to $\boldeta_k$ by requiring that $\Q\big(\der{\wt{\psi}_k}{\boldeta_k}\big)$ be asymptotically vanishing under the SVD constraint. Denote by $\wt{\U}_{-k}$ and $\wt{\V}_{-k}$ the submatrices of $\wt{\U}$ and $\wt{\V}$ after removing the $k$th columns, respectively. Then two propositions presented in the Supplementary Material give the constructions of $\M^{(k)}$ and $\W_k$ in SOFARI-R$_s$, respectively.

In light of Proposition \ref{prop:strong2:m} in Section \ref{sec:proof:prop2} of the Supplementary Material, given the properly chosen $\M^{(k)}$ and  consistent SOFAR estimator $\widetilde{\C}$ for $\C^*$, $\wt{\psi}_k$ would be approximately insensitive to nuisance parameter vector $\boldeta_k$ when constrained to be on the corresponding manifolds. We will employ the \textit{geodesic} to measure the distance on the manifold and utilize the Taylor expansion on the tangent space via the manifold gradient $\Q\big(\der{\wt{\psi}_k}{\boldeta_k}\big)$, so that the approximation error of $\wt{\psi}_k(\wt{\v}_k,{\wt{\boldeta}_k}) - \wt{\psi}_k(\wt{\v}_k,\boldeta_k^*)$ can be smaller than the root-$n$ order.

Proposition \ref{prop:strong3:w} in Section \ref{sec:proof:prop3} of the Supplementary Material gives an explicit construction of matrix $\W_k$. In fact, it can be verified that a key bias term takes the form of $\big[\I_q - \W_k ( \I_q - \M_k^u \wt{\u}_k\wt{\v}_k\trans + \M_k^u \sum_{i \neq k} \wt{\u}_i\wt{\v}_i\trans)\big](\wt{\v}_k - \v_k^*)$ in our debiased estimate $\wt{\v}_k - \W_k \wt{\psi}_k(\wt{\v}_k,\boldeta_k^*)$ after getting rid of the nuisance parameters. Hence, when $\W_k$ is the inverse of $\I_q - \M_k^u \wt{\u}_k\wt{\v}_k\trans + \M_k^u \sum_{i \neq k} \wt{\u}_i\wt{\v}_i\trans$, the bias term above can be removed. 

It is worth noting that $\M^{(k)}$ and $\W_k$ should not be considered separately because the former can affect the effectiveness of the latter.  A seemingly natural approach is to construct $\M^{(k)}$ such that the derivative  $\der{\wt{\psi}_k}{\boldeta_k}$ asymptotically vanishes in the Euclidean space.
However, this choice results in a loss of degrees of freedom, thereby leading to the nonexistence of a valid $\W_k$ matrix.

Based on $\M^{(k)}$ and $\W_k$, our SOFARI-R$_s$ estimate $\wh{\v}_k$ for $\v_k^*$ is finally defined as
\begin{align}\label{vk:strong}
	\wh{\v}_k & = \widetilde{\v}_k - \W_k\wt{\psi}_k(\widetilde{\v}_k,\widetilde{\boldeta}_k) 
	= \widetilde{\v}_k - \W_k\Big(\der{L}{\v_k}(\widetilde{\v}_k,\widetilde{\boldeta}_k) - \mathbf{M}^{(k)}\der{L}{\boldeta_k}(\widetilde{\v}_k,\widetilde{\boldeta}_k)\Big).
\end{align}
The SOFARI-R$_s$ procedure is summarized in Algorithm \ref{alg1} in Section \ref{new.sec.impleproc} of the Supplementary Material. In practice, we can calculate statistics $\{\wh{\v}_k\}_{k = 1}^{r^*}$ simultaneously and use parallel computing for inference in large-scale applications.

\subsection{SOFARI-R under weakly orthogonal factors}\label{sec:vw}

The SOFARI-R$_s$ inference procedure introduced in Section \ref{sec:vs} addresses cases involving strongly orthogonal latent factors, with the condition $\sum_{j \neq k} |\l_j\strans\wh{\bSigma}\l_k^*| = o(n^{-1/2})$. However, when correlations among latent factors diminish no faster than the root-$n$ rate, SOFARI-R$_s$ may not work, as the intrinsic bias induced by stronger correlations could compromise the asymptotic distribution. To tackle this problem, we now present the general form of SOFARI-R inference procedure for the case of weakly orthogonal factors, which accommodates a wider range of multi-response scenarios under a less stringent assumption regarding the factor correlations.

In contrast to SOFARI-R$_s$, which incorporates all unknown parameters within the constrained least-squares loss function \eqref{SVDc}, the general SOFARI-R infers $\v_k^*$ for a given $k$ by removing the other $r^* - 1$ layers from the response matrix via subtracting their estimates.
Specifically, this is achieved by subtracting signals of the other layers from the response matrix using the SOFARI \textit{debiased} estimates for latent left factor vectors and the SOFAR estimates for right factor vectors. After removing these layers, the intrinsic bias can be effectively controlled, even in the presence of weak orthogonality among factors. This is \textit{very different} from that in SOFARI \citep{sofari}, where the eigengap can also contribute to reducing the intrinsic bias. In contrast, we do not have such benefit here as the intrinsic bias is determined solely by the correlations among latent factors in view of \eqref{intribias}.

We begin with introducing the specific construction of the debiased estimate of $\v_k^*$ in SOFARI-R for each given $k$ with $1 \leq k \leq r^*$. To subtract the other $r^*-1$ layers $\X\C^*_{-k}$ from the response matrix, we propose constructing an estimate of $\C^*_{-k}$ using the SOFAR estimates $\wt{\v}_i$ for $\v_i^*$ and the debiased SOFARI estimates $\wh{\bmu}_i$ for $\bmu_i^*$, where $\bmu_i^* = d_i^* \l_i^* $ denotes the weighted left singular vector for each $1 \leq i \leq r^*$. The SOFARI debiased estimator $\wh{\bmu}_i$ is formally defined in Lemma \ref{lemm:sofari}, Section \ref{app:variance} of the Supplementary Material. Note that this debiased estimate $\wh{\bmu}_i$ is generally nonsparse, which can lead to accumulation of approximation errors in high-dimensional settings. 

Thus, we exploit the hard-thresholding technique on the debiased estimate $\wh{\bmu}_i$ and obtain the hard-thresholded debiased estimate 
$$\wh{\bmu}_i^t = (\wh{\mu}_{i1}^t, \ldots, \wh{\mu}_{i p}^t)\trans \ \text{ with } \wh{\mu}_{ij}^t  = \wh{\mu}_{ij} \mathbf{1}(\wh{\mu}_{ij} \geq   \frac{  \log n}{\sqrt{n}}), $$ where $\mathbf{1}(\cdot)$ denotes the indicator function. Since each component $\wh{\mu}_{ij}$ converges to $\mu^*_{ij}$ at the root-$n$ rate as implied by the asymptotic normality in Lemma \ref{lemm:sofari}, we can get
$ \operatorname{supp}(\wh{\bmu}_i^t) \subset  \operatorname{supp}({\bmu}_i^*).$
This reveals that $\wh{\bmu}_i^t$ is a sparse estimator as long as the true left factor ${\bmu}_i^*$ is sparse.
Based on $\wh{\bmu}_i^t$, we further define
\[ \wh{\u}_i^t =  (\wt{\bmu}_i\trans \wh{\bSigma}  \wt{\bmu}_i)^{-1/2} n^{-1/2} \X \wh{\bmu}_i^t. \]

With estimates $\wh{\u}_i^t$ and $\wt{\v}_i$, we have the surrogate for $\X\C^*_{-k}$ as $\sum_{i \neq k} \sqrt{n} \wh{\u}_i^t \wt{\v}_i\trans$. By subtracting this estimate of the other $r^* - 1$ layers from response matrix,  the new loss function can be formulated as 
\begin{align}\label{loss:weak}
	&L(\v_k,\boldsymbol{\eta}_k) = (2n)^{-1}\norm{\Y - \sqrt{n} \u_k\v_k\trans - \sum_{i \neq k} \sqrt{n} \wh{\u}_i^t \wt{\v}_i\trans}_F^2, \nonumber  \\
	&\text{subject to}
	~ \u_k\trans\u_k = 1, \  \v_k\trans\wt{\v}_i = 0,  \ i = 1, \ldots, r^* \text{ and } \ i \neq k,
\end{align}
where $\boldeta_k = \u_k$ is the nuisance parameter vector. Similarly, the modified score function for $\v_k$ can be defined as
\begin{align*}
	\wt{\psi}_k(\v_k,\boldeta_k) = \der{L}{\v_k} - \M_k\der{L}{\boldeta_k},
\end{align*}
where matrix $\mathbf{M}_k \in \mathbb{R}^{q \times n}$  will be chosen such that $\wt{\psi}_k(\v_k,\boldeta_k)$ is approximately insensitive to $\boldeta_k$. The following proposition characterizes the property of $\wt{\psi}_k$. 

\begin{proposition}\label{prop:weak1:psi}
    For an arbitrary $\M_k$, it holds that
	$\wt{\psi}_k(\wt{\v}_k,\boldeta^*_k)
	 = ( \I_q - \M_k \wt{\u}_k \wt{\v}_k\trans + \M_k\sum_{i \neq k} \wt{\u}_i \wt{\v}_i\trans ) $ $ (\wt{\v}_k - \v_k^* ) + \sum_{i \neq k} \wt{\v}_i (\wh{\u}_i^t  - \u_i^*)\trans\u_k^* + \sum_{i \neq k} (\wt{\v}_i  - \v_i^*)\u_i\strans\u_k^* 	+ \bdelta_k + \bepsilon_k$,  
where $\bepsilon_k =   - n^{-1/2}\E\trans\u_k^* + n^{-1/2} \M_k\E\trans\wt{\v}_k$ and 
	$\bdelta_k = - \M_k  \Big( \u_k^* \v_k\strans -\wt{\u}_k\wt{\v}_k\trans + \sum_{i \neq k} (\wt{\u}_i \wt{\v}_i\trans -\u_i^* \v_i\strans)\Big)  (\wt{\v}_k - \v_k^*)$.
\end{proposition}

In light of Proposition \ref{prop:weak1:psi} above, the score function vector $\wt{\psi}_k$ at the true parameter values $(\v_k^*, \boldeta^*_k)$ is
	$\wt{\psi}_k(\v_k^*,\boldeta^*_k)
	= \sum_{i \neq k} \wt{\v}_i (\wh{\u}_i^t  - \u_i^*)\trans\u_k^* + \sum_{i \neq k} (\wt{\v}_i  - \v_i^*)\u_i\strans\u_k^* + \bepsilon_k^*$, 
where $\bepsilon_k^*$ is obtained by replacing $\wt{\v}_k$ with $\v_k^*$ in $\bepsilon_k$. Although the expectation of $\bepsilon_k^*$ is zero, the expectation of $\wt{\psi}_k(\v_k^*,\boldeta^*_k)$ does not vanish due to the presence of the first two terms. The second term, $\sum_{i \neq k} (\wt{\v}_i  - \v_i^*)\u_i\strans\u_k^*$, is induced by the estimation errors of $\v_i^*$ and correlations between different layers $\u_i\strans\u_j^*$. It can be guaranteed to asymptotically vanish as long as the initial estimates are consistent and the correlations between different layers are not that strong. 

The main difficulty lies in the \textit{first} term as it does \textit{not} vanish asymptotically. Note that the hard-thresholded debiased estimate $\wh{\u}_i^t$ comes from the debiased estimate $\wh{\bmu}_i$, which enjoys asymptotic normality under weakly orthogonal factors (to be formally defined in Condition \ref{con:weak:orth}). Thus, after delicate analyses, we can show that the first term $\sum_{i \neq k} \wt{\v}_i (\wh{\u}_i^t  - \u_i^*)\trans\u_k^*$ can be decomposed into a main term following  asymptotically normal distribution and the remainder terms that  vanish asymptotically at the rate of $o(n^{-1/2})$ under mild conditions. Indeed, exploiting the SOFARI debiased estimates of left factor vectors is crucial for controlling the intrinsic bias associated with our inference problem under the case of weakly orthogonal factors. Then the following Propositions \ref{prop:weak2:m} and \ref{prop:weak3:w} provide constructions of matrices $\M_k$ and $\W_k$ in SOFARI-R.



\begin{proposition}\label{prop:weak2:m}
    When the construction of $\M_k$ is given by 
    $\M_k = -(\wt{\v}_k\trans\wt{\v}_k)^{-1} \wt{\v}_k\wt{\u}_k\trans$, 
the value of $\der{\wt{\psi}_k}{\boldeta_k\trans}$ at $(\wt{\v}_k, \wt{\boldeta}_k)$ is
     $\left(\der{\wt{\psi}_k}{\boldeta_k\trans}\right) = \left(\derr{L}{\v_k}{\boldeta_k\trans} - \M_k\derr{L}{\boldeta_k}{\boldeta_k\trans}\right) = \bDelta$, 
    where
    $\bDelta = \wt{\v}_k\wt{\u}_k\trans- \v_k^*\u_k\strans  + \sum_{i \neq k}(\wt{\v}_i\wh{\u}_i\trans - \v_i^*\u_i\strans) - n^{-1/2}\E\trans$.
\end{proposition}

\begin{proposition}\label{prop:weak3:w}   
For $\M_k$ given in Proposition \ref{prop:weak2:m} and
	$\W_k =   \I_q - 2^{-1} (\wt{\v}_k\trans\wt{\v}_k)^{-1} (  \wt{\v}_k \wt{\v}_k\trans -  \wt{\v}_k \wt{\u}_k\trans  \wt{\U}_{-k} \wt{\V}_{-k}\trans  )$, 
we have $ \W_k(\I_q - \M_k \wt{\u}_k \wt{\v}_k\trans + \M_k\sum_{i \neq k} \wt{\u}_i \wt{\v}_i\trans) = \I_q.$
\end{proposition}

Since the other $r^*-1$ layers have been removed from the response matrix, we have only one matrix $\M_k$ that corresponds to nuisance parameter vector $\u_k$, as opposed to construction of the full matrix $\M^{(k)}$ in Proposition \ref{prop:strong2:m} for SOFARI-R$_s$. Then matrix $\W_k$ can be constructed accordingly.
Based on $\M_k$ and $\W_k$, our SOFARI-R debiased estimate $\wh{\v}_k$ for $\v_k^*$ is defined as
\begin{align}\label{vk:weak}
	\wh{\v}_k & = \widetilde{\v}_k - \W_k\wt{\psi}_k(\widetilde{\v}_k,\widetilde{\boldeta}_k) 
	= \widetilde{\v}_k - \W_k\Big(\der{L}{\v_k}(\widetilde{\v}_k,\widetilde{\boldeta}_k) - \mathbf{M}_k\der{L}{\boldeta_k}(\widetilde{\v}_k,\widetilde{\boldeta}_k)\Big).
\end{align}
We summarize the SOFARI-R procedure in Algorithm \ref{alg2} in Section \ref{new.sec.impleproc} of the Supplementary Material.
In addition, the key differences between our new SOFARI-R and the previous SOFARI \citep{sofari} are delineated in Table \ref{tab:my_label}.

\section{Asymptotic properties} \label{sec4:theory}

We now provide theoretical justifications for both SOFARI-R$_s$ and SOFARI-R, which correspond to the settings of strongly orthogonal factors and weakly orthogonal factors, respectively. 

\subsection{Technical conditions}\label{new.Sec.4.1}
We first present some definitions and regularity conditions that will be used in our theoretical analyses.

\begin{definition}[SOFAR SVD estimates]\label{defi:sofar}
	A $p \times q$ matrix  $\wt{\C}$ with SVD components $ (\widetilde{\L}, \wt{\D},  \widetilde{\mathbf{R}})$ is called an acceptable estimator of matrix $\C^*$ if it satisfies that with probability at least $1 - \theta_{n,p,q}^{\prime}$ for some asymptotically vanishing $\theta_{n,p,q}^{\prime}$, the following estimation error bounds hold:
	\begin{align*}
		& (a) ~ \norm{\wt{\D}-\D^*}_F
		+ \norm{\wt{\L}\wt{\D} - \L^*\D^*}_F
		+ \norm{\wt{\mathbf{R}}\wt{\D} - \mathbf{R}^*\D^*}_F \leq  c \gamma_n, \\[5pt]
		& (b)~
		 \norm{\wt{\D}-\D^*}_0
		 + \norm{\wt{\L}\wt{\D} - \L^*\D^*}_0
		 + \norm{\wt{\mathbf{R}}\wt{\D} - \mathbf{R}^*\D^*}_0
		\leq (r^*+s_u +s_v)[1+o(1)], 
 	\end{align*}
	where $s_u = \norm{\L^*\D^*}_0$, $s_v = \norm{\mathbf{R}^*\D^*}_0$,  $ \gamma_n = ({r^*}+s_u+s_v)^{1/2}\eta_n^2\{n^{-1}\log(pq)\}^{1/2}$, $\eta_{n}=1+\delta^{-1 / 2} \left(\sum_{j=1}^{r^*}(d_{1}^{*} / d_{j}^{*})^{2} \right)^{1 / 2}$, and $c$ and $\delta$ are some positive constants. 
\end{definition}

\begin{definition}[Approximate Inverse]\label{defi:theta}
	A $p\times p$ matrix $\wh{\bTheta} = (\wh{\btheta}_1,\ldots,\wh{\btheta}_p)\trans$ is called an approximate inverse matrix of $\wh{\bSigma}$ if there exists some positive constant $C$ such that 1) $\norm{\I - \wh{\bTheta}\wh{\bSigma}}_{\max}\leq C\sqrt{(\log p)/n}$ and 
	2) $\max_{1\leq i\leq p}\norm{\wh{\btheta}_i}_{0} \leq s_{\max}$ and $\max_{1\leq i\leq p}\norm{\wh{\btheta}_i}_{2} \leq C$. %
\end{definition}
The above two definitions are defined similarly as in \cite{sofari}.
Definition \ref{defi:sofar} characterizes the properties of SOFAR SVD estimates, which can be established by Theorem 2 of \cite{uematsu2017sofar} under some mild conditions. 
Definition \ref{defi:theta} requires that the approximate inverse matrix $\wh{\bTheta}$ satisfies an entrywise approximation error bound of rate $\sqrt{(\log p)/n}$ and a rowwise sparsity level $s_{\max}$ with the length of each row  bounded above. 
It is worth pointing out that different from SOFARI, the approximate inverse matrix $\wh{\bTheta}$ is required \textit{solely} in the case of weakly orthogonal factors as we need to utilize the SOFARI debiased estimates of left factor vectors and guarantee its properties. This approximate inverse is \textit{not} needed under strongly orthogonal factors, \textit{nor} for correcting the bias of right factor vectors. 

\begin{condition}\label{con1:error}
	The error matrix $\E\sim\N(\0,\I_n\otimes\bSigma_e)$ and the maximum eigenvalue of $\bSigma_e$ is bounded from above.
\end{condition}

\begin{condition}\label{con2:re}
	There exist some sparsity level $s \geq  \max\{s_{\max}, 3(r^* + s_u + s_v)\}$ and positive constants $\rho_l$ and $\rho_u$  such that
 \begin{align*}
 \rho_l < \min_{\bdelta\in\R^p}\left\{ \frac{\norm{ \wh{\bSigma} \bdelta}_2}{  \norm{\bdelta}_2} : \norm{\bdelta}_0\leq s \right\} \leq \max_{\bdelta\in\R^p}\left\{ \frac{\norm{\wh{\bSigma}\bdelta}_2}{\norm{\bdelta}_2} : \norm{\bdelta}_0\leq s \right\} < \rho_u.
\end{align*}
\end{condition}

\begin{condition}\label{con3:eigend}
	The nonzero eigenvalues $d^{*2}_{i}$ of matrix $\C\strans\C^*$ satisfy that $d^{*2}_{i}  - d^{*2}_{i+1} \geq  \delta_1 d^{*2}_{i}$ for some positive constant $\delta_1 > 1 - \rho_l/\rho_u$ with $1 \leq i \leq r^*$. Also, $r^* \gamma_n = o(d^{*}_{r^*})$.
\end{condition}

\begin{condition}[Strong orthogonality]\label{con4:strong}
	The nonzero squared singular values $d^{*2}_{i}$ are at the constant level and $\sum_{j \neq k} | \l_j\strans\wh{\bSigma}\l_k^*| = o(n^{-1/2})$ for each given $k$ with $1 \leq k \leq r^*$.
\end{condition}

\begin{condition}[Weak orthogonality]\label{con:weak:orth}
	The nonzero squared singular values $d^{*2}_{i}$ and the latent factors jointly satisfy that $\sum_{j = k+1}^{r^*} (d_j^{*2}/d_k^*) |\l_j\strans \wh{\bSigma} \l_k^*| = o(n^{-1/2})$ for each given $k$ with $1 \leq k < r^*$. 
\end{condition}

\begin{condition}\label{con:threshold}
	For $\bmu_i^* = (\mu_{i1}^*, \ldots, \mu_{ip}^*)\trans$ with $ 1 \leq i \leq  r^*$, there exist some positive constants $C_u$ and $\alpha < 1/2$ such that $\mu_{ij}^*$ either belongs to set $\mathcal{S}_{\mu_i} = \{ j: | \mu_{ij}^*| \geq  C_u  {n^{-\alpha}} \}$ or jointly satisfies  $ \sum_{i \neq k} \norm{(\bmu_i^*)_{ \mathcal{S}_{\mu_i}^c}}_2 = o(\frac{ 1}{\sqrt{n}})$ for each $k$, $1 \leq k \leq  r^*$.
\end{condition}

Conditions \ref{con1:error}--\ref{con:weak:orth} are the same as those imposed in \cite{sofari} for inference on the left factor vectors. The Gaussian assumption in Condition \ref{con1:error} can be relaxed to accommodate non-Gaussian settings by leveraging a similar central limit theorem argument as in \cite{van2014asymptotically}. Condition \ref{con2:re} imposes a boundedness assumption on the $s$-sparse eigenvalues of $\wh{\bSigma}$. Notably, the sparsity level $s_{\max}$ for the precision matrix that appears in the lower bound of $s$ is necessary only under weakly orthogonal factors for the SOFARI estimator of left factors. 
Condition \ref{con3:eigend} requires distinction between the nonzero singular values so that different latent factors are separable. 
Conditions \ref{con4:strong} and \ref{con:weak:orth} correspond to the cases of strongly orthogonal factors and weakly orthogonal factors, respectively. 
Note that the bounded singular value assumption in Condition \ref{con4:strong} is imposed solely to simplify the technical analyses. Indeed, they can be diverging as long as the singular values and latent factors jointly satisfy a
similar requirement to that in Condition \ref{con:weak:orth}.

Condition 6 is needed in the case of weakly orthogonal factors, which imposes a weak sparsity pattern on the weighted left singular vectors. That is, except for a few identifiable signals above the order of $n^{-\alpha}$, the sum of the rest are asymptotically negligible in the sense that they are collectively lower than the root-$n$ order.

\subsection{Asymptotic theory of SOFARI-R\texorpdfstring{$_s$}{s}}

Define  
${\M}_k^*$ and $ {\W}_k^*$ as the population counterparts of ${\M}_k$ and $ {\W}_k$ suggested for the SOFARI-R$_s$ procedure, obtained by replacing $\wt{\U}, \wt{\V}$ with $\U^* = (\u_1^*, \ldots, \u_{r^*}^*)$ and $\V^* = (\v_1^*, \ldots, \v_{r^*}^*)$.
In addition, denote by
\[  \kappa_n = \max\{ (r^*+s_u+s_v)^{1/2}, \eta_n^2 \}(r^*+s_u+s_v)\eta_n^2\log(pq)/\sqrt{n},\]
which will be the key order of the error term. We now proceed to characterize the asymptotic distribution of the proposed estimator $\wh{\v}_k$ in the following theorem.
\begin{theorem}\label{theo:strong:vk}
		Assume that Conditions \ref{con1:error}--\ref{con4:strong} hold and $\wt{\C}$ satisfies Definition \ref{defi:sofar}. Then for $\a\in\mathcal{A}=\{\a\in\R^q:\norm{\a}_0\leq m,\norm{\a}_2=1\}$ satisfying $m^{1/2}\kappa_n = o(1)$, we have
		\begin{align*}
			\sqrt{n}\a\trans(\wh{\v}_k-\v_k^*) = h_k + t_k,
		\end{align*}
		where the distribution term $h_k =
		-\a\trans \W^{*}_k \M_{k}^* \E {\v}_k^{*} 
		+ \a\trans\W^*_k\E\trans\u_k^* \sim N(0,\nu_k^2)$ with 
		$\nu_k^2 = \a\trans\W_k^*(\bSigma_e + \v_k\strans\bSigma_e\v_k^* \M_{k}^* \M_{k}\strans - 2\M_{k}^*  \u_k^* \v_k\strans\bSigma_e)\W_k\strans\a$. 
		Moreover, the bias term $t_k = O_p(m^{1/2}\kappa_n)$ holds with probability at least $1 - \theta_{n,p,q}$,
		where 
			\begin{equation}\label{thetapro}
			\theta_{n,p,q} = \theta_{n,p,q}^{\prime} +  2(pq)^{1-c_0^2/2}
		\end{equation}
		with $\theta_{n,p,q}^{\prime}$ given in Definition \ref{defi:sofar} and some constant $c_0 > \sqrt{2
        }$.
	\end{theorem}

Theorem \ref{theo:strong:vk} establishes the asymptotic normal distribution for each latent right factor vector $\v_k^*$ with $1 \leq k \leq r^*$.
This theorem, along with Theorems 1 and 2 in \cite{sofari}, completes the manifold-based inference results for the SVD components of coefficient matrix $\C^*$ in multi-response regression. The main requirement for the validity of the asymptotic normal distribution is $m^{1/2}\kappa_n = o(1)$. Compared to Theorem 1 in \cite{sofari}, this requirement is \textit{weaker} by a factor of $s_{\max}$, the sparsity level of the precision matrix.
This is because the SOFARI-R$_s$ inference on latent right factor vectors is designed for response selection, which is from the prediction perspective. This contrasts with the SOFARI$_s$ inference for latent left factor vectors in \cite{sofari}, which is from the estimation point of view.
Thus, it does \textit{not} require estimation of the precision matrix, \textit{nor} does the error term impose any sparsity bound on the precision matrix. 

An important application of Theorem \ref{theo:strong:vk} is inference on the sparse PCA since the right singular vectors of a data matrix correspond to the eigenvectors of its sample covariance matrix. Compared to the debiased sparse PCA procedure developed in \cite{jankova2021biased}, which establishes the asymptotic normality of the first eigenvector, our manifold-based inference technique can provide asymptotic distributions for \textit{all} significant eigenvectors of the data matrix as demonstrated in Theorem \ref{theo:strong:vk}. Moreover, the debiased procedure in \cite{jankova2021biased} needs to estimate the inverse Hessian matrix and requires that the inverse Hessian matrix exhibits a certain level of sparsity. In contrast, our SOFARI-R$_s$ procedure does \textit{not} impose such requirement.

As the population variances $\nu_{k}^2$ presented in Theorem \ref{theo:strong:vk} are unknown in practice, we propose a surrogate using some consistent estimate of error covariance matrix $\bSigma_e$ along with initial SOFAR estimates. To this end, we introduce the following definition to characterize  estimation consistency of an estimate $\wt{\bSigma}_e$ for $\bSigma_e$,  which is attainable through existing covariance estimation techniques including the hard-thresholding \citep{bickel2008covariance} and adaptive thresholding \citep{cai2011adaptive}.
\begin{definition}\label{defi:error}
    A $q \times q$ matrix $\wt{\bSigma}_e$ is an acceptable estimator of ${\bSigma}_e$ if $\norm{\wt{\bSigma}_e - {\bSigma}_e}_2 = o_p(1). $
\end{definition}

Based on $\wt{\bSigma}_e$ and the SOFAR estimates, we define
\begin{align}\label{eqvarsk}
    \wt{\nu}^2_k = {\a}\trans{\W}_{k}(\wt{\bSigma}_e +  \wt{\v}_k\trans\wt{\bSigma}_e\wt{\v}_k {\M}_{k}{\M}_{k}\trans - 2{\M}_{k}\wt{\u}_k \wt{\v}_k\trans\wt{\bSigma}_e)
    {\W}_k\trans{\a}.
\end{align}

\begin{theorem}\label{theo:strong:var}
Assume that all the conditions of Theorem \ref{theo:strong:vk}  are satisfied. Then for each $k$, $1 \leq k \leq r^*$, with probability at least $1- \theta_{n,p,q}$, the estimation error bounds 
	$|\wt{\nu}_{k}^2 - {\nu}_{k}^2 | \leq  \widetilde{C}  \gamma_n$ hold,
where $\widetilde{C} > 0$ is some constant and $\theta_{n,p,q}$ is given in \eqref{thetapro}.
\end{theorem}
Theorem \ref{theo:strong:var} shows that the plug-in estimate $ \wt{\nu}^2_k$ converges to ${\nu}_{k}^2$ with the same rate as that of the SOFAR estimate in view of the first property in Definition \ref{defi:sofar}. Thus, Theorems \ref{theo:strong:vk} and \ref{theo:strong:var} together provide easy-to-use bias-corrected estimators for the SOFARI-R$_s$ procedure that enjoy asymptotic normality with estimable variance.

\subsection{Asymptotic theory of SOFARI-R}

We now turn to the theory for the SOFARI-R procedure. 
Denote by
\begin{align} \label{kappa}
	\kappa_n^{\prime} = \max\{s_{\max}^{1/2} , (r^*+s_u+s_v)^{1/2}, \eta_n^2\} (r^*+s_u+s_v)\eta_n^2\log(pq)/\sqrt{n}.
	\end{align}
Using matrices $\M_k$ and $\W_k$ constructed in Propositions \ref{prop:weak2:m} and \ref{prop:weak3:w}, the following theorem presents the asymptotic distribution for $\wh{\v}_k$ under the scenario of weakly orthogonal factors.
\begin{theorem}\label{theo:weak:uk}
	Assume that Conditions \ref{con1:error}--\ref{con3:eigend} and \ref{con:weak:orth}--\ref{con:threshold} hold, and $\wt{\C}$ and $\wh{\bTheta}$ satisfy Definitions \ref{defi:sofar} and \ref{defi:theta}, respectively. Then for each given $k$ with $1 \leq k \leq r^*$ and an arbitrary vector
	$\a\in\mathcal{A}=\{\a\in\R^q:\norm{\a}_0\leq m,\norm{\a}_2=1\}$,
	we have
	\begin{align*}
		\sqrt{n}\a\trans(\wh{\v}_k-\v_k^*) = h_k + t_k,
	\end{align*}
	where the distribution term $ h_k = h_{v_k} + \sum_{i \neq k} h_{u_i}^{\prime} \sim N(0, \nu_k^2)$ and the error term 
		$t_k = O_p\Big( m^{1/2} d_1^{*}    d_{r^*}^{*-2} \kappa_n  +    s_{u}^{1/2}r^*  ( d_1^{*}  d_{r^*}^{*-2} \gamma_n +  \kappa_n^{\prime}  \max\big\{1, d_{r^*}^{*-1}, d_{r^*}^{*-2}\big\})  \Big)$. 
	Here, the explicit formula of $h_k$ and $\nu_k^2$ are presented in Section \ref{app:variance} of the Supplementary Material.
\end{theorem}

Theorem \ref{theo:weak:uk} establishes the asymptotic normality of the SOFARI-R debiased estimator $\wh{\v}_k$ for each $k$ with $1 \leq k \leq r^*$. In comparison to Theorem \ref{theo:strong:vk} for SOFARI-R$_s$, the distribution term $h_k$ in this theorem consists of two parts. Specifically, the first part $h_{v_k}$ is similar to the distribution term in Theorem \ref{theo:strong:vk}, except for the distinct constructions of $\M_{k}^{*}$ and $\W_{k}^{*}$. It can be regarded as the main part for inferring the latent right factor vectors. While the second part comes from the summation of distribution terms for weighted left singular vectors across the other $r^*-1$ layers, it corresponds to expression $\sum_{i \neq k} \wt{\v}_i (\wh{\u}_i^t - \u_i^*)\trans \u_k^*$ in Proposition \ref{prop:weak1:psi}, suggesting that incorporating the debiased estimates $\wh{\bmu}_i$ is beneficial to mitigate the intrinsic bias and ensure the asymptotic normality.

Furthermore, we see that the order of error term $t_k$ in Theorem \ref{theo:weak:uk} is also distinct from that in Theorem \ref{theo:strong:vk}. To be specific, the first term is analogous but contains an extra term $d_1^{*} d_{r^*}^{*-2}$, which adds a mild constraint on singular values. The last two terms are induced by the additional approximation error when we replace the other $r^*-1$ layers with SOFAR estimates $\wt{\v}_i$ and SOFARI estimates $\wh{\u}_i$. 
In fact, the error term $d_1^{*}  d_{r^*}^{*-2} \gamma_n + \kappa_n^{\prime}  \max\big\{1, d_{r^*}^{*-1}, d_{r^*}^{*-2}\big\}$ is of the same order as that in Theorem 4 of \cite{sofari} since this error indeed arises from the debiased SOFARI estimate. Moreover, term $ s_{u}^{1/2}$ captures the sparsity level of the hard-thresholded debiased estimate, while term $r^*$ accounts for accumulation of the $r^*-1$ layers.

Similar to \eqref{eqvarsk}, we define $\wt{\nu}_{k}^2$ as the estimate of ${\nu}_{k}^2$ after plugging in the SOFAR estimates. The following theorem provides estimation accuracy of the variance estimate for SOFARI-R. 

\begin{theorem}\label{theo:weak:var}
	Assume that all the conditions of Theorem \ref{theo:weak:uk} are satisfied and $ \wt{\bSigma}_e$ is an acceptable estimator. Then for each $k$ with $1 \leq k \leq r^*$, the estimation error bound 
			$|\wt{\nu}_{k}^2 - {\nu}_{k}^2 | \leq  \widetilde{C}^{\prime} r^{*2} s_u  \gamma_n  d_1^*d_{r^*}^{*-2}$ 
		hold with probability at least $1- \theta_{n,p,q}$, where $\theta_{n,p,q}$ is given in \eqref{thetapro}, and $\widetilde{C}^{\prime}  > 0$ is some constant.
\end{theorem}

\section{Simulation studies}\label{new.Sec.5}

In this section, we evaluate the finite-sample performance of the SOFARI-R method in inferring the latent right factor vectors. The detailed simulation setup is presented in Section \ref{app:simusetup} of the Supplementary Material.
In addition, we consider two settings of different dimensionalities. In setting $1$, we set $(n, p, q) = (200, 25, 30)$, while $(n, p, q)$ are increased to $(200, 50, 60)$ in setting $2$.
It is noteworthy that both settings give rise to the high-dimensional regime since the total dimensionality due to both features and responses is $p*q$, significantly exceeding the available sample size $n$.

\begin{figure}[tp]
	\centering
	\includegraphics[width=0.6\linewidth]{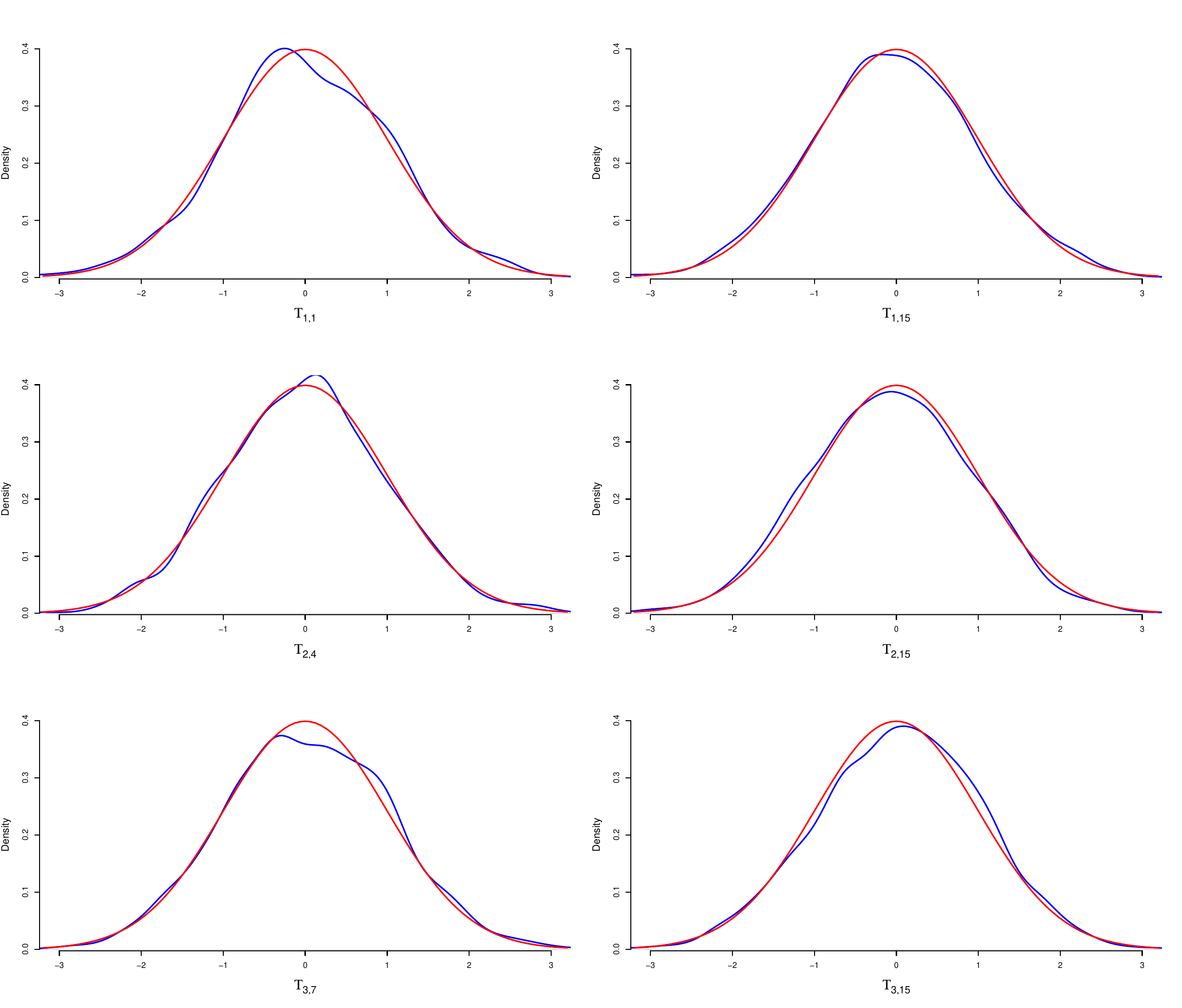}
	\caption{ \small{The kernel density estimates (KDEs) for the distributions of SOFARI-R estimates on the latent right factor vectors in different sparse SVD layers against the target standard normal density based on $1000$ replications for setting 1 in Section \ref{new.Sec.5}. Left panel: the KDEs of  $\mathrm{T}_{1,1}, \mathrm{T}_{2,4}$, and $\mathrm{T}_{3,7}$; right panel:  the KDEs of $\mathrm{T}_{1,15}, \mathrm{T}_{2,15}$, and $\mathrm{T}_{3,15}$, all viewed from top to bottom. The blue curves represent the KDEs for SOFARI-R estimators, whereas the red curves stand for the target standard normal density.}}\label{figure1}
\end{figure}

To implement the SOFARI-R inference procedure, we initially determine the rank of the multi-response regression model \eqref{model} using the self-tuning selection method outlined in \cite{Xin2019Adaptive}. The initial estimate $\wt{\C} = (\wt{\L}, \wt{\D}, \widetilde{\mathbf{R}})$ is obtained from the SOFAR procedure \citep{uematsu2017sofar} using the entrywise $L_1$-norm penalty (SOFAR-L). Moreover, the precision matrix of the covariates is estimated with the nodewise Lasso \citep{meinshausen2006high} as suggested in \cite{van2014asymptotically} and we exploit the adaptive thresholding method \citep{cai2011adaptive} for the covariance estimation of the random errors.

\begin{table}[t]
	\centering
	\caption{\label{table:uj1}
		The average performance measures of SOFARI-R on the individual components of the latent right factor vectors (i.e., the right singular vectors weighted by the corresponding variance-adjusted singular values) in different sparse SVD layers with squared singular values $(d_1^{*2}, d_2^{*2}, d_3^{*2} ) = (100^2, 15^2, 5^2)$ over $1000$ replications.}
	\smallskip
		\begin{tabular}{c|ccccccccc}
			\hline
			Setting& & ${\operatorname{CP}}$  & ${\operatorname{Len}}$ &    & ${\operatorname{CP}}$  & ${\operatorname{Len}}$ & & ${\operatorname{CP}}$  & ${\operatorname{Len}}$    \\
			\hline
			1&$v_{1,1}^*$
			& 0.946  &0.353  & $v_{2,4}^*$   & 0.953  & 0.356 & $v_{3,7}^*$    & 0.953  & 0.380 \\
			&$v_{1,2}^*$
			& 0.948   &0.353  & $v_{2,5}^*$   & 0.954  & 0.356 & $v_{3,8}^*$    & 0.938  & 0.379 \\
			&$v_{1,3}^*$
			& 0.947   &0.353  & $v_{2,6}^*$   & 0.942  & 0.357 & $v_{3,9}^*$    & 0.941  & 0.379 \\
			&$v_{1,q-2}^*$
			& 0.935   &0.355  & $v_{2,q-2}^*$   & 0.947  & 0.355 & $v_{3,q-2}^*$    & 0.953  & 0.355 \\
			&$v_{1,q-1}^*$
			& 0.943   &0.355  & $v_{2,q-1}^*$   & 0.951  & 0.355 & $v_{3,q-1}^*$    & 0.954  & 0.355 \\
			&$v_{1,q}^*$
			& 0.946   &0.354  & $v_{2,q}^*$   & 0.956  & 0.354 & $v_{3,q}^*$    & 0.954  & 0.354 \\
			\hline
			2&$v_{1,1}^*$
			& 0.935   &0.249  & $v_{2,4}^*$   & 0.939  & 0.252 & $v_{3,7}^*$    & 0.947  & 0.269 \\
			&$v_{1,2}^*$
			& 0.938   &0.249  & $v_{2,5}^*$   & 0.951  & 0.250 & $v_{3,8}^*$    & 0.941  & 0.268 \\
			&$v_{1,3}^*$
			& 0.947   &0.248  & $v_{2,6}^*$   & 0.953  & 0.252 & $v_{3,9}^*$    & 0.952  & 0.269 \\
			&$v_{1,q-2}^*$
			& 0.957   &0.250  & $v_{2,q-2}^*$   & 0.955  & 0.250 & $v_{3,q-2}^*$    & 0.958  & 0.250 \\
			&$v_{1,q-1}^*$
			& 0.956   &0.251  & $v_{2,q-1}^*$   & 0.957  & 0.251 & $v_{3,q-1}^*$    & 0.953  & 0.251 \\
			&$v_{1,q}^*$
			& 0.942   &0.250  & $v_{2,q}^*$   & 0.940  & 0.250 & $v_{3,q}^*$    & 0.947  & 0.250 \\
			\hline
		\end{tabular} 
\end{table}

We choose the significance level $\alpha = 0.05$ for statistical inference in both simulation examples and repeat the simulation $1000$ times for each setting.
Two performance measures are employed to evaluate the inference results: the average coverage probability (CP) and the average length (Len) of the $(1 - \alpha) 100\%$ (i.e., $95\%$) confidence intervals for the unknown population parameters over $1000$ replications.
To be specific, for each individual unknown parameter $v^*$, we denote by $\mathrm{CI}$ the corresponding $95\%$ confidence interval of $v^*$ constructed using SOFARI-R. Then the two performance measures are defined as 
${\operatorname{CP}} =  \wh{\mathbb{P}}\left[v^* \in \mathrm{CI}\right]$ and 
		${\operatorname{Len}} =  \operatorname{
			length}\left(\mathrm{CI}\right)$, 
respectively, where $\wh{\mathbb{P}}$ denotes the empirical probability measure. Note that CP is the empirical version of the expectation for the conditional coverage probability given both parameters and the covariate matrix. To verify the asymptotic normalities of the SOFARI-R estimates, we 
also define the following standardized quantities 
	$\mathrm{T}_{{k,j}} = \frac{\sqrt{n}(\wh{v}_{k,j} - {v}_{k,j}^*) }{\wt{\nu}_{{k,j}}}$ 
for each $k = 1, \ldots, r^*$ and $j = 1, \ldots, q$, where $\wt{\nu}_{{k,j}}^2$ is the corresponding variance estimate given in Theorem \ref{theo:weak:var}.

Now we proceed to evaluate the simulation results. First of all, the rank of the latent sparse SVD structure is correctly identified as $r = 3$ over both settings. Second, to examine the asymptotic normalities of different SOFARI-R estimates, we calculate the kernel density estimates (KDEs) for the standardized quantities $\mathrm{T}_{{k,j}}$  for both nonzero and zero components of the latent right factor vectors $\v^*_k = ({\l_i^*}\trans \wh{\bSigma}\l_i^*)^{1/2} d_k^* \r_k^*$. Since these KDEs are similar across the two settings, we only present in Figure \ref{figure1} the kernel density plots for setting 1 corresponding to the first nonzero component $v^*_{{k,3(k-1)+1}}$ and the last zero component $v^*_{{k,p}}$ in each latent sparse SVD layer for $1 \leq k \leq 3$.
By comparing the KDEs for SOFARI-R estimates to the standard normal density, Figure \ref{figure1} shows that the empirical distributions of the standardized SOFARI-R estimates for the representative parameters mimic the standard normal distribution closely, justifying our asymptotic normality theory. 

Third, we report the performance measures of SOFARI-R estimates for all three nonzero components and the last three zero components of $\v_k^*$ in each latent sparse SVD layer with $1 \leq k \leq 3$ over the two settings and summarize the results in Table \ref{table:uj1}.
It can be seen from Table \ref{table:uj1} that the average coverage probabilities of the corresponding confidence intervals constructed by SOFARI-R are all very close to the target level of $95\%$. Moreover, we observe that the average lengths of $95\%$ confidence intervals for different $v_{k,j}^*$ in each latent sparse SVD layer are relatively stable over $j$. These results verify the validity of our SOFARI-R inference procedure.

Beyond this simulation example, we have further assessed the robustness and effectiveness of SOFARI-R in another simulation study, where the correlations among latent factors violate our technical condition. Due to space constraints, these numerical results are provided in Section \ref{new.Sec.5.2} of the Supplementary Material.

\section{Real data application}\label{new.Sec.6} 
In this section, we demonstrate the practical performance of the SOFARI-R inference procedure through a monthly macroeconomic data set from the federal reserve economic database (FRED-MD) in \cite{mccracken2016fred}. This data set consists of 660 monthly observations from January 1960 to December 2014 for 134 macroeconomic variables, divided into different groups that broadly represent key aspects of economic activity and financial markets including labor market, housing, and stock market. Among those variables, we are interested in \textit{simultaneously} forecasting some key macroeconomic indicators such as the consumer price index (CPI), the unemployment rate, the stock market price index, and interest rates. Additionally, we select several typical macroeconomic variables as responses including the money supply, the personal income, exchange rates, and so on. This results in a total of 30 response variables from eight  distinct groups. We refer to Table \ref{table4} in Section \ref{appsec:D} of the Supplementary Material for the detailed list of the 30 selected responses, their descriptions, and the corresponding group classifications.

\begin{figure}[tp]
	\centering
	\includegraphics[width=0.7\linewidth]{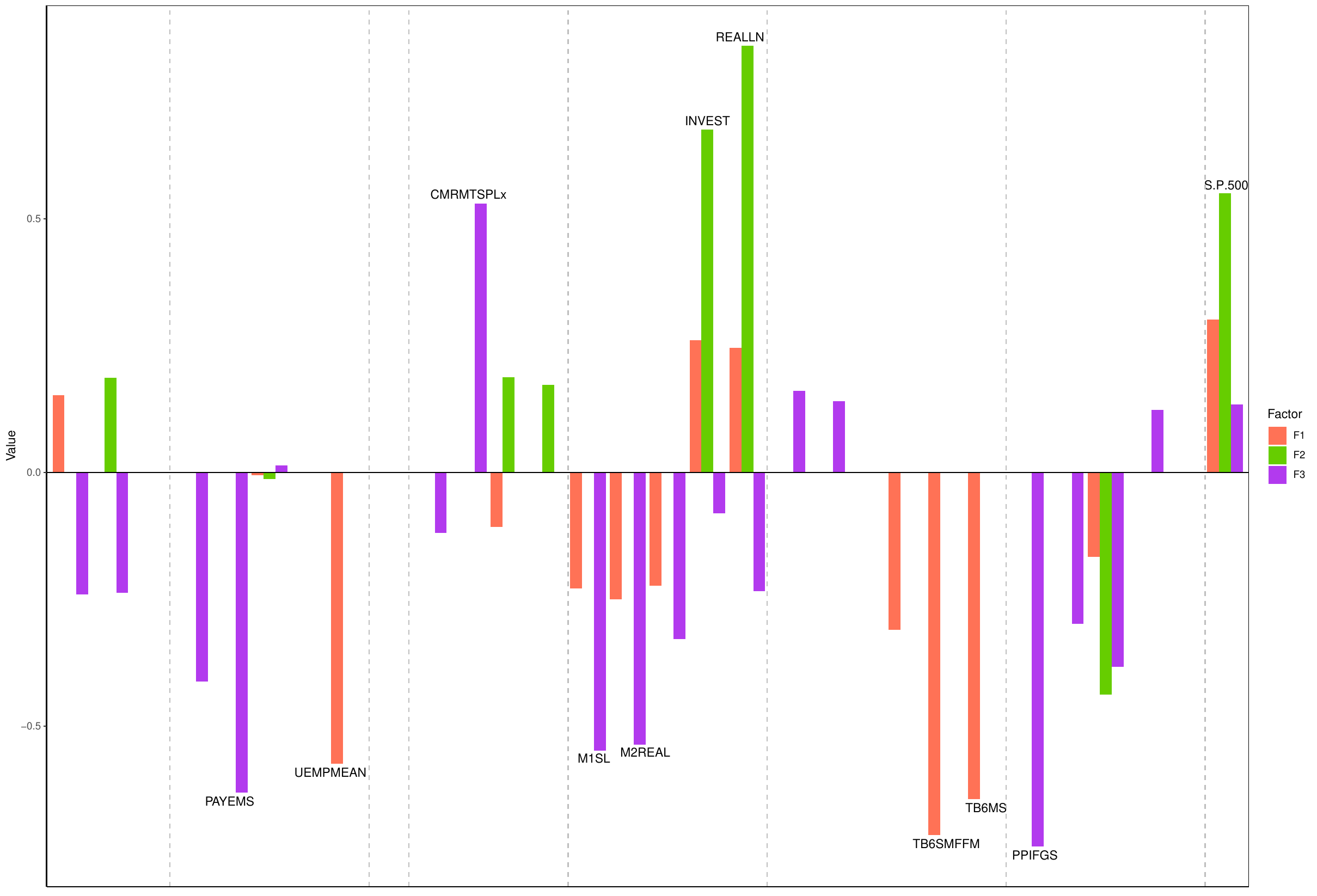}
	\caption{Bar charts of the significant components in the top three latent right factor vectors (i.e., the right singular vectors weighted by the corresponding variance-adjusted singular values). The significant components correspond to ones listed in Table 5. Different colors represent the three factors, F1, F2, and F3, ranked by their estimated singular values, respectively. The y-axis indicates the magnitude and sign of the corresponding coefficient in each factor. The dashed vertical lines indicate the boundaries between different groups of variables.}
    \label{figure2}
\end{figure}

 We use the remaining macroeconomic variables as covariates, excluding the four with missing values. Following \cite{zheng2021nonsparse}, 
 to address the issue of highly correlated covariates, we randomly select one representative covariate from those economic variables whose correlations in magnitude are over $0.9$, resulting in 70 representative covariates. In addition, to adapt to times series data, both responses and covariates are transformed through differencing and logarithmic transformation as in \cite{mccracken2016fred}. Furthermore, we include the first to fourth lags of the responses and covariates as additional predictors into the design matrix and standardize each column of both predictor and response matrices to have mean zero and standard deviation one, similarly as in \cite{tsknockoff2021}. After the preprocessing, we have $n = 654$ observations, $p = 400$ predictors, and $q = 30$ responses. This involves a high-dimensional multi-task learning problem in view of the unknown parameter matrix of dimensionality $p*q$. 

We begin our analysis by fitting the multi-response regression model \eqref{model} using the SOFAR estimator \citep{uematsu2017sofar} with the entrywise $L_1$-norm penalty (SOFAR-L) given its nice predictive performance as shown in Section \ref{appsec:D} of the Supplementary Material.
Then we implement the SOFARI-R inference procedure on the data similarly as in Section \ref{new.Sec.5}. First, the rank of model \eqref{model} is estimated as $r = 3$, indicating three significant latent factors in the SVD layers of the coefficient matrix. 
The estimated singular values for three layers are 53.997, 10.436, and 2.695, respectively.
Second, based on the SOFAR-L initial estimates, we choose the significance level $\alpha = 0.05$ to study the significance of compositions of the latent right factor vectors $\v_k^*$'s in different sparse SVD layers $k=1,2,3$.
As a result, the numbers of significant components in the three layers are 14, 8, and 19, respectively.
Moreover, we display those significant components of latent right factor vectors as a bar chart in Figure \ref{figure2}. As we can see from Figure \ref{figure2}, although there are significant gaps in the estimated singular values across the three layers, the signal magnitudes of the three layers become comparable after the variance adjustment in view of \eqref{uvtransform}. We provide some insights into the identified latent response factors in Section \ref{appsec:D}.

\section{Discussions}\label{sec:discu}

In this paper, we propose a new method SOFARI-R targeting at the  inference on the latent right singular vectors in high-dimensional multi-response regression.
The SOFARI-R procedure provides bias-corrected estimators for the latent right factor vectors that enjoy asymptotic normality with justified asymptotic variance estimates. Both the established theoretical analysis and empirical performance demonstrate the efficiency of our proposed method.
An interesting direction for future work is to extend our framework to nonlinear model settings, where the nonlinear effects can be captured through, e.g., the reproducing kernel Hilbert space (RKHS). Beyond this, it would also be valuable to explore extensions to causal inference and reinforcement learning. These extensions are beyond the scope of the current paper and will be interesting topics for future research.

\bibliographystyle{chicago}
\bibliography{references}

\newpage

\appendix
\setcounter{page}{1}
\setcounter{section}{0}
\renewcommand{\theequation}{A.\arabic{equation}}
\setcounter{equation}{0}

\begin{center}{\bf \Large Supplementary Material to ``SOFARI-R: High-Dimensional Manifold-Based Inference for Latent Responses''}

\bigskip

Zemin Zheng, Xin Zhou and Jinchi Lv
\end{center}

\noindent This Supplementary Material contains the proofs of Theorems \ref{theo:strong:vk}--\ref{theo:weak:var} and Propositions \ref{prop:deri2}--\ref{prop:strong3:w}, additional technical details, additional numerical results, and additional real data results. Throughout the proofs, we will use $c$ to denote a generic positive constant whose value may vary from line to line.

\section{Proofs of Theorems \ref{theo:strong:vk}--\ref{theo:weak:var} and Propositions \ref{prop:deri2}--\ref{prop:strong3:w}}

\subsection{Proof of Theorem \ref{theo:strong:vk}}

Denote by $\mathcal{E}_0$ the event on which the inequalities in Definition \ref{defi:sofar} hold. By Definition \ref{defi:sofar}, its probability is at least $1 - \theta_{n,p,q}^{\prime}$. Let us further define $\mathcal{E}_1 = \{ n^{-1}\norm{\X\trans\E}_{\max} \leq c_1[n^{-1}\log(pq)]^{1/2} \}$, where $c_1$ is some positive constant.
Since $\E\sim\N(\0,\I_n\otimes\bSigma_e)$ under Condition \ref{con1:error}, an application of similar arguments as in Step 2 of the proof of Theorem 1 in \cite{uematsu2017sofar} shows that event $\mathcal{E}_1$ holds  with probability at least $1- 2(pq)^{1-c_0^2/2}$, where $c_0 > \sqrt{2}$ is some positive constant.  Then we see that event $\mathcal{E} = \mathcal{E}_0 \cap \mathcal{E}_1$ holds with probability at least $1- \theta_{n,p,q}$ with $\theta_{n,p,q} = \theta_{n,p,q}^{\prime} + 2(pq)^{1-c_0^2/2}$. To ease the technical presentation, we will condition on event $\mathcal{E}$ throughout the proof. 

In view of Propositions \ref{prop:strong2:m} and \ref{prop:strong3:w}, it holds that $\M_k = - (\wt{\v}_k\trans\wt{\v}_k)^{-1} \wt{\V}_{-k} \wt{\U}_{-k}\trans$ and 
\begin{align}\label{eqwtheor1}
    \W_k &=\I_q -  (\wt{\v}_k\trans\wt{\v}_k)^{-1} ( \I_q + \wt{\V}_{-k} \wt{\U}_{-k}\trans \wt{\U}_{-k} \wt{\A}^{-1} \wt{\V}_{-k}\trans )  \wt{\V}_{-k} \wt{\U}_{-k}\trans  \wt{\u}_k \wt{\v}_k\trans \nonumber\\ & \quad + \wt{\V}_{-k} \wt{\U}_{-k}\trans \wt{\U}_{-k} \wt{\A}^{-1}\wt{\V}_{-k}\trans.
\end{align}
From Lemma \ref{lemma:wexist} in Section \ref{Sec.B.a5}, we see that $\W_k$ is well-defined. It follows from the definition of $\wh{\v}_k$ that
	\begin{align*}
	\wh{\v}_k & ={\psi}_k(\wt{\v}_k,\wt{\boldeta}_k) = \wt{\v}_k  - \W_k\wt{\psi}_k(\wt{\v}_k,\wt{\boldeta}_k) \\
	& = \wt{\v}_k  - \W_k\wt{\psi}_k(\wt{\v}_k,\boldeta_k^*) + \W_k(\wt{\psi}_k(\wt{\v}_k,\boldeta_k^*) -\wt{\psi}_k(\wt{\v}_k,\wt{\boldeta}_k)).
 \end{align*}
 
By Propositions \ref{prop:strong1:psi}--\ref{prop:strong3:w} and the initial estimates satisfying Definition \ref{defi:sofar}, it holds that
\begin{align*}
	\wt{\v}_k  - \W_k\wt{\psi}_k(\wt{\v}_k,\boldeta_k^*) & = \v_k^* - \W_k\bepsilon_k - \W_k\bdelta_k +  \W_k  \sum_{i \neq k} \v_i^*\u_i\strans \u_k^*\\
	&\quad + \Big[\I_q - \W_k( \I_q - \M_k \wt{\u}_k\wt{\v}_k\trans + \M_k \sum_{i \neq k} \wt{\u}_i\wt{\v}_i\trans )\Big] (\wt{\v}_k-\v_k^*),
	\end{align*}
where $\bepsilon_k =  n^{-1/2}\M_k \E\wt{\v}_k - n^{-1/2}\E\trans\u_k^*$ and 
\begin{align*}
	\bdelta_k =  \M_k ( \u_k^*\wt{\v}_k\trans -  \wt{\u}_k\wt{\v}_k\trans + \sum_{i \neq k} (\wt{\u}_i\wt{\v}_i\trans -  \u_i^*{\v}_i\strans)  )  (\v_k^* - \wt{\v}_k ). 
\end{align*}
Moreover, it follows from Proposition \ref{prop:strong3:w} that 
$\I_q - \W_k( \I_q - \M_k \wt{\u}_k\wt{\v}_k\trans + \M_k \sum_{i \neq k} \wt{\u}_i\wt{\v}_i\trans) = \mathbf{0}.$

Then for each given $\a\in\mathcal{A}=\{\a\in\R^q:\norm{\a}_0\leq m,\norm{\a}_2=1\}$, we can represent $\sqrt{n}\a\trans(\wh{\v}_k-\v_k^*) $ as 
	\begin{align}\label{eqafinal}
	\sqrt{n}\a\trans(\wh{\v}_k-\v_k^*)  =  &-\sqrt{n} \a\trans\W_k\bepsilon_k -\sqrt{n} \a\trans\W_k\bdelta_k + \sqrt{n}\a\trans\W_k \sum_{i \neq k} \v_i^*\u_i\strans \u_k^* \nonumber\\
	& - \sqrt{n}\a\trans\W_k( \wt{\psi}_k(\wt{\v}_k,\wt{\boldeta}_k) - \wt{\psi}_k(\wt{\v}_k,\boldeta^*_k)).
	\end{align}
Let us further define 
\begin{align}\label{eqh1}
	h_k =
	-\a\trans \W^{*}_k \M_{k}^* \E {\v}_k^{*} 
	+ \a\trans\W^*_k\E\trans\u_k^*,
\end{align}
where 
${\M}_k^* = - ({\v}_k\strans{\v}_k^*)^{-1} {\V}_{-k}^* {\U}_{-k}\strans$ and 
\begin{align*}
{\W}_k^* &=\I_q -  ({\v}_k\strans{\v}_k^*)^{-1} ( \I_q + {\V}_{-k}^* {\U}_{-k}\strans {\U}_{-k}^* {\A}^{*-1} {\V}_{-k}\strans )  {\V}_{-k}^* {\U}_{-k}\strans  {\u}_k^* {\v}_k\strans \\
&\quad + {\V}_{-k}^* {\U}_{-k}\strans {\U}_{-k}^* {\A}^{*-1}{\V}_{-k}\strans.
\end{align*}
We see from Lemma \ref{lemma:wexist} in Section \ref{Sec.B.a5} that ${\W}_k^*$ is well-defined.

The four lemmas below provide the upper bounds for the four terms given in \eqref{eqafinal}. Proofs of Lemmas \ref{lemm:taylorrank2:2}--\ref{lemma:1rk3} are presented in Sections \ref{new:sec:b1}--\ref{new.Sec.B.8}.

\begin{lemma}\label{lemm:taylorrank2:2}
	Assume that Conditions \ref{con1:error}--\ref{con4:strong} hold and $\wt{\C}$ satisfies Definition \ref{defi:sofar}. Then for an arbitrary
	$\a\in\mathcal{A}=\{\a\in\R^q:\norm{\a}_0\leq m,\norm{\a}_2=1\}$, with probability at least
	$1- \theta_{n,p,q}$ we have that 
	\begin{align*}
		& |\a\trans\W_k (\wt{\psi}_k(\wt{\v}_k,\wt{\boldeta}_k) - \wt{\psi}_k(\wt{\v}_k,\boldeta_k^*) )| \\[5pt]
		& \leq   c m^{1/2}\max\{(r^*+s_u+s_v)^{1/2}, \eta_n^2\}(r^*+s_u+s_v)\eta_n^2\{n^{-1}\log(pq)\},
	\end{align*}
	where $\theta_{n,p,q}$ is given in \eqref{thetapro} and $c$ is some positive constant.
\end{lemma}

\begin{lemma}\label{lemma:1rk2}
    	Assume that Conditions \ref{con2:re}--\ref{con4:strong} hold and $\wt{\C}$ satisfies Definition \ref{defi:sofar}. Then for any $\a\in\R^q$ satisfying $\norm{\a}_2 =1$, with probability at least
	$1- \theta_{n,p,q}$ we have that 
	\begin{align*}
		|\a\trans{\W_k} \bdelta_k | \leq c  (r^*+s_u+s_v)\eta_n^4\left\{n^{-1}\log(pq)\right\},
	\end{align*}
	where $\theta_{n,p,q}$ is given in \eqref{thetapro} and $c$ is some positive constant.
\end{lemma}

\begin{lemma}\label{lemma:1rk23}
	Assume that Conditions \ref{con2:re}--\ref{con4:strong} hold and $\wt{\C}$ satisfies Definition \ref{defi:sofar}.  For any $\a\in\R^q$ satisfying $\norm{\a}_2 =1$,
	with probability at least
	$	1- \theta_{n,p,q}$ for $\theta_{n,p,q}$ given in \eqref{thetapro}, it holds that
	\begin{align*}
		|\a\trans\W_k \sum_{i \neq k} \v_i^*\u_i\strans \u_k^* | = o( n^{-1/2}).
	\end{align*}
\end{lemma}

\begin{lemma}\label{lemma:1rk3}
	Assume that Conditions \ref{con1:error}--\ref{con4:strong} hold and $\wt{\C}$ satisfies Definition \ref{defi:sofar}. Then for any $\a\in\R^q$ satisfying $\norm{\a}_2 =1$,
	with probability at least
	$	1- \theta_{n,p,q}$ it holds that
	\begin{align*}
		\abs{-\a\trans{\W_k}\bepsilon_k - h_k / \sqrt{n}} \leq c (r^*+s_u + s_v)^{3/2}\eta_n^2\{ n^{-1}\log(pq)\},
	\end{align*}
	where $\theta_{n,p,q}$ is given in \eqref{thetapro} and $c$ is some positive constant.
\end{lemma}

Combining \eqref{eqafinal} and Lemmas \ref{lemm:taylorrank2:2}--\ref{lemma:1rk3}, we can obtain that 
\begin{align}
	\sqrt{n}\a\trans(\wh{\v}_k-\v_k^*)  =  h_k + t_k, \nonumber
	\end{align}
where $t_k = O\left[m^{1/2}\{ (r^*+s_u+s_v)^{1/2}, \eta_n^2 \}(r^*+s_u+s_v)\eta_n^2\log(pq)/\sqrt{n}\right]$.

Finally, let us derive the distribution of $h_k$.
For ease of presentation, denote by 
\begin{align*}
 \balpha_1 =-\M_{k}\strans \W_k\strans\a, ~ \bbeta_1  =  \v_k^*, \ \balpha_2 = \u_k^*, ~ \bbeta_2 =  \W_k\strans\a,
\end{align*}
where the four terms above are independent of $\E$.
Then $h_k$ can be represented as
\begin{align}\label{eqh2}
h_k & = \balpha_1\trans\E\bbeta_1 + \balpha_2\trans\E\bbeta_2 = (\balpha_1\otimes\bbeta_1)\trans\vect{\E} + (\balpha_2\otimes\bbeta_2)\trans\vect{\E},
\end{align}
where $\vect{\E}\in\R^{nq}$ stands for the vectorization of matrix $\E$.

Under Condition \ref{con1:error} that $\E\sim\N(\0,\I_n\otimes\bSigma_e)$, we see that $h_k$ is normally distributed.
In addition, it holds that $\mathbb{E}(h_k|\X) = 0$ and variance
\begin{align}\label{eqh3}
\text{var}(h_k|\X)
& = (\balpha_1\otimes\bbeta_1)\trans(\I_n\otimes\bSigma_e)(\balpha_1\otimes\bbeta_1)
+ (\balpha_2\otimes\bbeta_2)\trans(\I_n\otimes\bSigma_e)(\balpha_2\otimes\bbeta_2)   \nonumber          \\
& \quad\quad +2(\balpha_1\otimes\bbeta_1)\trans(\I_n\otimes\bSigma_e)(\balpha_2\otimes\bbeta_2),
\end{align}
which further leads to 
\begin{align}\label{eqh4}
\text{var}(h_k|\X)
& = \balpha_1\trans\I_n\balpha_1\bbeta_1\trans\bSigma_e\bbeta_1
+ \balpha_2\trans\I_n\balpha_2\bbeta_2\trans\bSigma_e\bbeta_2
+ 2\balpha_1\trans\I_n\balpha_2\bbeta_1\trans\bSigma_e\bbeta_2                    \nonumber      \\
& = \u_k\strans\u_k^* \cdot \a\trans\W_k^*\bSigma_e\W_k\strans\a +\v_k\strans\bSigma_e\v_k^* \cdot \a\trans\W_k^*\M_{k}^* \M_{k}\strans\W_k\strans\a  \nonumber\\
& \quad - 2\a\trans\W_k^*\M_{k}^*  \u_k^* \v_k\strans\bSigma_e\W_k\strans\a \nonumber \\
&=  \a\trans\W_k^*( \bSigma_e + \v_k\strans\bSigma_e\v_k^* \M_{k}^* \M_{k}\strans - 2\M_{k}^*  \u_k^* \v_k\strans\bSigma_e)\W_k\strans\a.
\end{align}
This completes the proof of Theorem \ref{theo:strong:vk}.

\subsection{Proof of Theorem \ref{theo:strong:var}}\label{sec:var1}
Observe that 
\begin{align*}
&  \wt{\nu}^2_k = {\a}\trans\W_{k}(\bSigma_e +  \wt{\v}_k\trans\bSigma_e\wt{\v}_k \M_{k}\M_{k}\trans - 2\M_{k}\wt{\u}_k \wt{\v}_k\trans\bSigma_e)
    \W_k\trans{\a}, \\
		&\nu_k^2 = \a\trans\W_k^*( \bSigma_e + \v_k\strans\bSigma_e\v_k^* \M_{k}^* \M_{k}\strans - 2\M_{k}^*  \u_k^* \v_k\strans\bSigma_e)\W_k\strans\a.
\end{align*}
Let us define 
$ \wt{\nu}^2_k =\varphi_1 + \varphi_2  - 2\varphi_3$ and $ {\nu}^2_k =\varphi_1^* + \varphi_2^*  - 2\varphi_3^*$, where
\begin{align*}
	&\varphi_1 =  {\a}\trans\W_{k}\bSigma_e\W_k\trans{\a},
	\ \ \varphi_1^* =  \a\trans\W_k^* \bSigma_e\W_k\strans\a, \\
	&\varphi_2 =   \wt{\v}_k\trans\bSigma_e\wt{\v}_k {\a}\trans\W_k\M_{k}\M_{k}\trans\W_k\trans{\a}, \ ~ ~ \varphi_2^* = \v_k\strans\bSigma_e\v_k^* \a\trans\W_k^*  \M_{k}^* \M_{k}\strans \W_k\strans\a, \\
	&\varphi_3 =  {\a}\trans\W_k\M_{k}\wt{\u}_k \wt{\v}_k\trans\bSigma_e\W_k\trans{\a}, \ \
	\varphi_3^* =   \a\trans\W_k^*  \M_{k}^*  \u_k^* \v_k\strans\bSigma_e\W_k\strans\a.
\end{align*}
It follows that 
\begin{align}\label{phiij20}
	|  \wt{\nu}^2_k -  {\nu}^2_k | &\leq  | \varphi_1 - \varphi_1^*| +  | \varphi_2 - \varphi_2^*| +  2| \varphi_3 - \varphi_3^*|   = A_1 + A_2 + 2 A_3.
\end{align}
We will bound the three terms on the right-hand side of (\ref{phiij20}) above separately based on Condition \ref{con4:strong} that the nonzero squared singular values $d^{*2}_{i}$ are at the constant level.

\noindent \textbf{(1). The upper bound on $A_1$}.
It holds that
\begin{align*}
    A_1 &= |{\a}\trans\W_{k}\bSigma_e\W_k\trans{\a} -  \a\trans\W_k^* \bSigma_e\W_k\strans\a| \\[5pt]
    &\leq |{\a}\trans\W_{k}\bSigma_e(\W_k\trans{\a} - \W_k\strans\a)| + 
|({\a}\trans\W_{k} -  \a\trans\W_k^* )\bSigma_e\W_k\strans\a| \\[5pt]
&\leq \norm{{\a}\trans\W_{k}}_2 \norm{\bSigma_e }_2 \norm{\W_k\trans{\a} - \W_k\strans\a }_2 + \norm{ {\a}\trans\W_{k} -  \a\trans\W_k^*}_2 \norm{ \bSigma_e }_2 \norm{\W_k\strans\a}_2.
\end{align*}
By Condition \ref{con1:error} and Lemma \ref{lemma:wr2bound} in Section \ref{new.Sec.B.5}, it can be easily seen that
\begin{align*}
    \norm{ \bSigma_e }_2 \leq c, \ \ \norm{{\a}\trans\W_{k}}_2 \leq c, \ \ \norm{ {\a}\trans\W_{k} -  \a\trans\W_k^*}_2 \leq c \gamma_n,
\end{align*}
which yield that 
\begin{align}
    A_1 \leq c \gamma_n. \label{a11}
\end{align}

\medskip

\noindent \textbf{(2). The upper bound on $A_2$}. Recall that 
\[ A_2 = | \wt{\v}_k\trans\bSigma_e\wt{\v}_k {\a}\trans\W_k\M_{k}\M_{k}\trans\W_k\trans{\a} - \v_k\strans\bSigma_e\v_k^* \a\trans\W_k^*  \M_{k}^* \M_{k}\strans \W_k\strans\a|. \]
From Condition \ref{con1:error} that $\norm{\bSigma_e }_2 \leq c$ and Lemma \ref{1} in Section \ref{new.Sec.B.3}, we have that $|  \wt{\v}_k\trans\bSigma_e\wt{\v}_k | \leq  \|  \wt{\v}_k\trans\|_2 \|\bSigma_e\|_2 \|\wt{\v}_k \|_2  \leq c$ and
\begin{align*}
	| \wt{\v}_k\trans\bSigma_e\wt{\v}_k - \v_k\strans\bSigma_e\v_k^* | &\leq   | \wt{\v}_k\trans\bSigma_e(\wt{\v}_k - \v_k^*) | +  | ( \wt{\v}_k\trans - \v_k\strans)\bSigma_e\v_k^* | \\
	&\leq  \| \wt{\v}_k\|_2 \|\bSigma_e\|_2 \|\wt{\v}_k - \v_k^* \|_2 +  \|  \wt{\v}_k - \v_k^*\|_2 \|\bSigma_e\|_2 \|\v_k^* \|_2 \\
	& \leq   c \gamma_n.
\end{align*}
Then an application of the triangle inequality leads to  
\begin{align}
	A_2 &\leq     | \wt{\v}_k\trans\bSigma_e\wt{\v}_k | |{\a}\trans\W_k\M_{k}\M_{k}\trans\W_k\trans{\a} -   \a\trans\W_k^*\M_{k}^* \M_{k}\strans\W_k\strans\a| \nonumber \\[5pt]
	&\quad + | \wt{\v}_k\trans\bSigma_e\wt{\v}_k - \v_k\strans\bSigma_e\v_k^* | |  \a\trans\W_k^*\M_{k}^* \M_{k}\strans\W_k\strans\a| \nonumber\\[5pt]
	&\leq c  |{\a}\trans\W_k\M_{k}\M_{k}\trans\W_k\trans{\a} -   \a\trans\W_k^*\M_{k}^* \M_{k}\strans\W_k\strans\a| \nonumber\\[5pt]
	&\quad +    c \gamma_n |  \a\trans\W_k^*\M_{k}^* \M_{k}\strans\W_k\strans\a|. \label{eqa22va}
\end{align}

Note that $\M_k = - (\wt{\v}_k\trans\wt{\v}_k)^{-1} \wt{\V}_{-k} \wt{\U}_{-k}\trans,
{\M}_k^* = - ({\v}_k\strans{\v}_k^*)^{-1} {\V}_{-k}^* {\U}_{-k}\strans.$
It follows from Lemma \ref{1} that
\begin{align}
    &\norm{\M_k}_2 \leq |\wt{\v}_k\trans\wt{\v}_k|^{-1} \norm{\wt{\V}_{-k}}_2 \norm{\wt{\U}_{-k}\trans}_2 \leq c, \label{m1wt} \\
    &\norm{{\M}_k^*}_2 \leq |{\v}_k\strans{\v}_k^*|^{-1} \norm{{\V}_{-k}^*}_2 \norm{{\U}_{-k}\strans}_2 \leq c. \label{m1true}
\end{align}
Further, by Lemma \ref{1} we can show that 
\begin{align}
    \norm{\M_k - &{\M}_k^*}_2 \leq  |\wt{\v}_k\trans\wt{\v}_k|^{-1} \norm{\wt{\V}_{-k} \wt{\U}_{-k}\trans - {\V}_{-k}^*{\U}_{-k}\strans }_2 \nonumber\\
    & + | (\wt{\v}_k\trans\wt{\v}_k)^{-1} -  ({\v}_k\strans{\v}_k^*)^{-1} | \norm{{\V}_{-k}^*}_2 \norm{{\U}_{-k}\strans}_2 \leq c \gamma_n. \label{m1diff}
\end{align}

We now bound the terms on the right-hand side of \eqref{eqa22va}.  It follows from Lemma \ref{lemma:wr2bound} in Section \ref{new.Sec.B.5}, \eqref{m1true}, and \eqref{m1diff} that 
\begin{align}
     |  \a\trans\W_k^*\M_{k}^* \M_{k}\strans\W_k\strans\a| & \leq  \norm{\a\trans\W_k^*}_2  \norm{\M_{k}^*}_2 \norm{\M_{k}\strans}_2 \norm{\W_k\strans\a}_2  \leq c, \label{m11111}\\[5pt]
     \|{\a}\trans\W_k\M_{k} -   \a\trans\W_k^*\M_{k}^*\|_2 & \leq   \|{\a}\trans\W_k(\M_{k} -   \M_{k}^*)\|_2 +   \|{\a}\trans(\W_k -   \W_k^*)\M_{k}^*\|_2 \nonumber\\
     &\leq  c \gamma_n. \label{m111112}
\end{align}
Moreover, by Lemma \ref{lemma:wr2bound} in Section \ref{new.Sec.B.5}, \eqref{m1wt}, \eqref{m1true}, and \eqref{m111112}, we can deduce that
\begin{equation}\label{dasklc}
    \begin{split}
        &|{\a}\trans\W_k\M_{k}\M_{k}\trans\W_k\trans\wt{\a} -   \a\trans\W_k^*\M_{k}^* \M_{k}\strans\W_k\strans\a| \\[5pt]
&\leq |{\a}\trans\W_k\M_{k}(\M_{k}\trans\W_k\trans{\a} -   \M_{k}\strans\W_k\strans\a)| \\[5pt]
&\ \ \ \ + |({\a}\trans\W_k\M_{k} -   \a\trans\W_k^*\M_{k}^*) \M_{k}\strans\W_k\strans\a| \\[5pt]
&\leq \|{\a}\trans\W_k \|_2 \|\M_{k}\|_2 \| \M_{k}\trans\W_k\trans{\a} -   \M_{k}\strans\W_k\strans\a\|_2 \\[5pt]
&\ \ \ \ + \|{\a}\trans\W_k\M_{k} -   \a\trans\W_k^*\M_{k}^*\|_2 \| \M_{k}\strans\|_2 \|\W_k\strans\a\|_2 \\[5pt]
&\leq c \| \M_{k}\trans\W_k\trans{\a} -   \M_{k}\strans\W_k\strans\a\|_2 + c \|{\a}\trans\W_k\M_{k} -   \a\trans\W_k^*\M_{k}^*\|_2 \\[5pt]
&\leq c \gamma_n.
    \end{split}
\end{equation}

Thus, combining \eqref{eqa22va}, \eqref{m11111}, and \eqref{dasklc} yields that
\begin{align}
    A_2 \leq c\gamma_n. \label{a22}
\end{align}
 
\medskip

\noindent \textbf{(3). The upper bound on $A_3$}. For term $A_3$, we have that 
\begin{align*}
	A_3 & =|{\a}\trans\W_k\M_{k}\wt{\u}_k \wt{\v}_k\trans\bSigma_e\W_k\trans{\a} -  \a\trans\W_k^*  \M_{k}^*  \u_k^* \v_k\strans\bSigma_e\W_k\strans\a| \\[5pt]
 &\leq   \| {\a}\trans\W_k\|_2 \|\M_{k} \|_2 \|\wt{\u}_k\wt{\v}_k\trans\bSigma_e\W_k\trans{\a} - \u_k^*\v_k\strans\bSigma_e\W_k\strans\a \|_2 \\[5pt]
	& \quad +   \| {\a}\trans\W_k\M_{k} -
	\a\trans\W_k^* \M_{k}^*\|_2 \| \u_k^* \v_k\strans\bSigma_e\W_k\strans\a\|_2  \\[5pt]
	& \leq c \|\wt{\u}_k\wt{\v}_k\trans\bSigma_e\W_k\trans{\a} - \u_k^*\v_k\strans\bSigma_e\W_k\strans\a\|_2 + c \gamma_n \| \u_k^* \v_k\strans\bSigma_e\W_k\strans\a\|_2,
\end{align*}
where the last step above has used Lemma \ref{lemma:wr2bound} in Section \ref{new.Sec.B.5}, \eqref{m1wt}, and \eqref{m111112}. Further, in light of Lemmas \ref{1} and \ref{lemma:wr2bound}, we can deduce that 
\begin{align*}
&\| \u_k^* \v_k\strans\bSigma_e\W_k\strans\a\|_2 \leq  \norm{\u_k^*}_2 \norm{\v_k\strans}_2 \norm{\bSigma_e}_2 \norm{\W_k\strans\a}_2 \leq c, \\[5pt]
    &\|\wt{\u}_k\wt{\v}_k\trans\bSigma_e\W_k\trans{\a} - \u_k^*\v_k\strans\bSigma_e\W_k\strans\a\|_2 \\[5pt]
    &\quad \leq \|\wt{\u}_k\wt{\v}_k\trans\bSigma_e(\W_k\trans - \W_k\strans)\a\|_2 + \|(\wt{\u}_k\wt{\v}_k\trans - \u_k^*\v_k\strans)\bSigma_e\W_k\strans\a\|_2 \\[5pt]
    &\quad \leq \norm{\wt{\u}_k}_2 \norm{\wt{\v}_k}_2 \norm{\bSigma_e}_2 \norm{(\W_k\trans - \W_k\strans)\a}_2 + \norm{\wt{\u}_k\wt{\v}_k\trans - \u_k^*\v_k\strans}_2 \norm{\bSigma_e}_2 \norm{\W_k\strans\a}_2 \\[5pt]
    &\quad \leq c \gamma_n.
\end{align*}
Hence, we can obtain that 
\begin{align}
    A_3 \leq c \gamma_n. \label{a33}
\end{align}

Therefore, combining \eqref{phiij20}, \eqref{a11}, \eqref{a22}, and \eqref{a33} leads to 
\begin{align*}
    |  \wt{\nu}^2_k -  {\nu}^2_k | \leq c \gamma_n. 
\end{align*}
This concludes the proof of Theorem \ref{theo:strong:var}.

\subsection{Proof of Theorem \ref{theo:weak:uk}}\label{sec:weaku}

By definition, it holds that 
\begin{align*}
	\wh{\v}_k & ={\psi}_k(\wt{\v}_k,\wt{\boldeta}_k) = \wt{\v}_k  - \W_k\wt{\psi}_k(\wt{\v}_k,\wt{\boldeta}_k) \\
	& = \wt{\v}_k  - \W_k\wt{\psi}_k(\wt{\v}_k,\boldeta_k^*) + \W_k(\wt{\psi}_k(\wt{\v}_k,\boldeta_k^*) -\wt{\psi}_k(\wt{\v}_k,\wt{\boldeta}_k)). 
\end{align*}
Using Propositions \ref{prop:weak1:psi}--\ref{prop:weak3:w},
we can deduce that 
\begin{align*}
	&\wt{\v}_k - \W_k\wt{\psi}_k(\wt{\v}_k,\boldeta^*_k)  = \v_k^* - \W_k\bepsilon_k - \W_k\bdelta_k - \sum_{i \neq k}\W_k   \wt{\v}_i (\wh{\u}_i^t  - \u_i^*)\trans\u_k^*   \\
	&- \sum_{i \neq k} \W_k (\wt{\v}_i  - \v_i^*)\u_i\strans\u_k^* + \Big[\I_q - \W_k( \I_q - \M_k \wt{\u}_k \wt{\v}_k\trans + \M_k\sum_{i \neq k}\wt{\u}_i \wt{\v}_i\trans )\Big] (\wt{\v}_k-\v_k^*),
\end{align*}
where 
\begin{align}
&\M_k = - ( \wt{\v}_k\trans\wt{\v}_k)^{-1}\widetilde{\v}_k\widetilde{\u}_k\trans, \ \ \ 
	\W_k =   \I_q - 2^{-1} (\wt{\v}_k\trans\wt{\v}_k)^{-1} (  \wt{\v}_k \wt{\v}_k\trans -  \wt{\v}_k \wt{\u}_k\trans  \wt{\U}_{-k} \wt{\V}_{-k}\trans  ), \nonumber\\[5pt] 
	&\bepsilon_k =   - n^{-1/2}\E\trans\u_k^* + n^{-1/2} \M_k\E\wt{\v}_k, \label{epe:weak} \\[5pt]
	&\bdelta_k = - \M_k  ( \u_k^* \v_k\strans -\wt{\u}_k\wt{\v}_k\trans + \sum_{i \neq k} (\wt{\u}_i \wt{\v}_i\trans -\u_i^* \v_i\strans))  (\wt{\v}_k - \v_k^*). \label{delta:weakrank2v1}
\end{align}

It follows from Proposition \ref{prop:weak3:w} that $$\I_q - \W_k( \I_q - \M_k \wt{\u}_k \wt{\v}_k\trans + \M_k\sum_{i \neq k}\wt{\u}_i \wt{\v}_i\trans) = \0. $$ 
Then for each given $\a\in\mathcal{A}=\{\a\in\R^q:\norm{\a}_0\leq m,\norm{\a}_2=1\}$, we can represent $\sqrt{n}\a\trans(\wh{\v}_k-\v_k^*) $ as 
	\begin{align}\label{eqafinal3az}
	&\sqrt{n}\a\trans(\wh{\v}_k-\v_k^*)  =   - \sqrt{n}\a\trans \W_k \sum_{i \neq k}   \wt{\v}_i (\wh{\u}_i^t  - \u_i^*)\trans\u_k^* -\sqrt{n} \a\trans\W_k\bepsilon_k  -\sqrt{n} \a\trans\W_k\bdelta_k \nonumber\\
	&  - \sqrt{n}\a\trans \W_k \sum_{i \neq k}  (\wt{\v}_i  - \v_i^*)\u_i\strans\u_k^* - \sqrt{n}\a\trans\W_k( \wt{\psi}_k(\wt{\v}_k,\wt{\boldeta}_k) - \wt{\psi}_k(\wt{\v}_k,\boldeta^*_k)).
	\end{align}

Denote by $ h_k =  \a\trans{\W}_k^* (n^{-1/2} \M_k^*\E\trans\v_k^*- n^{-1/2}\E\u_k^*  )$. The five terms given on the right-hand side of \eqref{eqafinal3az} above can be bounded by the following five lemmas. Proofs of Lemmas \ref{lemma:weak:delta}--\ref{lemma:disw} are provided in Sections \ref{new.Sec.B.61}--\ref{new.Sec.B.5a9}, respectively.
 
\begin{lemma}\label{lemma:weak:delta}
Assume that Conditions \ref{con2:re}, \ref{con3:eigend}, and \ref{con:weak:orth} hold, and $\wt{\C}$ satisfies Definition \ref{defi:sofar}. For $\bdelta_k$ defined in \eqref{delta:weakrank2v1} and an arbitrary vector
	$\a\in\mathcal{A}=\{\a\in\R^q:\norm{\a}_0\leq m,\norm{\a}_2=1\}$, with probability at least
	$1- \theta_{n,p,q}$ we have that 
	\begin{align*}
		|\a\trans{\W_k} \bdelta_k | \leq c (r^*+s_u+s_v)\eta_n^4\left\{n^{-1}\log(pq)\right\} d_k^{*-1},
 	\end{align*}
	where $\theta_{n,p,q}$ is given in \eqref{thetapro} and $c$ is some positive constant.
\end{lemma}

\begin{lemma}\label{lemma:weak:con}
	Assume that Conditions \ref{con2:re}, \ref{con3:eigend}, and \ref{con:weak:orth} hold, and $\wt{\C}$ satisfies Definition \ref{defi:sofar}.  For any $\a\in\mathcal{A}=\{\a\in\R^q:\norm{\a}_0\leq m,\norm{\a}_2=1\}$,
	with probability at least
	$	1- \theta_{n,p,q}$ it holds that
\begin{align*}
	\| \sqrt{n} \a\trans \W_k \sum_{i \neq k}  (\wt{\v}_i  - \v_i^*)\u_i\strans\u_k^* \|_2 \leq c  r^* d_1^{*}  d_{r^*}^{*-2} \gamma_n,
\end{align*}
	where $\theta_{n,p,q}$ is given in \eqref{thetapro} and $c$ is some positive constant.
\end{lemma}

\begin{lemma}\label{lemm:weak:taylor}
	Assume that Conditions \ref{con2:re}, \ref{con3:eigend}, and \ref{con:weak:orth} hold, and $\wt{\C}$ satisfies Definition \ref{defi:sofar}. Then for an arbitrary vector
	$\a\in\mathcal{A}=\{\a\in\R^q:\norm{\a}_0\leq m,\norm{\a}_2=1\}$, with probability at least
	$1- \theta_{n,p,q}$ we have that 
	\begin{align*}
		& |\a\trans\W_k (\wt{\psi}_k(\wt{\v}_k,\wt{\boldeta}_k) - \wt{\psi}_k(\wt{\v}_k,\boldeta_k^*) )| \\[5pt]
		&\quad  \leq  c m^{1/2} \max\{(r^*+s_u+s_v)^{1/2}, \eta_n^2\}(r^*+s_u+s_v)\eta_n^2\{n^{-1}\log(pq)\}d_k^{*-1},
	\end{align*}
	where $\theta_{n,p,q}$ is given in \eqref{thetapro} and $c$ is some positive constant.
\end{lemma}

\begin{lemma}\label{lemma:weak:ep}
	Assume that Conditions \ref{con2:re}, \ref{con3:eigend}, and \ref{con:weak:orth} hold, and $\wt{\C}$ satisfies Definition \ref{defi:sofar}.  For any $\a\in\mathcal{A}=\{\a\in\R^q:\norm{\a}_0\leq m,\norm{\a}_2=1\}$ and $\bepsilon_k$ given in \eqref{epe:weak},
	with probability at least
	$	1- \theta_{n,p,q}$ it holds that
	\begin{align*}
		\abs{-\a\trans\W_k\bepsilon_k - h_k / \sqrt{n}} \leq  c (r^*+s_u + s_v)^{3/2}\eta_n^2\{ n^{-1}\log(pq)\} d_1^{*} d_{k^*}^{*-2},
	\end{align*}
	where $\theta_{n,p,q}$ is given in \eqref{thetapro} and $c$ is some positive constant.
\end{lemma}

\begin{lemma}\label{lemma:disw}
	Assume that Conditions \ref{con2:re}, \ref{con3:eigend}, \ref{con:weak:orth}, and \ref{con:threshold} hold, and $\wt{\C}$ satisfies Definition \ref{defi:sofar}.
	For any $\a\in\mathcal{A}=\{\a\in\R^q:\norm{\a}_0\leq m,\norm{\a}_2=1\}$,  with probability at least
	$1- \theta_{n,p,q}$ for $\theta_{n,p,q}$ given in \eqref{thetapro}, we have that for all sufficiently large $n$, 
	\begin{align*}
		\sqrt{n} \a\trans\W_k \sum_{i \neq k}   \wt{\v}_i (\wh{\u}_i^t  - \u_i^*)\trans\u_k^* =  \sum_{i \neq k} \omega_{k,i} h_i((\wh{\bSigma} \bmu_k^*)^{t_i}) +  t_k^{\prime }.
	\end{align*}
	The distribution term is
$$h_i((\wh{\bSigma} \bmu_k^*)^{t_i}) = ((\wh{\bSigma} \bmu_k^*)^{t_i})\trans \W^{*}_{u_i}(\X\trans\E\r_i^* - \M_{u_i}^{*} \E\trans\X {\bmu}_i^{*})/\sqrt{n} \sim N(0,\nu_i((\wh{\bSigma} \bmu_k^*)^{t_i})^2),$$ where the variance is given by 
\begin{align*}
	\nu_i((\wh{\bSigma} \bmu_k^*)^{t_i})^2 = &((\wh{\bSigma} \bmu_k^*)^{t_i})\trans\W_{u_i}^*(\bmu_i\strans\wh{\bSigma}\bmu_i^* \M_{u_i}^* \bSigma_e \M_{u_i}^{* T} \\ & + \r_i\strans\bSigma_e\r_i^* \wh{\bSigma} - 2\wh{\bSigma} \bmu_i^*\r_i\strans\bSigma_e\M_{u_i}^{* T}) \W_{u_i}\strans(\wh{\bSigma} \bmu_k^*)^{t_i}.
\end{align*}
The error term is
\begin{align*}
	t_k^{\prime }  = O\Big(r^* s_{u}^{1/2}   (\kappa_n \max\big\{1, d_{r^*}^{*-1}, d_{r^*}^{*-2}\big\}  +  d_1^{*}  d_{r^*}^{*-2} \gamma_n)\Big)
\end{align*}
with $\kappa_n = \max\{s_{\max}^{1/2} , (r^*+s_u+s_v)^{1/2}, \eta_n^2\} (r^*+s_u+s_v)\eta_n^2\log(pq)/\sqrt{n}.$
\end{lemma}

Combining \eqref{eqafinal3az} and Lemmas \ref{lemma:weak:delta}--\ref{lemma:weak:ep} above yields that 
\begin{align*}
	\sqrt{n}\a\trans(\wh{\v}_k-\v_k^*)  =   - \sqrt{n}\a\trans \W_k \sum_{i \neq k}   \wt{\v}_i (\wh{\u}_i^t  - \u_i^*)\trans\u_k^* + h_k + t_k^{\prime \prime },
	\end{align*}
where 
$$ t_k^{\prime \prime } = O\Big( m^{1/2} d_1^{*}    d_{r^*}^{*-2} \max\{(r^*+s_u+s_v)^{1/2}, \eta_n^2\}(r^*+s_u+s_v)\eta_n^2\log(pq) / \sqrt{n} \Big).
$$
Further, with the aid of Lemma \ref{lemma:disw}, it holds that 
\begin{align*}
	\sqrt{n}\a\trans(\wh{\v}_k-\v_k^*)  =  h_k - \sum_{i \neq k} \omega_{k,i} h_i((\wh{\bSigma} \bmu_k^*)^{t_i})  + t_k,
\end{align*}
where the error term is 
$$t_k =   t_k^{\prime} + t_k^{\prime \prime} 
= O\Big( m^{1/2} d_1^{*}    d_{r^*}^{*-2} \kappa_n + r^* s_{u}^{1/2}   \kappa_n \max\big\{1, d_{r^*}^{*-1}, d_{r^*}^{*-2}\big\}  +   r^* s_{u}^{1/2} d_1^{*}  d_{r^*}^{*-2} \gamma_n \Big).
$$

In what follows, we will analyze the distribution term $h_k - \sum_{i \neq k} \omega_{k,i} h_i((\wh{\bSigma} \bmu_k^*)^{t_i})$.
For simplicity, denote by  $h_{u_i} =  h_i((\wh{\bSigma} \bmu_k^*)^{t_i})$ and 
$$\M_{k}^* =  \M_{v_k}^* n^{-1/2} \X\trans \text{ with }
\M_{v_k} = - (\l_k\strans \wh{\bSigma}\l_k^*)^{-1/2} ( {\v}_k\strans{\v}_k^*)^{-1}{\v}_k^*{\l}_k\strans.$$
Then the distribution term is $ h_k - \sum_{i \neq k} h_{u_i}$ with 
\begin{align*}
	&h_k =  \a\trans{\W}_k^* ( \M_{v_k}^* \X\trans \E  d_k^* (\l_k\strans \wh{\bSigma}\l_k^*)^{1/2} \r_k^* -   \E\trans \X  (\l_k\strans \wh{\bSigma}\l_k^*)^{-1/2} \l_k^*  )/\sqrt{n}, \\
	&h_{u_i} = ((\wh{\bSigma} \bmu_k^*)^{t_i})\trans \W^{*}_{u_i}(\X\trans\E\r_i^* - \M_{u_i}^{*} \E\trans\X d_i^* {\l}_i^{*})/\sqrt{n}.
\end{align*}
In addition, the variance is given by 
\begin{align*}
	\nu_k^2 =  \operatorname{cov} (h_{v_k},  h_{v_k} ) - 2 \sum_{i \neq k}  \omega_{k,i} \operatorname{cov} (h_{v_k},  h_{u_i} )   +  \sum_{i \neq k} \sum_{j \neq k} \omega_{k,i} \omega_{k,j}  \operatorname{cov} (  h_{u_i}, h_{u_j}), 
\end{align*}
We will deal with the three terms on the right-hand side above.

For each given $k$ and $i = 1, \ldots, r^*$, let us define 
\begin{align*}
& \balpha_1 = \X  (\l_k\strans \wh{\bSigma}\l_k^*)^{-1/2} \l_k^*, ~ \bbeta_1  = - \W_k\strans\a / \sqrt{n}, \\
& \balpha_{3i} = \X d_i^* \l_i^*, ~ \bbeta_{3i} = -\M_{u_i}\strans \W_{u_i}\strans(\wh{\bSigma} \bmu_k^*)^{t_i} / \sqrt{n}, \\
& \balpha_2 = \X\M_{v_k}\strans\W_k\strans\a / \sqrt{n}, ~ \bbeta_2 =  d_k^* (\l_k\strans \wh{\bSigma}\l_k^*)^{1/2} \r_k^*, \\
& \balpha_{4i} = \X\W_{u_i}\strans(\wh{\bSigma} \bmu_k^*)^{t_i} / \sqrt{n}, ~ \bbeta_{4i} = \r_i^*.
\end{align*}
Observe that all the quantities above are independent of $\E$.
Then we can rewrite $h_{v_k}$ and $h_{u_i}$ as
\begin{align}\label{edsqh2}
h_k & = \balpha_1\trans\E\bbeta_1 + \balpha_2\trans\E\bbeta_2 = (\balpha_1\otimes\bbeta_1)\trans\vect{\E} + (\balpha_2\otimes\bbeta_2)\trans\vect{\E}, \nonumber \\
h_{u_i}& = \balpha_{3i}\trans\E\bbeta_{3i} + \balpha_{4i}\trans\E\bbeta_{4i} = (\balpha_{3i}\otimes\bbeta_{3i})\trans\vect{\E} + (\balpha_{4i}\otimes\bbeta_{4i})\trans\vect{\E},
\end{align}
where $\vect{\E}\in\R^{nq}$ denotes the vectorization of matrix $\E$.
Similar to \eqref{eqh3} and \eqref{eqh4}, we can show that 
\begin{align*}
		\nu_k^2 = \a\trans\W_k^*( \bSigma_e + d_k^{*2} (\l_k\strans \wh{\bSigma}\l_k^*) \r_k\strans\bSigma_e\r_k^* \M_{v_k}^* \wh{\bSigma} \M_{v_k}\strans  - 2\M_{v_k}^* \wh{\bSigma} d_k^*  \l_k^* \r_k\strans \bSigma_e)\W_k\strans\a.
\end{align*}

Moreover, we can deduce that
\begin{align*}
&\operatorname{cov} (h_{v_k},  h_{u_i} )
 = (\balpha_1\otimes\bbeta_1)\trans(\I_n\otimes\bSigma_e)(\balpha_{3i}\otimes\bbeta_{3i})
+ (\balpha_2\otimes\bbeta_2)\trans(\I_n\otimes\bSigma_e)(\balpha_{3i}\otimes\bbeta_{3i})   \nonumber          \\
& \quad\quad +(\balpha_1\otimes\bbeta_1)\trans(\I_n\otimes\bSigma_e)(\balpha_{4i}\otimes\bbeta_{4i}) + (\balpha_2\otimes\bbeta_2)\trans(\I_n\otimes\bSigma_e)(\balpha_{4i}\otimes\bbeta_{4i}) \\
& = \balpha_1\trans\I_n\balpha_{3i}\bbeta_1\trans\bSigma_e\bbeta_{3i}
+ \balpha_2\trans\I_n\balpha_{3i}\bbeta_2\trans\bSigma_e\bbeta_{3i}
+ \balpha_1\trans\I_n\balpha_{4i}\bbeta_1\trans\bSigma_e\bbeta_{4i}     + \balpha_2\trans\I_n\balpha_{4i}\bbeta_2\trans\bSigma_e\bbeta_{4i} 
\end{align*}
\begin{align*}
&= d_i^* (\l_k\strans \wh{\bSigma}\l_k^*)^{-1/2} \l_i\strans \wh{\bSigma} \l_k^* \cdot \a\trans \W_k^* \bSigma_e \M_{u_i}\strans \W_{u_i}\strans(\wh{\bSigma} \bmu_k^*)^{t_i} \\
&\quad -   \a\trans \W_k^* \M_{v_k}^* \wh{\bSigma}  d_i^* \l_i^* \cdot d_k^* (\l_k\strans \wh{\bSigma}\l_k^*)^{1/2} \r_k\strans \bSigma_e \M_{u_i}\strans \W_{u_i}\strans(\wh{\bSigma} \bmu_k^*)^{t_i} \\
&\quad  - ((\wh{\bSigma} \bmu_k^*)^{t_i})\trans \W_{u_i}^* \wh{\bSigma} (\l_k\strans \wh{\bSigma}\l_k^*)^{-1/2} \l_k^* \cdot \r_i\strans \bSigma_e \W_k\strans\a \\
&\quad + \a\trans\W_k^* \M_{v_k}^* \wh{\bSigma} \W_{u_i}\strans(\wh{\bSigma} \bmu_k^*)^{t_i} \cdot d_k^* (\l_k\strans \wh{\bSigma}\l_k^*)^{1/2} \r_k\strans \bSigma_e \r_i^*.
\end{align*}

Similarly, it holds that 
\begin{align*}
    &\operatorname{cov} (h_{u_i},  h_{u_j} )
     \\
	 &= (\balpha_{3i}\otimes\bbeta_{3i})\trans(\I_n\otimes\bSigma_e)(\balpha_{3j}\otimes\bbeta_{3j})
    + (\balpha_{4i}\otimes\bbeta_{4i})\trans(\I_n\otimes\bSigma_e)(\balpha_{3j}\otimes\bbeta_{3j})   \nonumber          \\
    & \quad +(\balpha_{3i}\otimes\bbeta_{3i})\trans(\I_n\otimes\bSigma_e)(\balpha_{4j}\otimes\bbeta_{4j}) + (\balpha_{4i}\otimes\bbeta_{4i})\trans(\I_n\otimes\bSigma_e)(\balpha_{4j}\otimes\bbeta_{4j}) \\
    & = \balpha_{3i}\trans\I_n\balpha_{3j}\bbeta_{3i}\trans\bSigma_e\bbeta_{3j}
    + \balpha_{4i}\trans\I_n\balpha_{3j}\bbeta_{4i}\trans\bSigma_e\bbeta_{3j} \\ &
    \quad + \balpha_{3i}\trans\I_n\balpha_{4j}\bbeta_{3i}\trans\bSigma_e\bbeta_{4j}     + \balpha_{4i}\trans\I_n\balpha_{4j}\bbeta_{4i}\trans\bSigma_e\bbeta_{4j} \\
    &=     d_i^* d_j^* \l_i\strans \wh{\bSigma} \l_j^* \cdot ((\wh{\bSigma} \bmu_k^*)^{t_i})\trans \W_{u_i}^* \M_{u_i}^* \bSigma_e \M_{u_j}\strans \W_{u_j}\strans(\wh{\bSigma} \bmu_k^*)^{t_j} \\ 
    &\quad -   ((\wh{\bSigma} \bmu_k^*)^{t_i})\trans \W_{u_i}^* \wh{\bSigma}  d_j^* \l_j^* \cdot \r_i\strans \bSigma_e \M_{u_j}\strans \W_{u_j}\strans(\wh{\bSigma} \bmu_k^*)^{t_j} \\
    &\quad  - ((\wh{\bSigma} \bmu_k^*)^{t_j})\trans \W_{u_j}^* \wh{\bSigma} d_i^* \l_i^* \cdot \r_j\strans \bSigma_e 
    \M_{u_i}\strans \W_{u_i}\strans(\wh{\bSigma} \bmu_k^*)^{t_i} \\ 
    & \quad + ((\wh{\bSigma} \bmu_k^*)^{t_i})\trans\W_{u_i}^*  \wh{\bSigma} \W_{u_j}\strans(\wh{\bSigma} \bmu_k^*)^{t_j} \cdot  \r_i\strans \bSigma_e \r_j^*.
\end{align*}
This completes the proof of Theorem \ref{theo:weak:uk}.

\subsection{Proof of Theorem \ref{theo:weak:var}}\label{sec:weakvar}

To bound the estimated variance, note that 
\begin{align}\label{vareq3}
    |\wt{\nu}_k^2 - \nu_k^2| &\leq| \wt{\operatorname{cov}} (\wt{h}_k,  \wt{h}_{k} ) -  {\operatorname{cov}} (h_k,  h_{k} )| +  2 \sum_{i \neq k} | \wt{\omega}_{k,i} \wt{\operatorname{cov}} (\wt{h}_k,  \wt{h}_{u_i} ) - \omega_{k,i} {\operatorname{cov}} (h_k,  h_{u_i} )| \nonumber \\ 
    &+   \sum_{i \neq k} \sum_{j \neq k}| \wt{\omega}_{k,i} \wt{\omega}_{k,j}  \wt{\operatorname{cov}} (  \wt{h}_{u_i}, \wt{h}_{u_j})  - \omega_{k,i} \omega_{k,j}  {\operatorname{cov}} (  h_{u_i}, h_{u_j})  |,
\end{align}
where $\wt{\operatorname{cov}}(\cdot, \cdot)$ represents the covariance after plugging in the SOFAR initial estimates. 
The proof will be divided into three parts.

\noindent\textbf{(1). The upper bound on $| \wt{\operatorname{cov}} (\wt{h}_k,  \wt{h}_{k} ) -  {\operatorname{cov}} (h_k,  h_{k} )|$}. Notice that 
\begin{align*}
     &{\operatorname{cov}} (h_k,  h_{k} ) \\
	 &= \a\trans\W_k^*( \bSigma_e + d_k^{*2} (\l_k\strans \wh{\bSigma}\l_k^*) \r_k\strans\bSigma_e\r_k^* \M_{v_k}^* \wh{\bSigma} \M_{v_k}\strans  - 2\M_{v_k}^* \wh{\bSigma} d_k^*  \l_k^* \r_k\strans \bSigma_e)\W_k\strans\a, \\[5pt] 
     &\wt{\operatorname{cov}} (\wt{h}_k,  \wt{h}_{k} ) = {\a}\trans\W_{k}(\bSigma_e +   \wt{d}_k^{2} (\wt{\l}_k\trans \wh{\bSigma}\wt{\l}_k) \wt{\r}_k\trans\bSigma_e\wt{\r}_k\M_{v_k}\wh{\bSigma}\M_{v_k}\trans - 2\M_{v_k}\wt{d}_k\wt{\l}_k \wt{\r}_k\trans\bSigma_e)
    \W_k\trans{\a}.
\end{align*}
Let us define 
\begin{align*}
	&\varphi_{11} =  {\a}\trans\W_{k}\bSigma_e\W_k\trans{\a},
	\ \ \varphi_{11}^* =  \a\trans\W_k^* \bSigma_e\W_k\strans\a, \\
	&\varphi_{12} =    \wt{d}_k^{2} (\wt{\l}_k\trans \wh{\bSigma}\wt{\l}_k) \wt{\r}_k\trans\bSigma_e\wt{\r}_k \cdot {\a}\trans\W_{k} \M_{v_k}\wh{\bSigma}\M_{v_k}\trans,  \\
    & \varphi_{12}^* = d_k^{*2} (\l_k\strans \wh{\bSigma}\l_k^*) \r_k\strans\bSigma_e\r_k^* \cdot \a\trans\W_k^*  \M_{v_k}^* \wh{\bSigma} \M_{v_k}\strans , \\
	&\varphi_{13} =  {\a}\trans\W_k\M_{v_k}\wt{d}_k\wt{\l}_k \wt{\r}_k\trans\bSigma_e
    \W_k\trans{\a}, \ \
	\varphi_{13}^* =   \a\trans\W_k^*  \M_{v_k}^* \wh{\bSigma} d_k^*  \l_k^* \r_k\strans \bSigma_e\W_k\strans\a.
\end{align*}
It is easy to see that
\begin{align}\label{phiij201}
	| \wt{\operatorname{cov}} (\wt{h}_k,  \wt{h}_{k} ) - {\operatorname{cov}} (h_k,  h_{k} ) | &\leq  | \varphi_{11} - \varphi_{11}^*| +  | \varphi_{12} - \varphi_{12}^*| +  2| \varphi_{13} - \varphi_{13}^*|  \nonumber \\
	&: = A_{11} + A_{12} + 2 A_{13}.
\end{align}
Similar to the proof of Theorem \ref{theo:strong:var} in Section \ref{sec:var1}, we will bound the three terms on the right-hand side above separately.

First, for term $ A_{11}$, it holds that
\begin{align*}
     A_{11} &= |{\a}\trans\W_{k}\bSigma_e\W_k\trans{\a} -  \a\trans\W_k^* \bSigma_e\W_k\strans\a| \\[5pt]
    &\leq |{\a}\trans\W_{k}\bSigma_e(\W_k\trans{\a} - \W_k\strans\a)| + 
|({\a}\trans\W_{k}\bSigma_e -  \a\trans\W_k^* \bSigma_e)\W_k\strans\a| \\[5pt]
&\leq \norm{{\a}\trans\W_{k}}_2 \norm{\bSigma_e }_2 \norm{\W_k\trans{\a} - \W_k\strans\a }_2 + \norm{ {\a}\trans\W_{k} -  \a\trans\W_k^*}_2 \norm{ \bSigma_e }_2 \norm{\W_k\strans\a}_2.
\end{align*}
Using Condition \ref{con1:error} and Lemma \ref{lemma:weak:wbound} in Section \ref{new.Sec.B.59}, we can show that 
\begin{align*}
    \norm{ \bSigma_e }_2 \leq c, \ \ \norm{{\a}\trans\W_{k}}_2 \leq c, \ \ \norm{ {\a}\trans\W_{k} -  \a\trans\W_k^*}_2 \leq c \gamma_n d_1^* d_k^{*-2}.
\end{align*}
Hence, it follows that 
\begin{align}
     A_{11} \leq c \gamma_n d_1^* d_k^{*-2}. \label{a111}
\end{align}

Next, we provide the upper bound of term $A_{22}$. In light of the definitions of $\wt{\v}_k$, ${\v}_k^*$, $\M_{v_k}$, and ${\M}_{v_k}^*$, we can write term $ A_{12}$ as
\[ A_{12} = | \wt{\v}_k\trans\bSigma_e\wt{\v}_k {\a}\trans\W_k\M_{k}\M_{k}\trans\W_k\trans{\a} - \v_k\strans\bSigma_e\v_k^* \a\trans\W_k^*  \M_{k}^* \M_{k}\strans \W_k\strans\a|. \]
It follows from Condition \ref{con1:error} that $\norm{\bSigma_e }_2 \leq c$ and Lemma \ref{1} in Section \ref{new.Sec.B.3} that $|  \wt{\v}_k\trans\bSigma_e\wt{\v}_k | \leq  \|  \wt{\v}_k\trans\|_2 \|\bSigma_e\|_2 \|\wt{\v}_k \|_2  \leq c d_k^{*2}$ and
\begin{align*}
	| \wt{\v}_k\trans\bSigma_e\wt{\v}_k - \v_k\strans\bSigma_e\v_k^* | &\leq   | \wt{\v}_k\trans\bSigma_e(\wt{\v}_k - \v_k^*) | +  | ( \wt{\v}_k\trans - \v_k\strans)\bSigma_e\v_k^* | \\
	&\leq  \| \wt{\v}_k\|_2 \|\bSigma_e\|_2 \|\wt{\v}_k - \v_k^* \|_2 +  \|  \wt{\v}_k - \v_k^*\|_2 \|\bSigma_e\|_2 \|\v_k^* \|_2 \\
	& \leq   c \gamma_n d_k^*.
\end{align*}
Then an application of the triangle inequality gives that 
\begin{align}
	A_{12} &\leq     | \wt{\v}_k\trans\bSigma_e\wt{\v}_k | |{\a}\trans\W_k\M_{k}\M_{k}\trans\W_k\trans{\a} -   \a\trans\W_k^*\M_{k}^* \M_{k}\strans\W_k\strans\a| \nonumber \\[5pt]
	&\quad + | \wt{\v}_k\trans\bSigma_e\wt{\v}_k - \v_k\strans\bSigma_e\v_k^* | |  \a\trans\W_k^*\M_{k}^* \M_{k}\strans\W_k\strans\a| \nonumber\\[5pt]
	&\leq c  d_k^{*2} |{\a}\trans\W_k\M_{k}\M_{k}\trans\W_k\trans{\a} -   \a\trans\W_k^*\M_{k}^* \M_{k}\strans\W_k\strans\a| \nonumber\\[5pt]
	&\quad +    |  \a\trans\W_k^*\M_{k}^* \M_{k}\strans\W_k\strans\a| c  \gamma_n  d_k^{*}. \label{eqa122va}
\end{align}

For $\M_k$ and $\M_k^*$, it holds that
\begin{align}
 & \norm{\M_k^*}_2 =  \| ( {\v}_k\strans{\v}_k^*)^{-1}{\v}_k^*{\u}_k\strans\|_2 \leq |{\v}_k\strans{\v}_k^*|^{-1} \| {\v}_k^* \|_2 \|{\u}_k\strans\|_2 \leq c d_k^{*-1}, \label{m1} \\[5pt]
&\|\M_k\|_2 = \| ( \wt{\v}_k\trans\wt{\v}_k)^{-1}\widetilde{\v}_k\widetilde{\u}_k\trans\|_2 \leq |\wt{\v}_k\trans\wt{\v}_k|^{-1} \| \widetilde{\v}_k \|_2 \|\widetilde{\u}_k\trans\|_2 \leq c d_k^{*-1}. \label{m2}
\end{align}
In addition, we have that 
\begin{align}\label{m3}
	&\norm{ \M_k - \M_k^* }_2 = \| ( \wt{\v}_k\trans\wt{\v}_k)^{-1}\widetilde{\v}_k\widetilde{\u}_k\trans -  ( {\v}_k\strans{\v}^*_k)^{-1}{\v}^*_k{\u}_k\strans\|_2 \nonumber \\[5pt]
	&\leq | ( \wt{\v}_k\trans\wt{\v}_k)^{-1}| \| \widetilde{\v}_k\widetilde{\u}_k\trans - {\v}^*_k{\u}_k\strans\|_2 + | ( \wt{\v}_k\trans\wt{\v}_k)^{-1}  -  ( {\v}_k\strans{\v}^*_k)^{-1} | \| {\v}^*_k\|_2 \|{\u}_k\strans\|_2 \nonumber \\[5pt]
	&\leq c \gamma_n d_k^{*-2}.
\end{align}

Then by resorting to Lemma \ref{lemma:weak:wbound} and \eqref{m1}, we have that 
\begin{align*}
     |  \a\trans\W_k^*\M_{k}^* \M_{k}\strans\W_k\strans\a| \leq  \norm{\a\trans\W_k^*}_2  \norm{\M_{k}^*}_2 \norm{\M_{k}\strans}_2 \norm{\W_k\strans\a}_2  \leq c.
\end{align*}
From \eqref{m3}, it holds that 
\begin{align}\label{awmvv}
     \|{\a}\trans\W_k\M_{k} -   \a\trans\W_k^*  \M_{k}^*\|_2 
     &\leq   \|{\a}\trans\W_k(\M_{k} -   \M_{k}^*)\|_2 +   \|{\a}\trans(\W_k -   \W_k^*)\M_{k}^*\|_2 \nonumber\\
     &\leq  c \gamma_n d_k^{*-3} d_1^*.
\end{align}
Then we can deduce that 
\begin{align*}
&|{\a}\trans\W_k\M_{k}\M_{k}\trans\W_k\trans\wt{\a} -   \a\trans\W_k^*\M_{k}^* \M_{k}\strans\W_k\strans\a| \\
&\leq |{\a}\trans\W_k\M_{k}(\M_{k}\trans\W_k\trans{\a} -   \M_{k}\strans\W_k\strans\a)| + |({\a}\trans\W_k\M_{k} -   \a\trans\W_k^*\M_{k}^*) \M_{k}\strans\W_k\strans\a| \\
&\leq \|{\a}\trans\W_k \|_2 \|\M_{k}\|_2 \| \M_{k}\trans\W_k\trans{\a} -   \M_{k}\strans\W_k\strans\a\|_2 \\
&\quad + \|{\a}\trans\W_k\M_{k} -   \a\trans\W_k^*\M_{k}^*\|_2 \| \M_{k}\strans\|_2 \|\W_k\strans\a\|_2 \\
&\leq c \gamma_n d_k^{*-4} d_1^*.
\end{align*}
Hence, combining the above terms leads to 
\begin{align}
    A_{12} \leq c\gamma_n d_1^* d_k^{*-2}. \label{a122}
\end{align}

It remains to bound term $A_{13}$. Similar to $A_{12}$, we can bound $A_{13}$ as
\begin{align*}
	A_{13} & =|{\a}\trans\W_k\M_{k}\wt{\u}_k \wt{\v}_k\trans\bSigma_e\W_k\trans{\a} -  \a\trans\W_k^*  \M_{k}^*  \u_k^* \v_k\strans\bSigma_e\W_k\strans\a| \\[5pt]
 &\leq   \| {\a}\trans\W_k \|_2 \|\M_{k} \|_2 \|\wt{\u}_k\wt{\v}_k\trans\bSigma_e\W_k\trans{\a} - \u_k^*\v_k\strans\bSigma_e\W_k\strans\a \|_2 \\[5pt]
	& \quad +   \| {\a}\trans\W_k\M_{k} -
	\a\trans\W_k^* \M_{k}^*\|_2 \| \u_k^* \v_k\strans\bSigma_e\W_k\strans\a\|_2  \\[5pt]
	& \leq c d_k^{*-1}\|\wt{\u}_k\wt{\v}_k\trans\bSigma_e\W_k\trans{\a} - \u_k^*\v_k\strans\bSigma_e\W_k\strans\a\|_2 \\[5pt]
    & \quad+ c \gamma_n d_k^{*-3} d_1^* \| \u_k^* \v_k\strans\bSigma_e\W_k\strans\a\|_2.
\end{align*}
For the above two terms, it follows from Lemmas \ref{1} and \ref{lemma:weak:wbound} that
\begin{align*}
&\| \u_k^* \v_k\strans\bSigma_e\W_k\strans\a\|_2 \leq  \norm{\u_k^*}_2 \norm{\v_k\strans}_2 \norm{\bSigma_e}_2 \norm{\W_k\strans\a}_2 \leq c d_k^*, \\[5pt]
    &\|\wt{\u}_k\wt{\v}_k\trans\bSigma_e\W_k\trans{\a} - \u_k^*\v_k\strans\bSigma_e\W_k\strans\a\|_2 \\[5pt]
    &\quad \leq \|\wt{\u}_k\wt{\v}_k\trans\bSigma_e(\W_k\trans - \W_k\strans)\a\|_2 + \|(\wt{\u}_k\wt{\v}_k\trans - \u_k^*\v_k\strans)\bSigma_e\W_k\strans\a\|_2 \\[5pt]
    &\quad \leq \norm{\wt{\u}_k}_2 \norm{\wt{\v}_k}_2 \norm{\bSigma_e}_2 \norm{(\W_k\trans - \W_k\strans)\a}_2 + \norm{\wt{\u}_k\wt{\v}_k\trans - \u_k^*\v_k\strans}_2 \norm{\bSigma_e}_2 \norm{\W_k\strans\a}_2 \\[5pt]
    &\quad \leq c \gamma_n d_k^{*-1} d_1^*.
\end{align*}
Thus, we can obtain that 
\begin{align}
    A_{13} \leq c \gamma_n d_1^* d_k^{*-2}. \label{a133}
\end{align}

Combining \eqref{phiij201}, \eqref{a111}, \eqref{a122}, and \eqref{a133} yields that
\begin{align}\label{varpart1}
    | \wt{\operatorname{cov}} (\wt{h}_k,  \wt{h}_{k} ) - {\operatorname{cov}} (h_k,  h_{k} ) | \leq c \gamma_n d_1^* d_k^{*-2}. 
\end{align}

\noindent\textbf{(2). The upper bound on $\sum_{i \neq k} | \wt{\omega}_{k,i} \wt{\operatorname{cov}} (\wt{h}_k,  \wt{h}_{u_i} ) - \omega_{k,i} {\operatorname{cov}} (h_k,  h_{u_i} )|$}.

First, let us bound term $\omega_{k,i}$. It follows from Lemmas \ref{1} and \ref{lemma:weak:wbound} that
\begin{align}
     |\omega_{k,i}| & =  | ( \bmu_k\strans \wh{\bSigma}  \bmu_k^*)^{-1/2} \a\trans{\W}_k^* {\r}_i^*| \leq  | ( \bmu_k\strans \wh{\bSigma}  \bmu_k^*)^{-1/2}| \norm{\a\trans{\W}_k^* }_2 \norm{ {\r}_i^*}_2 \nonumber\\
     &\leq c d_k^{*-1}, \label{ome1} \\
	 |\wt{\omega}_{k,i}| & =  | ( \wt{\bmu}_k\trans \wh{\bSigma}  \wt{\bmu}_k)^{-1/2} \a\trans\W_k \wt{\r}_i| \leq  | ( \wt{\bmu}_k\trans \wh{\bSigma}  \wt{\bmu}_k)^{-1/2}| \norm{\a\trans\W_k }_2 \norm{ \wt{\r}_i}_2 \nonumber\\
     & \leq c d_k^{*-1}, \label{ome2} 
\end{align}	 
and 
\begin{align}
     &|\wt{\omega}_{k,i} - {\omega}_{k,i}| = | ( \wt{\bmu}_k\trans \wh{\bSigma}  \wt{\bmu}_k)^{-1/2} \a\trans\W_k \wt{\r}_i - ( \bmu_k\strans \wh{\bSigma}  \bmu_k^*)^{-1/2} \a\trans{\W}_k^* {\r}_i^*| \nonumber \\[5pt]
     &\ \  \leq  ( \wt{\bmu}_k\trans \wh{\bSigma}  \wt{\bmu}_k)^{-1/2} |\a\trans\W_k \wt{\r}_i - \a\trans{\W}_k^* {\r}_i^*| \\
     & \ \ \quad + | ( \wt{\bmu}_k\trans \wh{\bSigma}  \wt{\bmu}_k)^{-1/2} - ( \bmu_k\strans \wh{\bSigma}  \bmu_k^*)^{-1/2} | \|\a\trans{\W}_k^*\|_2 \| {\r}_i^*\|_2 \nonumber\\[5pt]
     &\ \ \leq c d_k^{*-1} ( \norm{\a\trans\W_k }_2  \norm{ \wt{\r}_i - {\r}_i^*}_2 +  \norm{\a\trans\W_k  - \a\trans{\W}_k^* }_2  \norm{ {\r}_i^*}_2 ) + c \gamma_n d_k^{*-2} \nonumber \\[5pt]
     &\ \ \leq c \gamma_n d_1^* d_k^{*-3}. \label{ome3}
\end{align}
Then with the aid of the triangle inequality, we can show that 
\begin{align}\label{covwcov}
    | \wt{\omega}_{k,i} \wt{\operatorname{cov}} (\wt{h}_k,  &\wt{h}_{u_i} ) - \omega_{k,i} {\operatorname{cov}} (h_k,  h_{u_i} )| \nonumber\\[5pt]
    &\leq | \wt{\omega}_{k,i} | |\wt{\operatorname{cov}} (\wt{h}_k,  \wt{h}_{u_i} ) -  {\operatorname{cov}} (h_k,  h_{u_i} )| + | \wt{\omega}_{k,i}  - \omega_{k,i} | |{\operatorname{cov}} (h_k,  h_{u_i} )| \nonumber\\[5pt]
    &\leq  c d_k^{*-1} |\wt{\operatorname{cov}} (\wt{h}_k,  \wt{h}_{u_i} ) -  {\operatorname{cov}} (h_k,  h_{u_i} )| + c \gamma_n d_1^* d_k^{*-3} |{\operatorname{cov}} (h_k,  h_{u_i} )|.
\end{align}

Next we bound terms $ |{\operatorname{cov}} (h_k,  h_{u_i} )|$ and $|\wt{\operatorname{cov}} (\wt{h}_k,  \wt{h}_{u_i} ) -  {\operatorname{cov}} (h_k,  h_{u_i} )|$. 
Recall that
\begin{align*}
&\operatorname{cov} (h_k,  h_{u_i} ) 
=        d_i^* (\l_k\strans \wh{\bSigma}\l_k^*)^{-1/2} \l_i\strans \wh{\bSigma} \l_k^* \cdot \a\trans \W_k^* \bSigma_e \M_{u_i}\strans \W_{u_i}\strans(\wh{\bSigma} {\bmu}_k^*)^t \\
&\quad -   \a\trans \W_k^* \M_{v_k}^* \wh{\bSigma}  d_i^* \l_i^* \cdot d_k^* (\l_k\strans \wh{\bSigma}\l_k^*)^{1/2} \r_k\strans \bSigma_e \M_{u_i}\strans \W_{u_i}\strans (\wh{\bSigma} {\bmu}_k^*)^t \\
&\quad  - ((\wh{\bSigma} {\bmu}_k^*)^t)\trans \W_{u_i}^* \wh{\bSigma} (\l_k\strans \wh{\bSigma}\l_k^*)^{-1/2} \l_k^* \cdot \r_i\strans \bSigma_e \W_k\strans\a \\
&\quad + \a\trans\W_k^* \M_{v_k}^* \wh{\bSigma} \W_{u_i}\strans (\wh{\bSigma} {\bmu}_k^*)^t \cdot d_k^* (\l_k\strans \wh{\bSigma}\l_k^*)^{1/2} \r_k\strans \bSigma_e \r_i^* =: \varphi^*_{21} + \varphi^*_{22} + \varphi^*_{23} + \varphi^*_{24}.
\end{align*}
Similarly, denote by $\varphi_{21}, \varphi_{22}, \varphi_{23}, \varphi_{24}$ the corresponding terms of $\wt{\operatorname{cov}} (\wt{h}_k,  \wt{h}_{u_i} )$. 

Similar to \eqref{phiij201}, it holds that 
\begin{align}\label{aaa444}
    |\wt{\operatorname{cov}} (\wt{h}_k,  \wt{h}_{u_i} ) -  {\operatorname{cov}} (h_k,  h_{u_i} )| &\leq |\varphi_{21} - \varphi_{21}^*  | + |\varphi_{22} - \varphi_{22}^*  | + |\varphi_{23} - \varphi_{23}^*  | + |\varphi_{24} - \varphi_{24}^*  | \nonumber \\
    & =: A_{21} + A_{22} + A_{23} + A_{24}.
\end{align}
For simplicity, let us define 
\[ \a_k = (\wh{\bSigma} \wt{\bmu}_k)^t, \ \ \a_k^* = (\wh{\bSigma} {\bmu}_k^*)^t. \] 
Then we have that 
\begin{align}
    \norm{\a_k^*}_0 \leq c s_u, \ \norm{\a_k}_0 \leq c s_u, \ \norm{\a_k^*}_2 \leq c d_k^*, \ \norm{\a_k}_2 \leq c d_k^*, \  \norm{\a_k - \a_k^*}_2 \leq c \gamma_n. \label{maa}
\end{align}

To bound the terms in \eqref{aaa444}, let us recall some preliminary results in \cite{sofari}.
In view of Lemmas 20 and 21 in \cite{sofari}, we can deduce that
\begin{align}
&\norm{ {\M}_{u_i}^* }_2  \leq c  d_i^{*-2} d_{i+1}^*, \ \norm{ \M_{u_i} }_2 \leq c  d_i^{*-2}d_{i+1}^*, \label{m1u}\\[5pt]
&\|\M_{u_i}  - {\M}_{u_i}^*  \|_2 
\leq  c \gamma_n d_i^{*-2}, \label{m2u}\\[5pt]
&\norm{\a_k\strans\W_{u_i}^*}_2 \leq c s_u^{1/2} d_k^*, \ \norm{\a_k\strans\W_{u_i}}_2 \leq c s_u^{1/2} d_k^*, \label{w0u} \\[5pt] 
&\norm{\a_k\trans\W_{u_i}^*}_2 \leq c s_u^{1/2} d_k^*, \ \norm{\a_k\trans\W_{u_i}}_2 \leq c s_u^{1/2} d_k^*,\label{w1u}\\[5pt]
&\norm{\a_k\strans(\W_{u_i}-\W_{u_i}^*)}_2 \leq c s_u^{1/2}  \gamma_n d_{i+1}^{*}d_i^{*-1}. \label{w2u}
\end{align}

For $\varphi^*_{21}$, it follows from Lemma \ref{1}, \eqref{m1u}, and \eqref{w1u} that
\begin{align*}
    |\varphi^*_{21}| &= | d_i^* (\l_k\strans \wh{\bSigma}\l_k^*)^{-1/2} \l_i\strans \wh{\bSigma} \l_k^* \cdot \a\trans \W_k^* \bSigma_e \M_{u_i}\strans \W_{u_i}\strans\a_k^*| \\ 
    &\leq d_i^* (\l_k\strans \wh{\bSigma}\l_k^*)^{-1/2} | \l_i\strans \wh{\bSigma} \l_k^*| \norm{ \a\trans \W_k^*}_2 \norm{\bSigma_e}_2  \norm{ \M_{u_i}\strans}_2  \norm{\W_{u_i}\strans \a_k^*}_2 \\
    &\leq c s_u^{1/2} d_i^{*-1} d_{i+1}^*  d_k^*.
\end{align*}
As for $\varphi^*_{22}$, similar to \eqref{m1} we have that 
\begin{align*}
    &|\varphi^*_{22}| = | \a\trans \W_k^* \M_{v_k}^* \wh{\bSigma}  d_i^* \l_i^* \cdot d_k^* (\l_k\strans \wh{\bSigma}\l_k^*)^{1/2} \r_k\strans \bSigma_e \M_{u_i}\strans \W_{u_i}\strans \a_k^*| \\
    &\leq d_i^* d_k^* (\l_k\strans \wh{\bSigma}\l_k^*)^{1/2} \norm{\a\trans \W_k^*}_2 \norm{\M_{k}^*}_2 \norm{n^{-1/2}\X \l_i^*}_2 \norm{\r_k\strans}_2 \norm{\bSigma_e}_2 \norm{\M_{u_i}\strans}_2 \norm{\W_{u_i}\strans \a_k^*}_2 \\
    &\leq c   s_u^{1/2} d_i^{*-1} d_{i+1}^* d_k^*.
\end{align*}

We can further show that 
\begin{align*}
     |\varphi^*_{23}| &= | \a_k\strans \W_{u_i}^* \wh{\bSigma} (\l_k\strans \wh{\bSigma}\l_k^*)^{-1/2} \l_k^* \cdot \r_i\strans \bSigma_e \W_k\strans\a| \\
     &\leq (\l_k\strans \wh{\bSigma}\l_k^*)^{-1/2} \norm{\a_k\strans \W_{u_i}^*}_2 \norm{\wh{\bSigma} \l_k^* }_2 \norm{ \r_i\strans}_2 \norm{ \bSigma_e}_2 \norm{\W_k\strans\a}_2 \\
     &\leq c s_u^{1/2} d_k^*.
\end{align*}
Moreover, it holds that
\begin{align*}
     |\varphi^*_{24}| &= |\a\trans\W_k^* \M_{v_k}^* \wh{\bSigma} \W_{u_i}\strans 
\a_k^* \cdot d_k^* (\l_k\strans \wh{\bSigma}\l_k^*)^{1/2} \r_k\strans \bSigma_e \r_i^* | \\
     &\leq  d_k^* (\l_k\strans \wh{\bSigma}\l_k^*)^{1/2} \norm{\a\trans\W_k^*}_2 \norm{\M_{v_k}^*}_2\norm{\wh{\bSigma} \W_{u_i}\strans\a_k^*}_2\norm{\r_k\strans}_2\norm{\bSigma_e}_2\norm{\r_i^*}_2 \\
     &\leq c   s_u^{1/2} d_k^*.
\end{align*}
Hence, combining the above terms gives that 
\begin{align}\label{covhh}
    \operatorname{cov} (h_k,  h_{u_i} )  \leq c   s_u^{1/2} d_k^*.
\end{align}

Next we bound term $|\wt{\operatorname{cov}} (\wt{h}_k,  \wt{h}_{u_i} ) -  {\operatorname{cov}} (h_k,  h_{u_i} )|$, which will be divided into four parts as shown in \eqref{aaa444}. For term $A_{21}$, it holds that 
\begin{align*}
    A_{21} &= | \wt{d}_i (\wt{\l}_k\trans \wh{\bSigma}\wt{\l}_k)^{-1/2} \wt{\l}_i\trans \wh{\bSigma} \wt{\l}_k \cdot \a\trans \W_k \bSigma_e \M_{u_i}\trans \W_{u_i}\trans\a_k \\[5pt]
    &\quad - d_i^* (\l_k\strans \wh{\bSigma}\l_k^*)^{-1/2} \l_i\strans \wh{\bSigma} \l_k^* \cdot \a\trans \W_k^* \bSigma_e \M_{u_i}\strans \W_{u_i}\strans\a_k^*| \\[5pt]
    &\leq | \wt{d}_i (\wt{\l}_k\trans \wh{\bSigma}\wt{\l}_k)^{-1/2} \wt{\l}_i\trans \wh{\bSigma} \wt{\l}_k | |  \a\trans \W_k \bSigma_e \M_{u_i}\trans \W_{u_i}\trans\a_k - \a\trans \W_k^* \bSigma_e \M_{u_i}\strans \W_{u_i}\strans\a_k^*| \\[5pt]
    &\quad +|\wt{d}_i (\wt{\l}_k\trans \wh{\bSigma}\wt{\l}_k)^{-1/2} \wt{\l}_i\trans \wh{\bSigma} \wt{\l}_k - d_i^* (\l_k\strans \wh{\bSigma}\l_k^*)^{-1/2} \l_i\strans \wh{\bSigma} \l_k^* | | \a\trans \W_k^* \bSigma_e \M_{u_i}\strans \W_{u_i}\strans\a_k^*|.
\end{align*}
Then by Lemma \ref{1}, we can obtain that $| \wt{d}_i (\wt{\l}_k\trans \wh{\bSigma}\wt{\l}_k)^{-1/2} \wt{\l}_i\trans \wh{\bSigma} \wt{\l}_k | \leq c d_i^*$ and
\begin{align*}
    &|\wt{d}_i (\wt{\l}_k\trans \wh{\bSigma}\wt{\l}_k)^{-1/2} \wt{\l}_i\trans \wh{\bSigma} \wt{\l}_k - d_i^* (\l_k\strans \wh{\bSigma}\l_k^*)^{-1/2} \l_i\strans \wh{\bSigma} \l_k^* | \\
    &\leq  (\wt{\bmu}_k\trans \wh{\bSigma}\wt{\bmu}_k)^{-1/2} |\wt{\bmu}_i\trans \wh{\bSigma} \wt{\bmu}_k - \bmu_i\strans \wh{\bSigma} \bmu_k^* | 
    + |(\wt{\bmu}_k\trans \wh{\bSigma}\wt{\bmu}_k)^{-1/2}  - (\bmu_k\strans \wh{\bSigma}\bmu_k^*)^{-1/2} | |\bmu_i\strans \wh{\bSigma} \bmu_k^* | \\
    &\leq  c\gamma_n d_i^{*-1} d_k^* + c \gamma_n d_k^{*-1} \max\{ d_i^{*}, d_k^*\} \\ 
    &\leq c\gamma_n d_i^{*} d_k^{*} \max\{  d_i^{*-2}, d_k^{*-2}\}.
\end{align*}

On the other hand, we have that 
\begin{align*}
    | \a\trans \W_k^* \bSigma_e \M_{u_i}\strans \W_{u_i}\strans\a_k^*| \leq \norm{ \a\trans \W_k^*}_2 \norm{\bSigma_e}_2 \norm{ \M_{u_i}\strans}_2 \norm{\W_{u_i}\strans\a_k^*}_2 \leq c s_u^{1/2}  d_i^{*-2} d_{i+1}^* d_k^*.
\end{align*}
It follows from \eqref{maa}--\eqref{w2u} that 
\begin{align}
    &\norm{\a_k\trans \W_{u_i}  -  \a_k\strans {\W}_{u_i}^*}_2 \leq  \norm{\a_k\trans (\W_{u_i}  -   {\W}_{u_i}^*)}_2 + \norm{(\a_k\trans  -  \a_k\strans) {\W}_{u_i}^*}_2 \nonumber\\
    & \leq c s_u^{1/2} d_k^*  \gamma_n d_{i+1}^{*}d_i^{*-2} + c s_u^{1/2} \gamma_n \leq c s_u^{1/2} \gamma_n \max\{d_k^*d_{i+1}^{*}d_i^{*-2}, 1 \}, \label{awaw}
\end{align}
and
\begin{align}\label{awmu}
    &\norm{\a_k\trans \W_{u_i} \M_{u_i} -  \a_k\strans {\W}_{u_i}^* {\M}_{u_i}^*}_2 \nonumber \\
    &\leq  \norm{\a_k\trans \W_{u_i}}_2 \norm{ \M_{u_i} -   {\M}_{u_i}^*}_2 +  \norm{\a_k\trans \W_{u_i}  -  \a_k\strans {\W}_{u_i}^*}_2 \norm{{\M}_{u_i}^*}_2 \nonumber\\
    &\leq  c s_u^{1/2}  \gamma_n d_i^{*-2} d_k^*  + c  s_u^{1/2} \gamma_n d_i^{*-2} d_{i+1}^* \max\{d_k^*d_{i+1}^{*}d_i^{*-2}, 1 \}  \nonumber\\
    &\leq c s_u^{1/2}  \gamma_n d_i^{*-2} \max\{d_k^*, d_{i+1}^{*}\}.
\end{align}

From Lemma \ref{lemma:weak:wbound} and \eqref{awmu}, we can deduce that 
\begin{align*}
     &|  \a\trans \W_k \bSigma_e \M_{u_i}\trans \W_{u_i}\trans\a_k - \a\trans \W_k^* \bSigma_e \M_{u_i}\strans \W_{u_i}\strans\a_k^*| \\ &\leq \|  \a\trans \W_k \|_2 \|\bSigma_e\|_2 \| \M_{u_i}\trans \W_{u_i}\trans\a_k -  \M_{u_i}\strans \W_{u_i}\strans\a_k^*\|_2 \\
     &\quad+ \|  \a\trans (\W_k  - \W_k^* )\|_2 \|\bSigma_e \|_2 \| \M_{u_i}\strans \|_2 \|\W_{u_i}\strans\a_k^*\|_2 \\
     &\leq c s_u^{1/2}  \gamma_n  d_i^{*-2} \max\{d_k^*, d_{i+1}^{*}\} + c s_u^{1/2}  \gamma_n d_k^{*-1}  d_i^{*-2} d_{i+1}^*  d_1^* \\
     &\leq c s_u^{1/2}  \gamma_n  d_i^{*-2} \max\{d_k^*, d_{i+1}^{*}, d_1^*  d_{i+1}^*d_k^{*-1}\}.
\end{align*}
Hence, it follows that 
\begin{align}\label{a21}
    A_{21} \leq c \gamma_n s_u^{1/2}  d_i^{*-1}\max\{  d_i^{*-2}  d_k^{*2}  d_{i+1}^*,  d_{i+1}^*,  d_k^*,  d_1^* d_{i+1}^*d_k^{*-1}  \}. 
\end{align}

For term $A_{22}$, it holds that
\begin{align*}
   & A_{22} = | \a\trans \W_k \M_{v_k} \wh{\bSigma}  \wt{d}_i \wt{\l}_i \cdot \wt{d}_k (\wt{\l}_k\trans \wh{\bSigma}\wt{\l}_k)^{1/2} \wt{\r}_k\trans \bSigma_e \M_{u_i}\trans \W_{u_i}\trans\a_k \\[5pt]
    &\quad \qquad -  \a\trans \W_k^* \M_{v_k}^* \wh{\bSigma}  d_i^* \l_i^* \cdot d_k^* (\l_k\strans \wh{\bSigma}\l_k^*)^{1/2} \r_k\strans \bSigma_e \M_{u_i}\strans \W_{u_i}\strans\a_k^*| \\[5pt]
    &\leq  | \a\trans \W_k \M_{v_k} \wh{\bSigma}  \wt{d}_i \wt{\l}_i | \\ & \qquad \cdot | \wt{d}_k (\wt{\l}_k\trans \wh{\bSigma}\wt{\l}_k)^{1/2} \wt{\r}_k\trans \bSigma_e \M_{u_i}\trans \W_{u_i}\trans\a_k  -  d_k^* (\l_k\strans \wh{\bSigma}\l_k^*)^{1/2} \r_k\strans \bSigma_e \M_{u_i}\strans \W_{u_i}\strans\a_k^*| \\[5pt]
    &\quad + | \a\trans \W_k \M_{v_k} \wh{\bSigma}  \wt{d}_i \wt{\l}_i -  \a\trans \W_k^* \M_{v_k}^* \wh{\bSigma}  d_i^* \l_i^*| | d_k^* (\l_k\strans \wh{\bSigma}\l_k^*)^{1/2} \r_k\strans \bSigma_e \M_{u_i}\strans \W_{u_i}\strans\a_k^*|.
\end{align*}
Then by Lemma \ref{lemma:weak:wbound},
we can obtain that
\begin{align}
&| \a\trans \W_k \M_{v_k} \wh{\bSigma}  \wt{d}_i \wt{\l}_i | \leq \norm{\a\trans \W_k}_2 \norm{ \M_{v_k}}_2 \norm{\wh{\bSigma}  \wt{d}_i \wt{\l}_i}_2 \leq c d_i^* d_k^{*-1}, \nonumber\\
	&\| \a\trans \W_k \M_{v_k}  -  \a\trans \W_k^* \M_{v_k}^*\|_2 \nonumber\\
	&\leq \| \a\trans \W_k\|_2 \| \M_{v_k}  - \M_{v_k}^*\|_2 + \| \a\trans (\W_k  - \W_k^*)\|_2 \| \M_{v_k}^*\|_2 \nonumber \\
	&\leq c   \gamma_n d_k^{*-2} + c \gamma_n d_1^* d_k^{*-3}  \leq c  \gamma_n d_1^* d_k^{*-3}. \label{awmv2}
\end{align}
Further, we can show that 
\begin{align*}
     &| \a\trans \W_k \M_{v_k} \wh{\bSigma}  \wt{d}_i \wt{\l}_i -  \a\trans \W_k^* \M_{v_k}^* \wh{\bSigma}  d_i^* \l_i^*| \\
     &\leq \| \a\trans \W_k \|_2 \|\M_{v_k} \|_2 \| \wh{\bSigma}  \wt{d}_i \wt{\l}_i -  \wh{\bSigma}  d_i^* \l_i^*\|_2  + \| \a\trans \W_k \M_{v_k}  -  \a\trans \W_k^* \M_{v_k}^*\|_2 \| \wh{\bSigma}  d_i^* \l_i^*\|_2 \\[5pt]
    &\leq c d_k^{*-1} \gamma_n + c d_i^* (\| \a\trans \W_k\|_2 \| \M_{v_k}  - \M_{v_k}^*\|_2 + \| \a\trans (\W_k  - \W_k^*)\|_2 \| \M_{v_k}^*\|_2) \\
    &\leq c d_k^{*-1} \gamma_n + c  d_i^* ( \gamma_n d_k^{*-2} + \gamma_n d_1^* d_k^{*-3} ) \leq c \gamma_n d_k^{*-2} \max\{ d_k^*, d_i^*, d_1^* d_i^{*} d_k^{*-1}\}.
\end{align*}
Observe that 
\begin{align*}
    &| d_k^* (\l_k\strans \wh{\bSigma}\l_k^*)^{1/2} \r_k\strans \bSigma_e \M_{u_i}\strans \W_{u_i}\strans\a_k^*| \\ &\leq  d_k^* (\l_k\strans \wh{\bSigma}\l_k^*)^{1/2} \norm{\r_k\strans}_2 \norm{\bSigma_e}_2 \norm{\M_{u_i}\strans}_2 \norm{ \W_{u_i}\strans\a_k^*}_2 \leq c s_u^{1/2} d_k^{*2}  d_i^{*-2}d_{i+1}^*.
\end{align*}

An application of Lemma \ref{1} and \eqref{awmu} leads to 
\begin{align*}
    & | \wt{d}_k (\wt{\l}_k\trans \wh{\bSigma}\wt{\l}_k)^{1/2} \wt{\r}_k\trans \bSigma_e \M_{u_i}\trans \W_{u_i}\trans \a_k  -  d_k^* (\l_k\strans \wh{\bSigma}\l_k^*)^{1/2} \r_k\strans \bSigma_e \M_{u_i}\strans \W_{u_i}\strans\a_k^*| \\[5pt]
    & \leq \| \wt{d}_k (\wt{\l}_k\trans \wh{\bSigma}\wt{\l}_k)^{1/2} \wt{\r}_k\trans \bSigma_e \|_2 \| \M_{u_i}\trans \W_{u_i}\trans\a_k  -   \M_{u_i}\strans \W_{u_i}\strans\a_k^*\|_2 \\[5pt]
    &\quad + \| \wt{d}_k (\wt{\l}_k\trans \wh{\bSigma}\wt{\l}_k)^{1/2} \wt{\r}_k\trans \bSigma_e   -  d_k^* (\l_k\strans \wh{\bSigma}\l_k^*)^{1/2} \r_k\strans \bSigma_e \|_2 \|\M_{u_i}\strans \W_{u_i}\strans\a_k^*\|_2 \\
    &\leq  c d_k^* \| \M_{u_i}\trans \W_{u_i}\trans\a_k  -   \M_{u_i}\strans \W_{u_i}\strans\a_k^*\|_2 + c \Big[  (\wt{\l}_k\trans \wh{\bSigma}\wt{\l}_k)^{1/2} \|\wt{d}_k\wt{\r}_k\trans   -  d_k^*  \r_k\strans \|_2  \\[5pt]
    &\ \
     + (  (\wt{\l}_k\trans \wh{\bSigma}\wt{\l}_k)^{1/2}   -   (\l_k\strans \wh{\bSigma}\l_k^*)^{1/2} )\|d_k^*\r_k\strans \|_2 \Big]\| \bSigma_e \|_2 \|\M_{u_i}\strans \W_{u_i}\strans\a_k^*\|_2 \\[5pt]
     &\leq c d_k^* \| \M_{u_i}\trans \W_{u_i}\trans\a_k  -   \M_{u_i}\strans \W_{u_i}\strans\a_k^*\|_2 + c \gamma_n \|\M_{u_i}\strans\|_2 \| \W_{u_i}\strans\a_k^*\|_2 \\[5pt]
     &\leq c d_k^* s_u^{1/2}  \gamma_n d_i^{*-2} \max\{d_k^*, d_{i+1}^{*}\} + c d_k^*  s_u^{1/2}  \gamma_n d_i^{*-2} d_{i+1}^* \\
     &\leq c  s_u^{1/2}  \gamma_n d_i^{*-2} d_k^* \max\{d_k^*, d_{i+1}^{*}\}.
\end{align*}
Thus, it holds that 
\begin{align}\label{a2222}
	A_{22} \leq c   s_u^{1/2}  \gamma_n d_i^{*-1}  \max\{d_k^*, d_{i+1}^{*}, d_{i+1}^* d_1^*  d_k^{*-1}\}.
\end{align}

For term $A_{23}$, it follows that
\begin{align*}
    A_{23} &= | \a_k\trans \W_{u_i} \wh{\bSigma} (\wt{\l}_k\trans \wh{\bSigma}\wt{\l}_k)^{-1/2} \wt{\l}_k \cdot \wt{\r}_i\trans \bSigma_e \W_k\trans\a \\
    &\quad -  (\a_k^*)\trans \W_{u_i}^* \wh{\bSigma} (\l_k\strans \wh{\bSigma}\l_k^*)^{-1/2} \l_k^* \cdot \r_i\strans \bSigma_e \W_k\strans\a| \\
    &\leq \| \a_k\trans \W_{u_i}\|_2 \| (\wt{\l}_k\trans \wh{\bSigma}\wt{\l}_k)^{-1/2} 
    \wh{\bSigma} \wt{\l}_k \wt{\r}_i\trans \bSigma_e \W_k\trans\a - (\l_k\strans \wh{\bSigma}\l_k^*)^{-1/2} \wh{\bSigma}  \l_k^*  \r_i\strans \bSigma_e \W_k\strans\a\|_2 \\[5pt]
    & + \| \a_k\trans \W_{u_i} -  \a_k\strans \W_{u_i}^* \|_2 \|\wh{\bSigma} (\l_k\strans \wh{\bSigma}\l_k^*)^{-1/2} \l_k^* \cdot \r_i\strans \bSigma_e \W_k\strans\a\|_2.
\end{align*}
It is easy to see that 
\[\|(\l_k\strans \wh{\bSigma}\l_k^*)^{-1/2} \wh{\bSigma}  \l_k^* \r_i\strans \bSigma_e \W_k\strans\a\|_2 \leq (\l_k\strans \wh{\bSigma}\l_k^*)^{-1/2} \norm{\wh{\bSigma}  \l_k^*}_2  \norm{ \r_i\strans}_2  \norm{\bSigma_e}_2  \norm{\W_k\strans\a}_2 \leq c. \]

By invoking Lemmas \ref{1} and \ref{lemma:weak:wbound}, we can deduce that 
\begin{align*}
     &\| (\wt{\l}_k\trans \wh{\bSigma}\wt{\l}_k)^{-1/2} 
    \wh{\bSigma} \wt{\l}_k \wt{\r}_i\trans \bSigma_e \W_k\trans\a - (\l_k\strans \wh{\bSigma}\l_k^*)^{-1/2} \wh{\bSigma}  \l_k^*  \r_i\strans \bSigma_e \W_k\strans\a\|_2 \\[5pt]
    &\leq \| (\wt{\l}_k\trans \wh{\bSigma}\wt{\l}_k)^{-1/2} 
    \wh{\bSigma} \wt{\l}_k \wt{\r}_i\trans\|_2 \| \bSigma_e \|_2 \| (\W_k\trans - \W_k\strans)\a\|_2 \\[5pt]
    &\ \ + \| (\wt{\l}_k\trans \wh{\bSigma}\wt{\l}_k)^{-1/2} 
    \wh{\bSigma} \wt{\l}_k \wt{\r}_i\trans  - (\l_k\strans \wh{\bSigma}\l_k^*)^{-1/2} \wh{\bSigma}  \l_k^*  \r_i\strans\|_2 \| \bSigma_e\|_2 \|\W_k\strans\a\|_2 \\[5pt]
    &\leq c \gamma_n d_1^* d_k^{*-2} + c  (\wt{\l}_k\trans \wh{\bSigma}\wt{\l}_k)^{-1/2} 
    \|\wh{\bSigma} \wt{\l}_k \wt{\r}_i\trans  - \wh{\bSigma}  \l_k^*  \r_i\strans\|_2 \\[5pt] 
    &\ \ + c  [(\wt{\l}_k\trans \wh{\bSigma}\wt{\l}_k)^{-1/2} 
     - (\l_k\strans \wh{\bSigma}\l_k^*)^{-1/2}]\| \wh{\bSigma}  \l_k^* \|_2 \| \r_i\strans\|_2 \\[5pt] 
    &\leq c \gamma_n d_1^* d_k^{*-2} + c \gamma_n d_k^{*-1} + c \gamma_n d_k^{*-1} \leq c \gamma_n d_1^* d_k^{*-2}.
\end{align*}
Then it follows from $ \| \a_k\trans \W_{u_i}\|_2 \leq c s_u^{1/2} \gamma_n$, \eqref{w1u}, and \eqref{awaw} that
\begin{align}\label{a23}
    A_{23} \leq c s_u^{1/2} \gamma_n \max\{d_k^*d_{i+1}^{*}d_i^{*-2}, 1 \} + c s_u^{1/2} \gamma_n d_1^* d_k^{*-2}.
\end{align}

For term $A_{24}$, it holds that
\begin{align*}
    A_{24} &= | \a\trans\W_k \M_{v_k} \wh{\bSigma} \W_{u_i}\trans\a_k \cdot \wt{d}_k (\wt{\l}_k\trans \wh{\bSigma}\wt{\l}_k)^{1/2} \wt{\r}_k\trans \bSigma_e \wt{\r}_i \\[5pt]
    &\quad \ - \a\trans\W_k^* \M_{v_k}^* \wh{\bSigma} \W_{u_i}\strans\a_k^* \cdot d_k^* (\l_k\strans \wh{\bSigma}\l_k^*)^{1/2} \r_k\strans \bSigma_e \r_i^*| \\[5pt]
    &\leq | \a\trans\W_k \M_{v_k} \wh{\bSigma} \W_{u_i}\trans\a_k - \a\trans\W_k^* \M_{v_k}^* \wh{\bSigma} \W_{u_i}\strans\a_k^* |  | d_k^* (\l_k\strans \wh{\bSigma}\l_k^*)^{1/2} \r_k\strans \bSigma_e \r_i^*| \\[5pt]
    &\quad \ + | \a\trans\W_k \M_{v_k} \wh{\bSigma} \W_{u_i}\trans\a_k | |  \wt{d}_k (\wt{\l}_k\trans \wh{\bSigma}\wt{\l}_k)^{1/2} \wt{\r}_k\trans \bSigma_e \wt{\r}_i -  d_k^* (\l_k\strans \wh{\bSigma}\l_k^*)^{1/2} \r_k\strans \bSigma_e \r_i^*|.
\end{align*}
Notice that $$| d_k^* (\l_k\strans \wh{\bSigma}\l_k^*)^{1/2} \r_k\strans \bSigma_e \r_i^*| \leq c d_k^*$$ and 
\begin{align*}
    &|  \wt{d}_k (\wt{\l}_k\trans \wh{\bSigma}\wt{\l}_k)^{1/2} \wt{\r}_k\trans \bSigma_e \wt{\r}_i -  d_k^* (\l_k\strans \wh{\bSigma}\l_k^*)^{1/2} \r_k\strans \bSigma_e \r_i^*| \\[5pt] &\leq |  \wt{d}_k (\wt{\l}_k\trans \wh{\bSigma}\wt{\l}_k)^{1/2} | |\wt{\r}_k\trans \bSigma_e \wt{\r}_i -  \r_k\strans \bSigma_e \r_i^*| + |  \wt{d}_k (\wt{\l}_k\trans \wh{\bSigma}\wt{\l}_k)^{1/2}  -  d_k^* (\l_k\strans \wh{\bSigma}\l_k^*)^{1/2}| | \r_k\strans \bSigma_e \r_i^*| \\[5pt]
    &\leq c d_k^* (\|\wt{\r}_k\trans\|_2 \|\bSigma_e\|_2 \|\wt{\r}_i -   \r_i^*\|_2 + \|\wt{\r}_k\trans  -  \r_k\strans\|_2 \| \bSigma_e \|_2 \| \r_i^*\|_2) + c  \gamma_n  \\[5pt]
    &\leq c \gamma_n  \max\{  d_i^{*-1} d_k^*, 1 \}.
\end{align*}

Using the arguments in part 2 of the proof of Theorem 6 in \cite{sofari}, we can show that
\begin{align*}
 &\norm{\wh{\bSigma} \W_{u_i}\trans\a_k}_2 \leq c s_u^{1/2} d_k^*, \\[5pt]
 &\|\wh{\bSigma}(\W_{u_i}\trans\a_k -  \W_{u_i}\strans\a^*_k)\|_2 \leq 
	\|\wh{\bSigma}\W_{u_i}\trans({\a}_k - \a^*_k)\|_2 + 	\|\wh{\bSigma}(\W_{u_i}\trans - {\W}_{u_i}\strans)\a_k^*\|_2 \\[5pt]
	&\qquad \qquad \qquad \qquad \leq   c s_u^{1/2}   \gamma_n \max\{1, d_{i+1}^*d_{i}^{*-2}d_k^*\}.
\end{align*}
Then it follows that 
\begin{align*}
    | \a\trans\W_k \M_{v_k} \wh{\bSigma} \W_{u_i}\trans\a_k | \leq \norm{ \a\trans\W_k}_2 \norm{ \M_{v_k}}_2  \norm{\wh{\bSigma} \W_{u_i}\trans\a_k}_2 \leq  c s_u^{1/2}.
\end{align*}

From \eqref{awmv2}, we can deduce that 
\begin{align*}
    &| \a\trans\W_k \M_{v_k} \wh{\bSigma} \W_{u_i}\trans\a_k - \a\trans\W_k^* \M_{v_k}^* \wh{\bSigma} \W_{u_i}\strans\a_k^* | \\[5pt]
    &\leq \norm{\a\trans\W_k}_2 \norm{\M_{v_k}}_2 \norm{\wh{\bSigma} (\W_{u_i}\trans\a_k - \W_{u_i}\strans\a_k^*)}_2 \\
	&\ \ + \norm{\a\trans\W_k \M_{v_k} - \a\trans\W_k^* \M_{v_k}^*}_2 \norm{\wh{\bSigma} \W_{u_i}\strans\a_k^*}_2 \\[5pt]
    &\leq  c  s_u^{1/2}   \gamma_n  d_{k}^{*-1}  \max\{1, d_{i+1}^*d_{i}^{*-2}d_k^*\}   +  c  s_u^{1/2}   \gamma_n   d_{1}^{*}  d_{k}^{*-2} \\
	&\leq c s_u^{1/2}   \gamma_n  d_{k}^{*-1}  \max\{1, d_{i+1}^*d_k^* d_{i}^{*-2},   d_{1}^{*}  d_{k}^{*-1}  \}.
\end{align*}
Hence, we can obtain that 
\begin{align}\label{a24}
	A_{24} \leq  c  s_u^{1/2}   \gamma_n   \max\{1,  d_k^*d_i^{*-1},   d_{1}^{*}  d_{k}^{*-1}  \}.
\end{align}

Combining \eqref{aaa444}, \eqref{a21}, \eqref{a2222}, \eqref{a23}, and \eqref{a24} yields that
\begin{align*}
    |\wt{\operatorname{cov}} (\wt{h}_k,  \wt{h}_{u_i} ) -  {\operatorname{cov}} (h_k,  h_{u_i} )| \leq c \gamma_n s_u^{1/2}  d_i^{*-1}\max\{ d_{i}^*,  d_k^*,   d_i^{*-2}  d_k^{*2}  d_{i+1}^*,   d_{1}^{*}  d_{i}^{*} d_{k}^{*-1} \}.
\end{align*}
This together with \eqref{ome2}, \eqref{ome3}, \eqref{covwcov}, and \eqref{covhh} results in 
\begin{align}\label{varpart2}
    \sum_{i \neq k}| \wt{\omega}_{k,i} \wt{\operatorname{cov}} (\wt{h}_k,  \wt{h}_{u_i} ) - \omega_{k,i} {\operatorname{cov}} (h_k,  h_{u_i} )| 
	\leq  c r^*  \gamma_n s_u^{1/2}
	\max\{     d_k^{*} d_{r^*}^{*-2}, d_{1}^{*} d_{k}^{*-2} \}.
\end{align}

\bigskip

\noindent\textbf{(3). The upper bound on $  \sum_{i \neq k} \sum_{j \neq k}| \wt{\omega}_{k,i} \wt{\omega}_{k,j}  \wt{\operatorname{cov}} (  \wt{h}_{u_i}, \wt{h}_{u_j})  - \omega_{k,i} \omega_{k,j}  {\operatorname{cov}} (  h_{u_i}, h_{u_j})  |$}. 
In view of \eqref{ome1}--\eqref{ome3}, it holds that 
\begin{align}\label{ome4}
     &| \wt{\omega}_{k,i} \wt{\omega}_{k,j} - \omega_{k,i} \omega_{k,j}| \leq | \wt{\omega}_{k,i}| | \wt{\omega}_{k,j} -  \omega_{k,j}| + | \wt{\omega}_{k,i} - \omega_{k,i} | |\omega_{k,j}| \leq c \gamma_n d_1^* d_k^{*-3},
\end{align}
Using \eqref{ome2}, we can deduce that 
\begin{align}\label{wwcovwwcov}
	&| \wt{\omega}_{k,i} \wt{\omega}_{k,j}  \wt{\operatorname{cov}} (  \wt{h}_{u_i}, \wt{h}_{u_j})  - \omega_{k,i} \omega_{k,j}  {\operatorname{cov}} (  h_{u_i}, h_{u_j})  | \nonumber \\
	& \leq  | \wt{\omega}_{k,i}| | \wt{\omega}_{k,j} | | \wt{\operatorname{cov}} (  \wt{h}_{u_i}, \wt{h}_{u_j})  - {\operatorname{cov}} (  h_{u_i}, h_{u_j})  | + | \wt{\omega}_{k,i} \wt{\omega}_{k,j}    - \omega_{k,i} \omega_{k,j} | | {\operatorname{cov}} (  h_{u_i}, h_{u_j})  | \nonumber \\
	&\leq c d_k^{*-2} | \wt{\operatorname{cov}} (  \wt{h}_{u_i}, \wt{h}_{u_j})  - {\operatorname{cov}} (  h_{u_i}, h_{u_j})  | + c \gamma_n d_1^* d_k^{*-3} | {\operatorname{cov}} (  h_{u_i}, h_{u_j})  |.
\end{align}

Let us define 
\begin{align*}
\operatorname{cov} (h_{u_i},  h_{u_j} )
&=     d_i^* d_j^* \l_i\strans \wh{\bSigma} \l_j^* \cdot \a_k\strans \W_{u_i}^* \M_{u_i}^* \bSigma_e \M_{u_j}\strans \W_{u_j}\strans\a_k^* \\ 
& \quad + \a_k\strans \W_{u_i}^*  \wh{\bSigma} \W_{u_j}\strans\a_k^* \cdot  \r_i\strans \bSigma_e \r_j^* \\
&\quad -   \a_k\strans \W_{u_i}^* \wh{\bSigma}  d_j^* \l_j^* \cdot \r_i\strans \bSigma_e \M_{u_j}\strans \W_{u_j}\strans\a_k^* \\
&\quad  - \a_k\strans \W_{u_j}^* \wh{\bSigma} d_i^* \l_i^* \cdot \r_j\strans \bSigma_e 
\M_{u_i}\strans \W_{u_i}\strans\a_k^* \\ 
&=: \varphi^*_{31} + \varphi^*_{32} + \varphi^*_{33} + \varphi^*_{34}.
\end{align*}
Similarly, denote by $\varphi_{31}, \varphi_{32}, \varphi_{33}, \varphi_{34}$ the corresponding terms of $\wt{\operatorname{cov}} (  \wt{h}_{u_i}, \wt{h}_{u_j})$. Then we have that 
\begin{align}\label{aaa444333}
    |\wt{\operatorname{cov}} (  \wt{h}_{u_i}, \wt{h}_{u_j}) - \operatorname{cov} (h_{u_i},  h_{u_j} )| &\leq |\varphi_{31} - \varphi_{31}^*  | + |\varphi_{32} - \varphi_{32}^*  | + |\varphi_{33} - \varphi_{33}^*  | + |\varphi_{34} - \varphi_{34}^*  | \nonumber \\
    & =: A_{31} + A_{32} + A_{33} + A_{34}.
\end{align}

It is easy to see that 
\begin{align*}
	|\varphi_{31}^*| &\leq  |\bmu_i\strans\wh{\bSigma}\bmu_j^*| \norm{\a_k\strans\W_{u_i}^* }_2 \norm{ \M_{u_i}^* }_2  \norm{\bSigma_e }_2  \norm{ \M_{u_j}\strans }_2  \norm{\W_{u_j}\strans\a_k^* }_2 \\
	&\leq  c s_u d_i^{*-1}d_{i+1}^*   d_j^{*-1}d_{j+1}^* d_k^{*2}.
\end{align*}
Similarly, we can show that 
\begin{align*}
	|\varphi_{33}^*| &\leq \norm{ \a_k\strans \W_{u_i}^* }_2  \norm{ \wh{\bSigma} \bmu_j^*}_2 \norm{ \r_i^* }_2 \norm{ \bSigma_e  }_2 \norm{ \M_{u_j}\strans }_2 \norm{ \W_{u_j}\strans  \a_k^* }_2 \\[5pt]
    &\leq c  s_u d_k^{*2} d_j^{*-1} d_{j+1}^*, \\[5pt]
	|\varphi_{34}^*| &\leq \norm{\a_k\strans \W_{u_j}^*  }_2  \norm{ \wh{\bSigma} \bmu_i^*}_2 \norm{ \r_j^* }_2 \norm{ \bSigma_e  }_2 \norm{ \M_{u_i}\strans }_2 \norm{ \W_{u_i}\strans  \a_k^* }_2  \\[5pt]
    &\leq c  s_u d_k^{*2} d_i^{*-1} d_{i+1}^*.
\end{align*}
Moreover, it holds that 
\begin{align*}
	|\varphi_{32}^*| &\leq \norm{\r_i\strans}_2 \|\bSigma_e\| \norm{\r_j^* }_2  \norm{ \a_k\strans\W_{u_i}^* }_2 \norm{\wh{\bSigma}\W_{u_j}\strans\a_k^* }_2  \leq  c  s_u d_k^{*2}.
\end{align*}
Then we can obtain that 
\begin{align}\label{phiijaa}
	|\operatorname{cov} (h_{u_i},  h_{u_j} )| \leq |\varphi_{31}^*| + |\varphi_{32}^*|  + |\varphi_{33}^*| + |\varphi_{34}^*|  \leq  c  s_u d_k^{*2}.
\end{align}

Next, we bound terms $ A_{31}, A_{32}, A_{33}, \text{ and } A_{34}$. 
For term $A_{31}$, it follows that
\begin{align}
		|A_{31}| \leq 
		&  |\wt{\bmu}_{i}\trans\wh{\bSigma}\wt{\bmu}_{j}| | \a_k\trans\W_{u_i}\M_{u_i}\bSigma_e\M_{u_j}\trans\W_{u_j}\trans\a_k -  \a_k\strans \W_{u_i}^* \M_{u_i}^* \bSigma_e \M_{u_j}\strans \W_{u_j}\strans\a_k^* | \nonumber\\[5pt]
		&\qquad ~ + |\wt{\bmu}_{i}\trans\wh{\bSigma}\wt{\bmu}_{j} -   \bmu_i\strans \wh{\bSigma} \bmu_j^*| | \a_k\strans \W_{u_i}^* \M_{u_i}^* \bSigma_e \M_{u_j}\strans \W_{u_j}\strans\a_k^* | \nonumber.
	\end{align}
Observe that $|\wt{\bmu}_{i}\trans\wh{\bSigma}\wt{\bmu}_{j}| \leq c d_i^* d_j^*$ and 
\begin{align*}
    |\wt{\bmu}_{i}\trans\wh{\bSigma}\wt{\bmu}_{j} -   \bmu_i\strans \wh{\bSigma} \bmu_j^*| &\leq  \|\wt{\bmu}_{i}\trans\|_2 \|\wh{\bSigma}(\wt{\bmu}_{j} -    \bmu_j^*)\|_2 + \|\wt{\bmu}_{i}\trans -   \bmu_i\strans \|_2 \| \wh{\bSigma} \bmu_j^*\|_2  \\ &\leq c \max\{  d_i^{*}, d_j^{*} \} \gamma_n.
\end{align*}
In addition, we have that 
\begin{align}
	| \a_k\strans \W_{u_i}^* \M_{u_i}^* \bSigma_e \M_{u_j}\strans \W_{u_j}\strans\a_k^* | & \leq \norm{\a_k\strans \W_{u_i}^* }_2  \norm{\M_{u_i}^* }_2  \norm{\bSigma_e }_2  \norm{\M_{u_j}\strans }_2 \norm{\W_{u_j}\strans\a_k^* }_2 \nonumber \\
	&\leq  c s_u d_k^{*2} d_i^{*-2} d_{i+1}^*   d_j^{*-2} d_{j+1}^*. \nonumber 
\end{align}

With the aid of \eqref{awmu}, we can deduce that 
\begin{align*}
	&| \a_k\trans\W_{u_i}\M_{u_i}\bSigma_e\M_{u_j}\trans\W_{u_j}\trans\a_k
 -  \a_k\strans \W_{u_i}^* \M_{u_i}^* \bSigma_e \M_{u_j}\strans \W_{u_j}\strans\a_k^* | \\[5pt]
 &\leq | \a_k\trans\W_{u_i}\M_{u_i}\bSigma_e(\M_{u_j}\trans\W_{u_j}\trans\a_k - \M_{u_j}\strans\W_{u_j}\strans\a_k^*)|  \\[5pt]
 &\quad +  |( \a_k\trans\W_{u_i}\M_{u_i} - \a_k\strans\W_{u_i}^*\M_{u_i}^*)\bSigma_e\M_{u_j}\strans\W_{u_j}\strans\wh{\bSigma} \a_k^*|  \\[5pt]
	&\leq  \|\a_k\trans\W_{u_i} \|_2 \|\M_{u_i}\|_2 \|\bSigma_e\|_2  \norm{\M_{u_j}\trans\W_{u_j}\trans\a_k - \M_{u_j}\strans\W_{u_j}\strans\a_k^*}_2 \\[5pt]
	&~ \qquad  +    \norm{ \a_k\trans\W_{u_i}\M_{u_i} - \a_k\strans\W_{u_i}^*\M_{u_i}^*}_2  \|\bSigma_e\|_2 \|\M_{u_j}\strans\|_2 \| \W_{u_j}\strans\a_k^*\|_2 \\[5pt]
	&\leq c s_u  \gamma_n  d_k^{*} d_{i+1}^* d_i^{*-2}  d_j^{*-2} \max\{d_k^*, d_{j+1}^{*}\}   + c s_u  \gamma_n  d_k^{*} d_{j+1}^* d_j^{*-2}   d_i^{*-2} \max\{d_k^*, d_{i+1}^{*}\}.
\end{align*}
Hence, it follows that 
\begin{align}\label{a31}
	A_{31} &\leq c s_u  \gamma_n  d_k^{*} d_{i+1}^* d_i^{*-1}  d_j^{*-1} \max\{d_k^*, d_{j+1}^{*}\}   + c s_u  \gamma_n  d_k^{*} d_{j+1}^* d_j^{*-1}   d_i^{*-1} \max\{d_k^*, d_{i+1}^{*}\} \nonumber \\
	&\quad  + c s_u \gamma_n    d_k^{*2} d_i^{*-2} d_{i+1}^*   d_j^{*-2} d_{j+1}^* \max\{  d_i^{*}, d_j^{*} \}.
\end{align}

For term $A_{32}$, it follows from \eqref{awaw} that 
	\begin{align}\label{a32}
		&|A_{32}| \leq     | \wt{\r}_i\trans\bSigma_e\wt{\r}_j | |\a_k\trans\W_{u_i}\wh{\bSigma}\W_{u_j}\trans\a_k -   \a_k\strans\W_{u_i}^*\wh{\bSigma}\W_{u_j}\strans\a^*_k| \nonumber \\
		&~ ~ + | \wt{\r}_i\trans\bSigma_e\wt{\r}_j - \r_i\strans\bSigma_e\r_j^* | |  \a_k\strans\W_{u_i}^*\wh{\bSigma}\W_{u_j}\strans\a^*_k| \nonumber\\
		&\leq c  (\|\a_k\trans\W_{u_i}\|_2 \|\wh{\bSigma}\W_{u_j}\trans\a_k -   \wh{\bSigma}\W_{u_j}\strans\a^*_k\|_2 + \|\a_k\trans\W_{u_i} -   \a_k\strans\W_{u_i}^*\|_2 \|\wh{\bSigma}\W_{u_j}\strans\a^*_k\|_2) \nonumber\\
		&\quad +  c \gamma_n \max\{|d_i^{*-1}|,|d_j^{*-1}|\} \|  \a_k\strans\W_{u_i}^*\|_2 \|\wh{\bSigma}\W_{u_j}\strans\a^*_k\|_2 \nonumber \\
  & \leq c s_u \gamma_n d_k^*  \max\{d_k^*d_i^{*-1},d_k^*d_j^{*-1}, 1 \}.  
	\end{align}

For term $A_{33}$, it holds that 
\begin{align*}
	|A_{33}| &\leq  |\a_k\trans\W_{u_i}\wh{\bSigma}\wt{\bmu}_j | |  \wt{\r}_i\trans\bSigma_e\M_{u_j}\trans\W_{u_j}\trans\a_k -  \r_i^* \bSigma_e \M_{u_j}\strans \W_{u_j}\strans  \a_k^*  | \\
	& ~ ~ +  |\a_k\trans\W_{u_i}\wh{\bSigma}\wt{\bmu}_j  -   \a_k\strans\W_{u_i}^*\wh{\bSigma} \bmu_j^*| | \r_i^* \bSigma_e \M_{u_j}\strans \W_{u_j}\strans  \a_k^*  |.
\end{align*}
Note that 
\begin{align*}
 &|\a_k\trans\W_{u_i}\wh{\bSigma}\wt{\bmu}_j | \leq \norm{\a_k\trans\W_{u_i}}_2 \norm{\wh{\bSigma}\wt{\bmu}_j}_2 \leq c s_u^{1/2} d_k^* d_j^*, \\
 &| \r_i^* \bSigma_e \M_{u_j}\strans \W_{u_j}\strans  \a_k^*  | \leq \norm{\r_i^*}_2 \norm{\bSigma_e }_2 \norm{\M_{u_j}\strans}_2  \norm{\W_{u_j}\strans  \a_k^*}_2  \leq c  s_u^{1/2} d_k^* d_j^{*-2} d_{j+1}^*.
\end{align*}
Further, by \eqref{awmu} we can show that 
\begin{align*}
	&|  \wt{\r}_i\trans\bSigma_e\M_{u_j}\trans\W_{u_j}\trans\a_k -  \r_i^* \bSigma_e \M_{u_j}\strans \W_{u_j}\strans  \a_k^*  | \\
 &\leq \|  \wt{\r}_i\trans \|_2 \|\bSigma_e \|_2 \| \M_{u_j}\trans\W_{u_j}\trans\a_k -  \M_{u_j}\strans \W_{u_j}\strans  \a_k^*  \|_2 + \|  \wt{\r}_i -  \r_i^* \|_2 \|\bSigma_e \|_2 \| \M_{u_j}\strans \W_{u_j}\strans  \a_k^*  \|_2 \\
	&\leq  c s_u^{1/2}  \gamma_n d_i^{*-2} \max\{d_k^*, d_{i+1}^{*}, d_j^{*-2} d_{j+1}^*  d_i^{*} d_k^* \}.
\end{align*}

In light of \eqref{awaw}, we have that 
\begin{align*}
	|\a_k\trans\W_{u_i}&\wh{\bSigma}\wt{\bmu}_j  -   \a_k\strans\W_{u_i}^*\wh{\bSigma} \bmu_j^*| \\&\leq
	  \|\a_k\trans\W_{u_i} \|_2  \|\wh{\bSigma}(\wt{\bmu}_j  -  \bmu_j^*)\|_2
	  + \|\a_k\trans\W_{u_i}  -   \a_k\strans\W_{u_i}^* \|_2  \|\wh{\bSigma} \bmu_j^*\|_2 \\
	  &\leq  c s_u^{1/2} \gamma_n  \max\{d_k^* d_j^* d_{i+1}^{*}d_i^{*-2}, d_j^*, d_k^* \}.
\end{align*}
Then it holds that 
\begin{align}\label{a333}
	A_{33} \leq &c s_u  \gamma_n d_k^* d_j^*  d_i^{*-2} \max\{d_k^*, d_{i+1}^{*}, d_j^{*-2} d_{j+1}^*  d_i^{*} d_k^* \} \nonumber \\
	& + c  s_u d_k^* d_j^{*-2} d_{j+1}^* \gamma_n  \max\{d_k^* d_j^* d_{i+1}^{*}d_i^{*-2}, d_j^*, d_k^* \}.
\end{align}
For term $A_{34}$, an application of similar arguments as for $A_{33}$ yields that 
\begin{align}\label{a444}
	A_{34} \leq &c s_u  \gamma_n d_k^* d_i^*  d_j^{*-2} \max\{d_k^*, d_{j+1}^{*}, d_i^{*-2} d_{i+1}^*  d_j^{*} d_k^* \} \nonumber \\
	& + c  s_u d_k^* d_i^{*-2} d_{i+1}^* \gamma_n  \max\{d_k^* d_i^* d_{j+1}^{*}d_j^{*-2}, d_i^*, d_k^* \}.
\end{align}

Combining \eqref{aaa444333}, \eqref{a31}, \eqref{a32}, \eqref{a333}, and \eqref{a444} gives that 
\begin{align*}
	|\wt{\operatorname{cov}} (  \wt{h}_{u_i}, \wt{h}_{u_j}) - \operatorname{cov} (h_{u_i},  h_{u_j} )| \leq c s_u  \gamma_n  d_1^*d_{r^*}^{*-2}d_k^{*2}.
\end{align*}
This along with \eqref{ome4}, \eqref{wwcovwwcov}, and  \eqref{phiijaa} leads to 
\begin{align}\label{varpart3}
	\sum_{i \neq k} \sum_{j \neq k} | \wt{\omega}_{k,i} \wt{\omega}_{k,j}  \wt{\operatorname{cov}} (  \wt{h}_{u_i}, \wt{h}_{u_j})  - \omega_{k,i} \omega_{k,j}  {\operatorname{cov}} (  h_{u_i}, h_{u_j})  | \leq c r^{*2} s_u  \gamma_n  d_1^*d_{r^*}^{*-2}.
\end{align}
Therefore, by \eqref{vareq3}, \eqref{varpart1}, \eqref{varpart2}, and \eqref{varpart3}, we can obtain that 
\begin{align*}
	|\wt{\nu}_k^2 - \nu_k^2| \leq c r^{*2} s_u  \gamma_n  d_1^*d_{r^*}^{*-2}.
\end{align*}
This concludes the proof of Theorem \ref{theo:weak:var}.

\subsection{Proof of Proposition \ref{prop:deri2}} \label{sec:proof:prop:deri2}

Under the constraint \eqref{SVDc}, we have  $\u_i\trans\u_i = 1$ for each $1 \leq i \leq r^*$. In view of Section H.2 in \cite{sofari}, for any matrix $\X \in \R^{n \times p}$ satisfying $\X\trans\X = \I_p$, it belongs to the Stiefel manifold $\text{St}(p,n) = \{\X \in \R^{n \times p}: \X\trans\X = \I_p \}$.  Then we see that all vectors $\u_i$ belong to the Stiefel manifold  $\text{St}(1, n) = \{ \u \in \mathbb{R}^n: \u\trans\u = 1   \}$. 
For function $\wt{\psi}_k$, denote by $\der{\wt{\psi}_k}{\u_i}$ with $1 \leq i \leq r^*$ the usual derivative vectors in the Euclidean space. Under the Stiefel manifold $\text{St}(1, n)$, an application of Lemma 30 in \cite{sofari} shows that the manifold gradient of $\wt{\psi}_k$ at $\u_i \in \text{St}(1, n) $ is given by $$(\I_n - \u_i{\u_i\trans})\der{\wt{\psi}_k}{\u_i}.$$ 

Moreover, for vectors $\v_j$ with $1 \leq j \leq r^*$ and $j \neq k$, it holds that 
$$\v_j\trans\v_j = d_j^2 \l_i\trans \wh{\bSigma}\l_i.$$ Since $d_j^2 \l_i\trans \wh{\bSigma}\l_i$ is unknown and its estimate varies across different estimation methods, there is no unit length constraint on ${\v}_i$ and we can take the gradient of $\wt{\psi}_k$ with respect to ${\v}_i$ directly in the Euclidean space $\mathbb{R}^q$ as $\der{\wt{\psi}_k}{\v_j}$. Recall that $\boldeta_k = \left(\u_1\trans,\ldots, \u_{r^*}\trans, \v_1\trans, \ldots, \v_{k-1}\trans, \v_{k+1}\trans, \ldots, \v_{r^*}\trans\right)\trans$. Therefore, the gradient of $\wt{\psi}_k$ on the manifold can be written as $$	\Q\big(\der{\wt{\psi}_k}{\boldeta_k}\big),$$ where $\Q = \diag{\I_n - \u_1\u_1\trans, \dots, \I_n - \u_{r^*}\u_{r^*}\trans, \I_{q(r^* - 1)}}$. This completes the proof of Proposition \ref{prop:deri2}.

\subsection{Proposition \ref{prop:strong1:psi} and its proof}\label{sec:proof:prop1}

\begin{proposition}\label{prop:strong1:psi}
   Under strongly orthogonal factors case of Section \ref{sec:vs}, for an arbitrary $\M^{(k)}$, it holds that
	\begin{align*}
	\wt{\psi}_k(\wt{\v}_k,\boldeta^*_k)
	& = ( \I_q - \M_k^u \wt{\u}_k\wt{\v}_k\trans + \M_k^u \sum_{i \neq k} \wt{\u}_i\wt{\v}_i\trans ) (\wt{\v}_k - \v_k^* ) \\
	&\quad ~ +   \sum_{j \neq k}  \M_j^v \sum_{i \neq j} \v_i^*\u_i\strans\u_j^* -  \sum_{i \neq k} \v_i^*\u_i\strans \u_k^*
	+ \bdelta_k + \bepsilon_k,
\end{align*}
where $\bepsilon_k =   n^{-1/2} \sum_{j \neq k} (\M_j^u \E\v_j^* + \M_j^v \E\trans\u_j^* ) + n^{-1/2}\M_k^u \E\wt{\v}_k - n^{-1/2}\E\trans\u_k^*$ and
\begin{align*}
	\bdelta_k =  \M_k^u ( \u_k^*\wt{\v}_k\trans -  \wt{\u}_k\wt{\v}_k\trans + \sum_{i \neq k} (\wt{\u}_i\wt{\v}_i\trans -  \u_i^*\v_i\strans)  )  (\v_k^* - \wt{\v}_k ).
\end{align*}
\end{proposition}

\noindent \textit{Proof}. Note that the loss function is 
\begin{align*}
	&L(\v_k,\boldeta_k) = (2n)^{-1}\norm{\Y - \sum_{i=1}^{r^*} \sqrt{n} \u_i\v_i\trans}_F^2.
\end{align*}
Under the orthogonality constraints $\v_i\trans\v_j=0$ for  $i, j \in \{1, \ldots, r^*\}$ and   $i \neq j$, it holds that
\begin{align*}
	\begin{aligned}
		L = (2 n)^{-1}\Big\{\|\mathbf{Y}\|_{F}^{2}
		&+2\langle\mathbf{Y},-\sqrt{n} \sum_{i \neq k} \u_i\v_i\trans \rangle
		+ n \boldsymbol{u}_{k}\trans \boldsymbol{u}_{k}\boldsymbol{v}_{k}\trans\boldsymbol{v}_{k}
		\\ &+\|\sum_{i \neq k} \sqrt{n} \u_i\v_i\trans\|_{F}^{2}
		-2 \sqrt{n} \boldsymbol{u}_{k}\trans\mathbf{Y} \boldsymbol{v}_{k}
		\Big\}.
	\end{aligned}
\end{align*}
After some calculations, we can obtain that 
\begin{align}
	& \der{L}{\u_k} = \u_k \v_k\trans\v_k - n^{-1/2}\Y\v_k,                                         \label{der1}       \\
	& \der{L}{\v_k} = \v_k\u_k\trans\u_k - n^{-1/2}\Y\trans\u_k. \label{der2} 
\end{align}
Similarly, for $j \neq k$, we also have that
\begin{align}
	& \der{L}{\u_j} = \u_j \v_j\trans\v_j - n^{-1/2}\Y\v_j,                \label{der3}        \\
	& \der{L}{\v_j} = \v_j\u_j\trans\u_j - n^{-1/2}\Y\trans\u_j. \label{der4}
\end{align}

Recall that $\boldeta_k^* = \left(\u_1\strans,  \ldots, \u_{r^*}\strans, \v_1\strans, \ldots, \v_{k-1}\strans, \v_{k+1}\strans, \v_{r^*}\strans\right)\trans$ and $\Y = \sqrt{n} \sum_{i = 1}^{r^*} \u_i^*\v_i\strans + \E.$
By some calculations with $\v_i\strans\v_j^* = 0 $ and $\u_i\strans\u_i^* = 1$, we can deduce that 
\begin{align*}
	 \der{L}{\v_j}\Big|_{\boldeta^*_k}
	&= \v_j^*\u_j\strans\u_j^* - \sum_{i = 1}^{r^*} \v_i^*\u_i\strans \u_j^*  - n^{-1/2}\E\trans\u_j^*
	 =  - \sum_{i \neq j} \v_i^*\u_i\strans \u_j^*  - n^{-1/2}\E\trans\u_j^*, \\
	 \der{L}{\v_k}\Big|_{\boldeta^*_k}
	 &= \v_k\u_k\strans\u_k^* - n^{-1/2}\Y\trans\u_k^* \\
	 &= \v_k\u_k\strans\u_k^* - \v_k^*\u_k\strans\u_k^* -  \sum_{i \neq k} \v_i^*\u_i\strans \u_k^* - n^{-1/2}\E\trans\u_k^* \\
	 &=   (\v_k - \v_k^*  )  -  \sum_{i \neq k} \v_i^*\u_i\strans \u_k^*  - n^{-1/2}\E\trans\u_k^*, \\
	  \der{L}{\u_j}\Big|_{\boldeta^*_k} & =  \u_j^* \v_j\strans\v_j^* - \sum_{i = 1}^{r^*} \u_i^*\v_i\strans \v_j^* - n^{-1/2}\E\v_j^* = - n^{-1/2}\E\v_j^*, 
	\end{align*}
\begin{align*}
	   \der{L}{\u_k}\Big|_{\boldeta^*_k} & =   \u_k^*\v_k\trans \v_k - n^{-1/2}\Y\v_k \\
	  &  =\u_k^*\v_k\trans\v_k - \u_k^*\v_k\strans\v_k - \sum_{i \neq k} \u_i^*\v_i\strans\v_k
	     - n^{-1/2}\E\v_k \\
	     & = \u_k^*\v_k\trans (\v_k - \v_k^*  ) - \sum_{i \neq k} \u_i^*\v_i\strans  (\v_k - \v_k^*  )    - n^{-1/2}\E\v_k \\
	     & = ( \u_k\v_k\trans - \sum_{i \neq k} \u_i\v_i\trans)  (\v_k - \v_k^*  )  +  \bdelta - n^{-1/2}\E\v_k,
\end{align*}
where $ \bdelta  =  ( \u_k^*\v_k\trans -  \u_k\v_k\trans + \sum_{i \neq k} (\u_i\v_i\trans -  \u_i^*\v_i\strans)  )  (\v_k - \v_k^*  )  $.

Since $\mathbf{M}=[\mathbf{M}_{1}^{u}, \ldots, \mathbf{M}_{r^*}^{u}, $ $\mathbf{M}_{1}^{v}, \ldots, \mathbf{M}_{k-1}^{v}, \mathbf{M}_{k+1}^{v}, \ldots,  \mathbf{M}_{r^*}^{v}]$, combining the above results yields that
\begin{align*}
	\wt{\psi}_k(\v_k,\boldeta^*_k)
	&= \der{L}{\v_k}\Big|_{\boldeta^*_k} - \M\der{L}{\boldeta_k}\Big|_{\boldeta^*_k} \nonumber \\
	&= \der{L}{\v_k}\Big|_{\boldeta^*_k} - \M_k^u\der{L}{\u_k}\Big|_{\boldeta^*_k} - \sum_{j \neq k} \M^u_j\der{L}{\u_j}\Big|_{\boldeta^*_k} - \sum_{j \neq k} \M_j^v\der{L}{\v_j}\Big|_{\boldeta^*_k}  \nonumber                 \\
	& = ( \I_q - \M_k^u \u_k\v_k\trans + \M_k^u \sum_{i \neq k} \u_i\v_i\trans ) (\v_k - \v_k^* ) \\
	&\quad ~ +   \sum_{j \neq k}  \M_j^v \sum_{i \neq j} \v_i^*\u_i\strans\u_j^* -  \sum_{i \neq k} \v_i^*\u_i\strans \u_k^*
	+ \bdelta_k + \bepsilon_k,
\end{align*}
where $\bepsilon_k =   n^{-1/2} \sum_{j \neq k} (\M_j^u \E\v_j^* + \M_j^v \E\trans\u_j^* ) + n^{-1/2}\M_k^u \E\v_k - n^{-1/2}\E\trans\u_k^*$ and 
\begin{align*}
	\bdelta_k =  \M_k^u ( \u_k^*\v_k\trans -  \u_k\v_k\trans + \sum_{i \neq k} (\u_i\v_i\trans -  \u_i^*\v_i\strans)  )  (\v_k^* - \v_k ).
\end{align*}
This concludes the proof of Proposition \ref{prop:strong1:psi}.

\subsection{Proposition \ref{prop:strong2:m} and its proof}\label{sec:proof:prop2}

\begin{proposition}\label{prop:strong2:m}
     Under strongly orthogonal factors case of Section \ref{sec:vs}, when the construction of $\M^{(k)}$ is given by 
\begin{align*}
    \M^u_k = - (\wt{\v}_k\trans\wt{\v}_k)^{-1} \wt{\V}_{-k} \wt{\U}_{-k}\trans,  \ \ \M_i^v = \0, \ \ \M_i^u = \0 \ \   \text{ for } 1 \leq i \leq r^* \ \text{and} \ i \neq k,
\end{align*}
    it holds that the value of $\big(\der{\wt{\psi}_k}{\boldeta_k\trans}\big)\Q$ at $(\wt{\v}_k, \wt{\boldeta}_k)$ is
    \begin{align*}
    \left(\der{\wt{\psi}_k}{\boldeta_k\trans}\right)\Q = \left(\derr{L}{\v_k}{\boldeta_k\trans} - \M^{(k)}\derr{L}{\boldeta_k}{\boldeta_k\trans}\right)\Q = \big(\0_{q \times n(k-1)}, \bDelta, \0_{q \times [n(r^* - k)+ q(r^* - 1)]}   \big),
    \end{align*}
    where
    $\bDelta = \left\{  n^{-1/2}(\wt{\C}-\C^*)\trans\X\trans - n^{-1/2}\E\trans\right\} (\I_n -  \wt{\u}_k\wt{\u}_k\trans )$.
\end{proposition}

\noindent \textit{Proof}. With derivatives \eqref{der1}--\eqref{der4},  for each $i, j \in \{1, \ldots, r^*\}$ with $i \neq j$ we have that
\begin{align*}
	&\derr{L}{\u_i}{\u_j\trans} = \0_{n\times n}, \ \derr{L}{\u_i}{\u_i\trans} = \v_i\trans\v_i \I_n, \\
	&\derr{L}{\u_i}{\v_j\trans} = \0_{n\times q}, \ \derr{L}{\u_i}{\v_i\trans} = 2 \u_i\v_i\trans -n^{-1/2}\Y, \\
	&\derr{L}{\v_i}{\u_j\trans} = \0_{q\times n}, \ \derr{L}{\v_i}{\u_i\trans} = 2\v_i\u_i\trans - n^{-1/2}\Y\trans, \\
	&\derr{L}{\v_i}{\v_j\trans} = \0_{q\times q}, \ \derr{L}{\v_i}{\v_i\trans} = \I_q.
\end{align*}
Observe that 
\begin{align*}
n^{-1/2} \Y & =  \sum_{i = 1}^{r^*}  \u_i\v_i\trans - \sum_{i = 1}^{r^*} ( \u_i\v_i\trans - \u_i^*\v_i\strans) + n^{-1/2} \E \\
& = 
\sum_{i = 1}^{r^*}  \u_i\v_i\trans - 
 n^{-1/2}\X(\C-\C^*) - n^{-1/2}\E.
\end{align*}
 
 Then the above derivatives can be written as
\begin{align*}
	\derr{L}{\u_i}{\u_j\trans} = \A^{uu}_{ij}  + \bDelta^{uu}_{ij}&, \ \derr{L}{\u_i}{\u_i\trans} = \A^{uu}_{ii}  + \bDelta^{uu}_{ii}, \\
	\derr{L}{\u_i}{\v_j\trans} = \A^{uv}_{ij}  + \bDelta^{uv}_{ij}&, \ \derr{L}{\u_i}{\v_i\trans} = \A^{uv}_{ii}  + \bDelta^{uv}_{ii}, \\
	\derr{L}{\v_i}{\u_j\trans} =  \A^{vu}_{ij}  + \bDelta^{vu}_{ij}&, \ \derr{L}{\v_i}{\u_i\trans} =  \A^{vu}_{ii}  + \bDelta^{vu}_{ii}, \\
	\derr{L}{\v_i}{\v_j\trans} =  \A^{vv}_{ij}  + \bDelta^{vv}_{ij}&, \ \derr{L}{\v_i}{\v_i\trans} =  \A^{vv}_{ii}  + \bDelta^{vv}_{ii},
\end{align*}
where
\begin{align*}
	&\A^{uu}_{ij} = \0_{n\times n}, \ \bDelta^{uu}_{ij} = \0_{n\times n}, \quad
	\A^{uu}_{ii} = \v_i\trans\v_i \I_n, \ \bDelta^{uu}_{ii} = \0_{n\times n}, \\
	& \A^{uv}_{ij} = \0_{n\times q}, \ \bDelta^{uv}_{ij}= \0_{n\times q},  \quad
	\A^{uv}_{ii}= \u_i\v_i\trans - \sum_{l \neq i}\u_l\v_l\trans, \ \bDelta^{uv}_{ii}=  \bdelta_{uv}, \\
	&  \A^{vu}_{ij} = \0_{q \times n}, \ \bDelta^{vu}_{ij}= \0_{q\times n}, \quad
	\A^{vu}_{ii} =  \v_i\u_i\trans - \sum_{l \neq i} \v_l\u_l\trans, \ \bDelta^{vu}_{ii} = \bdelta_{uv}\trans,   \\
	& \A^{vv}_{ij}  = \0_{q\times q}, \ \bDelta^{vv}_{ij}= \0_{q\times q}, \quad
	\A^{vv}_{ii} = \I_q, \ \bDelta^{vv}_{ii} = \0_{q\times q}
\end{align*}
with $\bdelta_{uv} = n^{-1/2}\X(\C-\C^*) - n^{-1/2}\E$.

We next calculate the term
\[ 	\left(\der{\wt{\psi}_k}{\boldeta_k\trans}\right)\Q = \left(\derr{L}{\v_k}{\boldeta_k\trans} - \M\derr{L}{\boldeta_k}{\boldeta_k\trans}\right)\Q,\]
where $\Q = \diag{\I_n - \u_1\u_1\trans, \dots, \I_n - \u_{r^*}\u_{r^*}\trans, \I_{q(r^* - 1)}}$.
It holds that
\begin{align*}
	\begin{aligned}
	\frac{\partial^{2} L}{\partial \boldsymbol{\eta}_k \partial \boldsymbol{\eta}_k^{T}} & =\left[\begin{array}{cc}
	\left(\mathbf{A}_{i j}^{u u}\right)_{\substack{1 \leq i \leq r^* \\
	1 \leq j \leq r^*}} & \left(\mathbf{A}_{i j}^{u v}\right)_{\substack{1 \leq i \leq r^* \\
	1 \leq j \leq r^*, j \neq k}} \\
	\left(\mathbf{A}_{i j}^{v u}\right)_{\substack{1 \leq i \leq r^*, i \neq k \\
	1 \leq j \leq r^*}} & \left(\mathbf{A}_{i j}^{v v}\right)_{\substack{1 \leq i \leq r^*, i \neq k \\
	1 \leq j \leq r^*, j \neq k}}
	\end{array}\right]
	\\[5pt]
 &\quad + \left[\begin{array}{cc}
	\left(\boldsymbol{\Delta}_{i j}^{u u}\right)_{\substack{1 \leq i \leq r^* \\
	1 \leq j \leq r^*}} & \left(\boldsymbol{\Delta}_{i j}^{u v}\right)_{\substack{1 \leq i \leq r^* \\
	1 \leq j \leq r^*, j \neq k}} \\
	\left(\boldsymbol{\Delta}_{i j}^{v u}\right)_{\substack{1 \leq i \leq r^*, i \neq k \\
	1 \leq j \leq r^*}} & \left(\boldsymbol{\Delta}_{i j}^{v v}\right)_{\substack{1 \leq i \leq r^*, i \neq k \\
	1 \leq j \leq r^*, j \neq k}}
	\end{array}\right],\\[5pt]
	\frac{\partial^{2} L}{\partial \boldsymbol{v}_{k} \partial \boldsymbol{\eta}_k^{T}} & =\left[\left(\mathbf{A}_{k j}^{v u}\right)_{1 \leq j \leq r^*},\left(\mathbf{A}_{k j}^{v v}\right)_{1 \leq j \leq r, j \neq k}\right]+\left[\left(\boldsymbol{\Delta}_{k j}^{v u}\right)_{1 \leq j \leq r^*},\left(\boldsymbol{\Delta}_{k j}^{v v}\right)_{1 \leq j \leq r^*, j \neq k}\right].
	\end{aligned}
\end{align*}

We aim to find matrix $\M$ that satisfies
\begin{align*}
	&\M \left[\begin{array}{cc}
	\left(\mathbf{A}_{i j}^{u u}\right)_{\substack{1 \leq i \leq r^* \\
	1 \leq j \leq r^*}} & \left(\mathbf{A}_{i j}^{u v}\right)_{\substack{1 \leq i \leq r^* \\
	1 \leq j \leq r^*, j \neq k}} \\
	\left(\mathbf{A}_{i j}^{v u}\right)_{\substack{1 \leq i \leq r^*, i \neq k \\
	1 \leq j \leq r^*}} & \left(\mathbf{A}_{i j}^{v v}\right)_{\substack{1 \leq i \leq r^*, i \neq k \\
	1 \leq j \leq r^*, j \neq k}}
	\end{array}\right] \Q   \\
 & = \left[\left(\mathbf{A}_{k j}^{v u}\right)_{1 \leq j \leq r^*},\left(\mathbf{A}_{k j}^{v v}\right)_{1 \leq j \leq r, j \neq k}\right] \Q.
\end{align*}
By some calculations, we can show that 
\begin{align*}
	& (\M^u_i  \v_i\trans\v_i  + \M^v_i ( \v_i\u_i\trans - \sum_{j \neq i} \v_j\u_j\trans ) )(\I_n  - \u_i\u_i\trans) = \0
	\ \  \text{for} \ i= 1, \ldots, r^* \ \text{with} \ i \neq k, \\
	& (  \M^u_i  ( \u_i\v_i\trans - \sum_{j \neq i} \u_j\v_j\trans ) + \M^v_i ) = \0
	\ \ \text{for} \ i= 1, \ldots, r^* \ \text{with} \ i \neq k, \\
	& (-\M^u_k \v_k\trans\v_k  + \v_k\u_k\trans - \sum_{j \neq k} \v_j\u_j\trans )(\I_n - \u_k\u_k\trans) = \0.
\end{align*}

Solving the above equations gives the choice of matrix $\M$ with
\begin{align*}
    \M^u_k = - (\v_k\trans\v_k)^{-1} \sum_{j \neq k} \v_j\u_j\trans, \   \M^u_i = \0,  \ \M^v_i = \0 \ \text{ for } \ i= 1, \ldots, r^* \ \text{with} \ i \neq k .
\end{align*}
Some further calculations lead to 
	\begin{align*}
		(\derr{L}{\v_k}{\boldeta\trans_k} - \M\derr{L}{\boldeta_k}{\boldeta\trans_k})\Q
		 = \big(\0_{q \times n(k-1)}, \bDelta, \0_{q \times [n(r^* - k)+ q(r^* - 1)]}   \big),
	\end{align*}
where $$\bDelta  =  \bDelta^{vu}_{kk} (\I_n - \u_k\u_k\trans) = \left\{  n^{-1/2} ({\C}-\C^*)\trans\X\trans - n^{-1/2}\E\trans \right\}  (\I_n - \u_k\u_k\trans).$$
This completes the proof of Proposition \ref{prop:strong2:m}.

\subsection{Proposition \ref{prop:strong3:w} and its proof}\label{sec:proof:prop3}

\begin{proposition}\label{prop:strong3:w}
 Under strongly orthogonal factors case of Section \ref{sec:vs}, when $\A =  (\wt{\v}_k\trans\wt{\v}_k) \I_{r^*-1} - \wt{\V}_{-k} \trans\wt{\V}_{-k} \wt{\U}_{-k}\trans \wt{\U}_{-k}$ is nonsingular, $\M^{(k)}$ is constructed as in Proposition \ref{prop:strong2:m}, and

	\begin{align*}
    \W_k =\I_q -  (\wt{\v}_k\trans\wt{\v}_k)^{-1} ( \I_{q} &+ \wt{\V}_{-k} \wt{\U}_{-k}\trans \wt{\U}_{-k} \A^{-1}\wt{\V}_{-k}\trans )  \wt{\V}_{-k} \wt{\U}_{-k}\trans  \wt{\u}_k \wt{\v}_k\trans\\
    &+ \wt{\V}_{-k} \wt{\U}_{-k}\trans \wt{\U}_{-k} \A^{-1}\wt{\V}_{-k}\trans,
\end{align*}
it holds that
$ \W_k ( \I_q - \M_k^u \wt{\u}_k\wt{\v}_k\trans + \M_k^u \sum_{i \neq k} \wt{\u}_i\wt{\v}_i\trans) = \I_q. $
\end{proposition}

\noindent \textit{Proof}. It is easy to see that the existence of $\W_k$ depends on the nonsingularity of $ \I_q - \M_k^u \u_k\v_k\trans + \M_k^u \sum_{i \neq k} \u_i\v_i\trans $. By the construction of $\M_k^u$ in Proposition \ref{prop:strong2:m}, we have that
\begin{align*}
	\I_q & + \M_k^u(- \u_k\v_k\trans +  \sum_{i \neq k} \u_i\v_i\trans ) =  \I_q  - (\v_k\trans\v_k)^{-1} \sum_{j \neq k} \v_j\u_j\trans (- \u_k\v_k\trans +  \sum_{i \neq k} \u_i\v_i\trans ) \\
	& =  \I_q +   (\v_k\trans\v_k)^{-1} (  \sum_{j \neq k} \v_j\u_j\trans \u_k \v_k\trans -   \sum_{j \neq k} \v_j\u_j\trans \cdot \sum_{i \neq k} \u_i\v_i\trans ) \\[5pt]
	& = \I_q + (\v_k\trans\v_k)^{-1}  [ \sum_{j \neq k} \v_j\u_j\trans \u_k   ,  - \sum_{j \neq k} \v_j\u_j\trans \U_{-k}  ]
	\left[\begin{array}{l}
		\boldsymbol{v}_{k}\trans\\
		 \V_{-k}\trans
	\end{array}\right] \\
	& =: \I_q  + \B_1 \B_2\trans,
\end{align*}
where $$\B_1 =    (\v_k\trans\v_k)^{-1}  [ \sum_{j \neq k} \v_j\u_j\trans \u_k   ,  - \sum_{j \neq k} \v_j\u_j\trans \U_{-k}  ] \in \mathbb{R}^{q \times r^*},   \B_2\trans =  \left[\begin{array}{l}
	\boldsymbol{v}_{k}\trans\\
 \V_{-k}\trans
\end{array}\right] \in \mathbb{R}^{r^* \times q}.$$

In light of $\v_i\strans\v_j^* = 0 $ for $i \neq j$, it holds that
\begin{align*}
	\I_{r^*} &+ \B_2\trans \B_1 =
	\I_{r^*} + (\v_k\trans\v_k)^{-1} \left[\begin{array}{l}
			\boldsymbol{v}_{k}\trans\\
		 \V_{-k}\trans
	\end{array}\right]  [ \sum_{j \neq k} \v_j\u_j\trans \u_k   ,  - \sum_{j \neq k} \v_j\u_j\trans \U_{-k}  ]
	\nonumber \\
 & =  \left[\begin{array}{cc}
		1 & \mathbf{0}\\
		(\v_k\trans\v_k)^{-1}  \V_{-k}\trans\V_{-k} \U_{-k}\trans\u_k  & \I_{r^*-1} - (\v_k\trans\v_k)^{-1}\V_{-k} \trans\V_{-k} \U_{-k}\trans \U_{-k}
	\end{array}\right].
\end{align*}
Under the assumption that $\A =  (\v_k\trans\v_k) \I_{r^*-1} - \V_{-k} \trans\V_{-k} \U_{-k}\trans \U_{-k}$ is nonsingular, we can see that the above matrix is also nonsingular. 

Denote by $\A_0 = (\v_k\trans\v_k)^{-1} \A $. By some calculations, we can show that 
\begin{align*}
	\begin{aligned}
	&(\I_{r^*} + \B_2\trans \B_1)^{-1} \\
	=	&  \left[\begin{array}{cc}
		1 & \mathbf{0}\\
		(\v_k\trans\v_k)^{-1}  \V_{-k}\trans\V_{-k} \U_{-k}\trans\u_k  & \I_{r^*-1} - (\v_k\trans\v_k)^{-1}\V_{-k} \trans\V_{-k} \U_{-k}\trans \U_{-k}
	\end{array}\right]^{-1} \\[5pt]
		&= \left(\begin{array}{cc}
			1 & 0\\
			-\mathbf{A}_0^{-1} (\v_k\trans\v_k)^{-1}  \V_{-k}\trans\V_{-k} \U_{-k}\trans\u_k  & \mathbf{A}_0^{-1}
		\end{array}\right).
	\end{aligned}
\end{align*}
It follows from the Sherman--Morrison--Woodbury formula that $(\I_q + \B_1\B_2\trans)^{-1} = \I_q - \B_1 (\I_{r^*} + \B_2\trans\B_1)^{-1} \B_2\trans  $. Therefore, we can set $\W_k$ as
\begin{align*}
    &\W_k = \I_q - \B_1 (\I_{r^*} + \B_2\trans\B_1)^{-1} \B_2\trans \\
    &=\I_q -  (\v_k\trans\v_k)^{-1} ( \I_q + \V_{-k} \U_{-k}\trans \U_{-k} \A^{-1} \V_{-k}\trans )  \V_{-k} \U_{-k}\trans  \u_k \v_k\trans \\
    &\quad + \V_{-k} \U_{-k}\trans \U_{-k} \A^{-1}\V_{-k}\trans.
\end{align*}
This concludes the proof of Proposition \ref{prop:strong3:w}.

\subsection{Proof of Proposition \ref{prop:weak1:psi}}\label{sec:proof:prop4}

In view of the loss function in \eqref{loss:weak}, it can be simplified as
\begin{align}
	L &= (2n)^{-1}\Big\{\norm{\Y}_F^2 + ( \sqrt{n}\u_k)\trans \sqrt{n}\u_k\v_k\trans\v_k + \sum_{i \neq k}\sqrt{n}(\wh{\u}_i^t)\trans \sqrt{n}\wh{\u}_i^t\wt{\v}_i\trans \wt{\v}_i \nonumber\\
&~ ~ - 2\sqrt{n}\u_k\trans\Y\v_k - \sum_{i \neq k}2\sqrt{n}(\wh{\u}_i^t)\trans\Y \wt{\v}_i\ + \sum_{i \neq k} 2 \sqrt{n} \u_k\trans \sqrt{n}\wh{\u}_i^t \v_k\trans \wt{\v}_i \Big\} \nonumber\\
 &  = (2n)^{-1}\norm{\Y}_F^2 + 2^{-1} \u_k\trans\u_k\v_k\trans\v_k + \sum_{i \neq k} 2^{-1} (\wh{\u}_i^t)\trans \wh{\u}_i^t\wt{\v}_i\trans \wt{\v}_i
	- \u_k\trans n^{-1/2}\Y\v_k \nonumber \\
 &\quad - (\wh{\u}_i^t)\trans n^{-1/2}\Y\wt{\v}_i  + \sum_{i \neq k}\u_k\trans\wh{\u}_i^t \v_k\trans \wt{\v}_i. \nonumber 
\end{align}
Some calculations yield that 
\begin{align}
	& \der{L}{\u_k} =  \u_k \v_k\trans\v_k  - n^{-1/2}\Y\v_k  + \sum_{i \neq k} \wh{\u}_i^t \wt{\v}_i\trans\v_k,                                              \label{1derisadq}      \\
	& \der{L}{\v_k} = \v_k\u_k\trans\u_k - n^{-1/2}\Y\trans\u_k + \sum_{i \neq k} \wt{\v}_i \u_k\trans\wh{\u}_i^t. \label{2derisadq2}
\end{align}

Notice that  $\boldeta_k = \u_k$ and $\boldeta_k^* = \u_k^*$. Using $\wt{\v}_i\trans\v_k = 0$ and $\v_i\strans\v_k^* = 0$, we can deduce that 
\begin{align*}
	\der{L}{\u_k}\Big|_{\boldeta_k^*}
	& = \u_k^* \v_k\trans\v_k  -  \u_k^* \v_k\trans\v_k^* -  \sum_{i \neq k} \u_i^* \v_i\strans\v_k - n^{-1/2}\E\v_k\\
	& =  \u_k^* \v_k\trans (\v_k - \v_k^*) - \sum_{i \neq k} \u_i^* \v_i\strans (\v_k - \v_k^* + \v_k^*) - n^{-1/2}\E\v_k \\
	& = (\u_k^* \v_k\trans -   \sum_{i \neq k} \u_i^* \v_i\strans) (\v_k - \v_k^*)  - n^{-1/2}\E\v_k \\
	& = {  ( \u_k \v_k\trans -  \sum_{i \neq k} \u_i \v_i\trans) (\v_k - \v_k^*) }  + \bdelta - n^{-1/2}\E\v_k,
\end{align*}
where
$\bdelta = ( \u_k^* \v_k\trans - \u_k\v_k\trans) (\v_k - \v_k^*) +   \sum_{i \neq k} (\u_i \v_i\trans -\u_i^* \v_i\strans)  (\v_k - \v_k^*).$
Further, it follows from $\u_k\strans\u_k^* = 1$ that
\begin{align*}
	\der{L}{\v_k}\Big|_{\boldeta^*_k}
	& =  \v_k\u_k\strans\u_k^* -   \sum_{i \neq k}\v_i^*\u_i\strans\u_k^*  - \v_k^*\u_k\strans\u_k^* - n^{-1/2}\E\trans\u_k^* +  \sum_{i \neq k} \wt{\v}_i \u_k\strans\wh{\u}_i^t \\
	& = (\v_k - \v_k^*)  + \sum_{i \neq k}(\wt{\v}_i  (\wh{\u}_i^t)\trans  - \v_i^*\u_i\strans)\u_k^*  - n^{-1/2}\E\trans\u_k^*        \\
	& =   (\v_k - \v_k^*)  + \sum_{i \neq k} \wt{\v}_i (\wh{\u}_i^t  - \u_i^*)\trans\u_k^* +  \sum_{i \neq k} (\wt{\v}_i  - \v_i^*)\u_i\strans\u_k^*  - n^{-1/2}\E\trans\u_k^*.
\end{align*}

Combining the above two terms and plugging in the SOFAR estimates $(\wt{\u}_k, \wt{\v}_k)$ lead to
\begin{align*}
	\wt{\psi}_k(\wt{\v}_k,\boldeta^*_k)
	&= \Big(\der{L}{\v_k}\Big|_{\boldeta^*_k} - \M_k \der{L}{\boldeta_k}\Big|_{\boldeta^*_k}\Big)\Big|_{(\wt{\u}_k, \wt{\v}_k)} \nonumber \\[5pt]
	& = ( \I_q - \M_k \wt{\u}_k \wt{\v}_k\trans + \M_k\sum_{i \neq k} \wt{\u}_i \wt{\v}_i\trans ) (\wt{\v}_k - \v_k^* )   \\[5pt]
	&\quad + \sum_{i \neq k} \wt{\v}_i (\wh{\u}_i  - \u_i^*)\trans\u_k^* + \sum_{i \neq k} (\wt{\v}_i  - \v_i^*)\u_i\strans\u_k^* 	+ \bdelta_k + \bepsilon_k,
\end{align*}
where $\bepsilon_k =   - n^{-1/2}\E\trans\u_k^* + n^{-1/2} \M_k\E\trans\wt{\v}_k$ and 
\begin{align*}
	\bdelta_k = - \M_k  ( \u_k^* \v_k\strans -\wt{\u}_k\wt{\v}_k\trans + \sum_{i \neq k}( \wt{\u}_i \wt{\v}_i\trans -\u_i^* \v_i\strans))  (\wt{\v}_k - \v_k^*).
\end{align*}
This completes the proof of Proposition \ref{prop:weak1:psi}.
	
\subsection{Proof of Proposition \ref{prop:weak2:m}}\label{sec:proof:prop5}

For the derivatives in \eqref{1derisadq} and \eqref{2derisadq2}, let us first simplify them using $\wt{\v}_i\trans\v_k = 0$ and $\u_k\trans\u_k = 1$. Subsequently, by taking derivatives with respect to $\boldeta_k = \u_k$, it holds that
	\begin{align*}
		\derr{L}{\v_k}{\boldeta_k\trans} &=  - n^{-1/2}\Y\trans + \sum_{i \neq k} \wt{\v}_i(\wh{\u}_i^t)\trans  \\
		&= -\v_k\u_k\trans + \Big[ (\v_k\u_k\trans- \v_k^*\u_k\strans)  + \sum_{i \neq k} (\wt{\v}_i(\wh{\u}_i^t)\trans - \v_i^*\u_i\strans) - n^{-1/2}\E\trans \Big], \\
		\derr{L}{\boldeta_k}{\boldeta_k\trans} &= \v_k\trans\v_k \I_n.
	\end{align*}
    
    With the initial SOFAR estimates $(\wt{\u}_k, \wt{\v}_k)$, we aim to find matrix $\M_k$ satisfying
$$ -\wt{\v}_k\wt{\u}_k\trans -  \M_k \wt{\v}_k\trans\wt{\v}_k  = \0.$$
 It is clear that the choice of $\M_k$ given by 
\begin{align*}
	 \M_k = -(\wt{\v}_k\trans\wt{\v}_k)^{-1} \wt{\v}_k\wt{\u}_k\trans
\end{align*}
satisfies the above equation.
Thus, we can obtain that
	\[  \Big(\derr{L}{\v_k}{\boldeta_k\trans} - \M_k \derr{L}{\boldeta_k}{\boldeta_k\trans}\Big)\Big|_{(\wt{\u}_k, \wt{\v}_k)} = \wt{\v}_k\wt{\u}_k\trans- \v_k^*\u_k\strans  + \sum_{i \neq k}(\wt{\v}_i(\wh{\u}_i^t)\trans - \v_i^*\u_i\strans) - n^{-1/2}\E\trans. \]
This concludes the proof of Proposition \ref{prop:weak2:m}.

\subsection{Proof of Proposition \ref{prop:weak3:w}}\label{sec:proof:prop6}

With $ \M_k = -(\wt{\v}_k\trans\wt{\v}_k)^{-1} \wt{\v}_k\wt{\u}_k\trans$ given in Proposition \ref{prop:weak2:m} and $\wt{\u}_k\trans\wt{\u}_k = 1$, we can deduce that
\begin{align*}
    \I_q - \M_k \wt{\u}_k \wt{\v}_k\trans + \M_k\sum_{i \neq k} \wt{\u}_i \wt{\v}_i\trans   &=	  \I_q + (\wt{\v}_k\trans\wt{\v}_k)^{-1} \wt{\v}_k \wt{\v}_k\trans -(\wt{\v}_k\trans\wt{\v}_k)^{-1} \wt{\v}_k \wt{\u}_k\trans \sum_{i \neq k}\wt{\u}_i \wt{\v}_i\trans \\
		& = 	\I_q + (\wt{\v}_k\trans\wt{\v}_k)^{-1}  [  \wt{\v}_k, -\wt{\v}_k\wt{\u}_k\trans\wt{\U}_{-k} ] \left[\begin{array}{l}
			\wt{\v}_{k}\trans\\
			\wt{\V}_{-k}\trans
		\end{array}\right]  \\
		& =: \I_q + \B_1\B_2\trans,
	\end{align*}
	where $$\B_1 = (\wt{\v}_k\trans\wt{\v}_k)^{-1}  [  \wt{\v}_k, -\wt{\v}_k\wt{\u}_k\trans\wt{\U}_{-k} ]  \in \R^{q \times r^*}, \B_2\trans =  \left[\begin{array}{l}
		\wt{\v}_{k}\trans\\
		\wt{\V}_{-k}\trans
	\end{array}\right] \in \R^{r^* \times q}.$$
    
	By some calculations, it holds that 
	\begin{align*}
		&\I_{r^*} + \B_2\trans \B_1 =
		\I_{r^*} +    \left[\begin{array}{l}
			\wt{\v}_{k}\trans\\
			\wt{\V}_{-k}\trans
		\end{array}\right]  [ (\wt{\v}_k\trans\wt{\v}_k)^{-1} \wt{\v}_{k}, - (\wt{\v}_k\trans\wt{\v}_k)^{-1}\wt{\v}_{k}\wt{\u}_k\trans\wt{\U}_{-k} ] 
		\\
		&=  \left[\begin{array}{cc}
			2 & -\wt{\u}_k\trans\wt{\U}_{-k} \\
			(\wt{\v}_k\trans\wt{\v}_k)^{-1} \wt{\V}_{-k}\trans\wt{\v}_{k} & \I_{r^*-1} - (\wt{\v}_k\trans\wt{\v}_k)^{-1}\wt{\V}_{-k}\trans\wt{\v}_{k}\wt{\u}_k\trans\wt{\U}_{-k}
		\end{array}\right] = \left[\begin{array}{cc}
			2 & -\wt{\u}_k\trans\wt{\U}_{-k} \\
			0 & \I_{r^*-1} 
		\end{array}\right],
	\end{align*}
 where the last step above is due to the fact of $\wt{\V}_{-k}\trans\wt{\v}_{k} = \mathbf{0}$. Hence, it is seen that matrix $\I_{r^*} + \B_2\trans \B_1$ is nonsingular. In addition, we can show that 
\begin{align*}
	\begin{aligned}
	(\I_{r^*} + \B_2\trans\B_1)^{-1} =	&  \left(\begin{array}{cc}
			2 & -\wt{\u}_k\trans\wt{\U}_{-k} \\
			0 & \I_{r^*-1}
		\end{array}\right)^{-1} 
		= \left(\begin{array}{cc}
			2^{-1} & 2^{-1} \wt{\u}_k\trans\wt{\U}_{-k}  \\
			0 & \I_{r^*-1}
		\end{array}\right).
	\end{aligned}
\end{align*}

Therefore, an application of the Sherman--Morrison--Woodbury formula yields that $(\I_q + \B_1\B_2\trans)^{-1} = \I_q - \B_1 (\I_{r^*} + \B_2\trans\B_1)^{-1} \B_2\trans  $ and consequently, \begin{align*}
	\W_k &= (  \I_q - \M_k \wt{\u}_k \wt{\v}_k\trans + \M_k\sum_{i \neq k} \wt{\u}_i \wt{\v}_i\trans )^{-1} = ( \I_q + \B_1 \B_2\trans)^{-1} \\
	&=   \I_q - 2^{-1} (\wt{\v}_k\trans\wt{\v}_k)^{-1} (  \wt{\v}_k \wt{\v}_k\trans -  \wt{\v}_k \wt{\u}_k\trans  \wt{\U}_{-k} \wt{\V}_{-k}\trans  ).
\end{align*}
This completes the proof of Proposition \ref{prop:weak3:w}.

\section{Explicit formulas in Theorem \ref{theo:weak:uk}} \label{app:variance}

In this section, we present explicit formulas of the distribution term and the associated variance in Theorem \ref{theo:weak:uk}. First, it is useful to provide the asymptotic distribution of $\wh{\bmu}_k$ derived in \cite{sofari}.
Denote by 
\begin{align*}
	\M_{\mu_k}^{*} & = -( \bmu_k\strans \wh{\bSigma}  \bmu_k^*)^{-1} \wh{\bSigma}\sum_{j = k+1}^{r^*}\bmu_j^*\r_j\strans, \\
	\W_{\mu_k}^{*} & = \widehat{\bTheta} \left\{  \I_p + ( \bmu_k\strans \wh{\bSigma}  \bmu_k^*)^{-1}  \wh{\bSigma}\L^{*(2)}_{d}(\I_{r^*-k} - ( \bmu_k\strans \wh{\bSigma}  \bmu_k^*)^{-1} (\L^{*(2)}_{d})\trans\wh{\bSigma} \L^{*(2)}_{d})^{-1}\right. \\
    &\quad \left.\cdot (\L^{*(2)}_{d})\trans\right\},
\end{align*}
where  $\L^{*(2)}_{d} = [d_{k+1}^* \l_{k+1}^*, \ldots, d_{r^*} \l_{r^*}]$. Further, let us define 
\begin{align*}
    \kappa_n^{(k)} =  \kappa_n^{\prime} \max\big\{1, d_k^{*-1}, d_k^{*-2}\big\}  + \gamma_n d_k^{*-3} d_{k+1}^{*} \big(\sum_{i=1}^{k-1}d_i^*\big)
\end{align*}
with $d_{r^*+1}^* = 0$.
The lemma below characterizes the asymptotic distribution of $\wh{\bmu}_k$.
\begin{lemma}[Theorem 3, \citet{sofari}]\label{lemm:sofari}
    Assume that Conditions \ref{con1:error}--\ref{con3:eigend} and \ref{con:weak:orth}  hold, and  $\wt{\C}$, $\wh{\bTheta}$ satisfy Definitions \ref{defi:sofar} and \ref{defi:theta}, respectively. Then for each given $k$ with $1 \leq k \leq r^*$ and an arbitrary vector
	$\a\in\mathcal{A}=\{\a\in\R^p:\norm{\a}_0\leq m,\norm{\a}_2=1\}$ satisfying $m^{1/2}\kappa_n^{(k)} = o(1) $,
	we have that 
	\begin{align*}
		\sqrt{n}\a\trans(\wh{\bmu}_k-\bmu_k^*) = h_{u_k} + t_{u_k},
	\end{align*}
	where the distribution term $h_{u_k} = \a\trans \W^{*}_{u_k}(\X\trans\E\r_k^* - \M_{u_k}^{*} \E\trans\X {\bmu}_k^{*})/\sqrt{n} \sim \N(0,\nu_{u_k}^2)$ with
	\begin{align*}
    \nu_{u_k}^2 = \a\trans\W_{u_k}^*(z_{kk}^{*}\M_{u_k}^{*}\bSigma_e\M_{u_k}\strans + \r_k\strans\bSigma_e\r_k^* \wh{\bSigma} - 2\wh{\bSigma}\bm_k^*  \r_k\strans\bSigma_e\M_{u_k}^{* T})\W_{u_k}\strans\a.
	\end{align*}
	Moreover, the error term $ t_{u_k} = O_p(m^{1/2}\kappa_n^{(k)} )$
 holds with probability at least $1 - \theta_{n,p,q}$ for $\theta_{n,p,q}$ given in \eqref{thetapro}.
\end{lemma}

Corresponding to matrices $\M_k$ and $\W_k$ suggested for SOFARI-R in Section \ref{sec:vw}, we define 
\begin{align*}
	&\M_{k}^* =  \M_{v_k}^* n^{-1/2} \X\trans \text{ with } \    \M_{v_k}^* = -  ( d_k^{*} \l_k\strans \wh{\bSigma}\l_k^*)^{-1} {\r}_k^*{\l}_k\strans, \\ 
	&\W_k^* =   \I_q - 2^{-1} (\v_k\strans\v_k^*)^{-1} (  \v_k^* \v_k\strans -  \v_k^* \u_k\strans  \U_{-k}^*  \V_{-k}\strans  ).
\end{align*}
In addition, denote by $ \omega_{k,i} =   ( \bmu_k\strans \wh{\bSigma}  \bmu_k^*)^{-1/2} \a\trans{\W}_k^* {\r}_i^*.$
For a vector $\b = (b_1, \ldots, b_p)\trans \in \mathbb{R}^p$, let us define  $\b^{t_i} = (b_1^{t_i}, \ldots, b_p^{t_i})\trans \in \mathbb{R}^p$ with $b_j^{t_i} = b_j$ if $j \in \mathcal{S}_{\mu_i}$ and $b_j^{t_i} = 0$ otherwise. 
Then the distribution term $h_k$ in Theorem \ref{theo:weak:uk} is 
\begin{align*}
	h_k = h_{v_k} + \sum_{i \neq k} h_{u_i}^{\prime} =  h_{v_k} -  \sum_{i \neq k} \omega_{k,i} h_{u_i}  \sim N(0, \nu_k^2)
\end{align*}
with 
\begin{align*}
	&h_{v_k} =  \a\trans{\W}_k^* ( \M_{v_k}^* \X\trans \E d_k^* (\l_k\strans \wh{\bSigma}\l_k^*)^{1/2} \r_k^*-   \E\trans \X  (\l_k\strans \wh{\bSigma}\l_k^*)^{-1/2} \l_k^*  )/\sqrt{n}, \\
	&h_{u_i} =  ((\wh{\bSigma} \bmu_k^*)^{t_i})\trans \W^{*}_{u_i}(\X\trans\E\r_i^* - \M_{u_i}^{*} \E\trans\X d_i^* {\l}_i^{*})/\sqrt{n}.
\end{align*}
The variance is given by 
\begin{align}\label{eq:variance}
	\nu_k^2 =  \operatorname{cov} (h_{v_k},  h_{v_k} ) - 2 \sum_{i \neq k}  \omega_{k,i} \operatorname{cov} (h_{v_k},  h_{u_i} )   +  \sum_{i \neq k} \sum_{j \neq k} \omega_{k,i} \omega_{k,j}  \operatorname{cov} (  h_{u_i}, h_{u_j}).
\end{align}

Specifically, we have that 
\begin{align*}
	&\operatorname{cov} (h_{v_k},  h_{v_k} ) \\&\ \ = \a\trans\W_k^*( \bSigma_e + d_k^{*2} (\l_k\strans \wh{\bSigma}\l_k^*) \r_k\strans\bSigma_e\r_k^* \M_{v_k}^* \wh{\bSigma} \M_{v_k}\strans  - 2\M_{v_k}^* \wh{\bSigma} d_k^*  \l_k^* \r_k\strans \bSigma_e)\W_k\strans\a.
\end{align*}
Moreover, it holds that 
\begin{align*}
\operatorname{cov} (h_{v_k},  h_{u_i} ) & =     d_i^* (\l_k\strans \wh{\bSigma}\l_k^*)^{-1/2} \l_i\strans \wh{\bSigma} \l_k^* \cdot \a\trans \W_k^* \bSigma_e \M_{u_i}\strans \W_{u_i}\strans(\wh{\bSigma} \bmu_k^*)^{t_i} \\[5pt]
&\quad -   \a\trans \W_k^* \M_{v_k}^* \wh{\bSigma}  d_i^* \l_i^* \cdot d_k^* (\l_k\strans \wh{\bSigma}\l_k^*)^{1/2} \r_k\strans \bSigma_e \M_{u_i}\strans \W_{u_i}\strans(\wh{\bSigma} \bmu_k^*)^{t_i} \\[5pt]
&\quad  - ((\wh{\bSigma} \bmu_k^*)^{t_i})\trans \W_{u_i}^* \wh{\bSigma} (\l_k\strans \wh{\bSigma}\l_k^*)^{-1/2} \l_k^* \cdot \r_i\strans \bSigma_e \W_k\strans\a \\[5pt]
&\quad + \a\trans\W_k^* \M_{v_k}^* \wh{\bSigma} \W_{u_i}\strans(\wh{\bSigma} \bmu_k^*)^{t_i} \cdot d_k^* (\l_k\strans \wh{\bSigma}\l_k^*)^{1/2} \r_k\strans \bSigma_e \r_i^*
\end{align*}
and 
\begin{align*}
   \operatorname{cov} (h_{u_i},  h_{u_j} )
   & =     d_i^* d_j^* \l_i\strans \wh{\bSigma} \l_j^* \cdot ((\wh{\bSigma} \bmu_k^*)^{t_i})\trans \W_{u_i}^* \M_{u_i}^* \bSigma_e \M_{u_j}\strans \W_{u_j}\strans(\wh{\bSigma} \bmu_k^*)^{t_j} \\ 
    &\quad -   ((\wh{\bSigma} \bmu_k^*)^{t_i})\trans \W_{u_i}^* \wh{\bSigma}  d_j^* \l_j^* \cdot \r_i\strans \bSigma_e \M_{u_j}\strans \W_{u_j}\strans(\wh{\bSigma} \bmu_k^*)^{t_j} \\
    &\quad  - ((\wh{\bSigma} \bmu_k^*)^{t_j})\trans \W_{u_j}^* \wh{\bSigma} d_i^* \l_i^* \cdot \r_j\strans \bSigma_e 
    \M_{u_i}\strans \W_{u_i}\strans(\wh{\bSigma} \bmu_k^*)^{t_i} \\ 
    & \quad + ((\wh{\bSigma} \bmu_k^*)^{t_i})\trans\W_{u_i}^*  \wh{\bSigma} \W_{u_j}\strans(\wh{\bSigma} \bmu_k^*)^{t_j} \cdot  \r_i\strans \bSigma_e \r_j^*.
\end{align*}

\section{Some key lemmas and their proofs} \label{new.Sec.lemma}

\subsection{Proof of Lemma \ref{lemm:taylorrank2:2}}\label{new:sec:b1}

For the nuisance parameter  $\boldeta_k = [  \u_1\trans,\ldots,  \u_{r^*}\trans, \v_1\trans, \ldots, \v_{k-1}\trans, \v_{k+1}\trans, \ldots, \v_{r^*}\trans]\trans$,
it follows from the definition of  $\wt{\psi}_k(\v_k,\boldeta_k)$ that 
\begin{align}\label{eqphikk}
	\wt{\psi}_k(\v_k,\boldeta_k)
	& = \der{L}{\v_k} - \M\der{L}{\boldeta_k} \nonumber\\
 &= \der{L}{\v_k} - \left(\M_k^u \der{L}{\u_k}+ \sum_{j \neq k}  \M_j^u \der{L}{\u_j} + \sum_{j \neq k}\M_j^v \der{L}{\v_j}\right).
\end{align}
From Proposition \ref{prop:strong2:m}, we have $\M_j^u = \0$ and $  \M_j^v = \0$ for $j \in \{ 1, \ldots, r^*\}$ with $j \neq k$. This implies that we need only to consider $\u_k$ as the nuisance parameter. By the derivatives \eqref{der1}, \eqref{der2}, and $\u_k\trans\u_k = 1$, it holds that 
\begin{align}
	\wt{\psi}_k(\v_k,\boldeta_k)
	 &= \der{L}{\v_k} - \M_k^u \der{L}{\u_k} \nonumber\\[5pt] 
       & = \v_k - n^{-1/2}\Y\trans\u_k - \M_k^u ( \u_k \v_k\trans\v_k - n^{-1/2}\Y\v_k ). \label{t11}
\end{align}

Since $\wt{\psi}_k(\u_k, \boldeta_k)$ depends only on $\u_k$ and $\v_k$ for any given $\M_k^u$, we need only to conduct the Taylor expansion of $\wt{\psi}_k(\u_k, \boldeta_k)$  with respect to $\v_k$. Since ${\u}_k$ and $\v_k^*$ belong to set $\{ \u \in \mathbb{R}^n: \u\trans\u = 1   \}$, we have that ${\u}_k, \u_k^* \in \text{St}(1, n)$ by the definition of the Stiefel manifold. Similar to the proof of Lemma 4 in \cite{sofari}, we can obtain the Taylor expansion of $\wt{\psi}_k(\v_k, \boldeta_k)$ that 
\begin{align*}
    \wt{\psi}_k&\left( \v_k,  {\boldeta}_k  \right) = \wt{\psi}_k\left( \v_k, \boldeta_k^* \right)
    +  \der{\wt{\psi}_k( \v_k, {\boldeta}_k  )}{{\u}_k\trans}\Big|_{\u_k^*}(\I_n - \u_k^*\u_k\strans) \exp^{-1}_{\u_k^*}({\u}_k)   + \r_{\u_k^*},
\end{align*}
where  the Taylor remainder term satisfies that
$ \|\r_{\u_k^*}\|_2 = O(\| \exp^{-1}_{\u_k^*}({\u}_k) \|_2^2). $ Here, $\exp^{-1}_{\u_k^*}({\u}_k)$ denotes the tangent vector in the tangent space of $\text{St}(1, n)$ such that ${\u}_k$ can be represented through the exponential map as ${\u}_k = \exp_{\u_k^*}(\bxi_1).$

In view of \eqref{t11}, it holds that 
\[ \der{\wt{\psi}_k( \v_k,  {\boldeta}_k   )}{{\u}_k\trans}\Big|_{\u_k^*} = - n^{-1/2}\Y\trans - \v_k\trans\v_k\M_k^u.\]
By Proposition \ref{prop:strong2:m} that $\M^u_k = - (\v_k\trans\v_k)^{-1} \sum_{j \neq k} \v_j\u_j\trans$  and the initial estimates in Definition \ref{defi:sofar}, we can deduce that 
\begin{align}\label{phiuuuu2}
    &\wt{\psi}_k(\wt{\v}_k,\wt{\boldeta}_k) - \wt{\psi}_k(\wt{\v}_k,\boldeta^*_k)= ( - n^{-1/2}\Y\trans + \sum_{j \neq k} \wt{\v}_j \wt{\u}_j\trans )(\I_n - \u_k^*\u_k\strans) \exp^{-1}_{\u_k^*}(\wt{\u}_k)  + \wt{\r}_{\u_k^*}  \nonumber\\
	 & =( \sum_{j \neq k}( \wt{\v}_j \wt{\u}_j\trans - \v_j^*\u_j\strans )- n^{-1/2}\E\trans)(\I_n - \u_k^*\u_k\strans) \exp^{-1}_{\u_k^*}(\wt{\u}_k)  + \wt{\r}_{\u_k^*},
\end{align}
where the Taylor remainder term is 
\begin{align}\label{tayttr}
    \|\wt{\r}_{\u_k^*}\|_2 = O(\| \exp^{-1}_{\u_k^*}(\wt{\u}_k) \|_2^2).
\end{align}

\noindent\textbf{(1). The upper bound on $\norm{ \exp^{-1}_{\u_k^*}(\wt{\u}_k)}_2$}. 
Let us define $\bxi_k =  \exp^{-1}_{\u_k^*}(\wt{\u}_k)$. An application of similar arguments as in the proof of Lemma 5 in \cite{sofari} shows that the geodesic starting from $\u_k^*$ with tangent vector $ \bxi_k$ is given by 
\begin{align*}
    \gamma(t ; \u_k^*, \bxi_k) = \u_k^* \cdot \cos( \norm{{\bxi}_k}_2 t) + \frac{{\bxi}_k}{\norm{{\bxi}_k}_2}  \cdot \sin(\norm{{\bxi}_k}_2 t).
\end{align*}
Similarly, it follows from the definition of the exponential map that 
\begin{align}\label{eqxi}
     \wt{\u}_k & = \exp_{\u_k^*}({\bxi}_k) = \gamma(1 ; \u_k^*, \bxi_k) = \u_k^* \cdot  \cos( \norm{{\bxi}_k}_2 ) + \frac{{\bxi}_k}{\norm{{\bxi}_k}_2} \cdot  \sin(\norm{{\bxi}_k}_2).
\end{align}

Recall that $\u_k =  (\l_k\trans \wh{\bSigma}\l_k)^{-1/2} n^{-1/2} \X \l_k$. 
Similar to (A.110) in \cite{sofari},  it holds that  $\norm{{\bxi}_k}_2/ \sin(\norm{{\bxi}_k}_2) \neq 0$ and 
\begin{align}
    {\bxi}_k &=  \left(  \wt{\u}_k -  \u_k^* \cos( \norm{{\bxi}_k}_2 ) \right) \cdot \frac{\norm{{\bxi}_k}_2}{\sin(\norm{{\bxi}_k}_2)} \nonumber \\
	& = n^{-1/2} \X \left( (\wt{\l}_k\trans \wh{\bSigma}\wt{\l}_k)^{-1/2}   \wt{\l}_k - (\l_k\strans \wh{\bSigma}\l_k^*)^{-1/2}   \l_k^* \cos( \norm{{\bxi}_k}_2 ) \right) \cdot \frac{\norm{{\bxi}_k}_2}{\sin(\norm{{\bxi}_k}_2)}. \nonumber
\end{align}
Then applying Lemma 3 in \cite{chen2012sparse} leads to 
$\norm{\bxi_k}_2  =   O(\norm{\wt{\u}_k - \u_k^*}_2   ).$ 
Further, Lemma \ref{1} shows that $	\norm{\wt{\u}_k - \u_k^*}_2 
	\leq c \gamma_n$,
which entails that 
\begin{align}\label{eqr11}
	\norm{\bxi_k}_2 \leq c\norm{\wt{\u}_k - \u_k^{*}}_{2} \leq c \gamma_n. 
\end{align}

On the other hand, denote by ${\bxi}_k =  n^{-1/2} \X \bxi_k^{\prime}$. 
In light of Definition \ref{defi:sofar},
 we can deduce that 
\begin{align}
    \norm{{\bxi}_k^{\prime} }_0 &= \norm{\left( (\wt{\l}_k\trans \wh{\bSigma}\wt{\l}_k)^{-1/2}   \wt{\l}_k - (\l_k\strans \wh{\bSigma}\l_k^*)^{-1/2}   \l_k^* \cos( \norm{{\bxi}_k}_2 ) \right) \cdot \frac{\norm{{\bxi}_k}_2}{\sin(\norm{{\bxi}_k}_2)}}_0 \nonumber \\ &= \norm{   (\wt{\l}_k\trans \wh{\bSigma}\wt{\l}_k)^{-1/2}   \wt{\l}_k - (\l_k\strans \wh{\bSigma}\l_k^*)^{-1/2}   \l_k^* \cos( \norm{{\bxi}_k}_2 ) }_0  \nonumber\\
    &\leq \norm{ \wt{\l}_k - \l_k^* }_0 + \norm{  \l_k^* }_0 
    \leq 3(r^*+s_u +s_v). \label{bxi0} 
\end{align}
Using the sparsity of ${\bxi}_k^{\prime}$ and Condition \ref{con2:re}, we have that $ \rho_l \norm{\bxi_k^{\prime}}_2 \leq  \norm{n^{-1/2} \X \bxi_k^{\prime}}_2 =  \norm{\bxi_k}_2 \leq c \gamma_n,$ which further implies that
\begin{align}\label{eqr1111}
   \norm{\bxi_k^{\prime}}_2   \leq c \gamma_n.
\end{align}

\noindent\textbf{(2). The upper bound on $|\a\trans \W_k (\wt{\psi}_k(\wt{\v}_k,\wt{\boldeta}_k) - \wt{\psi}_k(\wt{\v}_k,\boldeta^*_k) )|  $}.
Observe that 
\begin{align*}
	& |\a\trans \W_k (\wt{\psi}_k(\wt{\v}_k,\wt{\boldeta}_k) - \wt{\psi}_k(\wt{\v}_k,\boldeta^*_k) )| \\[5pt]
	&\leq |\a\trans\W_k(\sum_{j \neq k}( \wt{\v}_j \wt{\u}_j\trans - \v_j^*\u_j\strans )- n^{-1/2}\E\trans)(\I_n - \u_k^*\u_k\strans) \exp^{-1}_{\u_k^*}(\wt{\u}_k)|
 +	|\a\trans\W_k \wt{\r}_{\u_k^*}|. 
\end{align*}
By \eqref{eqr11} and $\|\u_k^*\|_2 = 1$, it holds that 
\begin{align*}
	\|(\I_n - \u_k^*\u_k\strans   )\bxi_k\|_2 \leq  \|\bxi_k\|_2 + |\u_k\strans  \bxi_k| \cdot \|\u_k^*\|_2 \leq \|\bxi_k\|_2 + \|\bxi_k\|_2 \|\u_k^*\|_2^2 \leq c\gamma_n.
\end{align*}
Then under Condition \ref{con4:strong} that $d_i^*$ is at the constant level, it follows from Lemmas \ref{1} and \ref{lemma:wr2bound} that 
\begin{align}
    |\a\trans &\W_k(\sum_{j \neq k}( \wt{\v}_j \wt{\u}_j\trans - \v_j^*\u_j\strans ))(\I_n - \u_k^*\u_k\strans) \exp^{-1}_{\u_k^*}(\wt{\u}_k)| \nonumber  \\
    &\leq   \|\a\trans \W_k \|_2  \norm{ \wt{\V}_{-k} \wt{\U}_{-k}\trans - \V_{-k}^*\U_{-k}\strans }_2\|(\I_n - \u_k^*\u_k\strans   )\bxi_k\|_2 \nonumber \\
    &\leq c ({r^*}+s_u+s_v)\eta_n^4\{n^{-1}\log(pq)\}. \label{sadz}
\end{align}

We next bound term
$\a\trans\W_k n^{-1/2}
\E\trans(\I_n - \u_k^*\u_k\strans   )\bxi_k$.
Denote by $\W_i\trans$ the $i$th row of $\W_k$. In view of Lemma \ref{lemma:wr2bound}, it holds that 
\begin{align}\label{wwww}
	\max_{1 \leq i \leq p}\norm{\W_i}_0 \leq c (r^*+s_u+s_v) \ \text{ and } \  \max_{1 \leq i \leq p}\norm{\W_i}_2 \leq c.
\end{align}
Since $\bxi =  n^{-1/2} \X \bxi_k^{\prime}$ and $\u_k^* =  (\l_k\strans \wh{\bSigma}\l_k^*)^{-1/2} n^{-1/2} \X \l_k^*$, we have 
\begin{align*}
    &| \W_i\trans n^{-1/2}
\E\trans(\I_n - \u_k^*\u_k\strans   )\bxi_k | \leq  
| \W_i\trans n^{-1/2}\E\trans \bxi_k| +  | \W_i\trans n^{-1/2}
\E\trans \u_k^* \u_k\strans   \bxi_k | \\[5pt]
&\quad \leq \norm{ \W_i\trans n^{-1}\E\trans \X }_{2,s} \norm{\bxi_k^{\prime}}_2 + \norm{ \W_i\trans n^{-1}\E\trans \X }_{2,s} \norm{ (\l_k\strans \wh{\bSigma}\l_k^*)^{-1/2} \l_k^* \u_k\strans   \bxi_k}_2,
\end{align*}
where $s = c(r^* + s_u + s_v)$.  Here, for an arbitrary vector $\x$, $\norm{\x}_{2,s}^2 \equiv \max_{\abs{S}\leq s}\sum_{i\in S}x_i^2$ with $S$ an index set. 

With the aid of \eqref{wwww} and the fact that $n^{-1}\norm{\X\trans\E}_{\max} \leq c\{n^{-1}\log(pq)\}^{1/2}$, it holds that
	\begin{align*}
		\|\W_i\trans n^{-1}\X\trans\E \|_{\max} &\leq  \norm{\W_i}_1 \| n^{-1}\X\trans\E \|_{\max} 
  \leq c (r^*+s_u+s_v)^{1/2}\{n^{-1}\log(pq)\}^{1/2},
	\end{align*}
which further entails that $	\|\W_i\trans n^{-1}\X\trans\E \|_{2,s}  \leq c(r^*+s_u+s_v)\{n^{-1}\log(pq)\}^{1/2}$.

For the second term above, it follows from \eqref{eqr11}, \eqref{daswqe}, and Lemma \ref{1} that 
\begin{align*}
    \norm{ (\l_k\strans \wh{\bSigma}\l_k^*)^{-1/2} \l_k^* \u_k\strans   \bxi_k}_2 \leq  (\l_k\strans \wh{\bSigma}\l_k^*)^{-1/2} \norm{  \l_k^*}_2 \| \u_k\strans  \|_2 \norm{ \bxi_k}_2  \leq c \gamma_n.
\end{align*}
Combining the above terms results in 
\begin{align}\label{czdwq}
    | \W_i\trans n^{-1/2}
\E\trans(\I_n - &\u_k^*\u_k\strans   )\bxi_k | \leq  
c(r^*+s_u+s_v)^{3/2} \eta_n^2 \{n^{-1}\log(pq)\}. 
\end{align}

Let us define $\wh{\bDelta} = \sum_{j \neq k}( \wt{\v}_j \wt{\u}_j\trans - \v_j^*\u_j\strans )- n^{-1/2}\E\trans$. In light of \eqref{sadz} and \eqref{czdwq}, we have that 
\begin{align*}
		&| \W_i\trans \wh{\bDelta} (\I_n - \u_k^*\u_k\strans   ){\bxi}_k| \\ & \leq  c\max\{ (r^*+s_u+s_v)^{1/2}, \eta_n^2\}(r^*+s_u+s_v)\eta_n^2\{n^{-1}\log(pq)\}.
	\end{align*}
Hence, for each vector $\a \in \mathbb{R}^q$ satisfying $\norm{\a}_0 =m$ and $\norm{\a}_2 =1$, it holds that
\begin{align}
	&|\a\trans\W_k\wh{\bDelta} (\I_n - \u_k^*\u_k\strans   ){\bxi}_k| \leq \norm{\a}_1 \norm{\W_k\wh{\bDelta} (\I_n - \u_k^*\u_k\strans   ){\bxi}_k}_{\max}
	\nonumber\\
 &\leq \norm{\a}_0^{1/2} \norm{\a}_2 \max_{1 \leq i \leq q}	| \W_i\trans \wh{\bDelta}  (\I_n - \u_k^*\u_k\strans   ){\bxi}_k| \nonumber\\
	&\leq { c m^{1/2} \max\{ (r^*+s_u+s_v)^{1/2}, \eta_n^2\}(r^*+s_u+s_v)\eta_n^2\{n^{-1}\log(pq)\}.} \nonumber
\end{align}

Moreover, using \eqref{tayttr}, \eqref{eqr11}, and Lemma \ref{lemma:wr2bound}, we can deduce that 
\begin{align*}
     |\a\trans\W_k \wt{\r}_{\u_k^*}| 
    \leq \| \a\trans\W_k\|_2 \|\wt{\r}_{\u_k^*}\|_2 \leq   c (r^*+s_u+s_v)\eta_n^4\{n^{-1}\log(pq)\}. 
\end{align*}
Therefore, for $\a\in\mathcal{A}=\{\a\in\R^q:\norm{\a}_0\leq m,\norm{\a}_2=1\}$, we can obtain that
\begin{align}
	&|\a\trans\W_k (\wt{\psi}_k\left(\wt{\v}_k, \wt{\boldeta}_k \right) - \wt{\psi}_k\left( \wt{\v}_k, \boldeta_k^* \right)) | \nonumber \\
	& \leq c m^{1/2}\max\{ (r^*+s_u+s_v)^{1/2}, \eta_n^2\}(r^*+s_u+s_v)\eta_n^2\{n^{-1}\log(pq)\}, \nonumber 
\end{align}
which completes the proof of Lemma  \ref{lemm:taylorrank2:2}.

\subsection{Proof of Lemma \ref{lemma:1rk2}} \label{new.Sec.B.6}

By the Cauchy--Schwarz inequality, we have that 
\begin{align*}
	|\a\trans{\W_k} \bdelta_k | &\leq  \|  \a\trans{\W_k} \|_2 \|   \M_k \|_2 \|  \u_k^*\wt{\v}_k\trans -  \wt{\u}_k\wt{\v}_k\trans + \sum_{i \neq k} (\wt{\u}_i\wt{\v}_i\trans -  \u_i^*{\v}_i\strans)  \|_2 \| \v_k^* - \wt{\v}_k \|_2.
\end{align*}
We will bound the above terms under Condition \ref{con4:strong} that the nonzero squared singular values $d^{*2}_{i}$ are at the constant level.
In view of parts (a), (c), and (e) of Lemma \ref{1}, it holds that 
\begin{align*}
	&\| \sum_{i \neq k} (\wt{\u}_i\wt{\v}_i\trans -  \u_i^*{\v}_i\strans)    \|_2 = \norm{\wt{\U}_{-k}\wt{\V}_{-k}\trans -  \U_{-k}^*{\V}_{-k}\strans  }_2 \leq c \gamma_n, \\
    &\norm{\u_k^* \wt{\v}_k\trans -\wt{\u}_k \wt{\v}_k\trans}_2 \leq \norm{\u_k^*  -\wt{\u}_k}_2 \norm{ \wt{\v}_k\trans}_2 \leq c\gamma_n, \\
	&\norm{\v_k^*  -\wt{\v}_k}_2 \leq c\gamma_n.
\end{align*}

Further, if follows from parts (b) and (d) of Lemma \ref{1} that
\begin{align*}
	&\|\M_k\|_2 = \| (\wt{\v}_k\trans\wt{\v}_k)^{-1} \wt{\V}_{-k} \wt{\U}_{-k}\trans \|_2  \leq  (\wt{\v}_k\trans\wt{\v}_k)^{-1}\| \wt{\V}_{-k} \|_2 \| \wt{\U}_{-k}\trans  \|_2 \leq c.
\end{align*}
This together with Lemma \ref{lemma:wr2bound} that $\norm{\a\trans{\W_k}}_2 \leq c$ leads to 
\begin{align*}
	|\a\trans{\W_k} \bdelta_k |   \leq  c  (r^*+s_u+s_v)\eta_n^4\left\{n^{-1}\log(pq)\right\}.
\end{align*}
This concludes the proof of Lemma \ref{lemma:1rk2}.

\subsection{Proof of Lemma \ref{lemma:1rk23}} \label{new.Sec.B.81}

Recall that 
$\u_i^* =  (\l_i\strans \wh{\bSigma}\l_i^*)^{-1/2} n^{-1/2} \X \l_i^* =  (\bmu_i\strans \wh{\bSigma}\bmu_i^*)^{-1/2} n^{-1/2} \X \bmu_i^*$ with $\bmu_i^* = d_i^* \l_i^*$. Under Condition \ref{con4:strong} that the nonzero squared singular values $d^{*2}_{i}$ are at the constant level,
it follows from Lemma \ref{lemma:wr2bound} that $\norm{\a\trans{\W_k}}_2 \leq c$ and consequently, 
\begin{align*}
	|\a\trans\W_k \sum_{i \neq k} \v_i^*\u_i\strans \u_k^* | &\leq \|\a\trans\W_k \|_2 \|\sum_{i \neq k} \v_i^*\u_i\strans \u_k^* \|_2  \\
	&\leq c \sum_{i \neq k} \|\v_i^* \|_2 (\bmu_i\strans \wh{\bSigma}\bmu_i^*)^{-1/2}(\bmu_k\strans \wh{\bSigma}\bmu_k^*)^{-1/2} d_i^* d_k^*|\l_i\strans \wh{\bSigma} \l_k^* | \\
	&\leq c \sum_{i \neq k}  |\l_i\strans \wh{\bSigma} \l_k^* | 
	= o( n^{-1/2}),
\end{align*}
where the last inequality above has used part (a) of Lemma \ref{1} and \eqref{daswqe}.
This completes the proof of Lemma \ref{lemma:1rk23}.

\subsection{Proof of Lemma \ref{lemma:1rk3}} \label{new.Sec.B.8}

Observe that 
\begin{align*}
	&\abs{-\a\trans\W_k\bepsilon_k - h_k / \sqrt{n}} \\[5pt]
	& \leq
	|\a\trans\W_k  \M_k  n^{-1/2}\E\wt{\v}_k -\a\trans \W^{*}_k \M_{k}^*n^{-1/2} \E {\v}_k^{*} |  + |\a\trans(\W_k -\W^*_k)n^{-1/2}\E\trans\u_k^*  | \\[5pt]
	&\leq |\a\trans\W_k  \M_k  n^{-1/2}\E(\wt{\v}_k - {\v}_k^{*})| + |(\a\trans\W_k  \M_k -\a\trans \W^{*}_k \M_{k}^*)n^{-1/2} \E{\v}_k^{*} | \\[5pt]
	&\quad + |\a\trans(\W_k -\W^*_k)n^{-1/2}\E\trans\u_k^*  |. 
\end{align*}
We will bound the above three terms under Condition \ref{con4:strong} that the nonzero squared singular values $d^{*2}_{i}$ are at the constant level. 

\noindent \textbf{(1). The upper bound on $|\a\trans\W_k  \M_k  n^{-1/2}\E(\wt{\v}_k - {\v}_k^{*})|. $}
Similar to the proof of Lemma \ref{1}, we define   $$\wt{\U} = n^{-1/2} \X \wt{\L}_0  \ \text{ with } 
 \wt{\L}_0 = ( (\wt{\l}_1\trans \wh{\bSigma}\wt{\l}_1)^{-1/2}\wt{\l}_1, \cdots, (\wt{\l}_{r^*}\trans \wh{\bSigma}\wt{\l}_{r^*})^{-1/2}\wt{\l}_{r^*}  ).$$ Similarly, denote by  $\wt{\U}_{-k} = n^{-1/2} \X \wt{\L}_{0, -k} $ and ${\U}_{-k}^* = n^{-1/2} \X {\L}_{0, -k}^*$.
Then it holds that
\begin{align*}
	&\M_k = - (\wt{\v}_k\trans\wt{\v}_k)^{-1} \wt{\V}_{-k} \wt{\U}_{-k}\trans = - (\wt{\v}_k\trans\wt{\v}_k)^{-1} \wt{\V}_{-k} \wt{\L}_{0,-k}\trans  n^{-1/2} \X\trans, \\[5pt]
	&{\M}_k^* = - ({\v}_k\strans{\v}_k^*)^{-1} {\V}_{-k}^* {\U}_{-k}\strans = - ({\v}_k\trans{\v}_k^*)^{-1} {\V}_{-k}^* {\L}_{0,-k}\strans  n^{-1/2} \X\trans.
\end{align*}

Let us define $s=c(r^*+s_u+s_v)$. 
It can be seen that
	$\norm{ \wt{\L}_{0,-k}}_0 \leq \norm{ \wt{\L}_0}_0 \leq s,$ which further entails that $\norm{ \wt{\L}_{0,-k} \b }_0 \leq s$ for any vector $\b \in \mathbb{R}^{r^*-1}$.
Hence, we have that $\| \a\trans\W_k   \wt{\V}_{-k} \wt{\L}_{0,-k}\trans \|_0 \leq s$ and 
\begin{align*}
	&|\a\trans\W_k  \M_k  n^{-1/2}\E(\wt{\v}_k - {\v}_k^{*})| 
	\\ &\leq | ( \wt{\v}_k\trans\wt{\v}_k)^{-1}|  \| \a\trans\W_k   \wt{\V}_{-k} \wt{\L}_{0,-k}\trans \|_2 \| n^{-1} \X\trans \E(\wt{\v}_k - {\v}_k^{*})  \|_{2,s}  \\
	&\leq | ( \wt{\v}_k\trans\wt{\v}_k)^{-1}|  \| \a\trans\W_k \|_2 \|  \wt{\V}_{-k} \|_2 \| \wt{\L}_{0,-k}\trans \|_2 \| n^{-1} \X\trans \E(\wt{\v}_k - {\v}_k^{*})  \|_{2,s} \\
	&\leq c \| n^{-1} \X\trans \E(\wt{\v}_k - {\v}_k^{*})  \|_{2,s},
\end{align*}
where the last inequality above is due to Lemma \ref{1}, \eqref{cxzkj}, and Lemma \ref{lemma:wr2bound}.

It remains to bound term $\| n^{-1} \X\trans \E(\wt{\v}_k - {\v}_k^{*})  \|_{2,s}$.
It follows from Definition \ref{defi:sofar} that 
$   \norm{\wt{\v}_k - {\v}_k^{*} }_{0} \leq  \norm{\wt{d}_k \wt{\r}_k}_0 + \norm{ {d}_k^* {\r}_k^{*} }_{0} \leq s.$
In addition, from Lemma \ref{1} we see that 
\begin{align*}
	\| n^{-1} \X\trans \E(\wt{\v}_k - {\v}_k^{*})  \|_{2,s} &\leq s^{1/2}  \| n^{-1} \X\trans \E(\wt{\v}_k - {\v}_k^{*})  \|_{\max}  \\
	&\leq c s^{1/2} \norm{ n^{-1} \X\trans \E }_{\max} \norm{\wt{\v}_k - {\v}_k^{*} }_{1} 
	\leq c s \{n^{-1}\log(pq)\}^{1/2} \gamma_n.
\end{align*}
Thus, we can obtain that 
\begin{align}
	|\a\trans\W_k  \M_k  n^{-1/2}\E(\wt{\v}_k - {\v}_k^{*})| \leq c   s \{n^{-1}\log(pq)\}^{1/2} \gamma_n. \label{cxznkffq}
\end{align}

\noindent \textbf{(2). The upper bound on $|(\a\trans\W_k  \M_k -\a\trans \W^{*}_k \M_{k}^*)n^{-1/2} \E{\v}_k^{*} |$}.
It holds that 
\begin{align*}
	&|(\a\trans\W_k  \M_k -\a\trans \W^{*}_k \M_{k}^*)n^{-1/2} \E{\v}_k^{*} | \\[5pt] 
	& =
	|(\a\trans\W_k   (\wt{\v}_k\trans\wt{\v}_k)^{-1} \wt{\V}_{-k} \wt{\L}_{0,-k}\trans - \a\trans \W^{*}_k ({\v}_k\trans{\v}_k^*)^{-1} {\V}_{-k}^* {\L}_{0,-k}\strans  )  n^{-1}\X\trans \E {\v}_k^{*}| \\[5pt]
	&\leq  \|\a\trans\W_k   (\wt{\v}_k\trans\wt{\v}_k)^{-1} \wt{\V}_{-k} \wt{\L}_{0,-k}\trans - \a\trans \W^{*}_k ({\v}_k\trans{\v}_k^*)^{-1} {\V}_{-k}^* {\L}_{0,-k}\strans \|_2  \|n^{-1}\X\trans \E {\v}_k^{*}\|_{2,s}.
\end{align*}
Here, the last step above has used the sparsity of ${\L}_{0,-k}^*$ and $\wt{\L}_{0,-k}$ in \eqref{l0001} and \eqref{l000} such that
\begin{align*}
	\|(\a\trans\W_k   (\wt{\v}_k\trans\wt{\v}_k)^{-1} \wt{\V}_{-k}) \cdot \wt{\L}_{-k}\trans & -( \a\trans \W^{*}_k ({\v}_k\trans{\v}_k^*)^{-1} {\V}_{-k}^*) \cdot {\L}_{-k}\strans \|_0  \\&\leq \| \wt{\L}_{0,-k} \|_0 + \| {\L}_{0,-k}^*\|_0 \leq s.
\end{align*}

Further, we can deduce that 
\begin{align*}
	&\| (\wt{\v}_k\trans\wt{\v}_k)^{-1} \a\trans\W_k   \wt{\V}_{-k} \wt{\L}_{-k}\trans - ({\v}_k\trans{\v}_k^*)^{-1}\a\trans \W^{*}_k  {\V}_{-k}^* {\L}_{-k}\strans \|_2 \\[5pt]
	&\leq |\wt{\v}_k\trans\wt{\v}_k|^{-1} \|  \a\trans\W_k   \wt{\V}_{-k} \wt{\L}_{-k}\trans - \a\trans \W^{*}_k  {\V}_{-k}^* {\L}_{-k}\strans \|_2  \\[5pt] 
    &\quad \quad + |(\wt{\v}_k\trans\wt{\v}_k)^{-1}  - ({\v}_k\trans{\v}_k^*)^{-1}|\| \a\trans \W^{*}_k \|_2 \| {\V}_{-k}^* \|_2 \|{\L}_{-k}\strans \|_2 \\[5pt]
	&\leq c \Big[  \|  \a\trans\W_k \|_2 \|  \wt{\V}_{-k} \wt{\L}_{-k}\trans -   {\V}_{-k}^* {\L}_{-k}\strans \|_2  + \|  \a\trans\W_k - \a\trans \W^{*}_k \|_2 \| {\V}_{-k}^*\|_2 \| {\L}_{-k}\strans \|_2  \Big]  \\[5pt] 
    &\qquad  + |(\wt{\v}_k\trans\wt{\v}_k)^{-1}  - ({\v}_k\trans{\v}_k^*)^{-1}|\| \a\trans \W^{*}_k \|_2 \| {\V}_{-k}^* \|_2 \|{\L}_{-k}\strans \|_2 \\
	&\leq c    \|  \wt{\V}_{-k} (\wt{\L}_{-k}\trans -    {\L}_{-k}\strans) \|_2  + c \| ( \wt{\V}_{-k}  -   {\V}_{-k}^*) {\L}_{-k}\strans \|_2  + c  \gamma_n \\[5pt]
	&\leq c  \gamma_n,
\end{align*}
where we have applied Lemmas \ref{1} and  \ref{lemma:wr2bound}, \eqref{uuszfaa22}, and \eqref{sdzxc}.
Moreover, it holds that 
\begin{align*}
	\| n^{-1} \X\trans \E {\v}_k^{*}  \|_{2,s} &\leq s^{1/2}  \| n^{-1} \X\trans \E {\v}_k^{*}  \|_{\max}  \\
	&\leq c s^{1/2} \norm{ n^{-1} \X\trans \E }_{\max} \norm{{\v}_k^{*} }_{1} 
	\leq c s \{n^{-1}\log(pq)\}^{1/2}.
\end{align*}
Hence, it follows that 
\begin{align}\label{czkjge}
	|(\a\trans\W_k  \M_k -\a\trans \W^{*}_k \M_{k}^*)n^{-1/2} \E {\v}_k^{*} |  \leq c s \{n^{-1}\log(pq)\}^{1/2} \gamma_n.
\end{align}

\noindent \textbf{(3). The upper bound on  $ |\a\trans(\W_k -\W^*_k)n^{-1/2}\E\trans\u_k^*  |$}.
By Lemma \ref{lemma:wr2bound}, it holds that 
$\|\a\trans(\W_k-{\W}_k^*)\|_0  \leq s$ and $\|\a\trans(\W_k-{\W}_k^*)\|_2  \leq c \gamma_n.$	
Similarly, we can obtain that
\begin{align}
	|\a\trans(\W_k -\W^*_k)n^{-1/2}\E\trans\u_k^*  | & \leq \|\a\trans(\W_k -\W^*_k)\|_2 \|n^{-1/2}\E\trans\u_k^*  \|_{2,s} \nonumber\\[5pt]
	& \leq c  \gamma_n \|n^{-1}\E\trans\X (\l_k\strans \wh{\bSigma}\l_k^*)^{-1/2}  \l_k^*  \|_{2,s}  \nonumber\\[5pt]
	&\leq  c  \gamma_n  s^{1/2} |\bmu_k\strans \wh{\bSigma}\bmu_k^*|^{-1/2}  \|n^{-1}\E\trans\X   \|_{\max}  \norm{\bmu_k^* }_0^{1/2}\norm{\bmu_k^* }_2 \nonumber\\[5pt]
	&\leq c   s \{n^{-1}\log(pq)\}^{1/2} \gamma_n, \label{sdzxchh}
\end{align}
where the last step above has used \eqref{daswqe}, $\norm{\bmu_k^* }_0 \leq s_u$, and $\norm{\bmu_k^* }_2 \leq d_i^*$.

Therefore, combining \eqref{cxznkffq}, \eqref{czkjge}, and \eqref{sdzxchh} yields that 
\begin{align*}
		\abs{-\a\trans{\W_k}\bepsilon_k - h_k / \sqrt{n}} \leq c (r^*+s_u + s_v)^{3/2}\eta_n^2\{ n^{-1}\log(pq)\}.
\end{align*}
This concludes the proof of Lemma \ref{lemma:1rk3}.

\subsection{Proof of Lemma \ref{lemma:weak:delta}} \label{new.Sec.B.61}

In view of the definition of $\bdelta_k$ in \eqref{delta:weakrank2v1}, it holds that  
\begin{align*}
	&|\a\trans{\W_k}\bdelta_k| = |\a\trans{\W_k} \M_k  ( \u_k^* \v_k\strans -\wt{\u}_k\wt{\v}_k\trans + \sum_{i \neq k}( \wt{\u}_i \wt{\v}_i\trans -\u_i^* \v_i\strans))  (\wt{\v}_k - \v_k^*)|\\
 &\leq  \norm{\a\trans{\W_k} }_2 \|\M_k\|_2 ( \norm{\u_k^* \v_k\strans -\wt{\u}_k\wt{\v}_k\trans }_2  + \sum_{i \neq k}\norm{  \wt{\u}_i \wt{\v}_i\trans -\u_i^* \v_i\strans }_2) \norm{\wt{\v}_k - \v_k^*}_2  \\
 &\leq c \gamma_n  \|\M_k\|_2  \sum_{i = 1}^{r^*}\norm{  \wt{\u}_i \wt{\v}_i\trans -\u_i^* \v_i\strans }_2,
\end{align*}
where the last inequality above is due to Lemmas \ref{1} and \ref{lemma:weak:wbound}. Moreover, it follows from Lemma \ref{1} that 
\begin{align*}
	\|\M_k\|_2 = \| ( \wt{\v}_k\trans\wt{\v}_k)^{-1}\widetilde{\v}_k\widetilde{\u}_k\trans\|_2 \leq |\wt{\v}_k\trans\wt{\v}_k|^{-1} \| \widetilde{\v}_k \|_2 \|\widetilde{\u}_k\trans\|_2 \leq c d_k^{*-1}.
\end{align*}

For term $\sum_{i = 1}^{r^*}\norm{  \wt{\u}_i \wt{\v}_i\trans -\u_i^* \v_i\strans }_2$, from \eqref{uvuvzxa} we can show that 
\begin{align*}
	\sum_{i = 1}^{r^*}\| \wt{\u}_i \wt{\v}_i\trans -\u_i^* \v_i\strans\|_2 \leq c \gamma_n.
\end{align*}
Thus, combining above terms leads to 
\begin{align*}
	|\a\trans{\W_k}\bdelta_k| \leq c   (r^*+s_u+s_v)\eta_n^4\left\{n^{-1}\log(pq)\right\} d_k^{*-1},
\end{align*}
which completes the proof of Lemma \ref{lemma:weak:delta}.

\subsection{Proof of Lemma \ref{lemma:weak:con}} \label{new.Sec.B.8conz}

Note that $\u_i^* =  (\l_i\strans \wh{\bSigma}\l_i^*)^{-1/2} n^{-1/2} \X \l_i^*$.  By Lemma \ref{lemma:weak:wbound} that $\| \a\trans \W_k \|_2  \leq c$, we can deduce that 
\begin{align*}
	&\| \sqrt{n} \a\trans \W_k \sum_{i \neq k}  (\wt{\v}_i  - \v_i^*)\u_i\strans\u_k^* \|_2 \leq  \| \a\trans \W_k \|_2 \|  \sqrt{n} \sum_{i \neq k}  (\wt{\v}_i  - \v_i^*)\u_i\strans\u_k^* \|_2 \\ 
	&\leq c \|   \sqrt{n} \sum_{i \neq k}  (\wt{\v}_i  - \v_i^*)\u_i\strans\u_k^* \|_2    \\
	&\leq 	  \sqrt{n} \sum_{i \neq k}  \|\wt{\v}_i  - \v_i^*\|_2 d_i^* d_k^* (\bmu_i\strans \wh{\bSigma}\bmu_i^*)^{-1/2} (\bmu_k\strans \wh{\bSigma}\bmu_k^*)^{-1/2} |\l_i\strans \wh{\bSigma} \l_k^* | \\
	&\leq c \gamma_n  \sqrt{n} \sum_{i \neq k}  |\l_i\strans \wh{\bSigma} \l_k^* |,
\end{align*}
where the last step above has applied Lemma \ref{1} that $\norm{\wt{\v}_i - \v_i^* }_2 \leq c \gamma_n$ and \eqref{daswqe}.

Further, it follows from Condition \ref{con:weak:orth} that
\begin{align}
	&\sqrt{n} \sum_{i \neq k} |\l_i\strans \wh{\bSigma} \l_k^* | \nonumber \\&=  \sqrt{n} \sum_{i < k} (d_k^{*2}/d_i^*) |\l_i\strans \wh{\bSigma} \l_k^* | \cdot (d_i/d_k^{*2}) +  \sqrt{n} \sum_{i > k} (d_i^{*2}/d_k^*) |\l_i\strans \wh{\bSigma} \l_k^* | \cdot  (d_k/d_i^{*2})  \nonumber\\
	&\leq \sqrt{n} \sum_{i < k} (d_k^{*2}/d_i^*) |\l_i\strans \wh{\bSigma} \l_k^* | \cdot (d_i/d_k^{*2}) + (d_k/d_{r^*}^{*2}) \sqrt{n} \sum_{i > k} (d_i^{*2}/d_k^*) |\l_i\strans \wh{\bSigma} \l_k^* |   \nonumber\\
	&=  o (\sum_{i < k} d_i/d_k^{*2} ) + o( d_k/d_{r^*}^{*2} ) \leq c r^* d_1^* d_{r^*}^{*-2}. \label{weakceq}
\end{align}
Therefore, combining the above terms gives that
\begin{align*}
	\| \sqrt{n} \a\trans \W_k \sum_{i \neq k}  (\wt{\v}_i  - \v_i^*)\u_i\strans\u_k^* \|_2 \leq c r^* d_1^{*}  d_{r^*}^{*-2} \gamma_n.
\end{align*}
This concludes the proof of Lemma \ref{lemma:weak:con}.

\subsection{Proof of Lemma \ref{lemm:weak:taylor}}

For function $\wt{\psi}_k({\v}_k,{\boldeta}_k)$ with respect to ${\boldeta}_k = \u_k$,
the first-order Taylor expansion at $\u_k^*$ is given by 
\begin{align*}
	\wt{\psi}_k({\v}_k,{\u}_k) = \wt{\psi}_k(\wt{\v}_k,\u_k^*) +   \der{\wt{\psi}_k}{{\u}_k\trans}\Big|_{\u_k^*}({\u}_k - {\u}_k^*  ) + \r_{\u_k},
\end{align*}
where $\r_{\u_k}$ is the Taylor remainder term such that $ \|\r_{\u_k}\|_2 = O( \|{\u}_k - {\u}_k^* \|_2^2 )  $.
By the derivatives in \eqref{1derisadq} and \eqref{2derisadq2}, we can obtain that
\begin{align*}
	\wt{\psi}_k(\v_k,{\boldeta_k}) &= \der{L}{\v_k} - \M_k\der{L}{{\u_k}} \\
 &= \v_k - n^{-1/2}\Y\trans\u_k + \sum_{i \neq k} \wt{\v}_i \u_k\trans\wh{\u}_i^t - \M_k(  \u_k \v_k\trans\v_k  - n^{-1/2}\Y\v_k).
\end{align*}
Then we can see that 
	$\der{\wt{\psi}_k}{{\u}_k\trans}\Big|_{\u_k^*} = - n^{-1/2}\Y\trans + \sum_{i \neq k} \wt{\v}_i (\wh{\u}_i^t)\trans - \v_k\trans\v_k\M_k.$
 
 With the initial estimator in Definition \ref{defi:sofar} and $\M_k =  -(\wt{\v}_k\trans\wt{\v}_k)^{-1} \wt{\v}_k\wt{\u}_k\trans$, it holds that 
\begin{align}
    &\big|\a\trans\W_k \der{\wt{\psi}_k}{{\u}_k\trans}\Big|_{\u_k^*} (\wt{\u}_k - {\u}_k^*  ) \big| =  |\a\trans\W_k( - n^{-1/2}\Y + \sum_{i \neq k} \wt{\v}_i (\wh{\u}_i^t)\trans + \wt{\v}_k\wt{\u}_k\trans) (\wt{\u}_k - {\u}_k^* )|  \nonumber\\
    & = |\a\trans\W_k(  \sum_{i \neq k} (\wt{\v}_i (\wh{\u}_i^t)\trans - {\v}_i^*\u_i\strans) + \wt{\v}_k\wt{\u}_k\trans - {\v}_k^*{\u}_k\strans  -  n^{-1/2}\E) (\wt{\u}_k - {\u}_k^* )| \nonumber\\
    & \leq \|\a\trans\W_k\|_2 \Big(\| \sum_{i \neq k} (\wt{\v}_i (\wh{\u}_i^t)\trans - {\v}_i^*\u_i\strans )\|_2 + \| \wt{\v}_k\wt{\u}_k\trans - {\v}_k^*{\u}_k\strans \|_2 \Big) \|\wt{\u}_k - {\u}_k^* \|_2 \nonumber\\
	&\quad  + | \a\trans\W_k n^{-1/2}\E (\wt{\u}_k - {\u}_k^*) | \nonumber\\
    &=: A_1 + A_2. \label{zcxasdqq}
\end{align}
We will bound the above two terms $A_1$ and $A_2$ separately. In light of Lemma \ref{1}, we have that 
\begin{align*}
    \|\wt{\u}_k - {\u}_k^* \|_2 &\leq c \gamma_n d_k^{*-1}, \\[5pt]
    \| \wt{\u}_k\wt{\v}_k\trans -\u_k^*\v_k\strans\|_2 &\leq 	 \| (\wt{\u}_k- \u_k^*)\v_k\strans \|_2+\| \wt{\u}_k(\wt{\v}_k- \v_k^*)\trans \|_2 \nonumber\\[5pt]	
    &\leq  \|\wt{\u}_k- \u_k^*\|_2 \|\v_k^* \|_2 +  \|\wt{\u}_k\|_2 \|\wt{\v}_k- \v_k^*\|_2  \leq c \gamma_n.
\end{align*}

For term $\| \sum_{i \neq k} (\wt{\v}_i (\wh{\u}_i^t)\trans - {\v}_i^*\u_i\strans )\|_2$,
it follows that
\begin{align*}
	\|  (\wt{\v}_i (\wh{\u}_i^t)\trans - {\v}_i^*\u_i\strans )\|_2 &\leq 		\|  \wt{\v}_i \|_2 \|\wh{\u}_i^t - \u_i^* \|_2  + 	\|  \wt{\v}_i - {\v}_i^*\|_2 \|\u_i^* \|_2 \\[5pt]
	&\quad \leq  c d_i^* \|\wh{\u}_i^t - \u_i^* \|_2 + c \gamma_n,
\end{align*}
where we have used Lemma \ref{1} and {\eqref{whueqz2} in the proof of Lemma \ref{lemma:disw} that }
\begin{align}
	\|\wh{\u}_i^t  -\u_i^* \|_2 \leq  c  \gamma_n d_i^{*-1}. \nonumber
\end{align}
Moreover, Lemma \ref{lemma:weak:wbound} implies that $\|\a\trans\W_k \|_2 \leq c$.
Hence, combining above terms leads to 
\begin{align}
    A_1 \leq c \gamma_n^2 d_k^{*-1}. \label{sdacaq}
\end{align}

We next bound term $A_2$. Denote by  $\w_i\trans$ the $i$th row of $\W_k$. It follows from Lemma \ref{lemma:weak:wbound} that
\begin{align}
	\norm{\w_i}_0 \leq c(r^* + s_u + s_v), \ \ \norm{\w_i}_2 \leq c. \label{www2}
\end{align}
Then for $\a\in\mathcal{A}=\{\a\in\R^q:\norm{\a}_0\leq m,\norm{\a}_2=1\}$, we can deduce that 
\begin{align*}
	&| \a\trans\W_k n^{-1/2}\E\trans (\wt{\u}_k - {\u}_k^*) |  \leq \norm{\a}_1 \norm{\W_kn^{-1/2}\E\trans (\wt{\u}_k - {\u}_k^*)}_{\max} \\[5pt]
	&\leq \norm{\a}_0^{1/2} \norm{\a}_2 \max_{1 \leq i \leq q}	| \w_i\trans n^{-1}\E\trans  \X  (  (\wt{\bmu}_k\trans \wh{\bSigma}\wt{\bmu}_k)^{-1/2} \wt{\bmu}_k -  (\bmu_k\strans \wh{\bSigma}\bmu_k^*)^{-1/2} {\bmu}_k^*)| \\[5pt]
	&\leq m^{1/2} \max_{1 \leq i \leq q}	| \w_i\trans n^{-1}\E\trans  \X  (  (\wt{\bmu}_k\trans \wh{\bSigma}\wt{\bmu}_k)^{-1/2} \wt{\bmu}_k -  (\bmu_k\strans \wh{\bSigma}\bmu_k^*)^{-1/2} {\bmu}_k^*)|.
\end{align*}

From \eqref{www2} and the fact that $n^{-1}\norm{\X\trans\E}_{\max} \leq c\{n^{-1}\log(pq)\}^{1/2}$, it holds that
	\begin{align*}
		\|\w_i\trans n^{-1}\E\trans\X \|_{\max} &\leq  \norm{\W_i}_1 \| n^{-1}\X\trans\E \|_{\max} \\
  &\leq c (r^*+s_u+s_v)^{1/2}\{n^{-1}\log(pq)\}^{1/2},
	\end{align*}
which further leads to 
	\begin{align*}
		\|\w_i\trans n^{-1}\X\trans\E \|_{2,s}  \leq c(r^*+s_u+s_v)\{n^{-1}\log(pq)\}^{1/2}.
	\end{align*}
    
Further, by \eqref{daswqe}, \eqref{sdczcsa}, and Definition \ref{defi:sofar}, we can show that 
\begin{align*}
    &\norm{ (\wt{\bmu}_k\trans \wh{\bSigma}\wt{\bmu}_k)^{-1/2} \wt{\bmu}_k -  (\bmu_k\strans \wh{\bSigma}\bmu_k^*)^{-1/2} {\bmu}_k^* }_2 \\[5pt]
	&\leq  (\wt{\bmu}_k\trans \wh{\bSigma}\wt{\bmu}_k)^{-1/2}  \norm{ \wt{\bmu}_k -   {\bmu}_k^*}_2  +  \abs{    (\wt{\bmu}_k\trans \wh{\bSigma}\wt{\bmu}_k)^{-1/2}  -  (\bmu_k\strans \wh{\bSigma}\bmu_k^*)^{-1/2}  }  \|{\bmu}_k^*\|_2
	\leq   c \gamma_n d_k^{*-1},
\end{align*}
where $ \gamma_n = ({r^*}+s_u+s_v)^{1/2}\eta_n^2\{n^{-1}\log(pq)\}^{1/2}$. Then it follows that 
\begin{align}\label{dasxzc}
	| \a\trans\W_k n^{-1/2}\E\trans (\wt{\u}_k - {\u}_k^*) | \leq c m^{1/2} (r^*+s_u+s_v)^{3/2}\eta_n^2\{n^{-1}\log(pq)\}  d_k^{*-1}.    
\end{align}
Hence, combining \eqref{zcxasdqq}, \eqref{sdacaq}, and \eqref{dasxzc} yields that 
\begin{align*}
    &\big|\a\trans\W_k \der{\wt{\psi}_k}{{\u}_k\trans}\Big|_{\u_k^*} (\wt{\u}_k - {\u}_k^*  ) \big| \\ & \leq  c m^{1/2} \max\{(r^*+s_u+s_v)^{1/2}, \eta_n^2\}(r^*+s_u+s_v)\eta_n^2\{n^{-1}\log(pq)\}d_k^{*-1}.
\end{align*}

Moreover, for the Taylor remainder terms, since $ \|\wt{\r}_{\u_k}\|_2 = O( \|\wt{\u}_k - {\u}_k^* \|_2^2 )  $ it can be seen that 
\begin{align*}
     |\a\trans\W_k \wt{\r}_{\u_k}| \leq \norm{\a\trans\W_k}_2 \norm{\wt{\r}_{\u_k}}_2 
    \leq  c (r^*+s_u+s_v)\eta_n^4\{n^{-1}\log(pq)\}. 
\end{align*}
Therefore, combining the above terms gives that 
\begin{align*}
	& |\a\trans\W_k (\wt{\psi}_k(\wt{\v}_k,\wt{\boldeta}_k) - \wt{\psi}_k(\wt{\v}_k,\boldeta_k^*) )| \\[5pt]
	& \leq   c m^{1/2} \max\{(r^*+s_u+s_v)^{1/2}, \eta_n^2\}(r^*+s_u+s_v)\eta_n^2\{n^{-1}\log(pq)\}d_k^{*-1},
\end{align*}
which completes the proof of Lemma  \ref{lemm:weak:taylor}.

\subsection{Proof of Lemma \ref{lemma:weak:ep}} \label{new.Sec.B.8z}

Recall in Proposition \ref{prop:weak2:m} that 
\begin{align*}
	&\M_k = - ( \wt{\v}_k\trans\wt{\v}_k)^{-1} \wt{d}_k \widetilde{\r}_k\widetilde{\l}_k\trans n^{-1/2} \X\trans, \ \
	{\M}_k^* = - ( {\v}_k\strans{\v}_k^*)^{-1} d_k^* {\r}_k^*{\l}_k\strans n^{-1/2} \X\trans. 
\end{align*}
Denote by  $\M_k = \M_{v_k} n^{-1/2} \X\trans$ and $\M_{k}^* = {\M}_{v_k}^* n^{-1/2} \X\trans$, where
\begin{align*}
	&\M_{v_k} = - ( \wt{\v}_k\trans\wt{\v}_k)^{-1} \wt{d}_k \widetilde{\r}_k\widetilde{\l}_k\trans, \ \
	{\M}_{v_k}^* = - ( {\v}_k\strans{\v}_k^*)^{-1} d_k^* {\r}_k^*{\l}_k\strans. 
\end{align*}
Then for $\bepsilon_k =   - n^{-1/2}\E\trans\u_k^* + n^{-1/2} \M_k\E\wt{\v}_k$, we can deduce that 
\begin{align}
	&\abs{-\a\trans\W_k\bepsilon_k - h_k / \sqrt{n}} \nonumber \\[5pt]
	& \leq
	|\a\trans\W_k  \M_k  n^{-1/2}\E\wt{\v}_k -\a\trans \W^{*}_k \M_{k}^*n^{-1/2} \E {\v}_k^{*} |  + |\a\trans(\W_k -\W^*_k)n^{-1/2}\E\trans\u_k^*  | \nonumber \\[5pt]
	&\leq |\a\trans\W_k  \M_k  n^{-1/2}\E(\wt{\v}_k - {\v}_k^{*})| + |(\a\trans\W_k  \M_k -\a\trans \W^{*}_k \M_{k}^*)n^{-1/2} \E{\v}_k^{*} | \nonumber \\[5pt]
	&\quad + |\a\trans(\W_k -\W^*_k)n^{-1/2}\E\trans\u_k^*  | \nonumber \\[5pt]
	&= |\a\trans\W_k  \M_{v_k}  n^{-1}\X\trans\E(\wt{\v}_k - {\v}_k^{*})| + |(\a\trans\W_k  \M_{v_k} -\a\trans \W^{*}_k {\M}_{v_k}^*)n^{-1/2}\X\trans \E{\v}_k^{*} | \nonumber \\[5pt]
	&\quad + |\a\trans(\W_k -\W^*_k)n^{-1/2}\E\trans\u_k^*  |.
    \label{randomsdca}
\end{align}
The above three terms can be bounded similarly as in the proof of Lemma \ref{lemma:1rk3}. We will first show the sparsity of $\a\trans\W_k  \M_{v_k}$, $ (\a\trans\W_k  \M_{v_k}-\a\trans \W^{*}_k {\M}_{v_k}^*)$, and $\a\trans(\W_k -\W^*_k)$. 

Let us define $s = c (r^* + s_u + s_v)$.
Since both $\widetilde{\l}_k$ and ${\l}_k^*$ are $s$-sparse, we can show that 
\begin{align*}
	&\norm{\a\trans\W_k  \M_{v_k}}_0 = \norm{(\a\trans\W_k  ( \wt{\v}_k\trans\wt{\v}_k)^{-1}  \wt{d}_k \widetilde{\r}_k) \cdot \widetilde{\l}_k\trans}_0 \leq s, \\ 
	&\norm{\a\trans\W_k  \M_{v_k} -\a\trans \W^{*}_k {\M}_{v_k}^*}_0 \leq   \norm{\widetilde{\l}_k\trans}_0 + \norm{{\l}_k\strans }_0 \leq s.
\end{align*}
It follows from Lemma \ref{lemma:weak:wbound} that  
$\norm{\a\trans(\W_k -\W^*_k)}_0 \leq  s.$ For $\M_{v_k}$ and ${\M}_{v_k}^*$, by Lemma \ref{1} it holds that 
\begin{align}
	&\|\M_{v_k}\|_2 \leq  ( \wt{\v}_k\trans\wt{\v}_k)^{-1} |\wt{d}_k| \|  \widetilde{\r}_k \|_2 \|\widetilde{\l}_k\trans\|_2 \leq c d_k^{*-1}, \label{mzxx1} \\[5pt]
	&\norm{{\M}_{v_k}^*}_2 \leq  ( {\v}_k\strans{\v}_k^*)^{-1} |d_k^*| \| {\r}_k^* \|_2 \|{\l}_k\strans\|_2 \leq c d_k^{*-1},  \label{mzxx2}
\end{align}
\begin{align}
	\norm{\M_{v_k} - {\M}_{v_k}^*}_2 &\leq ( \wt{\v}_k\trans\wt{\v}_k)^{-1} \norm{ \wt{d}_k \widetilde{\r}_k\widetilde{\l}_k\trans  -  d_k^* {\r}_k^*{\l}_k\strans}_2 \nonumber\\ &\qquad+ |( \wt{\v}_k\trans\wt{\v}_k)^{-1}   -  ( {\v}_k\strans{\v}_k^*)^{-1}| \norm{ d_k^* {\r}_k^*{\l}_k\strans}_2 \leq c \gamma_n  d_k^{*-2}.  \label{mzxx3}
\end{align}

We are now ready to bound the three terms in \eqref{randomsdca}. Using the sparsity of $\a\trans\W_k  \M_{v_k}$, we have that 
\begin{align*}
	|\a\trans\W_k  \M_k  n^{-1/2}\E(\wt{\v}_k - {\v}_k^{*})| &\leq \|\a\trans\W_k  \M_{v_k} \|_2 \| n^{-1} \X\trans \E(\wt{\v}_k - {\v}_k^{*})\|_{2,s} \\
	&\leq \|\a\trans\W_k \|_2 \| \M_{v_k} \|_2  s^{1/2} \norm{ n^{-1} \X\trans \E }_{\max} \norm{\wt{\v}_k - {\v}_k^{*} }_{1} \\
	&\leq c   d_k^{*-1}  s   \{n^{-1}\log(pq)\}^{1/2} \gamma_n \\
	&\leq  c (r^*+s_u + s_v)^{3/2}\eta_n^2\{ n^{-1}\log(pq)\}    d_k^{*-1},
\end{align*}
where we have used \eqref{mzxx1} and Lemmas \ref{1} and \ref{lemma:weak:wbound}.

Similarly, it follows from \eqref{mzxx2}, \eqref{mzxx3}, and Lemmas \ref{1} and \ref{lemma:weak:wbound} that 
\begin{align*}
	&|(\a\trans\W_k  \M_k -\a\trans \W^{*}_k \M_{k}^*)n^{-1/2} \E{\v}_k^{*} | \\ & \leq \|\a\trans\W_k  \M_{v_k} -\a\trans \W^{*}_k {\M}_{v_k}^*\|_2 \|n^{-1} \X\trans \E {\v}_k^{*} \|_{2,s} \\[5pt]
	&\leq (\|\a\trans\W_k \|_2 \|  \M_{v_k} - {\M}_{v_k}^*\|_2  + \|\a\trans(\W_k  - \W^{*}_k) \|_2 \| {\M}_{v_k}^*\|_2  )   s^{1/2} \norm{ n^{-1} \X\trans \E }_{\max} \norm{{\v}_k^{*} }_{1} \\[5pt]
	&\leq  c(   d_k^{*-2} \gamma_n + d_1^{*} d_{k^*}^{*-2}   d_k^{*-1} \gamma_n ) s \{n^{-1}\log(pq)\}^{1/2}  d_{k}^* \\[5pt]
	& \leq  c (r^*+s_u + s_v)^{3/2}\eta_n^2\{ n^{-1}\log(pq)\}   d_1^{*}  d_k^{*-2}.
\end{align*}

Moreover, by resorting to Lemma \ref{lemma:weak:wbound}, we can deduce that 
\begin{align*}
	&|\a\trans(\W_k -\W^*_k)n^{-1/2}\E\trans\u_k^*  |  \leq \|\a\trans(\W_k -\W^*_k)\|_2 \| n^{-1/2}\E\trans\u_k^*  \|_{2,s} \\
	&\leq {\|\a\trans(\W_k -\W^*_k)\|_2}   s^{1/2} \norm{ n^{-1} \X\trans \E }_{\max} \norm{\u_k^*}_{1}  \\
	&\leq c d_1^{*} d_{k^*}^{*-2}   \gamma_n  s \{n^{-1}\log(pq)\}^{1/2}   \leq  c (r^*+s_u + s_v)^{3/2}\eta_n^2\{ n^{-1}\log(pq)\}  d_1^{*} d_{k^*}^{*-2}.
\end{align*}
Thus, combining the above terms yields that 
\begin{align*}
	\abs{-\a\trans\W_k\bepsilon_k - h_k / \sqrt{n}} \leq  c (r^*+s_u + s_v)^{3/2}\eta_n^2\{ n^{-1}\log(pq)\}   d_1^{*} d_{k^*}^{*-2}.
\end{align*}
This concludes the proof of Lemma  \ref{lemma:weak:ep}.

\subsection{Proof of Lemma \ref{lemma:disw}} \label{new.Sec.B.5a9}

Observe that 
\begin{align}\label{1223zz}
	&\sqrt{n} \a\trans\W_k \sum_{i \neq k}   \wt{\v}_i (\wh{\u}_i^t  - \u_i^*)\trans\u_k^*   \nonumber\\
	& =
	\sqrt{n}\a\trans{\W}_k^* \sum_{i \neq k}   \wt{\v}_i (\wh{\u}_i^t  - \u_i^*)\trans\u_k^* + \sqrt{n} \a\trans(\W_k  - \W_k^*)\sum_{i \neq k}   \wt{\v}_i (\wh{\u}_i^t  - \u_i^*)\trans\u_k^* \nonumber\\
	&=   \sqrt{n} \a\trans{\W}_k^* \sum_{i \neq k} {\v}_i^* (\wh{\u}_i^t  -\u_i^*)\trans\u_k^* +  \sqrt{n} \a\trans{\W}_k^*  \sum_{i \neq k}( \wt{\v}_i -  {\v}_i^*) (\wh{\u}_i^t  -\u_i^*)\trans\u_k^* \nonumber\\ 
	&\quad  +  \sqrt{n} \a\trans(\W_k  - \W_k^*) \sum_{i \neq k} \wt{\v}_i (\wh{\u}_i^t  -\u_i^*)\trans\u_k^* \nonumber \\
	&=: A_1 + A_2 + A_3.
\end{align}
To bound the above three terms, we will first analyze term $\wh{\u}_i^t  -\u_i^*$.

Since $\u_i^* = (\bmu_i\strans \wh{\bSigma}  \bmu_i^*)^{-1/2} n^{-1/2} \X \bmu_i^*$ and 
$ \wh{\u}_i^t =  (\wt{\bmu}_i\trans \wh{\bSigma}  \wt{\bmu}_i)^{-1/2} n^{-1/2} \X \wh{\bmu}_i^t$,  we see that $\wh{\u}_i^t  -\u_i^*$ can be represented as 
\begin{align}
	&\wh{\u}_i^t  -\u_i^* =  (\wt{\bmu}_i\trans \wh{\bSigma}  \wt{\bmu}_i)^{-1/2} n^{-1/2} \X \wh{\bmu}_i^t -  (\bmu_i\strans \wh{\bSigma}  \bmu_i^*)^{-1/2} n^{-1/2} \X \bmu_i^* \nonumber \\
	& = (\bmu_i\strans \wh{\bSigma}  \bmu_i^*)^{-1/2}  n^{-1/2} \X (\wh{\bmu}_i^t - \bmu_i^* ) + [   (\wt{\bmu}_i\trans \wh{\bSigma}  \wt{\bmu}_i)^{-1/2} -   (\bmu_i\strans \wh{\bSigma}  \bmu_i^*)^{-1/2}]  n^{-1/2} \X \bmu_i^*  \nonumber \\
	&\quad + [   (\wt{\bmu}_i\trans \wh{\bSigma}  \wt{\bmu}_i)^{-1/2} -   (\bmu_i\strans \wh{\bSigma}  \bmu_i^*)^{-1/2}] n^{-1/2} \X (\wh{\bmu}_i^t - \bmu_i^* ). \label{zcpaisd}
\end{align}
Based on this and $\u_i^* =  ( \bmu_i\strans \wh{\bSigma}  \bmu_i^*)^{-1/2} n^{-1/2} \X  \bmu_i^*$,
term $A_1$ can be decomposed as 
\begin{align}\label{1223zzz}
	&A_1 = \sqrt{n} \a\trans{\W}_k^* \sum_{i \neq k} {\v}_i^* (\wh{\u}_i^t  -\u_i^*)\trans\u_k^* \nonumber \\[5pt]
	&= \sqrt{n} \a\trans{\W}_k^* \sum_{i \neq k} {\v}_i^* ( \bmu_k\strans \wh{\bSigma}  \bmu_k^*)^{-1/2} (  (\wt{\bmu}_i\trans \wh{\bSigma}  \wt{\bmu}_i)^{-1/2} \wh{\bmu}_i^t  -  ( \bmu_i\strans \wh{\bSigma}  \bmu_i^*)^{-1/2} \bmu_i^*)\trans \wh{\bSigma} \bmu_k^*  \nonumber\\[5pt]
	&= \sqrt{n} \a\trans{\W}_k^*  \sum_{i \neq k} {\v}_i^*  ( \bmu_k\strans \wh{\bSigma}  \bmu_k^*)^{-1/2} (\bmu_i\strans \wh{\bSigma}  \bmu_i^*)^{-1/2}   (\wh{\bmu}_i^t - \bmu_i^* )\trans   \wh{\bSigma} \bmu_k^* \nonumber\\[5pt]
	&~ +  \sqrt{n} \a\trans{\W}_k^*  \sum_{i \neq k} {\v}_i^* ( \bmu_k\strans \wh{\bSigma}  \bmu_k^*)^{-1/2}
	[   (\wt{\bmu}_i\trans \wh{\bSigma}  \wt{\bmu}_i)^{-1/2} -   (\bmu_i\strans \wh{\bSigma}  \bmu_i^*)^{-1/2}]    \bmu_i\strans  \wh{\bSigma}  \bmu_k^* \nonumber \\[5pt]
	&~ + \sqrt{n} \a\trans{\W}_k^*  \sum_{i \neq k} {\v}_i^* ( \bmu_k\strans \wh{\bSigma}  \bmu_k^*)^{-1/2} [   (\wt{\bmu}_i\trans \wh{\bSigma}  \wt{\bmu}_i)^{-1/2} -   (\bmu_i\strans \wh{\bSigma}  \bmu_i^*)^{-1/2}]  (\wh{\bmu}_i^t - \bmu_i^* )\trans \wh{\bSigma}  \bmu_k^* \nonumber \\[5pt]
	&=: A_{11} + A_{12} + A_{13}.
\end{align}
We will prove that the first term $A_{11}$ above is normally distributed, and the last two terms $A_{12}$ and $A_{13}$ are asymptotically negligible.

\noindent\textbf{(1). The asymptotic distribution of  $A_{11} $}. For term 
$$A_{11} =\sqrt{n}   \sum_{i \neq k} \a\trans{\W}_k^* {\v}_i^*  (\bmu_i\strans \wh{\bSigma}  \bmu_i^*)^{-1/2}  ( \bmu_k\strans \wh{\bSigma}  \bmu_k^*)^{-1/2}  (\wh{\bmu}_i^t - \bmu_i^* )\trans   \wh{\bSigma} \bmu_k^*,$$ 
let us define 
$	\omega_{k,i} =  (\bmu_i\strans \wh{\bSigma}  \bmu_i^*)^{-1/2}  ( \bmu_k\strans \wh{\bSigma}  \bmu_k^*)^{-1/2} \a\trans{\W}_k^* {\v}_i^*   =  ( \bmu_k\strans \wh{\bSigma}  \bmu_k^*)^{-1/2} \a\trans{\W}_k^* {\r}_i^*.$
By invoking Lemmas \ref{1} and \ref{lemma:weak:wbound}, we have that 
\begin{align}
	|\omega_{k,i} | \leq \norm{ \a\trans{\W}_k^*}_2  \norm{{\r}_i^*}_2  \abs{( \bmu_k\strans \wh{\bSigma}  \bmu_k^*)^{-1/2}} \leq  c d_k^{*-1}.
\end{align}

By Lemma \ref{lemma:threuk}  that $\operatorname{supp}(\wh{\bmu}_i^t) =  \mathcal{S}_{\mu_i}$,
term $A_{11}$ can be written as 
\begin{align*}
	A_{11} &=  \sum_{i \neq k} \sqrt{n} \omega_{k,i} (\wh{\bmu}_i^t - \bmu_i^* )\trans   \wh{\bSigma} \bmu_k^* \\ &=  \sum_{i \neq k} \sqrt{n} \omega_{k,i} (\wh{\bmu}_i^t - \bmu_i^* )\trans  (\wh{\bSigma} \bmu_k^*)^{t_i} +  \sum_{i \neq k} { \sqrt{n} \omega_{k,i} (\bmu_i\strans)_{ \mathcal{S}_{\mu_i}^c}   (\wh{\bSigma} \bmu_k^*)_{\mathcal{S}_{\mu_i}^c}}.
\end{align*}
Here, for a vector $\b = (b_1, \ldots, b_p)\trans \in \mathbb{R}^p$, denote by  $\b^{t_i} = (b_1^{t_i}, \ldots, b_p^{t_i})\trans \in \mathbb{R}^p$ with $b_j^{t_i} = b_j$ if $j \in \mathcal{S}_{\mu_i}$ and $b_j^{t_i} = 0$ otherwise. 

For the second term above, note that $\sum_{i \neq k} \| (\bmu_i\strans)_{ \mathcal{S}_{\mu_i}^c} \|_2  = o(n^{-1/2})$ in Condition \ref{con:threshold}. It holds that 
\begin{align*}
	\sum_{i \neq k} |\sqrt{n}& \omega_{k,i} (\bmu_i\strans)_{ \mathcal{S}_{\mu_i}^c}   (\wh{\bSigma} \bmu_k^*)_{\mathcal{S}_{\mu_i}^c} | \leq \sum_{i \neq k}\sqrt{n} |\omega_{k,i}|  \|(\bmu_i\strans)_{ \mathcal{S}_{\mu_i}^c} \|_2 \|  (\wh{\bSigma} \bmu_k^*)_{\mathcal{S}_{\mu_i}^c} \|_2 \\
	&\leq c \sum_{i \neq k} \sqrt{n} d_k^{*-1}   \|(\bmu_i\strans)_{ \mathcal{S}_{\mu_i}^c} \|_2  \|  \wh{\bSigma} \bmu_k^* \|_2 \leq c \sqrt{n}\sum_{i \neq k} \|(\bmu_i\strans)_{ \mathcal{S}_{\mu_i}^c} \|_2
	= o( 1 ),
\end{align*}
where we have used $\|  \wh{\bSigma} \bmu_k^* \|_2 \leq  c d_k^*$.
For the first term $\sum_{i \neq k} \sqrt{n} \omega_{k,i} (\wh{\bmu}_i - \bmu_i^* )\trans  (\wh{\bSigma} \bmu_k^*)^{t_i}$, notice that 
\begin{align*}
	\| (\wh{\bSigma} \bmu_k^*)^{t_i}\|_0 \leq s_u, \ \  \| (\wh{\bSigma} \bmu_k^*)^{t_i}\|_2 \leq  \| \wh{\bSigma} \bmu_k^*\|_2 \leq c d_k^*.
\end{align*}

Then in view of Lemma \ref{lemm:sofari}, replacing $\a$ with $(\wh{\bSigma} \bmu_k^*)^{t_i}$ leads to 
\begin{align}
	\sum_{i \neq k} \sqrt{n} \omega_{k,i} (\wh{\bmu}_i - \bmu_i^* )\trans   (\wh{\bSigma} \bmu_k^*)^{t_i} = \sum_{i \neq k} \omega_{k,i} h_i((\wh{\bSigma} \bmu_k^*)^{t_i}) + \sum_{i \neq k} \omega_{k,i} t_i((\wh{\bSigma} \bmu_k^*)^{t_i}). \label{1223zzzz}
\end{align}
The distribution term is
$$h_i((\wh{\bSigma} \bmu_k^*)^{t_i}) = ((\wh{\bSigma} \bmu_k^*)^{t_i})\trans \W^{*}_{u_i}(\X\trans\E\r_i^* - \M_{u_i}^{*} \E\trans\X {\bmu}_i^{*})/\sqrt{n} \sim N(0,\nu_i((\wh{\bSigma} \bmu_k^*)^{t_i})^2),$$ 
where the variance is given by 
\begin{align*}
	\nu_i((\wh{\bSigma} \bmu_k^*)^{t_i})^2 &= ((\wh{\bSigma} \bmu_k^*)^{t_i})\trans\W_{u_i}^*(\bmu_i\strans\wh{\bSigma}\bmu_i^* \M_{u_i}^* \bSigma_e \M_{u_i}^{* T} \\ & ~~ + \r_i\strans\bSigma_e\r_i^* \wh{\bSigma} - 2\wh{\bSigma} \bmu_i^*\r_i\strans\bSigma_e\M_{u_i}^{* T}) \W_{u_i}\strans(\wh{\bSigma} \bmu_k^*)^{t_i}.
\end{align*}
The error term is 
\begin{align*}
	\sum_{i \neq k} \omega_{k,i} t_i((\wh{\bSigma} \bmu_k^*)^{t_i}) &\leq c \sum_{i \neq k} s_{u}^{1/2}  ( \kappa_n \max\big\{1, d_i^{*-1}, d_i^{*-2}\big\}  + c   \gamma_n  d_i^{*-3} d_{i+1}^{*} \big(\sum_{j=1}^{i-1}d_j^*\big)) \\
	& \leq c r^* s_{u}^{1/2}  ( \kappa_n \max\big\{1, d_i^{*-1}, d_i^{*-2}\big\}  +   \gamma_n  d_1^{*} d_{r^*}^{*-2}),
\end{align*}
where 
$\kappa_n = \max\{s_{\max}^{1/2} , (r^*+s_u+s_v)^{1/2}, \eta_n^2\} (r^*+s_u+s_v)\eta_n^2\log(pq)/\sqrt{n}.$

\noindent\textbf{(2). Upper bounds on $A_{12}, A_{13}, A_2, A_3$}. For term $A_{12}$, it follows from Lemmas \ref{1} and \ref{lemma:weak:wbound}, \eqref{weakceq}, \eqref{daswqe0}, \eqref{daswqe}, and \eqref{sdczcsa} that 
\begin{align}\label{a1212}
	&|A_{12}| = |\sqrt{n} \a\trans{\W}_k^*  \sum_{i \neq k} {\v}_i^*
	[   (\wt{\bmu}_i\trans \wh{\bSigma}  \wt{\bmu}_i)^{-1/2} -   (\bmu_i\strans \wh{\bSigma}  \bmu_i^*)^{-1/2}]  ( \bmu_k\strans \wh{\bSigma}  \bmu_k^*)^{-1/2}  \bmu_i\strans  \wh{\bSigma}  \bmu_k^*| \nonumber \\[5pt]
	&\leq \sqrt{n} \norm{\a\trans{\W}_k^* }_2  \sum_{i \neq k} \norm{\v_i}_2 \big|   (\wt{\bmu}_i\trans \wh{\bSigma}  \wt{\bmu}_i)^{-1/2} -   (\bmu_i\strans \wh{\bSigma}  \bmu_i^*)^{-1/2}\big|  | \bmu_k\strans \wh{\bSigma}  \bmu_k^*|^{-1/2}  |  \bmu_i\strans  \wh{\bSigma}  \bmu_k^* | \nonumber \\[5pt]
	&\leq c \sqrt{n}  \sum_{i \neq k}  \gamma_n d_i^{*-2} d_k^{*-1} d_i^* d_i^* d_k^* | \l_i\strans  \wh{\bSigma}  \l_k^* | \nonumber  \\
	&\leq c    r^*  \gamma_n d_1^* d_{r^*}^{*-2}.
\end{align}

To bound terms $A_{13}, A_2, \text{ and } A_3$, we will first derive the upper bounds on $\|\sqrt{n}(\wh{\bmu}_k^t- \bmu_k^*)\|_2 $, $\|\wh{\u}_i^t  -\u_i^*\|_2$, and  $|\sqrt{n}(\wh{\u}_i^t  -\u_i^*)\trans\u_k^*|$, separately. 

\textbf{(2.1). The upper bound on $\|\sqrt{n}(\wh{\bmu}_k^t- \bmu_k^*)\|_2 $}.
In light of the definition of $\wh{\bmu}_k^t$, we have that $\wh{\mu}_{kj}^t = \wh{\mu}_{kj}$ if $ |\wh{\mu}_{kj}| \geq \frac{ \log n}{ \sqrt{n}}$ and $\wh{\mu}_{kj}^t = 0$ otherwise.
From Lemma \ref{lemma:threuk}, we see that $\operatorname{supp}(\wh{\bmu}_k^t) =  \mathcal{S}_{\mu_k}$.
Then for $j \notin \mathcal{S}_{\mu_k}$, it holds that $\wh{\mu}_{kj}^t = 0$. Also, from Condition \ref{con:threshold} we have 
$  \norm{(\bmu_k^*)_{ \mathcal{S}_{\mu_k}^c}}_2 = o(\frac{ 1}{\sqrt{n}})$. Hence, for sufficiently large $n$, it follows that 
\begin{align}\label{czlkad}
	\sum_{j \notin \mathcal{S}_{\mu_k}} | \sqrt{n}(\wh{\mu}_{kj}^t -  {\mu}_{{kj}}^* ) |^2 = \sum_{j \notin \mathcal{S}_{\mu_k}} | \sqrt{n}  {\mu}_{{kj}}^*  |^2 =  \norm{\sqrt{n}  (\bmu_k^*)_{ \mathcal{S}_{\mu_k}^c}}_2^2 = o(1).
\end{align}

For  $j \in  \mathcal{S}_{\mu_k}$, it holds that $\wh{\mu}_{kj}^t = \wh{\mu}_{kj}$. 
Applying Lemma \ref{lemm:sofari} by replacing $\a$ with $\e_j$, for each $j = 1, \ldots, p$ we can obtain that 
\begin{align}
	\sqrt{n} \e_j\trans (\wh{\bmu}_k-\bmu_k^*) = h_{k,j} + t_{k,j}, \nonumber 
\end{align}
where the distribution term $h_{k,j} = \e_j\trans \W^{*}_k(\X\trans\E\v_k^* - \M_{k}^{*} \E\trans\X {\u}_k^{*})/\sqrt{n} \sim \N(0,\nu_{k,j}^2)$ with
\begin{align*}
	\nu_{k,j}^2 = \e_j\trans\W_k^*(z_{kk}^{*}\M_{k}^{*}\bSigma_e\M_{k}\strans + \v_k\strans\bSigma_e\v_k^* \wh{\bSigma} - 2\wh{\bSigma}\u_k^*  \v_k\strans\bSigma_e\M_{k}^{* T})\W_k\strans\e_j.
\end{align*}

With the aid of Theorem 6 in \cite{sofari}, we can show that 
$ \nu_{k,j} \leq c.$ Since $ t_{k,j} = o(1)$, it holds that $$|\sqrt{n} \e_j\trans (\wh{\bmu}_k-\bmu_k^*)| = | \sqrt{n}(\wh{\mu}_{kj}^t -  {\mu}_{{kj}}^*) | \leq  c| h_{k,j}   | \leq c \nu_{k,j} \leq c.$$
This further entails that 
\begin{align*}
	\sum_{j \in \mathcal{S}_{\mu_k}} | \sqrt{n}(\wh{\mu}_{kj}^t -  {\mu}_{{kj}}^*) |^2 \leq c |\mathcal{S}_{\mu_k}| \leq c s_u.
\end{align*}
For sufficiently large $n$, this along with \eqref{czlkad} yields that 
\begin{align}\label{whu1463}
	\norm{\sqrt{n}(\wh{\bmu}_k^t - {\bmu}_k^*)}_2 &=\Big(\sum_{j \in \mathcal{S}_{\mu_k}} | \sqrt{n}\wh{\mu}_{kj}^t -  {\mu}_{{kj}}^*) |^2  +(\sum_{j \notin \mathcal{S}_{\mu_k}} | \sqrt{n}(\wh{\mu}_{kj}^t -  {\mu}_{{kj}}^* ) |^2 \Big)^{1/2} \nonumber\\
	& \leq  c s_{u}^{1/2}.
\end{align}

\textbf{(2.2). The upper bound on $\|\wh{\u}_i^t  -\u_i^*\|_2$}.
Similarly, we bound $\|\wh{\u}_i^t  -\u_i^*\|_2$, which follows the decomposition in \eqref{zcpaisd}.
Note that $ \norm{ \sqrt{n}(\wh{\bmu}_i^t - \bmu_i^* ) }_2  \leq c s_u^{1/2}$ in \eqref{whu1463} and $\operatorname{supp}(\wh{\bmu}_k^t) =  \mathcal{S}_{u_k}$ with  $|\mathcal{S}_{u_k}| \leq s_u$ in Lemma \ref{lemma:threuk}, which implies that $\norm{ \wh{\bmu}_i^t - \bmu_i^*  }_0 \leq 2 s_u$. By Condition \ref{con2:re}, it holds that 
\begin{align*}
   \sqrt{n} \norm{ n^{-1/2} \X (\wh{\bmu}_i^t - \bmu_i^* ) }_2  \leq \norm{ \sqrt{n} (\wh{\bmu}_i^t - \bmu_i^* ) }_2  \leq c s_u^{1/2}.
\end{align*}
Based on this and Lemma \ref{1}, we can deduce that
\begin{align*}
	\|\sqrt{n} (\bmu_i\strans \wh{\bSigma}  \bmu_i^*)^{-1/2}  n^{-1/2} \X (\wh{\bmu}_i^t - \bmu_i^* ) \|_2 &\leq c (\bmu_i\strans \wh{\bSigma}  \bmu_i^*)^{-1/2}  \norm{ \sqrt{n}(\wh{\bmu}_i^t - \bmu_i^* ) }_2 \\ &\leq c d_i^{*-1} s_u^{1/2}.
\end{align*}

Further, we can show that 
\begin{align*}
	&\|\sqrt{n} [   (\wt{\bmu}_i\trans \wh{\bSigma}  \wt{\bmu}_i)^{-1/2} -   (\bmu_i\strans \wh{\bSigma}  \bmu_i^*)^{-1/2}]  n^{-1/2}\X \bmu_i^* \|_2 \leq c \sqrt{n} \gamma_n d_i^{*-1}, \\[5pt]
	&\| \sqrt{n} [   (\wt{\bmu}_i\trans \wh{\bSigma}  \wt{\bmu}_i)^{-1/2} -   (\bmu_i\strans \wh{\bSigma}  \bmu_i^*)^{-1/2}] n^{-1/2} \X (\wh{\bmu}_i^t - \bmu_i^* )\|_2 \leq  c s_u^{1/2}\gamma_n d_i^{*-2}.
\end{align*}
Thus, for sufficiently large $n$, combining the above three terms with \eqref{zcpaisd} leads to 
\begin{align}
	\|\wh{\u}_i^t  -\u_i^* \|_2 \leq c  \gamma_n d_i^{*-1}. \label{whueqz2}
\end{align}

\textbf{(2.3). The upper bound on $|\sqrt{n}(\wh{\u}_i^t  -\u_i^*)\trans\u_k^*|$}.
Similar to \eqref{zcpaisd}, it holds that 
\begin{align*}
	& \sqrt{n}(\wh{\u}_i^t  -\u_i^*)\trans\u_k^* \\[5pt]
	&=   \sqrt{n}(\bmu_i\strans \wh{\bSigma}  \bmu_i^*)^{-1/2}  ( \bmu_k\strans \wh{\bSigma}  \bmu_k^*)^{-1/2}  (\wh{\bmu}_i^t - \bmu_i^* )\trans   \wh{\bSigma} \bmu_k^* \\[5pt]
	&\quad +  
	\sqrt{n} [   (\wt{\bmu}_i\trans \wh{\bSigma}  \wt{\bmu}_i)^{-1/2} -   (\bmu_i\strans \wh{\bSigma}  \bmu_i^*)^{-1/2}]  ( \bmu_k\strans \wh{\bSigma}  \bmu_k^*)^{-1/2}  \bmu_i\strans  \wh{\bSigma}  \bmu_k^* \\[5pt]
	&\quad + \sqrt{n}  [   (\wt{\bmu}_i\trans \wh{\bSigma}  \wt{\bmu}_i)^{-1/2} -   (\bmu_i\strans \wh{\bSigma}  \bmu_i^*)^{-1/2}] ( \bmu_k\strans \wh{\bSigma}  \bmu_k^*)^{-1/2}   (\wh{\bmu}_i^t - \bmu_i^* )\trans \wh{\bSigma}  \bmu_k^*.
\end{align*}
By \eqref{whu1463}, we have $ \norm{ \sqrt{n}(\wh{\bmu}_i^t - \bmu_i^* ) }_2  \leq c s_u^{1/2}.$ Then from Lemma \ref{1}, we can deduce that 
\begin{align*}
	&|\sqrt{n}(\bmu_i\strans \wh{\bSigma}  \bmu_i^*)^{-1/2}  ( \bmu_k\strans \wh{\bSigma}  \bmu_k^*)^{-1/2}  (\wh{\bmu}_i^t - \bmu_i^* )\trans   \wh{\bSigma} \bmu_k^*| \\[5pt]
	&\leq c (\bmu_i\strans \wh{\bSigma}  \bmu_i^*)^{-1/2}  ( \bmu_k\strans \wh{\bSigma}  \bmu_k^*)^{-1/2} \norm{ \sqrt{n}(\wh{\bmu}_i^t - \bmu_i^* ) }_2 \norm{ \wh{\bSigma} \bmu_k^*}_2 \\[5pt]
	&\leq c d_i^{*-1} s_u^{1/2}.
\end{align*}

Moreover, we can show that
\begin{align*}
	&|\sqrt{n} [   (\wt{\bmu}_i\trans \wh{\bSigma}  \wt{\bmu}_i)^{-1/2} -   (\bmu_i\strans \wh{\bSigma}  \bmu_i^*)^{-1/2}]  ( \bmu_k\strans \wh{\bSigma}  \bmu_k^*)^{-1/2}  \bmu_i\strans  \wh{\bSigma}  \bmu_k^*| \leq c \sqrt{n} \gamma_n d_i^{*-1}  | \l_i\strans  \wh{\bSigma}  \l_k^* |, \\[5pt]
	&| \sqrt{n}  [   (\wt{\bmu}_i\trans \wh{\bSigma}  \wt{\bmu}_i)^{-1/2} -   (\bmu_i\strans \wh{\bSigma}  \bmu_i^*)^{-1/2}] ( \bmu_k\strans \wh{\bSigma}  \bmu_k^*)^{-1/2}  (\wh{\bmu}_i^t - \bmu_i^* )\trans \wh{\bSigma}  \bmu_k^*| \leq  c s_u^{1/2}\gamma_n d_i^{*-2}.
\end{align*}
Hence, for sufficiently large $n$, it follows that 
\begin{align}
	|\sqrt{n}(\wh{\u}_i^t  -\u_i^*)\trans\u_k^*| \leq c d_i^{*-1} s_u^{1/2}. \label{whueqz}
\end{align}

We are now ready to bound terms $A_{13}, A_2, \text{ and } A_3$.
For term $A_{13}$, we have that 
\begin{align}\label{a131313}
	&|A_{13}| \nonumber \\&= |\sqrt{n} \a\trans{\W}_k^*  \sum_{i \neq k} {\v}_i^*  ( \bmu_k\strans \wh{\bSigma}  \bmu_k^*)^{-1/2} [   (\wt{\bmu}_i\trans \wh{\bSigma}  \wt{\bmu}_i)^{-1/2} -   (\bmu_i\strans \wh{\bSigma}  \bmu_i^*)^{-1/2}]  (\wh{\bmu}_i^t - \bmu_i^* )\trans \wh{\bSigma}  \bmu_k^*| \nonumber \\[5pt] 
	&\leq   \sum_{i \neq k}  \norm{{\v}_i^* }_2   |(\wt{\bmu}_i\trans \wh{\bSigma}  \wt{\bmu}_i)^{-1/2} -   (\bmu_i\strans \wh{\bSigma}  \bmu_i^*)^{-1/2} | \norm{\sqrt{n} (\wh{\bmu}_i^t - \bmu_i^* ) }_2  \nonumber \\
	&\qquad \times  ( \bmu_k\strans \wh{\bSigma}  \bmu_k^*)^{-1/2} \|\a\trans{\W}_k^*\|_2 \norm{  \wh{\bSigma}  \bmu_k^*}_2 \nonumber\\
	&\leq c  \sum_{i \neq k} d_i^*  \gamma_n d_i^{*-2} s_u^{1/2}   \leq  c r^*  s_u^{1/2} \gamma_n    d_{r^*}^{*-1}.
\end{align}

It remains to bound terms $A_2, A_3$. Using \eqref{whueqz} and Lemmas \ref{1} and \ref{lemma:weak:wbound}, it holds that 
\begin{align}\label{a2222222}
	|A_2| &= |\sqrt{n} \a\trans{\W}_k^*  \sum_{i \neq k}( \wt{\v}_i -  {\v}_i^*) (\wh{\u}_i^t  -\u_i^*)\trans\u_k^*| \nonumber \\
	&\leq \norm{\a\trans{\W}_k^*}_2 \sum_{i \neq k} \norm{  \wt{\v}_i -  {\v}_i^*}_2 |\sqrt{n} (\wh{\u}_i^t  -\u_i^*)\trans\u_k^*| \nonumber\\
	&\leq c  r^* s_u^{1/2}  d_{r^*}^{*-1} \gamma_n.
\end{align}
With the aid of \eqref{whueqz} and Lemmas \ref{1} and \ref{lemma:weak:wbound},  we can deduce that 
\begin{align}\label{a32222222}
	| A_3| &= |\sqrt{n} \a\trans(\W_k  - \W_k^*) \sum_{i \neq k} \wt{\v}_i (\wh{\u}_i^t  -\u_i^*)\trans\u_k^*| \nonumber \\
	&\leq \norm{ \a\trans(\W_k  - \W_k^*)}_2  \sum_{i \neq k} \norm{\wt{\v}_i  }_2 |\sqrt{n} (\wh{\u}_i^t  -\u_i^*)\trans\u_k^*| \nonumber \\ 
	&\leq c r^* s_u^{1/2} d_1^{*} d_{k^*}^{*-2} \gamma_n. 
\end{align}
Thus, it follows from \eqref{a1212}, \eqref{a131313}, \eqref{a2222222}, and \eqref{a32222222} that the error bound for $A_{12}, A_{13}, A_2, A_3$ is given by 
\begin{align}\label{1223zzzzz}
	t^{\prime} = O(r^* s_u^{1/2} d_1^{*} d_{r^*}^{*-2} \gamma_n ).
\end{align}

Therefore, combining \eqref{1223zz}, \eqref{1223zzz}, \eqref{1223zzzz}, and \eqref{1223zzzzz} yields that 
\begin{align*}
	\sqrt{n} \a\trans\W_k \sum_{i \neq k}   \wt{\v}_i (\wh{\u}_i^t  - \u_i^*)\trans\u_k^* =  \sum_{i \neq k} \omega_{k,i} h_i((\wh{\bSigma} \bmu_k^*)^{t_i}) + \sum_{i \neq k}\omega_{k,i}  t_i((\wh{\bSigma} \bmu_k^*)^{t_i}) + t^{\prime}.
\end{align*}
Here, the distribution term is
$$h_i((\wh{\bSigma} \bmu_k^*)^{t_i}) = ((\wh{\bSigma} \bmu_k^*)^{t_i})\trans \W^{*}_{u_i}(\X\trans\E\r_i^* - \M_{u_i}^{*} \E\trans\X {\bmu}_i^{*})/\sqrt{n} \sim N(0,\nu_i((\wh{\bSigma} \bmu_k^*)^{t_i})^2),$$ where the variance is given by 
\begin{align*}
	\nu_i((\wh{\bSigma} \bmu_k^*)^{t_i})^2 &= ((\wh{\bSigma} \bmu_k^*)^{t_i})\trans\W_{u_i}^*(\bmu_i\strans\wh{\bSigma}\bmu_i^* \M_{u_i}^* \bSigma_e \M_{u_i}^{* T} \\ &~ + \r_i\strans\bSigma_e\r_i^* \wh{\bSigma} - 2\wh{\bSigma} \bmu_i^*\r_i\strans\bSigma_e\M_{u_i}^{* T}) \W_{u_i}\strans(\wh{\bSigma} \bmu_k^*)^{t_i}.
\end{align*}
The error term is 
\begin{align*}
	\sum_{i \neq k}\omega_{k,i}  t_i((\wh{\bSigma} \bmu_k^*)^{t_i})  = r^* s_{u}^{1/2}  ( \kappa_n \max\big\{1, d_i^{*-1}, d_i^{*-2}\big\}  + \sum_{i \neq k} s_{u}^{1/2}   \gamma_n d_i^{*-3} d_{i+1}^{*} \big(\sum_{i=1}^{i-1}d_i^*\big))
\end{align*}
with 
$\kappa_n = \max\{s_{\max}^{1/2} , (r^*+s_u+s_v)^{1/2}, \eta_n^2\} (r^*+s_u+s_v)\eta_n^2\log(pq)/\sqrt{n}.$
Moreover, it holds that 
	$t^{\prime} = O(r^* s_u^{1/2} d_1^{*}  d_{r^*}^{*-2} \gamma_n ).$
This completes the proof of Lemma \ref{lemma:disw}.

\subsection{Lemma \ref{1} and its proof} \label{new.Sec.B.3}

\begin{lemma}\label{1}
	Assume that Conditions \ref{con2:re} and \ref{con3:eigend}  hold, and $\wt{\C}$ satisfies Definition \ref{defi:sofar}. Then with probability at least
$1- \theta_{n,p,q}$ for $\theta_{n,p,q}$ given in \eqref{thetapro}, we have that for sufficiently large $n$ and each $i=1, \ldots, r^*$, 

\noindent	(a) $\|\v_i^*\|_2 \leq c d_i^*, \ \|\wt{\v}_i\|_2 \leq c d_i^*,$ $\norm{\wt{\v}_i - \v_i^* }_2 \leq c \gamma_n$,
    
\noindent    (b)  $| \v_i\strans \v_i^*|^{-1} \leq c d_i^{*-2}$,  $ | (\wt{\v}_i\trans \wt{\v}_i)^{-1} -  (\v_i\strans \v_i^*)^{-1} | \leq c \gamma_n d_i^{*-3}$,
    
 \noindent    (c) $\|\u_i^*\|_2 \leq c, \ \|\wt{\u}_i\|_2 \leq c,$ \ $\norm{\wt{\u}_i - \u_i^* }_2 \leq c \gamma_n d_i^{*-1}$, 
     
\noindent	(d) $\|\U^*_{-k}\|_2 \leq c, \ \| \wt{\U}_{-k}\|_2 \leq c, \ \|\V^*_{-k}\|_2 \leq c d_1^*, \  \|\wt{\V}_{-k}\|_2 \leq c d_1^*,$
    
\noindent	(e) $\| \wt{\U}_{-k} -	\U_{-k}^* \|_2 \leq  c \gamma_n d_1^* d_{r^*}^{*-2}, \ \ 
	\| \wt{\V}_{-k} -	\V_{-k}^* \|_2 \leq   c \gamma_n d_1^* d_{r^*}^{*-1},$ \\[5pt]
\qquad \qquad $\| \wt{\U}_{-k}\wt{\V}_{-k}\trans  -	\U_{-k}^*\V_{-k}\strans \|_2 \leq  c \gamma_n$, 

\noindent where $\gamma_n = (r^*+s_u+s_v)^{1/2} \eta_n^2\{n^{-1}\log(pq)\}^{1/2}$ and $c$ is some positive constant.
\end{lemma}

\noindent \textit{Proof}. Let us first prove parts (a) and (b). Observe that $\v_i^* =   (\l_i\strans \wh{\bSigma}\l_i^*)^{1/2} d_i^* \r_i^*$ and $\wt{\v}_i =   (\wt{\l}_i\trans \wh{\bSigma}\wt{\l}_i)^{1/2} \wt{d}_i \wt{\r}_i$.
By Condition \ref{con2:re}, $\norm{\l_i^*}_0 = s_u$, and $\norm{\l_i^*}_2 = 1$, we have that $|\l_i\strans \wh{\bSigma}\l_i^*| \leq \norm{\l_i\strans}_2 \norm{\wh{\bSigma}\l_i^*}_2 \leq c  $. This together with  $\norm{\r_i^*}_2 = 1$ leads to 
\begin{align}\label{eqcvvc}
	\norm{\v_i^* }_2 \leq (\l_i\strans \wh{\bSigma}\l_i^*)^{1/2}  d_i^* \norm{\r_i^*}_2 \leq c  d_i^*.
\end{align}
Since $\bmu_i^* = d_i^* \l_i^*$ and $\wt{\bmu}_i = \wt{d}_i \wt{\l}_i$, we can deduce that 
\begin{align}
	\norm{\wt{\v}_i - \v_i^* }_2 &\leq \norm{ (\wt{\l}_i\trans \wh{\bSigma}\wt{\l}_i)^{1/2} \wt{d}_i \wt{\r}_i -   (\l_i\strans \wh{\bSigma}\l_i^*)^{1/2} d_i^* \r_i^*}_2 \nonumber \\
	&= \norm{ (\wt{\bmu}_i\trans \wh{\bSigma}\wt{\bmu}_i)^{1/2} \wt{\r}_i -   (\bmu_i\strans \wh{\bSigma}\bmu_i^*)^{1/2}  \r_i^*}_2 \nonumber\\ 
	&\leq |\bmu_i\strans \wh{\bSigma}\bmu_i^* |^{1/2} \norm{  \wt{\r}_i  -  \r_i^* }_2 + | (\wt{\bmu}_i\trans \wh{\bSigma}\wt{\bmu}_i) ^{1/2} - (\bmu_i\strans \wh{\bSigma}\bmu_i^*)^{1/2} | \norm{\wt{\r}_i }_2 \nonumber\\
	&\leq  c  |\bmu_i\strans \wh{\bSigma}\bmu_i^* |^{1/2} \gamma_n d_i^{*-1} + | (\wt{\bmu}_i\trans \wh{\bSigma}\wt{\bmu}_i) ^{1/2} - (\bmu_i\strans \wh{\bSigma}\bmu_i^*)^{1/2} |, \label{xzcadf}
\end{align}
where the last step above has used $\norm{\wt{\r}_i }_2 = 1$ and part (a) of Lemma 6 in \cite{sofari} that $\norm{  \wt{\r}_i  -  \r_i^* }_2 \leq c \gamma_n d_i^{*-1}$. 

For term $|\bmu_i\strans \wh{\bSigma}\bmu_i^* |$, it follows that 
\begin{align}\label{daswqe0}
     |\bmu_i\strans \wh{\bSigma}\bmu_i^* | = d_i^{*2} |\l_i\strans \wh{\bSigma}\l_i^*| \leq c d_i^{*2}. 
\end{align}
Based on Definition 2, an application of similar arguments gives that 
\begin{align}\label{daswqe01}
     |\wt{\bmu}_i\trans \wh{\bSigma}\wt{\bmu}_i | \leq c d_i^{*2}. 
\end{align}
Further, from part (c) of Lemma 6 in \cite{sofari}, it holds that 
\begin{align}
    &({\bmu}_i\strans\wh{\bSigma}{\bmu}_i^{*})^{-1} \leq  cd_i^{*-2},   \ (\wt{\bmu}_i\trans \wh{\bSigma}\wt{\bmu}_i)^{-1} \leq  cd_i^{*-2}, \label{daswqe}\\[5pt]
    &| \wt{\bmu}_i\trans \wh{\bSigma}\wt{\bmu}_i -\bmu_i\strans \wh{\bSigma}\bmu_i^* | \leq c\gamma_n d_i^*, \ | (\wt{\bmu}_i\trans \wh{\bSigma}\wt{\bmu}_i)^{-1} -(\bmu_i\strans \wh{\bSigma}\bmu_i^*)^{-1} | \leq c\gamma_n d_i^{*-3}. \label{daswqe2}
\end{align}
Hence, by some calculations, we can obtain that
\begin{align}\label{sdhajhc}
	| (\wt{\bmu}_i\trans \wh{\bSigma}\wt{\bmu}_i )^{1/2} - (\bmu_i\strans \wh{\bSigma}\bmu_i^*)^{1/2}  | = \frac{ | \wt{\bmu}_i\trans \wh{\bSigma}\wt{\bmu}_i -\bmu_i\strans \wh{\bSigma}\bmu_i^* | }{(\wt{\bmu}_i\trans \wh{\bSigma}\wt{\bmu}_i )^{1/2} + (\bmu_i\strans \wh{\bSigma}\bmu_i^*)^{1/2} } \leq c \gamma_n.
\end{align}

Combining \eqref{xzcadf}, \eqref{daswqe0}, and \eqref{sdhajhc} leads to 
\begin{align*}
	\norm{\wt{\v}_i - \v_i^* }_2  \leq  c \gamma_n.
\end{align*}
This along with \eqref{eqcvvc} yields that for sufficiently large $n$, 
\begin{align*}
	\norm{\wt{\v}_i  }_2 \leq \norm{ \v_i^* }_2 +  \norm{\wt{\v}_i - \v_i^* }_2  \leq c d_i^*,
\end{align*}
which completes the proof of part (a).

For the proof of part (b),
observe that $ \v_i\strans \v_i^*  =   \bmu_i\strans \wh{\bSigma}\bmu_i^*$ and $\wt{\v}_i\trans \wt{\v}_i = \wt{\bmu}_i\trans \wh{\bSigma} \wt{\bmu}_i$. Then the results of part (b) follow from \eqref{daswqe} and \eqref{daswqe2}.

We next prove part (c).
Notice that $$\u_i^* =  (\l_i\strans \wh{\bSigma}\l_i^*)^{-1/2} n^{-1/2} \X \l_i^* = (\bmu_i\strans \wh{\bSigma}\bmu_i^*)^{-1/2} n^{-1/2} \X \bmu_i^* .$$ Since $\norm{\bmu_i^*}_0 \leq s_u$ and $\|\bmu_i^*\|_2 = d_i^*\|\l_i^*\|_2 = d_i^*$, by Condition \ref{con2:re} it holds that 
\begin{align*}
  \|n^{-1/2} \X \bmu_i^*\|_2^2 = |\bmu_i\strans \wh{\bSigma}\bmu_i^*| \leq  \|\bmu_i^*\|_2 \|\wh{\bSigma} \bmu_i^*\|_2 \leq  c d_i^{*2}.
\end{align*}
This together with \eqref{daswqe} leads to 
\begin{align*}
	\norm{\u_i^* }_2 \leq |\bmu_i\strans \wh{\bSigma}\bmu_i^*|^{-1/2}   \|n^{-1/2} \X \bmu_i^*\|_2 \leq c.
\end{align*}
Further, it follows from Definition \ref{defi:sofar} that $\|\wt{\bmu}_i - \bmu_i^*\|_0 \leq 2(r^* + s_u + s_v)$ and $ \|\wt{\bmu}_i - \bmu_i^*\|_2 \leq c \gamma_n$. Then under Condition \ref{con2:re}, we have that 
\begin{align*}
    \|n^{-1/2} \X (\wt{\bmu}_i - \bmu_i^*)\|_2 \leq c \gamma_n.
\end{align*}

In view of \eqref{daswqe} and \eqref{sdhajhc}, it holds that 
\begin{align}
	\Big|(\wt{\bmu}_i\trans \wh{\bSigma}\wt{\bmu}_i)^{-1/2} - (\bmu_i\strans \wh{\bSigma}\bmu_i^*)^{-1/2}\Big| = \frac{|(\wt{\bmu}_i\trans \wh{\bSigma}\wt{\bmu}_i)^{1/2} - (\bmu_i\strans \wh{\bSigma}\bmu_i^*)^{1/2}|}{(\wt{\bmu}_i\trans \wh{\bSigma}\wt{\bmu}_i)^{1/2} (\bmu_i\strans \wh{\bSigma}\bmu_i^*)^{1/2}} \leq c \gamma_n d_i^{*-2}. \label{sdczcsa}
\end{align}
Hence, for term $\norm{\wt{\u}_i - \u_i^* }_2$, combining the above results gives that 
\begin{align*}
	\norm{\wt{\u}_i - \u_i^* }_2 & =  \|(\wt{\bmu}_i\trans \wh{\bSigma}\wt{\bmu}_i)^{-1/2} n^{-1/2} \X \wt{\bmu}_i - (\bmu_i\strans \wh{\bSigma}\bmu_i^*)^{-1/2} n^{-1/2} \X \bmu_i^*\|_2 \\[5pt]
	& \leq  |(\wt{\bmu}_i\trans \wh{\bSigma}\wt{\bmu}_i)^{-1/2} - (\bmu_i\strans \wh{\bSigma}\bmu_i^*)^{-1/2}| \| n^{-1/2} \X \wt{\bmu}_i \|_2 \\
    &\quad+  \|n^{-1/2} \X (\wt{\bmu}_i - \bmu_i^*)\|_2 |\bmu_i\strans \wh{\bSigma}\bmu_i^*|^{-1/2} \\[5pt]
	& \leq c \gamma_n d_i^{*-1}. 
\end{align*}

We now proceed to prove parts (d) and (e).
For matrix $\U^* = (\u_1^*, \ldots, \u_{r^*}^*)$, denote by 
$\U^* = n^{-1/2} \X \mathbf{L}^*_0 $ with
$$ \L^*_0 =: ( (\l_1\strans \wh{\bSigma}\l_1^*)^{-1/2}\l_1^*, \cdots, (\l_{r^*}\strans \wh{\bSigma}\l_{r^*}^*)^{-1/2}\l_{r^*}^*  ).$$
Since $\norm{\L^*}_0 \leq s_u$ with $\L^* = ( \l_1^*, \cdots, \l_{r^*}^*  ) $,  it holds that 
\begin{align}
    \norm{ \L^*_0}_0 \leq s_u, \label{l0001}
\end{align}
which further leads to
$\norm{\L^*_0\b}_0 \leq s_u$ for any vector $\b \in \mathbb{R}^{r^*}$. In addition, under Condition \ref{con3:eigend} we have $\norm{\wh{\bSigma} \L_0^*\b}_2 \leq \norm{ \L_0^*\b}_2 $. 

It follows from the definition of the induced $2$-norm that 
\begin{align}\label{uuszfaa22}
	\norm{\U^*}_2^2 = \sup_{\b\trans\b = 1} \norm{n^{-1/2} \X \L_0^*\b}_2^2 = \sup_{\b\trans\b = 1} \b\trans\L_0\trans \wh{\bSigma} \L_0^*\b \leq c \sup_{\b\trans\b = 1}\norm{ \L_0^*\b}_2^2.
\end{align}
Observe that $\L_0\strans\L^*_0  = \diag{ (\l_1\strans \wh{\bSigma}\l_1^*)^{-1}, \cdots,  (\l_{r^*}\strans \wh{\bSigma}\l_{r^*}^*)^{-1} } $. 
From \eqref{daswqe}, we have $|\l_i\strans \wh{\bSigma}\l_i^*|^{-1} = |{\bmu}_i\strans\wh{\bSigma}{\bmu}_i^{*}|^{-1} d_i^{*2}  \leq  c$.
Then it holds that 
\begin{align}
	\sup_{\b\trans\b = 1} \norm{ \L^*_0\b}_2^2 &= \sup_{\b\trans\b = 1} \b\trans\L_0\strans\L^*_0\b \nonumber\\
 &= \sup_{\b\trans\b = 1}\b\trans \diag{ (\l_1\strans \wh{\bSigma}\l_1^*)^{-1}, \cdots,  (\l_{r^*}\strans \wh{\bSigma}\l_{r^*}^*)^{-1} } \b \leq c, \label{lllzq}
\end{align} 
which yields $\norm{ \L^*_0}_2 \leq c.$ Moreover, it also implies that  
\[\norm{\U^*}_2 \leq c.\]

Similarly, denote by $\wt{\U} = n^{-1/2} \X \wt{\L}_0 $ with 
\begin{align}
    \wt{\L}_0 = ( (\wt{\l}_1\trans \wh{\bSigma}\wt{\l}_1)^{-1/2}\wt{\l}_1, \cdots, (\wt{\l}_{r^*}\trans \wh{\bSigma}\wt{\l}_{r^*})^{-1/2}\wt{\l}_{r^*}  ). \label{llzlcas}
\end{align}
In light of Definition \ref{defi:sofar}, it can be easily seen that 
\begin{align}
     \norm{\wt{\L}_0}_0 \leq \sum_{i=1}^{r^*} \norm{\wt{\l}_i}_0 \leq  \sum_{i=1}^{r^*} \norm{\wt{\l}_i - \l^*_i}_0 + \sum_{i=1}^{r^*} \norm{{\l}_i^*}_0 \leq 3 (r^* + s_u + s_v). \label{l000}
\end{align}
Define 
$\wt{\U}_d = (\wt{d}_1 \wt{\l}_1, \ldots, \wt{d}_{r^*} \wt{\l}_{r^*}), \, \wt{\mathbf{D}}_l = \diag{(\wt{\bmu}_1\trans \wh{\bSigma}\wt{\bmu}_1)^{-1/2}, \cdots, (\wt{\bmu}_{r^*}\trans \wh{\bSigma}\wt{\bmu}_{r^*})^{-1/2}   }  $, and $\U_d^*, \, {\mathbf{D}}_l^*$ analogously. 
Similar to \eqref{lllzq}, we can show that $\norm{\U_d^*}_2  \leq c d_1^*$. By \eqref{daswqe}, we have $\norm{ \wt{\mathbf{D}}_l}_2 \leq cd_{r^*}^{*-1}$. Furthermore, it holds that
\begin{align*}
\norm{\wt{\L}_0 - \L^*_0}_0
	&= \norm{\wt{\U}_d \wt{\mathbf{D}}_l - \U_d^* {\mathbf{D}}_l^*}_0 \\
	&\leq  \norm{(\wt{\U}_d - \U_d^*) \wt{\mathbf{D}}_l}_0 + \norm{\U_d^* ( \wt{\mathbf{D}}_l -  {\mathbf{D}}_l^*)}_0  \\
	&\leq  \norm{\wt{\U}_d - \U_d^*}_0  + \norm{\U_d^*}_0   \\
	&\leq  3 (r^* + s_u + s_v),
\end{align*}
where the last step above holds due to Definition \ref{defi:sofar}. 

Based on the sparsity of $\wt{\L}_0 - \L^*_0$, similar to \eqref{uuszfaa22},  we can deduce that 
\[
\norm{n^{-1/2} \X (\wt{\L}_0 - \L^*_0)}_2  \leq \norm{\wt{\L}_0 - \L^*_0}_2.
\]
It follows from \eqref{sdczcsa} that 
\begin{align}\label{sadqqead}
    \norm{ \wt{\mathbf{D}}_l -  {\mathbf{D}}_l^*}_2  \leq  c \gamma_n d_{r^*}^{*-2}.
\end{align}
Then for term $\norm{\wt{\U} - \U^*}_2$, we can obtain that
\begin{align}
	&\norm{\wt{\U} - \U^*}_2 = \norm{n^{-1/2} \X (\wt{\L}_0 - \L^*_0)}_2  \nonumber\\ & \leq \norm{\wt{\L}_0 - \L^*_0}_2 
	= \norm{\wt{\U}_d \wt{\mathbf{D}}_l - \U_d^* {\mathbf{D}}_l^*}_2  \nonumber\\
	&\leq  \norm{(\wt{\U}_d - \U_d^*) \wt{\mathbf{D}}_l}_2 + \norm{\U_d^* ( \wt{\mathbf{D}}_l -  {\mathbf{D}}_l^*)}_2  \nonumber\\
    & \leq  \norm{\wt{\U}_d - \U_d^*}_2 \norm{ \wt{\mathbf{D}}_l}_2 + \norm{\U_d^*}_2  \norm{ \wt{\mathbf{D}}_l -  {\mathbf{D}}_l^*}_2 \nonumber  \\
	&\leq c \gamma_n d_1^*  d_{r^*}^{*-2}, \label{sdzxc}
\end{align}
where the last inequality above has used \eqref{sadqqead}, $\norm{\U_d^*}_2  \leq c d_1^*$, $\norm{ \wt{\mathbf{D}}_l}_2 \leq cd_{r^*}^{*-1}$ and Definition \ref{defi:sofar}.
Hence, for sufficiently large $n$, we have that 
\begin{align}
\norm{ \wt{\mathbf{L}}_0}_2 \leq  \norm{ {\mathbf{L}}^*_0}_2 + \norm{\wt{\L}_0 - \L^*_0}_2 \leq c,
  \  \norm{ \wt{\mathbf{U}}}_2 \leq  \norm{ {\mathbf{U}}^*}_2 + \norm{\wt{\U} - \U^*}_2 \leq c.  \label{cxzkj}
\end{align}

Next we analyze matrix $\V$ following similar analysis as for $\U$. Note that 
$$ \V^* = (\v_1^*, \ldots, \v_{r^*}^* ) =   \Big( (\bmu_1\strans \wh{\bSigma}\bmu_1^*)^{1/2} \r_1^*, \cdots, (\bmu_{r^*}\strans \wh{\bSigma}\bmu_{r^*}^*)^{1/2}\r_{r^*}^*  \Big).$$
Similar to \eqref{uuszfaa22} and \eqref{lllzq}, we can show that 
$\norm{\V^*}_2 \leq c d_1^*$. Similarly, we have that $\norm{\wt{\V}}_2 \leq c d_1^*$.
Further, let us define 
\begin{align*}
	\mathbf{R}^* = ( \r_1^*, \ldots,  \r_{r^*}^*), \ {\mathbf{D}}_{v}^* = \diag{({\bmu}_1\strans \wh{\bSigma}{\bmu}_1^*)^{1/2}, \cdots, ({\bmu}_{r^*}\strans \wh{\bSigma}{\bmu}_{r^*}^*)^{1/2}   },
\end{align*}
and $ \wt{\mathbf{R}}_d, \wt{{\mathbf{D}}}_v$ 
analogously. Then we see that $\V^* =  \mathbf{R}^* {\mathbf{D}}_v^*$ and $\wt{\V} =  \wt{\mathbf{R}} \wt{{\mathbf{D}}}_v$. In view of \eqref{daswqe0} and \eqref{sdhajhc}, it holds that 
\begin{align*}
	\norm{ \wt{\mathbf{D}}_v}_2 \leq c d_1^*, \ \ 
	\norm{ \wt{\mathbf{D}}_v -  {\mathbf{D}}_v^*}_2 \leq c \gamma_n.
\end{align*}
By definition, we have $\norm{\mathbf{R}^*}_2 = 1$. 
Thus, an application of part (a) of Lemma 6 in \cite{sofari} yields that $\norm{  \wt{\r}_i  -  \r_i^* }_2 \leq c \gamma_n d_i^{*-1}$, which leads to $\norm{\wt{\mathbf{R}} - \mathbf{R}^*}_2 \leq c \gamma_n d_{r^*}^{*-1}$.

Combining the above results gives that 
\begin{align*}
	&\norm{  \wt{\V} - \V^*   }_2 
  =  \norm{\wt{\mathbf{R}} \wt{\mathbf{D}}_v - \mathbf{R}^* {\mathbf{D}}_v^*}_2  \\
  	&\leq  \norm{(\wt{\mathbf{R}} - \mathbf{R}^*) \wt{\mathbf{D}}_v}_2 + \norm{\mathbf{R}^* ( \wt{\mathbf{D}}_v -  {\mathbf{D}}_v^*)}_2 \\
    &\leq  \norm{\wt{\mathbf{R}} - \mathbf{R}^*}_2 \norm{ \wt{\mathbf{D}}_v}_2 + \norm{\mathbf{R}^*}_2  \norm{ \wt{\mathbf{D}}_v -  {\mathbf{D}}_v^*}_2  \\
	&\leq c \gamma_n d_1^* d_{r^*}^{*-1}.
\end{align*}
Observe that 
$  \U\V\trans = \sum_{i = 1}^{r^*} n^{-1/2}\X d_i \l_i \r_i\trans = n^{-1/2}\X \C.$
It follows from Definition \ref{defi:sofar} that 
\begin{align*}
	\sum_{i =1}^{r^*}\norm{\wt{d}_k\wt{\r}_k - d_k^*\r^*_k}_2 \leq  c\gamma_n \ \text{ and } \  \sum_{i =1}^{r^*}\abs{d_k^*-\wt{d}_k} \leq c\gamma_n,
\end{align*}
Notice that $\sum_{i =1}^{r^*} d_k^{*}(\wt{\r}_k - \r_k^*) = \sum_{i =1}^{r^*} (\wt{d}_k\wt{\r}_k - d_k^*\r_k^*) + \sum_{i =1}^{r^*} (d_k^*-\wt{d}_k)\wt{\r}_k$. Since $\norm{\wt{\r}_k}_2=1$, we have that
	\begin{align*}
	\sum_{i =1}^{r^*} \norm{d_k^{*}(\wt{\r}_k - \r_k^*)}_2 &\leq \sum_{i =1}^{r^*}\norm{\wt{d}_k\wt{\r}_k - d_k^*\r^*_k}_2 + \sum_{i =1}^{r^*} \abs{d_k^*-\wt{d}_k}\norm{\wt{\r}_k}_2 
		 \leq  c \gamma_n.	
   \end{align*}
   
Further, we can deduce that 
\begin{align}
    &\norm{ \wt{\U}\wt{\V}\trans  -   \U^*  \V\strans}_2  = \norm{   \sum_{i =1}^{r^*} (\wt{\u}_k \wt{\v}_k\trans   -  \u_k^* \v_k\strans) }_2 =  \norm{ \sum_{i =1}^{r^*} (  \wt{d}_k \wt{\l}_k \wt{\r}_k\trans   - {d}_k^* \l_k^* \r_k\strans)}_2 \nonumber \\
    &\leq \| \sum_{i =1}^{r^*}   \wt{\l}_k \wt{d}_k (  \wt{\r}_k\trans   -   \r_k\strans) \|_2 + \| \sum_{i =1}^{r^*}  (  \wt{d}_k \wt{\l}_k    -  d_k^* \l_k^*)  \r_k\strans \|_2 \nonumber\\
     &\leq  \sum_{i =1}^{r^*}  \| \wt{\l}_k \|_2 \| \wt{d}_k (  \wt{\r}_k\trans   -   \r_k\strans) \|_2 +  \sum_{i =1}^{r^*} \|  \wt{d}_k \wt{\l}_k    -  d_k^* \l_k^* \|_2 \|\r_k\strans \|_2 \leq c \gamma_n. \label{uvuvzxa}
\end{align}
Using similar arguments, we can also obtain that 
\begin{align*}
    \norm{ \wt{\U}_{-k}\wt{\V}_{-k}\trans  -   \U^*_{-k}  \V_{-k}\strans}_2 \leq c \gamma_n.
\end{align*}
This concludes the proof of Lemma \ref{1}.

\subsection{Lemma \ref{lemma:wexist} and its proof} \label{Sec.B.a5}
\begin{lemma}\label{lemma:wexist}
	Assume that all the conditions of Theorem \ref{theo:strong:vk} are satisfied.  Then for each given $k$ with $1 \leq k \leq r^*$, with probability at least
	$1- \theta_{n,p,q}$ for $\theta_{n,p,q}$ given in \eqref{thetapro}, $\W_k $ and $\W_k^{*}$  are well-defined.
\end{lemma}

\noindent \textit{Proof}. 
Let us define $\wt{\A} = (\wt{\v}_k\trans\wt{\v}_k) \I_{r^*-1} - \wt{\V}_{-k} \trans\wt{\V}_{-k} \wt{\U}_{-k}\trans \wt{\U}_{-k}$ and $\wt{\A} = (a_{ij})$ with $i, j \in \mathcal{A} = \{1 \leq \ell \leq r^*: \ell \neq k\}$.
In view of the definition of $\W_k$ in \eqref{eqwtheor1}, to prove that $\W_k $ is well-defined, it is sufficient to show that $\wt{\A}$ is nonsingular.
By some calculations, it holds that 
\begin{align}\label{a0aa}
	a_{ij} = \left\{\begin{array}{l}
		\wt{\bmu}_k\trans \wh{\bSigma}\wt{\bmu}_k -  \wt{\bmu}_i\trans \wh{\bSigma}\wt{\bmu}_i, \quad \text{if} \ i=j, \\[5pt]
		- \wt{\bmu}_i\trans \wh{\bSigma}\wt{\bmu}_j ({\wt{\bmu}_i\trans \wh{\bSigma}\wt{\bmu}_i})^{1/2} ({\wt{\bmu}_j\trans \wh{\bSigma}\wt{\bmu}_j})^{-1/2},		   \quad \text{if} \ i \neq j.
		\end{array}\right.
\end{align} 
We will prove that $ \sum_{j \in \mathcal{A}, j \neq i}|a_{ij}| = o( |a_{\ell \ell}| ) $ for any $i, \ell \in \mathcal{A}$.

In light of Conditions \ref{con3:eigend} and \ref{con4:strong}, by Lemma 7 of \cite{sofari}, we have $|a_{\ell \ell}| \geq c$. 
For term $a_{ij}$ with $i \neq j$, under Condition \ref{con4:strong} that nonzero $d_i^*$ is at the constant level, it follows from \eqref{daswqe01} and \eqref{daswqe} in Lemma \ref{1} that $({\wt{\bmu}_i\trans \wh{\bSigma}\wt{\bmu}_i})^{1/2} \leq c $ and $({\wt{\bmu}_j\trans \wh{\bSigma}\wt{\bmu}_j})^{-1/2} \leq c$. Further, under Condition \ref{con4:strong}, by Definition \ref{defi:sofar} and part (b) of Lemma 6 in \cite{sofari}, it holds that
\begin{align*}
	|\wt{\bmu}_i\trans \wh{\bSigma}\wt{\bmu}_j| &\leq |{\bmu}_i\strans \wh{\bSigma}{\bmu}_j^*| + |\wt{\bmu}_i\trans \wh{\bSigma}\wt{\bmu}_j - {\bmu}_i\strans \wh{\bSigma}{\bmu}_j^*| \\
 &\leq c  |{\l}_i\strans \wh{\bSigma}{\l}_j^*| + \norm{\wt{\bmu}_i\trans}_2\norm{\wh{\bSigma}(\wt{\bmu}_j - {\bmu}_j^*)}_2 + \norm{\wt{\bmu}_i - {\bmu}_i^*}_2 \norm{\wh{\bSigma}{\bmu}_j^*}_2 \\
 &\leq o(n^{-1/2}) + c \gamma_n \leq c \gamma_n.
\end{align*}
Hence, it follows that $|a_{ij}| \leq c \gamma_n$.
Since $r^* \gamma_n = o(1) $ under Condition \ref{con3:eigend}, we can obtain that $ \sum_{j \in  \mathcal{A}, \, j \neq i} |a_{ij}| = o(|a_{\ell \ell}|)$ for any $i, \ell \in \mathcal{A}$, which leads to
\[ |a_{ii}| > \sum_{j \in  \mathcal{A}, \, j \neq i}|a_{ij}| \ \text{ for all } i \in  \mathcal{A}.   \]
This shows that matrix $\wt{\A} $ is strictly diagonally dominant.

An application of the Levy--Desplanques Theorem  \citep{horn2012matrix} shows that matrix $\wt{\A} = (\wt{\v}_k\trans\wt{\v}_k) \I_{r^*-1} - \wt{\V}_{-k} \trans\wt{\V}_{-k} \wt{\U}_{-k}\trans \wt{\U}_{-k}$  is nonsingular. Moreover, using similar arguments we can also show that matrix $ ({\v}_k\strans{\v}_k^*) \I_{r^*-1} - {\V}_{-k} \strans{\V}_{-k}^* {\U}_{-k}\strans {\U}_{-k}^*$ is strictly diagonally dominant and thus is nonsingular. Therefore, we see that both matrices $\W_k $ and $\W_k^{*}$ are well-defined, which completes the proof of Lemma  \ref{lemma:wexist}.

\subsection{Lemma \ref{lemma:wr2bound} and its proof} \label{new.Sec.B.5}

	\begin{lemma}\label{lemma:wr2bound}
		Assume that all the conditions of Theorem \ref{theo:strong:vk} are satisfied. Denote by $\W_i\trans$ the $i$th row of $\W_k$ with $\W_k$ defined in \eqref{eqwtheor1}. Then with probability at least
		$1- \theta_{n,p,q}$ for $\theta_{n,p,q}$ given in \eqref{thetapro}, it holds that 
\begin{align}
	\max_{1 \leq i \leq q}\norm{\W_i}_0 \leq c (r^*+s_u+s_v) \ \text{ and } \  \max_{1 \leq i \leq q}\norm{\W_i}_2 \leq c. \nonumber
\end{align}
Moreover, for any $\a\in\R^q$ satisfying $\norm{\a}_2 =1$,  with probability at least
		$1- \theta_{n,p,q}$ for $\theta_{n,p,q}$ given in \eqref{thetapro}, we have that
		\begin{align*}
			&\norm{\a\trans\W^*_k}_2 \leq c, \  \norm{\a\trans\W_k}_2 \leq c, \\[5pt]
   &\norm{\a\trans(\W_k-\W_k^*)}_0 \leq c(r^* + s_u + s_v), \  \norm{\a\trans(\W_k-\W_k^*)}_2 \leq c \gamma_n,
		\end{align*}
		where $\gamma_n = (r^*+s_u+s_v)^{1/2} \eta_n^2\{n^{-1}\log(pq)\}^{1/2}$ and $c$ is some positive constant.
	\end{lemma}

	\noindent \textit{Proof}. Let $\e_i \in \mathbb{R}^q$ be the unit vector with the $i$th component $1$ and other components $0$. It holds that $\W_i\trans = \e_i\trans\W_k$ and 
\begin{align}
   \norm{\W_i}_0 & \leq \|\e_i \|_0 + \norm{\e_i\trans  (\wt{\v}_k\trans\wt{\v}_k)^{-1} ( \I_q + \wt{\V}_{-k} \wt{\U}_{-k}\trans \wt{\U}_{-k} \wt{\A}^{-1} \wt{\V}_{-k}\trans )  \wt{\V}_{-k} \wt{\U}_{-k}\trans  \wt{\u}_k \cdot \wt{\v}_k\trans }_0 \nonumber \\[5pt]
    &\quad + \norm{  \e_i\trans  \wt{\V}_{-k} \wt{\U}_{-k}\trans \wt{\U}_{-k} \wt{\A}^{-1} \cdot \wt{\V}_{-k}\trans}_0 \nonumber \\
    &\leq 1 + \norm{\wt{\v}_k }_0 + \norm{ \wt{\V}_{-k}}_0 \leq c(r^* + s_u + s_v), \label{szcardq}
\end{align}
 where we have used the fact that $\norm{ \wt{\V}_{-k} \mathbf{b} }_0 \leq \norm{ \wt{\V}_{-k} }_0$ for any $\mathbf{b} \in \mathbb R^{r^*-1}$, and  Definition \ref{defi:sofar}. Moreover, it is easy to see that 
 $\max_{1 \leq i \leq q}\norm{\W_i}_0 \leq c (r^*+s_u+s_v)$.
 
 To prove $ \max_{1 \leq i \leq q}\norm{\W_i}_2 \leq c$, it suffices to show that 
 $\norm{\a\trans\W_k}_2 \leq c$.
By definition, we have that 
\begin{align}
    \|\a\trans\W_k\|_2 & \leq \|\a \|_2 + \norm{\a\trans  (\wt{\v}_k\trans\wt{\v}_k)^{-1} ( \I_q + \wt{\V}_{-k} \wt{\U}_{-k}\trans \wt{\U}_{-k} \wt{\A}^{-1} \wt{\V}_{-k}\trans )  \wt{\V}_{-k} \wt{\U}_{-k}\trans  \wt{\u}_k \wt{\v}_k\trans }_2 \nonumber \\[5pt]
    &\quad + \norm{  \a\trans  \wt{\V}_{-k} \wt{\U}_{-k}\trans \wt{\U}_{-k} \wt{\A}^{-1}\wt{\V}_{-k}\trans}_2. \label{awadasdw}
\end{align}
From the proof of Lemma \ref{lemma:wexist},
we see that $\wt{\A} =  (a_{ij})$ is symmetric and strictly diagonally dominant.  Let us define
	$$\alpha_1 = \min_i(|a_{ii}| - \sum_{j \neq i}|a_{ij}| ) \ \text{ and } \   \alpha_2 = \min_i(|a_{ii}| - \sum_{j \neq i}|a_{ji}| ).$$
It holds that $\alpha_1 = \alpha_2 \asymp  \min_{i} |a_{ii}| = \min_{i \neq k}|\wt{\bmu}_k\trans \wh{\bSigma}\wt{\bmu}_k -  \wt{\bmu}_i\trans \wh{\bSigma}\wt{\bmu}_i|$. Using similar argument as for (A.181) in Section B.14 of \cite{sofari}, we can obtain that $\alpha_1 = \alpha_2 \leq c$.
Then by Corollary 2 in \cite{varah1975lower}, it follows that 
\begin{align}\label{rreq32}
		\norm{ \wt{\A}^{-1}  }_2  \leq  \frac{1}{\sqrt{\alpha_1 \alpha_2} }  \leq c.
\end{align}

We proceed to bound terms in \eqref{awadasdw}.
Under Condition \ref{con4:strong} that nonzero $d_i^*$ is at the constant level, with some calculations we can deduce that $\norm{ \I_q + \wt{\V}_{-k} \wt{\U}_{-k}\trans \wt{\U}_{-k} \wt{\A}^{-1} \wt{\V}_{-k}\trans }_2 \leq 1 + \|\wt{\V}_{-k} \|_2 \|\wt{\U}_{-k}\trans \|_2 \| \wt{\U}_{-k} \|_2 \|\wt{\A}^{-1} \|_2 \| \wt{\V}_{-k}\trans\|_2 \leq c$ and 
\begin{align*}
  &\norm{\a\trans  (\wt{\v}_k\trans\wt{\v}_k)^{-1} ( \I_q + \wt{\V}_{-k} \wt{\U}_{-k}\trans \wt{\U}_{-k} \wt{\A}^{-1} \wt{\V}_{-k}\trans )  \wt{\V}_{-k} \wt{\U}_{-k}\trans  \wt{\u}_k \wt{\v}_k\trans }_2\\[5pt]
  &\leq  \norm{\a }_2 |\wt{\v}_k\trans\wt{\v}_k|^{-1} \norm{ \I_q + \wt{\V}_{-k} \wt{\U}_{-k}\trans \wt{\U}_{-k} \wt{\A}^{-1} \wt{\V}_{-k}\trans }_2 \norm{\wt{\V}_{-k} }_2 \norm{\wt{\U}_{-k}\trans }_2 \norm{ \wt{\u}_k \wt{\v}_k\trans}_2 \\[5pt]
  &\leq c,
\end{align*}
where the last step above has applied Lemma \ref{1} and \eqref{rreq32}.
Further, applying Lemma \ref{1} and \eqref{rreq32} again leads to 
\begin{align*}
	\norm{  \a\trans  \wt{\V}_{-k} \wt{\U}_{-k}\trans \wt{\U}_{-k} \wt{\A}^{-1}\wt{\V}_{-k}\trans}_2 \leq 
	\norm{\a}_2 \norm{ \wt{\V}_{-k}}_2 \norm{\wt{\U}_{-k}\trans}_2 \norm{\wt{\U}_{-k}}_2 \norm{\wt{\A}^{-1}}_2\norm{\wt{\V}_{-k}\trans}_2 \leq c.
\end{align*}
Hence, combining the above terms yields that 
\begin{align}
	\norm{\a\trans\W_k}_2 \leq c. \label{czxkljcsa}
\end{align}

Next we analyze term $\a\trans(\W_k-\W_k^*)$. It follows from the definitions of $\V^*$ and $\wt{\V}$ that $\norm{\V^*}_0 \leq s_v \leq c(r^* + s_u + s_v)$ and $\norm{\wt{\V}}_0 \leq c(r^* + s_u + s_v) $. 
Similar to \eqref{szcardq}, we can deduce that 
\begin{align*}
		&\norm{\a\trans(\W_k-\W_k^*)}_0 \\& \leq     \Big\| (\wt{\v}_k\trans\wt{\v}_k)^{-1} \Big\{ \a\trans( \I_q + \wt{\V}_{-k} \wt{\U}_{-k}\trans \wt{\U}_{-k} \wt{\A}^{-1} \wt{\V}_{-k}\trans )  \wt{\V}_{-k} \wt{\U}_{-k}\trans  \wt{\u}_k \Big\} \cdot \wt{\v}_k\trans \nonumber \\[5pt]
 &-     (\v_k\strans\v_k^*)^{-1} \Big\{ \a\trans( \I_q + \V_{-k}^* \U_{-k}\strans \U_{-k}^* \A^{* -1} \V_{-k}\strans )  \V_{-k}^* \U_{-k}\strans  \u_k^* \Big\} \cdot \v_k\strans \Big\|_0 \nonumber \\[5pt]
 &+ \Big\| \Big\{ \a\trans\wt{\V}_{-k} \wt{\U}_{-k}\trans \wt{\U}_{-k} \wt{\A}^{-1} \Big\} \cdot \wt{\V}_{-k}\trans -  \Big\{\a\trans\V_{-k}^* \U_{-k}\strans \U_{-k}^* \A^{* -1} \Big\} \cdot \V_{-k}\strans \Big\|_0 \\[5pt]
 &\leq \norm{\wt{\v}_k  }_0 +  \norm{\v_k^* }_0 + \norm{\wt{\V}_{-k} }_0 + \norm{ \V_{-k}^*}_0 \\[5pt]
 &\leq  c (r^* + s_u + s_v).
\end{align*}

For term $\norm{\a\trans(\W_k-\W_k^*)}_2 $, since $\| \a \|_2 = 1$ it holds that 
\begin{align}
&\norm{\a\trans(\W_k-\W_k^*)}_2  \leq    \Big\| (\wt{\v}_k\trans\wt{\v}_k)^{-1} ( \I_q + \wt{\V}_{-k} \wt{\U}_{-k}\trans \wt{\U}_{-k} \wt{\A}^{-1} \wt{\V}_{-k}\trans )  \wt{\V}_{-k} \wt{\U}_{-k}\trans  \wt{\u}_k \wt{\v}_k\trans \nonumber \\[5pt]
 &\qquad -     (\v_k\strans\v_k^*)^{-1} ( \I_q + \V_{-k}^* \U_{-k}\strans \U_{-k}^* \A^{* -1} \V_{-k}\strans )  \V_{-k}^* \U_{-k}\strans  \u_k^* \v_k\strans \Big\|_2 \nonumber \\[5pt]
 & \qquad+ \norm{ \wt{\V}_{-k} \wt{\U}_{-k}\trans \wt{\U}_{-k} \wt{\A}^{-1}\wt{\V}_{-k}\trans -  \V_{-k}^* \U_{-k}\strans \U_{-k}^* \A^{* -1}\V_{-k}\strans }_2. \label{woeaios}
\end{align}
To bound the above terms in \eqref{woeaios}, let us first analyze terms $\norm{  {\A}^{*-1}  }_2,  \norm{  \A^* - \wt{\A}  }_2$ and  $ \norm{ \wt{\A}^{-1} - \A^{* -1} }_2$.
Observe that $ {\A}^{*} = (\v_k\strans\v_k^*) \I_{r^*-1} - \V_{-k} \strans\V_{-k}^* \U_{-k}\strans \U_{-k}^*$. An application of similar arguments as for \eqref{rreq32} gives that 
\begin{align}
    \norm{  {\A}^{*-1}  }_2 \leq c. \label{rreq323}
\end{align}

For term $ \norm{  \A^* - \wt{\A}  }_2$, we have that
\begin{align*}
		&\norm{ \wt{\A} - \A^*   }_2 \\&= \norm{
  [(\wt{\v}_k\trans\wt{\v}_k) \I_{r^*-1} - \wt{\V}_{-k} \trans\wt{\V}_{-k} \wt{\U}_{-k}\trans \wt{\U}_{-k} ]
  -[(\v_k\strans\v_k^*) \I_{r^*-1} - \V_{-k} \strans\V_{-k}^* \U_{-k}\strans \U_{-k}^*] }_2 \\[5pt]
		& \leq |\wt{\v}_k\trans\wt{\v}_k - \v_k\strans\v_k^*| + \norm{\wt{\V}_{-k} \trans\wt{\V}_{-k} \wt{\U}_{-k}\trans \wt{\U}_{-k} -  \V_{-k} \strans\V_{-k}^* \U_{-k}\strans \U_{-k}^*}_2.
\end{align*}
Note that nonzero $d_i^*$ is at the constant level under Condition \ref{con4:strong}. It follows from Lemma \ref{1} that 
\begin{align*}
    |\wt{\v}_k\trans\wt{\v}_k - \v_k\strans\v_k^*|  \leq  \norm{\wt{\v}_k }_2 \norm{\wt{\v}_k - \v_k^* }_2 + \norm{\wt{\v}_k - \v_k^* }_2 \norm{{\v}_k^* }_2 \leq c \gamma_n.
\end{align*}

Moreover, an application of Lemma \ref{1} leads to
\begin{align*}
	&\norm{\wt{\V}_{-k} \trans\wt{\V}_{-k} \wt{\U}_{-k}\trans \wt{\U}_{-k} -  \V_{-k} \strans\V_{-k}^* \U_{-k}\strans \U_{-k}^*}_2 \\[5pt]
	&\leq \norm{\wt{\V}_{-k} \trans\wt{\V}_{-k} (\wt{\U}_{-k}\trans \wt{\U}_{-k} -  \U_{-k}\strans \U_{-k}^*)}_2 + \norm{(\wt{\V}_{-k} \trans\wt{\V}_{-k} -  \V_{-k} \strans\V_{-k}^*) \U_{-k}\strans \U_{-k}^*}_2\\[5pt]
	&\leq  \norm{\wt{\V}_{-k} \trans}_2 \norm{\wt{\V}_{-k}}_2 \norm{ \wt{\U}_{-k}\trans \wt{\U}_{-k} -  \U_{-k}\strans \U_{-k}^*}_2 + \norm{\wt{\V}_{-k} \trans\wt{\V}_{-k} -  \V_{-k} \strans\V_{-k}^*}_2 \norm{ \U_{-k}\strans }_2 \norm{\U_{-k}^*}_2 \\[5pt]
	&\leq c \norm{\wt{\U}_{-k}\trans}_2  \norm{(\wt{\U}_{-k} -  \U_{-k}\strans)  }_2 + c \norm{ \wt{\U}_{-k} -  \U_{-k}\strans}_2\norm{\U_{-k}^* }_2   \\[5pt]
	&\quad + c \norm{\wt{\V}_{-k}\trans }_2 \norm{\wt{\V}_{-k} -  \V_{-k}\strans  }_2 + c\norm{ \wt{\V}_{-k} -  \V_{-k}\strans}_2 \norm{ \V_{-k}^* }_2   \\[5pt]
	& \leq c \gamma_n. 
\end{align*}
Thus, it follows that 
\begin{align}\label{rreq324}
		\norm{ \wt{\A} - \A^*   }_2 \leq c \gamma_n.
\end{align}

For term $\norm{\wt{\A}^{-1} - \A^{* -1}}_2$, combining \eqref{rreq32}, \eqref{rreq323}, and \eqref{rreq324} yields that 
\begin{align}
	&  \norm{ \wt{\A}^{-1} - \A^{* -1} }_2 = \norm{ \wt{\A}^{-1}( \A^* - \wt{\A}  )\A^{*-1}  }_2 \nonumber\\
	&\leq \norm{ \wt{\A}^{-1}}_2 \norm{  \A^* - \wt{\A}   }_2 \norm{ \A^{*-1}  }_2 \leq c \gamma_n. \label{rreq325}
\end{align}

We next bound the two terms in \eqref{woeaios}.
Let us define
\begin{align*}
	&\boldsymbol{\pi}_1 = ( \I_q + \wt{\V}_{-k} \wt{\U}_{-k}\trans \wt{\U}_{-k} \wt{\A}^{-1} \wt{\V}_{-k}\trans )  \wt{\V}_{-k} \wt{\U}_{-k}\trans  \wt{\u}_k \wt{\v}_k\trans, \\
	&\boldsymbol{\pi}_1^* =  ( \I_q + \V_{-k}^* \U_{-k}\strans \U_{-k}^* \A^{* -1} \V_{-k}\strans )  \V_{-k}^* \U_{-k}\strans  \u_k^* \v_k\strans.
\end{align*}
For the first term on the right-hand of \eqref{woeaios}, it holds that 
\begin{align}
	 &  \|(\wt{\v}_k\trans\wt{\v}_k)^{-1}\boldsymbol{\pi}_1 -  (\v_k\strans\v_k^*)^{-1} \boldsymbol{\pi}_1^* \|_2 \nonumber\\
   &\leq  |\wt{\v}_k\trans\wt{\v}_k|^{-1} \norm{ \boldsymbol{\pi}_1 -   \boldsymbol{\pi}_1^* }_2 + | (\wt{\v}_k\trans\wt{\v}_k)^{-1} - (\v_k\strans\v_k^*)^{-1}  |  \| \boldsymbol{\pi}_1^* \|_2 \nonumber\\
   &\leq c \norm{ \boldsymbol{\pi}_1 -   \boldsymbol{\pi}_1^* }_2 + c \gamma_n \| \boldsymbol{\pi}_1^* \|_2, \label{pipi2}
  \end{align}
  where we have applied part (b) of Lemma \ref{1}. 
  
  The upper bounds for $\| \boldsymbol{\pi}_1^* \|_2$ and $\norm{ \boldsymbol{\pi}_1 -   \boldsymbol{\pi}_1^* }_2 $ will be based on Lemma \ref{1}, \eqref{rreq32}, \eqref{rreq323}, \eqref{rreq324}, and \eqref{rreq325}.
  For term $\| \boldsymbol{\pi}_1^* \|_2$, it follows from Lemma \ref{1} and \eqref{rreq323} that 
\begin{align}
	\| \boldsymbol{\pi}_1^* \|_2 & \leq \| \I_q + \V_{-k}^* \U_{-k}\strans \U_{-k}^* \A^{* -1} \V_{-k}\strans \|_2 \| \V_{-k}^* \U_{-k}\strans \|_2 \|  \u_k^* \v_k\strans \|_2 \nonumber \\
	&\leq (1 + \| \V_{-k}^* \|_2 \| \U_{-k}\strans \|_2 \| \U_{-k}^*\|_2 \| \A^{* -1} \|_2 \| \V_{-k}\strans \|_2  ) \| \V_{-k}^* \|_2 \| \U_{-k}\strans \|_2 \|  \u_k^*\|_2 \| \v_k\strans \|_2 \nonumber \\
	&\leq c. \label{pipi}
\end{align}

For $\norm{ \boldsymbol{\pi}_1 -   \boldsymbol{\pi}_1^* }_2$, it holds that 
\begin{align*}
 &\norm{ \boldsymbol{\pi}_1 -   \boldsymbol{\pi}_1^* }_2  \leq \norm{( \I_q + \V_{-k}^* \U_{-k}\strans \U_{-k}^* \A^{* -1} \V_{-k}\strans )  (\wt{\V}_{-k} \wt{\U}_{-k}\trans  \wt{\u}_k \wt{\v}_k\trans -   \V_{-k}^* \U_{-k}\strans  \u_k^* \v_k\strans) }_2 \\[5pt]
 & \ \ + \|(  \wt{\V}_{-k} \wt{\U}_{-k}\trans \wt{\U}_{-k} \wt{\A}^{-1} \wt{\V}_{-k}\trans -
 \V_{-k}^* \U_{-k}\strans \U_{-k}^* \A^{* -1} \V_{-k}\strans ) \wt{\V}_{-k} \wt{\U}_{-k}\trans  \wt{\u}_k \wt{\v}_k\trans \|_2 \\[5pt]
 & \leq   (1 + \| \V_{-k}^* \|_2 \| \U_{-k}\strans \|_2 \| \U_{-k}^*\|_2 \| \A^{* -1} \|_2 \| \V_{-k}\strans \|_2  ) \norm{ \wt{\V}_{-k} \wt{\U}_{-k}\trans  \wt{\u}_k \wt{\v}_k\trans -   \V_{-k}^* \U_{-k}\strans  \u_k^* \v_k\strans }_2  \\[5pt]
 & \ \ + \|  \wt{\V}_{-k} \wt{\U}_{-k}\trans \wt{\U}_{-k} \wt{\A}^{-1} \wt{\V}_{-k}\trans -
 \V_{-k}^* \U_{-k}\strans \U_{-k}^* \A^{* -1} \V_{-k}\strans \|_2 \|  \wt{\V}_{-k} \|_2 \| \wt{\U}_{-k}\trans \|_2 \|  \wt{\u}_k\|_2 \| \wt{\v}_k\trans \|_2 \\[5pt]
 &\leq c \norm{ \wt{\V}_{-k} \wt{\U}_{-k}\trans  \wt{\u}_k \wt{\v}_k\trans -   \V_{-k}^* \U_{-k}\strans  \u_k^* \v_k\strans }_2 \\[5pt]
 &\ \ + c \|  \wt{\V}_{-k} \wt{\U}_{-k}\trans \wt{\U}_{-k} \wt{\A}^{-1} \wt{\V}_{-k}\trans -
 \V_{-k}^* \U_{-k}\strans \U_{-k}^* \A^{* -1} \V_{-k}\strans \|_2,
 \end{align*}
 where the last step above has used Lemma \ref{1} and \eqref{rreq323}.
For the first term above, by  Lemma \ref{1} we have that 
\begin{align*}
	&\norm{ \wt{\V}_{-k} \wt{\U}_{-k}\trans  \wt{\u}_k \wt{\v}_k\trans -   \V_{-k}^* \U_{-k}\strans  \u_k^* \v_k\strans }_2 \\ 
	&\leq \norm{ \wt{\V}_{-k} \wt{\U}_{-k}\trans  (\wt{\u}_k \wt{\v}_k\trans -  \u_k^* \v_k\strans )}_2 + \norm{ (\wt{\V}_{-k} \wt{\U}_{-k}\trans   -   \V_{-k}^* \U_{-k}\strans)  \u_k^* \v_k\strans }_2 \\ 
	&\leq \norm{ \wt{\V}_{-k}}_2  \norm{\wt{\U}_{-k}\trans}_2 \norm{ \wt{\u}_k \wt{\v}_k\trans -  \u_k^* \v_k\strans }_2 + \norm{ \wt{\V}_{-k} \wt{\U}_{-k}\trans   -   \V_{-k}^* \U_{-k}\strans }_2 \norm{  \u_k^*}_2  \norm{\v_k\strans }_2 \\
	&\leq c \gamma_n.
\end{align*}

Further, using Lemma \ref{1} and \eqref{rreq323}, we can deduce that 
\begin{align*}
	 &\|  \wt{\V}_{-k} \wt{\U}_{-k}\trans \wt{\U}_{-k} \wt{\A}^{-1} \wt{\V}_{-k}\trans -
	\V_{-k}^* \U_{-k}\strans \U_{-k}^* \A^{* -1} \V_{-k}\strans \|_2 \\[5pt]
	&\leq  \|  \wt{\V}_{-k} \wt{\U}_{-k}\trans( \wt{\U}_{-k} \wt{\A}^{-1} \wt{\V}_{-k}\trans -
	 \U_{-k}^* \A^{* -1} \V_{-k}\strans) \|_2 \\[5pt]
  & \qquad \qquad+ \| ( \wt{\V}_{-k} \wt{\U}_{-k}\trans  -
	\V_{-k}^* \U_{-k}\strans) \U_{-k}^* \A^{* -1} \V_{-k}\strans \|_2 \\[5pt]
	&\leq \|  \wt{\V}_{-k} \|_2 \| \wt{\U}_{-k}\trans \|_2 \| \wt{\U}_{-k} \wt{\A}^{-1} \wt{\V}_{-k}\trans -
	\U_{-k}^* \A^{* -1} \V_{-k}\strans \|_2 \\[5pt]
 &\qquad \qquad + \|  \wt{\V}_{-k} \wt{\U}_{-k}\trans  -
   \V_{-k}^* \U_{-k}\strans \|_2 \| \U_{-k}^* \|_2 \| \A^{* -1} \|_2 \| \V_{-k}\strans \|_2 \\[5pt]
   &\leq c \| \wt{\U}_{-k} \wt{\A}^{-1} \wt{\V}_{-k}\trans -
   \U_{-k}^* \A^{* -1} \V_{-k}\strans \|_2 + c \gamma_n.
\end{align*}

It remains to bound term $\| \wt{\U}_{-k} \wt{\A}^{-1} \wt{\V}_{-k}\trans -
   \U_{-k}^* \A^{* -1} \V_{-k}\strans \|_2$. It follows that
\begin{align*}
	&\| \wt{\U}_{-k} \wt{\A}^{-1} \wt{\V}_{-k}\trans -
	\U_{-k}^* \A^{* -1} \V_{-k}\strans \|_2 \\
	& \leq \| \wt{\U}_{-k} (\wt{\A}^{-1} \wt{\V}_{-k}\trans -
	 \A^{* -1} \V_{-k}\strans) \|_2  + \| (\wt{\U}_{-k} -
	\U_{-k}^*) \A^{* -1} \V_{-k}\strans \|_2  \\
	&\leq \| \wt{\U}_{-k}\|_2 \Big(\| \wt{\A}^{-1} \|_2 \| \wt{\V}_{-k}\trans -
	\V_{-k}\strans \|_2 + \| \wt{\A}^{-1} -
	\A^{* -1} \|_2 \| \V_{-k}\strans \|_2 \Big) \\ &\qquad + \| \wt{\U}_{-k} -
	\U_{-k}^* \|_2 \| \A^{* -1} \|_2 \| \V_{-k}\strans \|_2 
	\leq c \gamma_n,
\end{align*}
where the last step above has applied Lemma \ref{1},  \eqref{rreq32}, \eqref{rreq323}, and \eqref{rreq325}.

Therefore, combining the above results yields that
 $   \norm{ \boldsymbol{\pi}_1 -  \boldsymbol{\pi}_1^* }_2 \leq c \gamma_n. $
This along with \eqref{woeaios}, \eqref{pipi2}, and \eqref{pipi} entails that 
\begin{align*}
	\norm{\a\trans(\W_k-\W_k^*)}_2 \leq c \gamma_n.
\end{align*}
Moreover, by the triangle inequality, for sufficiently large $n$ it holds that
\begin{align*}
	\norm{\a\trans \W_k^*}_2  \leq \norm{\a\trans\W_k}_2 + \norm{\a\trans(\W_k-\W_k^*)}_2 \leq c.
\end{align*}
This concludes the proof of Lemma \ref{lemma:wr2bound}.

\subsection{Lemma \ref{lemma:weak:wbound} and its proof} \label{new.Sec.B.59}

	\begin{lemma}\label{lemma:weak:wbound}
		Assume that Conditions \ref{con2:re}, \ref{con3:eigend}, and \ref{con:weak:orth} hold, and $\wt{\C}$ satisfies Definition \ref{defi:sofar}.
		For  $\W_k^* =   \I_q - 2^{-1} (\v_k\strans\v_k^*)^{-1} (  \v_k^* \v_k\strans -  \v_k^* \u_k\strans  \U_{-k}^*  \V_{-k}\strans  )  $ and $\W_k =   \I_q - 2^{-1} (\wt{\v}_k\trans\wt{\v}_k)^{-1} (  \wt{\v}_k \wt{\v}_k\trans -  \wt{\v}_k \wt{\u}_k\trans  \wt{\U}_{-k} \wt{\V}_{-k}\trans  )$,  with probability at least
		$1- \theta_{n,p,q}$ for $\theta_{n,p,q}$ given in \eqref{thetapro}, we have that for sufficiently large $n$, 
		\begin{align}
			\max_{1 \leq i \leq q}\norm{\w_i}_0 \leq c (r^*+s_u+s_v) \ \text{ and } \  \max_{1 \leq i \leq q}\norm{\w_i}_2 \leq c, \nonumber
		\end{align}
		where ${\w}_{i}\trans$ is the $i$th row of ${\W}_k$ with $i = 1, \ldots, p$. Moreover, for any $\a\in\R^q$ satisfying $\norm{\a}_2 =1$,  with probability at least
				$1- \theta_{n,p,q}$ for $\theta_{n,p,q}$ given in \eqref{thetapro}, we have that
				\begin{align*}
					&\norm{\a\trans\W^*_k}_2 \leq c, \  \norm{\a\trans\W_k}_2 \leq c, \\[5pt]
		   &\norm{\a\trans(\W_k-\W_k^*)}_0 \leq c(r^* + s_u + s_v), \  \norm{\a\trans(\W_k-\W_k^*)}_2 \leq c \gamma_n d_1^* d_k^{*-2},
				\end{align*}
				where $\gamma_n = (r^*+s_u+s_v)^{1/2} \eta_n^2\{n^{-1}\log(pq)\}^{1/2}$ and $c$ is some positive constant.
	\end{lemma}
	
	\noindent \textit{Proof}.
	Denote by $\e_i \in \mathbb{R}^q$ a unit vector with the $i$th component $1$ and other components $0$. It is easy to see that $\w_i\trans = \e_i\trans\W_k$ and 
	\begin{align}
		\norm{\w_i}_0 & \leq \|\e_i \|_0 + \norm{\e_i\trans  2^{-1} (\wt{\v}_k\trans\wt{\v}_k)^{-1} (  \wt{\v}_k \wt{\v}_k\trans -  \wt{\v}_k \wt{\u}_k\trans  \wt{\U}_{-k} \wt{\V}_{-k}\trans   )}_0 \nonumber \\[5pt]
		 &\leq 1 + \norm{\wt{\v}_k }_0 + \norm{ \wt{\V}_{-k}}_0 \leq c(r^* + s_u + s_v), \nonumber
	 \end{align}
	 where we have used the fact that $\norm{ \wt{\V}_{-k} \mathbf{b} }_0 \leq \norm{ \wt{\V}_{-k} }_0$ for any $\mathbf{b} \in \mathbb R^{r^*-1}$, and  Definition \ref{defi:sofar}. Also, it holds that 
	 $\max_{1 \leq i \leq q}\norm{\w_i}_0 \leq c (r^*+s_u+s_v)$.

 To prove $ \max_{1 \leq i \leq q}\norm{\w_i}_2 \leq c$, it suffices to show that
 $\norm{\a\trans\W_k}_2 \leq c$.
First, for term $\norm{\a\trans\W^*_k}_2 $, it follows from the triangle inequality and Lemma \ref{1} that 
\begin{align*}
	\norm{\a\trans\W^*_k}_2 &\leq \norm{\a}_2 + \norm{\a\trans  2^{-1} (\v_k\strans\v_k^*)^{-1} (  \v_k^* \v_k\strans -  \v_k^* \u_k\strans  \U_{-k}^*  \V_{-k}\strans  ) }_2 \\[5pt]
 & \leq 1 + c | \v_k\strans\v_k^*|^{-1} \norm{\v_k^*}_2 \norm{\v_k\strans}_2 + c \norm{  \sum_{i \neq k}(\v_k\strans\v_k^*)^{-1} \boldsymbol{v}_{k}^*\u_k\strans\u_i^*\v_i\strans }_2 \\
	&\leq  1 + c d_k^{*-2}  d_k^{*2}  + c \norm{  \sum_{i \neq k}(\v_k\strans\v_k^*)^{-1} \boldsymbol{v}_{k}^*\u_k\strans\u_i^*\v_i\strans }_2 \\[5pt]
	&\leq c  + c \norm{  \sum_{i \neq k}(\v_k\strans\v_k^*)^{-1} \boldsymbol{v}_{k}^*\u_k\strans\u_i^*\v_i\strans }_2.
\end{align*}
In addition, by the definitions of $\v_k^*$ and $\u_k^*$, we can deduce that
\begin{align*}
    &\norm{  \sum_{i \neq k}(\v_k\strans\v_k^*)^{-1} \boldsymbol{v}_{k}^*\u_k\strans\u_i^*\v_i\strans }_2  = \norm{ \sum_{i \neq k} (\v_k\strans\v_k^*)^{-1} d_k^* \r_k^* \l_k\strans \wh{\bSigma} \l_i^* d_i^{*} \r_i\strans}_2 \\
    &\leq  (\v_k\strans\v_k^*)^{-1} \| \sum_{i \neq k} \l_k\strans \wh{\bSigma} \l_i^*  d_k^* \r_k^*  d_i^{*} \r_i\strans  \|_2 \leq c  d_k^{*-2}  \sum_{i \neq k} | \l_k\strans \wh{\bSigma} \l_i^*  | d_k^*  d_i^{*} \| \r_k^*  \|_2  \|\r_i\strans  \|_2 \\ &\leq c  d_k^{*-1} d_1^{*}  \sum_{i \neq k} |\l_k\strans \wh{\bSigma} \l_i^* | 
    \leq  c r^* d_1^{*2}  d_k^{*-1} d_{r^*}^{-2}  /\sqrt{n},
\end{align*}
{where the last step above is due to \eqref{weakceq}.}
Hence, for sufficiently large $n$, we have that
\begin{align*}
    \norm{\a\trans\W^*_k}_2 \leq c.
\end{align*}

We next bound term $\norm{\a\trans(\W_k - \W^*_k)}_2 $. By definition, it holds that
\begin{align*}
    &\norm{\a\trans(\W_k - \W^*_k)}_2 \\
    &\leq  \norm{\a}_2 \|2^{-1} (\wt{\v}_k\trans\wt{\v}_k)^{-1} (  \wt{\v}_k \wt{\v}_k\trans -  \wt{\v}_k \wt{\u}_k\trans  \wt{\U}_{-k} \wt{\V}_{-k}\trans  )\\ & \qquad-   2^{-1} (\v_k\strans\v_k^*)^{-1} (  \v_k^* \v_k\strans -  \v_k^* \u_k\strans  \U_{-k}^*  \V_{-k}\strans  )\|_2\\
    &\leq c | (\wt{\v}_k\trans\wt{\v}_k)^{-1} - (\v_k\strans\v_k^*)^{-1} |   \norm{ \v_k^* \v_k\strans -  \v_k^* \u_k\strans  \U_{-k}^*  \V_{-k}\strans  }_2 \\
    &\quad  + c | \wt{\v}_k\trans\wt{\v}_k|^{-1} \norm{ (  \wt{\v}_k \wt{\v}_k\trans -  \wt{\v}_k \wt{\u}_k\trans  \wt{\U}_{-k} \wt{\V}_{-k}\trans  ) -  (  \v_k^* \v_k\strans -  \v_k^* \u_k\strans  \U_{-k}^*  \V_{-k}\strans  )}_2.
\end{align*}
By invoking Lemma \ref{1}, we can show that
\begin{align*}
    &| (\wt{\v}_k\trans\wt{\v}_k)^{-1} - (\v_k\strans\v_k^*)^{-1} |  \leq c \gamma_n d_k^{*-3}, \ \
    | \wt{\v}_k\trans\wt{\v}_k|^{-1} \leq  c d_k^{*-2},\\
    &\norm{ \v_k^* \v_k\strans -  \v_k^* \u_k\strans  \U_{-k}^*  \V_{-k}\strans  }_2 \leq \| \v_k^*\|_2 \|  \v_k\strans\|_2 +  \|  \v_k^*\|_2 \| \u_k\strans \|_2 \| \U_{-k}^* \|_2 \|\V_{-k}\strans \|_2 \leq c d_1^* d_k^*.
\end{align*}
Then it follows that
\begin{align}\label{zcjald}
    &\norm{\a\trans(\W_k - \W^*_k)}_2  \\
	&\leq c \gamma_n d_1^* d_k^{*-2} + c d_k^{*-2} \norm{ (  \wt{\v}_k \wt{\v}_k\trans -  \wt{\v}_k \wt{\u}_k\trans  \wt{\U}_{-k} \wt{\V}_{-k}\trans  ) -  (  \v_k^* \v_k\strans -  \v_k^* \u_k\strans  \U_{-k}^*  \V_{-k}\strans  )}_2. \nonumber
\end{align}

It remains to bound the last term above. Notice that
\begin{align*}
     &\norm{ (  \wt{\v}_k \wt{\v}_k\trans -  \wt{\v}_k \wt{\u}_k\trans  \wt{\U}_{-k} \wt{\V}_{-k}\trans  ) -  (  \v_k^* \v_k\strans -  \v_k^* \u_k\strans  \U_{-k}^*  \V_{-k}\strans  )}_2 \\
     &\leq \norm{   \wt{\v}_k \wt{\v}_k\trans  -   \v_k^* \v_k\strans }_2  + \norm{  \wt{\v}_k \wt{\u}_k\trans  \wt{\U}_{-k} \wt{\V}_{-k}\trans  -  \v_k^* \u_k\strans  \U_{-k}^*  \V_{-k}\strans  }_2.
\end{align*}
In view of Lemma \ref{1}, we can obtain that 
\begin{align*}
	&\norm{   \wt{\v}_k \wt{\v}_k\trans  -   \v_k^* \v_k\strans }_2 \leq  \norm{   \wt{\v}_k}_2 \norm{\wt{\v}_k\trans  -   \v_k\strans }_2 + \norm{   \wt{\v}_k   -   \v_k^*}_2 \norm{ \v_k\strans }_2 
	\leq c \gamma_n d_k^*, \nonumber \\[5pt]
	&\norm{   \wt{\U}_{-k} \wt{\V}_{-k}\trans  -   \U_{-k}^*  \V_{-k}\strans  }_2 \leq c \gamma_n.
\end{align*}

Similarly, we can show that $ \norm{\wt{\v}_k \wt{\u}_k\trans   -  \v_k^* \u_k\strans }_2 \leq c \gamma_n$.
Thus, it holds that
\begin{align}
	&\norm{  \wt{\v}_k \wt{\u}_k\trans  \wt{\U}_{-k} \wt{\V}_{-k}\trans  -  \v_k^* \u_k\strans  \U_{-k}^*  \V_{-k}\strans  }_2  \nonumber \\[5pt]
	&\leq \norm{  \wt{\v}_k \wt{\u}_k\trans  (\wt{\U}_{-k} \wt{\V}_{-k}\trans  -   \U_{-k}^*  \V_{-k}\strans)  }_2 + \norm{  (\wt{\v}_k \wt{\u}_k\trans   -  \v_k^* \u_k\strans)  \U_{-k}^*  \V_{-k}\strans  }_2  \nonumber \\[5pt]
	&\leq \norm{  \wt{\v}_k}_2  \norm{ \wt{\u}_k\trans}_2  \norm{ \wt{\U}_{-k} \wt{\V}_{-k}\trans  -   \U_{-k}^*  \V_{-k}\strans}_2 +  \norm{\wt{\v}_k \wt{\u}_k\trans   -  \v_k^* \u_k\strans }_2 \norm{ \U_{-k}^* }_2 \norm{  \V_{-k}\strans}_2  \nonumber \\[5pt]
	&\leq c  \gamma_n  d_1^*. \label{jkjkzza}
\end{align}
Combining \eqref{zcjald} and \eqref{jkjkzza} yields that 
\begin{align*}
    \norm{\a\trans(\W_k - \W^*_k)}_2 
    \leq c   \gamma_n d_1^* d_k^{*-2}.
\end{align*}

Finally, for term $\norm{\a\trans \W_k }_2$, it follows from the triangle inequality that for sufficiently large $n$, 
\begin{align*}
	\norm{\a\trans \W_k }_2 \leq \norm{\a\trans(\W_k - \W^*_k)}_2  + \norm{\a\trans \W^*_k}_2  \leq  c.
\end{align*}
This completes the proof of Lemma \ref{lemma:weak:wbound}.

\subsection{Lemma \ref{lemma:threuk} and its proof}
	\begin{lemma}\label{lemma:threuk}
	Assume that all the conditions of Theorem \ref{theo:weak:uk} hold.
	For the  hard-thresholded debiased estimate
$\wh{\bmu}_i^t = (\wh{\mu}_{i1}^t, \ldots, \wh{\mu}_{i p}^t)\trans \text{ with } \wh{\mu}_{ij}^t  = \wh{\mu}_{ij} \mathbf{1}(\wh{\mu}_{ij} \geq   \frac{  \log n}{\sqrt{n}}), $ $i = 1, \ldots, r^*$,
	with probability at least $1 - \theta_{n,p,q}$ for $\theta_{n,p,q}$ defined in \eqref{thetapro}, for sufficiently large $n$ we have 
	$\operatorname{supp}(\wh{\bmu}_k^t) =  \mathcal{S}_{u_k}$.
\end{lemma}

\noindent \textit{Proof}. 
Under Conditions \ref{con1:error}--\ref{con3:eigend} and \ref{con:weak:orth}, for  $j \in  \{1, \ldots, p\}$, by Lemma \ref{lemm:sofari} we have that 
\begin{align}
& \sqrt{n}(\wh{\bmu}_{kj} -  {\bmu}_{{kj}}^*) =  h_{k,j} + t_{k,j},  \ \  h_{k,j}   \sim \N(0,\nu_{\mu_k,j}^2), \ \ 
t_{k,j} = o(1). \nonumber
\end{align}
An application of the proof of Theorem 6 in \cite{sofari} shows that $\nu_{\mu_k,j} \leq c$.
Since $ t_{k,j} = o(1)$ and  $ h_{k,j}   \sim \N(0,\nu_{\mu_k,j}^2)$, for sufficiently large $n$
it follows that  
\begin{align}\label{unbau1bound}
    | \sqrt{n}(\wh{\mu}_{kj} -  {\mu}_{{kj}}^*) | \leq  c| h_{k,j}   | \leq c \nu_{\mu_k,j} \leq c.
\end{align}

On one hand, for $j \in \mathcal{S}_{\mu_k}^c$, it follows from the definition of $\mathcal{S}_{\mu_k}$ in Condition \ref{con:threshold} that $\mu_{kj}^* = 0 $ or $\mu_{kj}^* = o(\frac{1}{\sqrt{n}})$, which leads to  $\wh{\mu}_{kj}  \leq c n^{-1/2} $ for sufficiently large $n$.
	On the other hand, for $j \in \mathcal{S}_{\mu_k}$, by Condition \ref{con:threshold} there exist some positive constants $C_u$ and $\alpha < 1/2$ such that
	$\min_{j \in \mathcal{S}_{\mu_k}} |\mu_{kj}^*| \geq C_u  n^{-\alpha}.$ 
	Then for any $j \in \mathcal{S}_{\mu_k}$ and sufficiently large $n$, we have that 
	\begin{align}\label{u1jthre}
	|\wh{\mu}_{kj}|  \geq C_u  n^{-\alpha} - c_1  n^{-1/2} \geq c_2 n^{-\alpha} \geq  \frac{  \log n}{\sqrt{n}},
    \end{align}
	where $ c_1 $ and $c_2$ are some positive constants.	
	Note that $\wh{\bmu}_k^t = (\wh{\mu}_{k1}^t, \ldots, \wh{\mu}_{k p}^t)\trans \text{ with } \wh{\mu}_{kj}^t  = \wh{\mu}_{kj} \mathbf{1}(\wh{\mu}_{kj} \geq   \frac{  \log n}{\sqrt{n}}).$
	Hence, \eqref{u1jthre} implies that $\mathcal{S}_{\mu_k} \subset \operatorname{supp}(\wh{\bmu}_k^t)$.

	We next show that $\operatorname{supp}(\wh{\bmu}_k^t) \subset \mathcal{S}_{\mu_k}$, which can be proved by contradiction. Suppose that $\operatorname{supp}(\wh{\bmu}_k^t) \not\subset \mathcal{S}_{\mu_k}$. Then there exists some $i \in \{1, \ldots, p\} $ such that $i \in \operatorname{supp}(\wh{\bmu}_k^t)$ and $i \notin \mathcal{S}_{\mu_k}$. For such $i$, it holds that $|\wh{\mu}_{ki}| \geq   \frac{  \log n}{\sqrt{n}}$ and $\mu_{ki}^* = o(\frac{1}{\sqrt{n}})$. For sufficiently large $n$, we can obtain that 
	\begin{align*}
		| \wh{\mu}_{ki} - \mu_{ki}^*  |  \geq  | \wh{\mu}_{ki}| - |\mu_{ki}^*  | \geq    \frac{  \log n}{\sqrt{n}} -  o(\frac{1}{\sqrt{n}}) \geq   \frac{  \log n}{\sqrt{n}} > \frac{c}{\sqrt{n}}.
	\end{align*}
	This is a contradiction with \eqref{unbau1bound}. Thus, we have $\operatorname{supp}(\wh{\bmu}_k^t) \subset  \mathcal{S}_{\mu_k}$.

	Therefore, for sufficiently large $n$, we have that $\operatorname{supp}(\wh{\bmu}_k^t) =  \mathcal{S}_{\mu_k}$   with probability at least $1 - \theta_{n,p,q}$ for $\theta_{n,p,q}$ defined in \eqref{thetapro}. This concludes the proof of Lemma \ref{lemma:threuk}.

\section{Implementation procedures} \label{new.sec.impleproc}

We provide in Algorithms \ref{alg1} and \ref{alg2} the implementation procedures for SOFARI-R\textsubscript{s} and SOFARI-R, respectively.

\begin{algorithm}[!t]
\caption{SOFARI-R\textsubscript{s}}
\label{alg1}
\begin{algorithmic}[1]
\Require Data $\mathbf{X} \in \mathbb{R}^{n \times p}$, $\mathbf{Y} \in \mathbb{R}^{n \times q}$
\Ensure $\{ \wh{\v}_k, \wt{\nu}^2_k \}_{k=1}^{\wh{r}}$

\State \textbf{Initialization:}
\State \quad Determine the rank $\wh{r}$.
\State \quad Compute initial SOFAR estimates $\big\{\wt{d}_i, \wt{\boldsymbol{\ell}}_i, \wt{\boldsymbol{r}}_i \big\}_{i=1}^{\wh{r}}$.
\State \quad Let
\State \quad \[
    \wt{\u}_i = (\wt{\l}_i\trans \wh{\bSigma} \wt{\l}_i)^{-1/2} n^{-1/2} \mathbf{X} \wt{\l}_i, \quad \wt{\v}_i = (\wt{\l}_i\trans \wh{\bSigma} \wt{\l}_i)^{1/2} \wt{d}_i \wt{\r}_i.
\]

\For{$k = 1, \ldots, \wh{r}$}
    \State \textbf{$\mathbf{M}$-step:} Compute 
    \[
        {\mathbf{M}}^{(k)} = \left[\mathbf{0}_{q \times n(k-1)}, {\mathbf{M}}_{k}, \mathbf{0}_{q \times [n(\wh{r} - k) + q(\wh{r} - 1)]} \right]
    \]
    \State \quad with ${\mathbf{M}}_k = - (\wt{\v}_k\trans \wt{\v}_k)^{-1} \wt{\mathbf{V}}_{-k} \wt{\mathbf{U}}_{-k}\trans$.
    
    \State \textbf{$\mathbf{W}$-step:} Compute 
    \[
        {\mathbf{A}} = (\wt{\v}_k\trans \wt{\v}_k) \mathbf{I}_{\wh{r}-1} - \wt{\mathbf{V}}_{-k}\trans \wt{\mathbf{V}}_{-k} \wt{\mathbf{U}}_{-k}\trans \wt{\mathbf{U}}_{-k}.
    \]
    \State \quad Then compute
    \begin{align*}
        {\mathbf{W}}_k = &\mathbf{I}_q - (\wt{\v}_k\trans \wt{\v}_k)^{-1} \left( \mathbf{I}_q + \wt{\mathbf{V}}_{-k} \wt{\mathbf{U}}_{-k}\trans \wt{\mathbf{U}}_{-k} {\mathbf{A}}^{-1} \wt{\mathbf{V}}_{-k}\trans \right) \wt{\mathbf{V}}_{-k} \wt{\mathbf{U}}_{-k}\trans \wt{\u}_k \wt{\mathbf{v}}_k\trans \\ & + \wt{\mathbf{V}}_{-k} \wt{\mathbf{U}}_{-k}\trans \wt{\mathbf{U}}_{-k} {\mathbf{A}}^{-1} \wt{\mathbf{V}}_{-k}\trans.
    \end{align*}

    \State \textbf{Debiased Estimate:} For $\boldeta_k = \left(\u_1\trans, \ldots, \u_{\wh{r}}\trans, \v_1\trans, \ldots, \v_{k-1}\trans, \v_{k+1}\trans, \ldots, \v_{\wh{r}}\trans\right)\trans$, compute
    \[
        \wh{\v}_k = \wt{\v}_k - {\mathbf{W}}_k \left( \frac{\partial L}{\partial \v_k} - {\mathbf{M}}^{(k)} \frac{\partial L}{\partial \boldsymbol{\eta}_k} \right) \Bigg|_{(\wt{\v}_k, \wt{\boldsymbol{\eta}}_k)}.
    \]
    
    \State \textbf{Variance Estimate:} 
    \[
        \wt{\nu}^2_k = \a\trans {\mathbf{W}}_{k} \left( \bSigma_e + \wt{\v}_k\trans \bSigma_e \wt{\v}_k {\mathbf{M}}_{k} {\mathbf{M}}_{k}\trans - 2 {\mathbf{M}}_{k} \wt{\u}_k \wt{\v}_k\trans \bSigma_e \right) {\mathbf{W}}_k\trans \a.
    \]
\EndFor

\State \textbf{Output:} $\{ \wh{\v}_k, \wt{\nu}^2_k \}_{k=1}^{\wh{r}}$
\end{algorithmic}
\end{algorithm}

\begin{algorithm}[!t]
\caption{SOFARI-R}
\label{alg2}
\begin{algorithmic}[1]
\Require Data $\mathbf{X} \in \mathbb{R}^{n \times p}$, $\mathbf{Y} \in \mathbb{R}^{n \times q}$
\Ensure Debiased estimates and variance estimates $\{ \wh{\v}_k, \wt{\nu}^2_k \}_{k=1}^{\wh{r}}$

\State \textbf{Initial Step:}
\State \quad Determine the rank $\wh{r}$.
\State \quad Compute initial SOFAR estimates $\big\{\wt{d}_i, \wt{\boldsymbol{\ell}}_i, \wt{\boldsymbol{r}}_i \big\}_{i=1}^{\wh{r}}$.
\State \quad Let
\State \quad \[
    \wt{\u}_i =  (\wt{\l}_i\trans \wh{\bSigma} \wt{\l}_i)^{-1/2} n^{-1/2} \X \wt{\l}_i, \quad \wt{\v}_i = (\wt{\l}_i\trans \wh{\bSigma} \wt{\l}_i)^{1/2} \wt{d}_i \wt{\r}_i.
\]

\For{$i = 1, \ldots, \wh{r}$}
    \State Run the SOFARI algorithm in \cite{sofari} to obtain the debiased estimate $\wh{\bmu}_i$.
    \State Compute $\wh{\bmu}_i^t = (\wh{\mu}_{i1}^t, \ldots, \wh{\mu}_{i p}^t)\trans$ with $\wh{\mu}_{ij}^t = \wh{\mu}_{ij} \cdot \mathbf{1}\left( \wh{\mu}_{ij} \geq \frac{\log n}{\sqrt{n}} \right)$.
\EndFor

\For{$k = 1, \ldots, \wh{r}$}
    \State \textbf{$\mathbf{M}$-step:} Compute $ \M_k = -(\wt{\v}_k\trans \wt{\v}_k)^{-1} \wt{\v}_k \wt{\u}_k\trans$.
    \State \textbf{$\mathbf{W}$-step:} Compute
    \[
        {\mathbf{W}}_k = \I_q - \frac{1}{2} (\v_k\trans \v_k)^{-1} \left( \wt{\v}_k \wt{\v}_k\trans - \wt{\v}_k \wt{\u}_k\trans \wt{\U}_{-k} \wt{\V}_{-k}\trans \right).
    \]
    \State \textbf{Debiased estimate:} For $\boldeta_k = \u_k$, compute
    \[
        \wh{\v}_k = \wt{\v}_k - {\mathbf{W}}_k \left( \frac{\partial L}{\partial \v_k} - {\mathbf{M}}_k \frac{\partial L}{\partial \boldsymbol{\eta}_k} \right) \Bigg|_{(\wt{\v}_k,\wt{\boldsymbol{\eta}}_k)}.
    \]
    \State \textbf{Variance estimate:} Compute estimated variance $\wt{\nu}^2_k$ defined in \eqref{eq:variance}.
\EndFor

\State \textbf{Output:} Debiased estimates and variance estimates $\{ \wh{\v}_k, \wt{\nu}^2_k \}_{k=1}^{\wh{r}}$
\end{algorithmic}
\end{algorithm}

\section{Additional simulation results} \label{new.sec.addisimuresu}

\subsection{simulation setup}\label{app:simusetup}

We consider a setup similar to that in \cite{mishra2017} where the latent factors are weakly orthogonal to each other, meaning that there are certain correlations among the latent factors. 
Specifically, the true regression coefficient matrix $\mathbf{C}^{*}=\sum_{k=1}^{r^*} d_{k}^{*} \l_{k}^{*} \mathbf{r}_{k}\strans$ satisfies that $r^* = 3$, $d_{1}^{*}=100, d_{2}^{*}=15, d_{3}^{*}=5$, and
\begin{align*}
	\l_{k}^{*}=\check{\l}_{k} /\|\check{\l}_{k}\|_{2}  & \ \text{ with }
	\check{\l}_{k}=\big(\operatorname{rep}(0,s_1(k-1)), \operatorname{unif}\left(S_{1}, s_1\right), \operatorname{rep}(0, p - ks_1)\big)\trans, \\
	\mathbf{r}_{k}^{*}=\check{\mathbf{r}}_{k} /\left\|\check{\mathbf{r}}_{k}\right\|_{2} & \ \text{ with }  \check{\mathbf{r}}_{k}=\big(\operatorname{rep}(0,s_2(k-1)), \operatorname{unif}\left(S_{2}, s_2\right), \operatorname{rep}(0, q - ks_2)\big)\trans.
\end{align*}
Here, $\operatorname{unif}\left({S}, s\right)$ represents an $s$-dimensional random vector with independent and identically distributed (i.i.d.) components from the uniform distribution on set ${S}$, $\operatorname{rep}(a,s)$ is an $s$-dimensional vector with identical components $a$, $S_{1} = \{-1, 1\}$, $S_2 = [-1,-0.3] \cup[0.3,1]$, $s_1 = 3$, and $s_2 = 3$.

Given the matrix consisting of left singular vectors $\L^* = (\l_1^*, \ldots, \l_{r^*}^*)$, we can choose a matrix $\L_{\perp}^{*} \in \mathbb{R}^{p \times\left(p-r^{*}\right)}$ such that $\mathbf{P}=\left[\L^{*}, \L_{\perp}^{*}\right] \in \mathbb{R}^{p \times p}$ is nonsingular. The design matrix $\mathbf{X}$ is generated according to the three steps below. First, a matrix $\mathbf{X}_{1} \in \mathbf{R}^{n \times r^{*}}$ is created by drawing a random sample from $N\left(\mathbf{0}, \mathbf{I}_{r *}\right)$ of size $n$.
Second, denote by $\bSigma_X=\left(0.3^{|i-j|}\right)_{1 \leq i, j \leq p}$ the population covariance matrix, $\wt{\x} \sim {N}(\mathbf{0}, \bSigma_X)$,  $\wt{\x}_1 = \L^{* {T}} \wt{\x}$, and $\wt{\x}_{2}=\L_{\perp}^{* {T}} \wt{\x}$. We then generate $\mathbf{X}_{2} \in \mathbb{R}^{n \times\left(p-r^{*}\right)}$ by drawing a random sample from the conditional distribution of $\wt{\x}_{2}$ given $\wt{\x}_{1}$ of size $n$. 
Finally, design matrix $\mathbf{X}$ is set as $\mathbf{X}=\left[\mathbf{X}_{1}, \mathbf{X}_{2}\right] \mathbf{P}^{-1}$ 
so that the latent factors $n^{-1/2}\X\l_i^*$ are weakly orthogonal to each other.

For the random error matrix $\E$, we assume that the rows of $\E$ are i.i.d. copies from $N\left(\mathbf{0}, \sigma^{2} \boldsymbol{\Sigma}_E\right)$ with $\boldsymbol{\Sigma}_E=\left(0.3^{|i-j|}\right)_{1 \leq i,j \leq q}$, which is independent of design matrix $\mathbf{X}$. The noise level $\sigma^{2}$ is chosen such that the signal-to-noise ratio (SNR) $\left\| \mathbf{X} (d_{r^*}^{*}\l_{r^*}^{*} \mathbf{r}_{r^*}\strans)\right\|_{F} /\|\mathbf{E}\|_{F}$ is equal to $1$.

\begin{table}[t]
	\centering
	
	\smallskip
\caption{\label{table:uj2sofar}
		The average performance measures of SOFARI-R on the individual components of the latent right factor vectors (i.e., the right singular vectors weighted by the corresponding variance-adjusted singular values) in different sparse SVD layers with squared singular values $ (d_1^{*2}, d_2^{*2}, d_3^{*2} ) = (200^2, 15^2, 5^2)$ over $1000$ replications for simulation example 2 in Section \ref{new.Sec.5.2}.}
		\begin{tabular}{c|ccccccccc}
			\hline	
			Setting& & ${\operatorname{CP}}$  & ${\operatorname{Len}}$ &    & ${\operatorname{CP}}$  & ${\operatorname{Len}}$ & & ${\operatorname{CP}}$  & ${\operatorname{Len}}$    \\
			\hline
			3
&$v_{1,1}^*$
& 0.949   &0.264  & $v_{2,6}^*$   & 0.947  & 0.289 & $v_{3,11}^*$    & 0.949  & 0.310 \\
&$v_{1,2}^*$
& 0.952   &0.265  & $v_{2,7}^*$   & 0.948  & 0.291 & $v_{3,12}^*$    & 0.955  & 0.309 \\
&$v_{1,3}^*$
& 0.943   &0.264  & $v_{2,8}^*$   & 0.938  & 0.290 & $v_{3,13}^*$    & 0.951  & 0.310 \\
&$v_{1,4}^*$
& 0.942  &0.263  & $v_{2,9}^*$   & 0.948  & 0.290 & $v_{3,14}^*$    & 0.948  & 0.311 \\
&$v_{1,5}^*$
& 0.946   &0.264  & $v_{2,10}^*$   & 0.946  & 0.291 & $v_{3,15}^*$    & 0.941  & 0.311 \\
&$v_{1,q-4}^*$
& 0.951   &0.266  & $v_{2,q-4}^*$   & 0.941  & 0.266 & $v_{3,q-4}^*$    & 0.948  & 0.266 \\
&$v_{1,q-3}^*$
& 0.955   &0.266  & $v_{2,q-3}^*$   & 0.953  & 0.266 & $v_{3,q-3}^*$    & 0.955  & 0.266 \\
&$v_{1,q-2}^*$
& 0.954   &0.266  & $v_{2,q-2}^*$   & 0.950  & 0.266 & $v_{3,q-2}^*$    & 0.954  & 0.266 \\
&$v_{1,q-1}^*$
& 0.944   &0.266  & $v_{2,q-1}^*$   & 0.948  & 0.266 & $v_{3,q-1}^*$    & 0.947  & 0.266 \\
&$v_{1,q}^*$
& 0.950   &0.265  & $v_{2,q}^*$   & 0.943  & 0.265 & $v_{3,q}^*$    & 0.947  & 0.265 \\
\hline
4&$v_{1,1}^*$
& 0.950   &0.186  & $v_{2,6}^*$   & 0.943  & 0.204 & $v_{3,11}^*$    & 0.949  & 0.220 \\
&$v_{1,2}^*$
& 0.936  &0.186  & $v_{2,7}^*$   & 0.945  & 0.205 & $v_{3,12}^*$    & 0.949  & 0.220 \\
&$v_{1,3}^*$
& 0.947   &0.186  & $v_{2,8}^*$   & 0.945  & 0.205 & $v_{3,13}^*$    & 0.947  & 0.220 \\
&$v_{1,4}^*$
& 0.938   &0.186  & $v_{2,9}^*$   & 0.951  & 0.205 & $v_{3,14}^*$    & 0.941  & 0.220 \\
&$v_{1,5}^*$
& 0.948   &0.186  & $v_{2,10}^*$   & 0.953  & 0.205 & $v_{3,15}^*$    & 0.954  & 0.220 \\
&$v_{1,q-4}^*$
& 0.950   &0.186  & $v_{2,q-4}^*$   & 0.946  & 0.186 & $v_{3,q-4}^*$    & 0.943  & 0.186 \\
&$v_{1,q-3}^*$
& 0.958   &0.187  & $v_{2,q-3}^*$   & 0.947  & 0.187 & $v_{3,q-3}^*$    & 0.948  & 0.187 \\
&$v_{1,q-2}^*$
& 0.950   &0.187  & $v_{2,q-2}^*$   & 0.952  & 0.187 & $v_{3,q-2}^*$    & 0.938  & 0.187 \\
&$v_{1,q-1}^*$
& 0.940   &0.187  & $v_{2,q-1}^*$   & 0.960  & 0.187 & $v_{3,q-1}^*$    & 0.942  & 0.187 \\
&$v_{1,q}^*$
& 0.952   &0.187  & $v_{2,q}^*$   & 0.951  & 0.187 & $v_{3,q}^*$    & 0.948  & 0.187 \\
\hline
		\end{tabular} 

\end{table}

\subsection{Additional simulation example}\label{new.Sec.5.2}

The setting of this second simulation example is similar to that in \cite{uematsu2017sofar}. The major difference with the first example in Section \ref{new.Sec.5} is that we now do \textit{not} assume any particular form of the orthogonality constraint on the latent factors, so that this setup allows for \textit{stronger} correlations among the latent factors. Thus, the technical assumptions in Conditions \ref{con4:strong} and \ref{con:weak:orth} can be violated here. This challenging setup is mainly designed to test the robustness of the SOFARI-R inference procedure when some of the orthogonality conditions are \textit{not} satisfied.
Specifically, the rows of design matrix $\X$ are i.i.d. and drawn directly from $N(\mathbf{0}, \bSigma_X)$ with covariance matrix  $\bSigma_X = (0.3^{|i-j|})_{p \times p}$. The true regression coefficient matrix $\mathbf{C}^{*}$ follows the same latent sparse SVD structure as that in simulation example 1, except that now $d_{1}^{*}$ increases from $100$ to $200$ and both $s_1$ and $s_2$ increase from $3$ to $5$.
Similarly, we consider two settings in this second example, and the other setups for settings 3 and 4 are the same as those for settings 1 and 2 in Section \ref{new.Sec.5}, respectively.

Table \ref{table:uj2sofar} summarizes the average performance measures of various SOFARI-R estimates in simulation example 2. First, similar to simulation example 1, the rank of the latent sparse SVD structure is identified consistently as $r = 3$. Second, from Table \ref{table:uj2sofar}, we can see that the average coverage probabilities of the confidence intervals constructed by SOFARI-R for the representative parameters remain very close to the target level of $95\%$. Third, it is clear that the average lengths of the $95\%$ confidence intervals for different components of the latent right factor vectors across different settings are stable over both $j$ and $k$. It demonstrates that the suggested SOFARI-R inference procedure can still perform well even when the correlations among the latent factors are no longer weak, provided that the eigengap among the nonzero singular values are sufficiently large.

\section{Additional real data results} \label{appsec:D}
In this section, we provide in Table \ref{table4} the list of $q = 30$ selected responses along with their descriptions for the real data application in Section \ref{new.Sec.6}. Additionally, we compare the prediction performance of different methods on this dataset. Furthermore, we provide a detailed interpretation of Figure \ref{figure2} to demonstrate its underlying economic implications.

\begin{table}[t]
    \centering
    \caption{The list of $30$ selected responses from eight groups for the real data application in Section \ref{new.Sec.6}. Group 1: Output and income; Group 2: Labor market; Group 3: Housing; Group 4: Consumption, orders, and inventories; Group 5: Money and credit; Group 6: Interest and exchange rates; Group 7: Prices; Group 8: Stock market.}\label{table4}
    \smallskip
     \resizebox{\textwidth}{!}{
    \begin{tabular}{lll}
    \toprule
       Variable & Description & Group \\
    \midrule
            RPI & Real personal income & 1 \\
         INDPRO & Total industrial production & 1 \\
      IPMANSICS & Industrial production of Manufacturing & 1 \\
         UNRATE & Civilian unemployment rate & 2 \\
         PAYEMS & Total number of employees on non-agricultural payrolls & 2 \\
        CLAIMSx & Initial claims & 2 \\
  CES0600000008 & Avg hourly earnings: goods-producing & 2 \\
       UEMPMEAN & Average duration of unemployment & 2 \\
        HOUST & Housing starts & 3 \\
DPCERA3M086SBEA & Real personal consumption expenditures & 4 \\
      CMRMTSPLx & Real manufacturing and trade industries sales & 4 \\
        AMDMNOx & New orders for durable goods & 4 \\
        BUSINVx & Total business inventories & 4 \\
           M1SL & M1 money stock & 5 \\
         M2REAL & Real M2 money stock & 5 \\
         CONSPI & Nonrevolving consumer credit to personal income & 5 \\
         INVEST & Securities in bank credit at all commercial banks & 5 \\
         REALLN & Real estate loans at all commercial banks & 5 \\
       FEDFUNDS & Effective federal funds rate & 6 \\
        EXUSUKx & U.S.-U.K. exchange rate & 6 \\
           GS10 & 10-year treasury rate & 6 \\
      COMPAPFFx & 3-month commercial paper minus FEDFUNDS & 6 \\
       TB6SMFFM & 6-month treasury c minus FEDFUNDS & 6 \\
          TB6MS & 6-month treasury bill & 6 \\
         PPIFGS & Producer price index for finished goods & 7 \\
         PPICMM & Producer price index for commodities & 7 \\
       CPIAUCSL & Consumer price index for all items & 7 \\
          PCEPI & Personal consumption expenditure implicit price deflator & 7 \\
      OILPRICEx & Crude oil, spliced WTI and cushing & 7 \\
      S.P.500 & S\&P's common stock price index: composite & 8 \\
    \bottomrule
    \end{tabular}}
\end{table}

\begin{table}[tp]
	\centering
	\caption{Prediction errors of different methods for the real data application in Section \ref{new.Sec.6}.}\label{tablereal01}
	\smallskip
	\begin{tabular}{lccccc}
		\hline	
		& SOFAR-L &  SOFAR-GL &RRR& RRSVD & SRRR \\
		\hline
		Prediction error& 0.876 &  0.884 &1.525& 1.152 & 0.927 \\
		\hline
	\end{tabular} 
\end{table}

We fit the multi-response regression model \eqref{model} using various methods. These include the SOFAR \citep{uematsu2017sofar} with the entrywise $L_1$-penalty (SOFAR-L) or the rowwise $(2,1)$-norm penalty (SOFAR-GL), reduced rank regression (RRR), sparse reduced rank regression (SRRR) \citep{chen2012sparse}, and reduced rank regression with sparse SVD (RSSVD) \cite{chen2012reduced}. By splitting the data into a training set consisting of the first 474 observations and a testing set comprising the remaining $n_1 = 180$ observations, we fit model \eqref{model} using the five methods on the training set and then calculate the prediction error $\norm{\Y - \X\widehat{\C}}_F^2/(n_1 q)$ based on the testing set. The prediction errors of all the methods are summarized in Table \ref{tablereal01}, which shows that the sparse learning methods exhibit much better prediction performance compared to the regular reduced rank regression. Furthermore, SOFAR-L achieves the highest prediction accuracy, closely followed by SOFAR-GL. This indicates that the SOFAR initial estimate can be a good approximation to the latent sparse SVD structure of the underlying data.

In addition, we have annotated eleven responses in Figure \ref{figure2} with their abbreviated names, as these coefficients are significantly larger in magnitude than those of the remaining responses. These responses represented by three factors are categorized by different groups of economic variables. Specifically, the first factor loading primarily corresponds to Group 2: labor market (UEMPMEAN: average duration of unemployment) and Group 6: interest and exchange rates (TB6SMFFM and TB6MS: 6-month treasury), which captures key dynamics in employment and monetary policy. Furthermore, to gain a comprehensive perspective of the relationships between responses and covariates, we also examine the estimation results of left singular vectors, revealing that variables such as M2REAL (real M2 money stock) and CPIAUCSL (the overall CPI) play pivotal roles in the first layer. In light of the two positive factor coefficients and three negative factor loadings, this suggests that higher liquidity (M2REAL) and increased inflation expectations (CPIAUCSL) typically stimulate economic activity, thereby reducing unemployment (UEMPMEAN) and easing monetary conditions (TB6SMFFM and TB6MS).
 
The second factor loading is associated with Group 5: money and credit (INVEST: investment securities, REALLN: real estate Loans) and Group 8 (stock market: S.P.500), with the coefficients being positive. Additionally, considering the estimated significant left factor coefficients such as M2SL (real M2 money stock) being positive, this reflects a positive relationship, where an increase in money supply (M2SL) corresponds to higher credit availability (INVEST, REALLN) and stronger stock market performance (S.P.500). 
The third factor loading relates to Group 2: labor market (PAYEMS: payroll employment), Group 4: consumption, orders, and inventories (CMRMTSPLx: trade sales), Group 5: money \& credit (M1SL, M2REAL: money stock), and Group 7: prices (PPIFGS: producer prices). This factor includes a broader range of economic activities, providing a comprehensive view of economic performance from consumer behavior to financial stability and price levels.

\end{document}